\documentclass{emulateapj}
\usepackage{pdflscape,amsmath,rotating,color}

\shorttitle{FGKM Stars}
\shortauthors{Rayner et al.}

\begin{document}


\title{The Infrared Telescope Facility (IRTF) Spectral Library: Cool Stars}


\author{John T. Rayner}
\affil{Institute for Astronomy, University of Hawai`i, 2680 Woodlawn Drive, Honolulu, HI 96822}
\email{rayner@ifa.hawaii.edu}

\author{Michael C. Cushing\altaffilmark{1}}
\affil{Institute for Astronomy, University of Hawai'i, 2680 Woodlawn Drive, Honolulu, HI 96822}
\email{michael.cushing@gmail.com}

\and

\author{William D. Vacca\altaffilmark{1}}
\affil{SOFIA-USRA, NASA Ames Research Center MS N211-3, Moffett Field, CA 94035} 
\email{wvacca@sofia.usra.edu}

\altaffiltext{1}{Visiting Astronomer at the Infrared Telescope Facility,
  which is operated by the University of Hawai`i under cooperative
  Agreement no. NCC 5-538 with the National Aeronautics and Space
  Administration, Office of Space Science, Planetary Astronomy Program.}

\begin{abstract}
  We present a 0.8$-$5 \micron~spectral library of 210 cool stars
  observed at a resolving power of $R\equiv\lambda/\Delta\lambda\sim$
  2000 with the medium-resolution infrared spectrograph, SpeX, at the
  3.0 m NASA Infrared Telescope Facility (IRTF) on Mauna Kea, Hawaii.
  The stars have well established MK spectral classifications and are
  mostly restricted to near-solar metallicities. The sample contains the
  F, G, K, and M spectral types with luminosity classes between I and V,
  but also includes some AGB, carbon, and S stars. In contrast to some
  other spectral libraries, the continuum shape of the spectra are
  measured and preserved in the data reduction process.  The spectra are
  absolutely flux calibrated using Two Micron All Sky Survey (2MASS)
  photometry.  Potential uses of the library include studying the
  physics of cool stars, classifying and studying embedded young
  clusters and optically obscured regions of the Galaxy, evolutionary
  population synthesis to study unresolved stellar populations in
  optically-obscured regions of galaxies, and synthetic photometry.  The
  library is available in digital form from the IRTF website.

\end{abstract}


\section{\label{Intro} Introduction}

Spectral libraries play an important role in attempts to understand and
classify individual stellar sources as well as to decompose the
integrated spectrum of an aggregate system, such as a galaxy, into its
various stellar populations. For example, the most widely used stellar
classification process, as originally developed by
\citet{1943QB881.M6.......}, consists of comparing the spectrum of a
star against a set of reference stellar spectra \citep[for a review
see][]{1994ASPC...60....3G}.  Infrared spectral libraries are
particularly useful for studying the physics of cool stars
\citep[e.g.][]{1998AJ....116.2520J, 2004A&A...422..289G}, classifying
and studying stars in nearby embedded young clusters \citep[e.g.,
][]{1995ApJ...450..233G,2008ApJ...685..313P} and optically obscured
regions of the Galaxy \citep[e.g.,][for evolved, globular, and young
clusters respectively] {1995ApJ...447L..29F, 2001AJ....122.1896F,
  2007A&A...475..209K}, and studying the unresolved stellar populations
of optically obscured extra-galactic regions using evolutionary
population synthesis (EPS) \citep[e.g.,
][]{2007A&A...468..205L,2008MNRAS.388..803R}.  EPS techniques attempt to
simulate observed galaxy spectra by combining individual stellar spectra
from a library and thereby derive the chemical and evolutionary
properties of the unresolved stellar populations \citep[e.g.,
][]{1997A&A...326..950F,1999ApJS..123....3L,2003MNRAS.344.1000B,2005MNRAS.362..799M,
  2007ASPC..374..303B, 2009ApJ...694.1379R}.

The niche of near-infrared (NIR $\sim$ 1-5~$\micron$) spectral
classification is clear. While stars earlier than roughly M0 ($\sim$3800
K) are brighter at optical wavelengths, unobscured stars later than
about M0 are brighter in the NIR and are thus better characterized at
these wavelengths. Furthermore it makes sense to use infrared
diagnostics only when optical extinction compromises optical
diagnostics.  The optimum infrared wavelengths for observation depend on
the amount of extinction.  For some objects, such as young stellar
objects or evolved stars, the presence of circumstellar dust can result
in significant excess continuum emission longward of 2~$\micron$.  For
this reason the $J$ and $H$ bands are perhaps best to characterize
embedded young stars since they avoid the veiling due to warm dust in
the $K$ band, while at the same time taking advantage of the reduced
extinction relative to the optical \citep[e.g.][]{1998ApJ...508..397M}.
On the other hand, heavily obscured objects without veiling are better
characterized in the $K$ band, or even the $L^{\prime}$ band in extreme
cases.  Consequently, the ideal infrared spectral library should contain
spectra covering a wide range of wavelengths to satisfy a variety of
possible applications.

With the maturing of NIR spectrographs and detector arrays, it has
become possible to generate increasingly sophisticated NIR libraries of
stellar spectra. \citet{2004ApJS..151..387I} presented a compilation of
NIR spectral libraries available at that time. In Table
\ref{tab:History} we revise and update this list. (The list does not
include spectral libraries covering mostly L and T dwarfs. For our
purposes a `library' is assumed to contain more than ten objects.)  All
of these libraries have shortcomings since none of them contains a large
sample of stars, with a range of metallicities, covering all spectral
types and luminosity classes, with spectra spanning a large wavelength
range.  In light of this, we have undertaken a project to construct an
improved spectral library using the facility NIR spectrograph, SpeX, at
the 3.0~m NASA Infrared Telescope Facility (IRTF) on Mauna Kea, Hawaii.
The result of this work is the IRTF Spectral Library, which we are
presenting in a series of papers.  In the first paper of this series,
\citet{2005ApJ...623.1115C} presented the spectra of M, L, and T dwarfs.
The current paper presents 210 spectra of F, G, K, and M stars with
luminosity classes between I and V (with mostly near-solar
metallicities), and includes some asymptotic giant branch (AGB) stars,
carbon stars, and S stars.  The spectra of all of these stars, including
the 13 L dwarfs and 2 T dwarfs from \citet{2005ApJ...623.1115C}, and the
gas giant planets (spectra summed along the central meridian), are
available in digital form on the IRTF
website\footnote{\url{http://irtfweb.ifa.hawaii.edu/~spex/IRTF\_Spectral\_Library}}. Additional
papers on hot stars are currently in preparation.

There are several important features of the IRTF Spectral Library.  The
wide wavelength range of $\sim$0.8-5~$\micron$ (with a larger subset at
0.8-2.4~$\micron$) is covered in only two cross-dispersed instrument
settings. For each setting, several spectral orders are simultaneously
recorded during a single exposure. In addition, most of the spectral
orders in each setting have significant wavelength overlap with the
adjacent spectral orders.  These instrumental aspects minimize potential
calibration problems posed by stitching together multiple
non-overlapping wavelength ranges observed at different times (e.g.,
sequentially), a situation typically encountered with observations
obtained with non-cross-dispersed (single-order) spectrographs. The
signal-to-noise (S/N) is better than $\sim$100 across most of this range
(except for the regions of poor atmospheric transmission and for
$\lambda > 4~\micron$) and the resolving powers of
$R\equiv\lambda/\Delta\lambda \approx$ 2000 at 0.8-2.4~$\micron$, and
$R\approx$ 2500 at 2.4-5~$\micron$, enable the accurate measurement of
spectral type and luminosity class using established equivalent width
and line ratio criteria (see \S\ref{Spectra}).  In contrast to some other
NIR spectral libraries, the continuum shape is preserved during data
reduction (for details see \S\ref{Red}) which is particularly useful
for characterizing cool stars with strong molecular absorption bands
that have been observed at low-resolution $R\sim$100.  Preserving the
continuum shape also allows for absolute flux calibration by scaling the
spectra to published Two Micron All Sky Survey photometry \citep[2MASS,
][]{2006AJ....131.1163S} and for the computation of synthetic colors
(e.g. $Y-J$, $J-H$, $H-K$, and $K-L^\prime$).

\section{\label{ODR}OBSERVATIONS AND DATA REDUCTION}

\subsection{\label{Sample_Selection}Sample Selection}

As described by \citet{1973ARAA..11...29M}, the MK spectral
classification system ``is a phenomenology of spectral lines, blends,
and bands, based on a general progression of color index (abscissa) and
luminosity (ordinate). It is defined by an array of standard stars
located on a two-dimensional spectral type versus luminosity-class
diagram. These standard reference points do not depend on specific line
intensities or ratios of intensities; they have come to be defined by
the {\em totality} of lines, blends, and bands in the ordinary {\em
  photographic} region'' (emphasis added).  In the MK system, the
classification gives the spectral sub-type and luminosity class
(e.g. K0~III); this is the observational analogue to the projection on
the luminosity-temperature plane (H-R diagram) for stars of a particular
composition.  Abundance adds a third dimension to the two-dimensional MK
diagram and is represented by additional symbols determined by the
relative intensities of lines or bands that reveal compositional
differences from the Sun.  For example, as a means of distinguishing a
solar metallicity Population I giant K0~III star from one with a lower
metal/hydrogen abundance, the classification of the latter becomes
K0~III~CN-1, or K0~III~CN-2 \citep[e.g.][]{1973ARAA..11...29M}.  With
better quality spectra increased precision in spectral classification is
possible. For example, giants can often be subdivided into luminosity
subclasses IIIa, IIIab, and IIIb.  A fundamental characteristic of the
MK system is that a finite array of discrete cells (spectral types)
represents a continuum i.e. spectra of stars of a given spectral subtype
(e.g. K5~V) are not all identical. The precision attainable with MK
classification has been estimated to be $\pm$0.6 spectral subtypes for B
and A dwarfs by \citet{1973IAUS...50...43J}, and $\pm$0.65 spectral
subtypes for G and K dwarfs by \citet{1971VeARI..24....1G}. This
precision depends upon observational dispersion (heterogeneous group of
observers and instruments) and cosmic dispersion (e.g. chemical
composition effects, and rotation effects).

We attempted to construct a sample of stars with undisputed spectral
types, traceable to the original developers of the MK classification
system. The original MK standard stars \citep{1953ApJ...117..313J,
  1973ARAA..11...29M} are generally too bright for us to observe.  To
that end, for the majority of the sample, we chose stars with
classifications given by \citet{1973AJ.....78..386M},
\citet{1973ARAA..11...29M}, \citet{1978rmsa.book.....M},
\citet{1989ApJS...71..245K},
\cite{2000webKN}\footnote{\url{http://www.astronomy.ohio-state.edu/MKCool}},
which in several cases were supplemented by stars taken from
compilations of MK standard stars by \citet{1989BICDS..36...27G} and
\citet{1978BICDS..15..121J}.  Whenever references gave conflicting
classifications, we chose the most recent revision.  For F stars, we
supplemented the lists generated from the aforementioned references with
stars whose spectral types are given by \citet{1989ApJS...69..301G},
\citet{2001AJ....121.2148G}, and \citet{1995ApJS...99..135A}.
Additionally, for M stars, we included objects with classifications
given by \citet{1991ApJS...77..417K}, \citet{1994AJ....108.1437H}, and
\cite{1997AJ....113.1421K}, again deferring to the latest revised
classifications whenever conflicting or multiple spectral types were
found in the various sources.  In a few instances stars with less
certain classifications were observed in order to fill gaps in our
coverage of spectral types due to observing limitations.

Despite their known variability in spectral type, and relative rarity,
we included asymptotic giant branch (AGB) stars in our sample because
their high luminosity makes them important in EPS studies of galaxies.
Population synthesis and star counts in clusters indicate that
AGB stars contribute more than 50\% of the $K$-band light of stellar
populations at 0.1 to 1 Gyr after an instantaneous burst of star
formation \citep{1999A&A...344L..21L,1999IAUS..191..579L}. The AGB
phase is also important in providing feedback in the chemical
evolution of galaxies.  AGB stars are intermediate mass stars
($\sim$0.8-8$M_\odot$) which ascend the asymptotic giant branch in the
HR diagram when helium and hydrogen ignite in shells surrounding their
cores (this phase lasts about $2\times10^5$~yr). Shell burning in young
AGB stars is stable but becomes increasingly unstable as the stars
become more luminous which leads to thermal pulsations. These stars are
known as thermally pulsating AGB (TPAGB) stars.  TPAGB stars are
recognizable by a variety of observational criteria by which they are
variously named: characteristic spectra (late-M, S, and C stars),
pulsating variability (Mira variables, long-period variables), mass loss
and maser emission (OH/IR stars). In our sample TPAGB stars are
identified by their variability types (L: irregular, SR: semiregular,
and M: Mira) given in the General Catalog of Variable Stars
\citep[GCVS,][]{1998GCVS4.C......0K}\footnote{\url{http://www.sai.msu.su/groups/cluster/gcvs/gcvs}}.
About 40 TPAGB stars are included in our sample.

Mass loss eventually removes the hydrogen-rich stellar envelope,
effectively terminating the TPAGB phase. The central star subsequently
evolves to higher temperatures while the circumstellar envelope expands
and cools, exposing the star. Ionizing wind and radiation from the star
quickly form a planetary nebula (PN). The transition from TPAGB to PN is
known as the post-AGB (PAGB) or protoplanetary nebula phase and lasts a
few thousand years.  Although PAGB stars were not targeted in our
sample, several supergiants that were observed have some of the
characteristics of PAGB stars (see \S\ref{Unusual}).  \citep[For a
comprehensive review of AGB stars see][]{2003agbs.conf.....H}.

In order to obtain high S/N out into the thermal infrared
($\sim$2.3-5~\micron), we selected relatively bright stars.
Consequently, they tend to be local and therefore of mostly solar
composition.  Figure \ref{fig:FeH}a shows the distribution of
metallicities for stars in our sample with spectroscopic measurements of
[Fe/H] \citep{1997A&AS..124..299C}. The distribution is typical for
stars in the solar neighborhood \citep{2004A&A...418..989N}.

The object name, spectral classification and associated reference, GCVS
variable type, $V$, $B-V$, and 2MASS ($J$, $H$, and $K_{S}$)
magnitudes for each star in the sample are given in Table
\ref{tab:Sample}.  All objects in the sample have declinations
$-30>\delta>+70$ degrees, a range set by the latitude of IRTF, and an
airmass $<$~2 for good telluric correction.  The stars have $K$-band
magnitudes of $\sim$11~$>K>~$0; the faint limit was set by the desire to
obtain high S/N spectra in less than about 30 minutes of integration
time, and the bright limit corresponds to detector saturation in the
minimum exposure time of 0.1~s (although several brighter targets were
observed using ad hoc methods).  The total number of stars in the sample
is 210. Table \ref{tab:Composition} gives the composition of the sample
by spectral type and luminosity class.  Due to practical limitations of
observing time there is no multi-epoch coverage of variable stars or
large numbers of stars with non-solar metallicity. However, we
anticipate future observing campaigns with SpeX at IRTF will add to the
sample.

\subsection{\label{Obs}Observations}

The observations were carried out over a period of eight years using
SpeX at the IRTF.  A detailed description of SpeX is given by
\cite{2003PASP..115..362R}.  Briefly, SpeX is a 0.8$-$5.4~$\mu$m,
medium-resolution, cross-dispersed spectrograph equipped with a 1024
$\times$ 1024 Aladdin 3 InSb array.  The entire 0.8 to 5.4~$\mu$m
wavelength range can be covered with two cross-dispersed modes, the
short-wavelength cross-dispersed mode (SXD), and the long-wavelength
cross-dispersed mode (LXD).  The SXD mode provides simultaneous coverage
of the 0.8$-$2.42~$\mu$m wavelength range, except for a 0.06 $\mu$m gap
between the $H$ and $K$ bands, while the LXD1.9, LXD2.1, and LXD2.3
modes cover the 1.9$-$4.2, 2.20$-$5.0 and 2.38$-$5.4~$\mu$m wavelength
ranges, respectively.  For nearly all stars, the 0$\farcs$3 (2 pixel)
slit was used for both the SXD and LXD modes, providing resolving powers
of 2000 and 2500, respectively. Measurements of arc lines
obtained with the internal calibration unit indicate that the FWHM is
2 pixels at all wavelengths for the 0$\farcs$3 slit.  (The resolving
power R varies by $\sim$20\% across a spectral order since it depends
on the changing grating diffraction angle.)  The length of the slit
for these modes is 15$\farcs$0 and the spatial scale is
0$\farcs$15/pixel.  The spectrograph also includes a high-throughput
low-resolution $R\sim$~200 prism mode and a single-order 60$\arcsec$
long-slit $R\sim$~2000 mode.  An autonomous infrared slit viewer
employing a 512 $\times$ 512 Aladdin 2 InSb array is used for object
acquisition, guiding, and imaging photometry. The slit viewer covers a
60$\arcsec$ $\times$ 60$\arcsec$ field-of-view at a spatial scale of
0$\farcs$12/pixel. An internal K-mirror image rotator enables the field
to be rotated on the slit. Calibration observations are
obtained using the internal calibration unit consisting of flat field
and arc lamps, integrating sphere, and illumination optics which
reproduce the beam from the telescope.  A log of the observations
including the object name, spectral type, UT date of observation,
spectroscopic mode, resolving power, exposure time, associated telluric
standard star, and sky conditions, is presented in Table
\ref{tab:ObsLog}.

To facilitate subtraction of the additive components of the total signal
(electronic bias level, dark current, sky and background emission)
during the reduction process, the observations were obtained in a series
of exposures in which the target was nodded along the slit between two
positions separated by 7$\farcs$5, and a sequence of nodded pairs was
taken to build up S/N. A minimum of three pairs was taken (six spectra)
to allow noisy pixels (due mainly to cosmic ray hits) to be rejected by
a sigma clipping algorithm.  Guiding was done on spill-over from the
science target in the slit using the infrared slit-viewing camera. In
the SXD mode, where atmospheric dispersion is significant compared to
the the slit width of 0$\farcs$3 (see Figure \ref{fig:atmosdisp}), the
image rotator was set to the parallactic angle prior to each
observation. As discussed in \S\ref{Red}, observing at the parallactic angle
minimized spectral slope variations. This is not as important in the LXD
mode where atmospheric dispersion is an order a magnitude smaller.
 
An A0~V star was observed before or after each science object to correct
for absorption due to the Earth's atmosphere (see Figure
\ref{fig:AtmosSpec}) and to flux calibrate the science object spectra.
The airmass difference between the object and ``telluric standard'' was
almost always less than 0.1 and usually less than 0.05.  However, in a
few cases where there was a paucity of nearby A0~V stars, the airmass
difference was as large as 0.15.  Standard stars were also chosen to be
located within 10 degrees of the science object whenever possible, to
minimize the effects of any differential flexure in the instrument
between observations of the object and standard.  This limit on the
angular distance provides a good compromise between the requirements to
match airmass, minimize flexure, and find suitably bright standard
stars.  On those few occasions when it was necessary to observe a
telluric standard more than 10 degrees away from the object due to a
lack of A0~V stars in certain parts of the sky, we found that
telluric CO$_2$ (predominantly at 2.01~$\mu$m) features were sometimes
not adequately removed by the standard star despite a good airmass match
and good correction of telluric H$_2$O. (See, for example, Figure
\ref{fig:F_IIIK}, where the F7~III and F8~III stars were corrected with
telluric standard stars at separations and airmass differences of
14~degrees and 0.09, and 21~degrees and 0.05, respectively.)  We attribute
this to the possibility that telluric H$_2$O and CO$_2$ are not well
mixed and to patchy CO$_2$ distribution.  Finally, a set of internal
flat field exposures and argon arc lamp exposures were taken after each
object/standard pair for flat fielding and wavelength calibration
purposes.

\subsection{\label{Red}Data Reduction}

We reduced the data using Spextool \citep{2004PASP..116..362C}, the
facility IDL-based data reduction package for SpeX.  The initial image
processing consisted of correcting each science frame for non-linearity,
subtracting the pairs of images taken at the two different slit
positions, and dividing the pair-subtracted images by a normalized flat
field.  In each frame, the spectra in the individual orders were then
optimally extracted \citep[e.g.,][]{1986PASP...98..609H} and wavelength
calibrated. (All wavelengths are given in vacuum.) The extracted spectra
in each order from the set of frames for a given object were then
combined using the median. This resulted in a single spectrum in each
order for a given object.

The spectra in the individual orders were then corrected for telluric
absorption and flux calibrated using the extracted A0~V spectra and the
technique described in \citet{2003PASP..115..389V}.  In addition
to correcting for the absorption due to the atmosphere, this process
also removes the signature of the instrumental throughput and restores
the intrinsic (i.e., above the atmosphere) spectrum of each science
object in each spectral order.  Briefly, this technique scales a
theoretical model spectrum of Vega to the observed visual magnitude of
the observed standard star, convolves it to the observed resolution,
and adjusts the \ion{H}{1} line strengths to match the observed
strengths of the standard star. The ratio of the adjusted model
spectrum to the observed spectrum of the A0V star gives the telluric
correction spectrum (which also includes correction for the instrument
throughput) in each order. The science object spectrum is then divided
by the telluric spectrum. The resulting flux calibration is accurate
to about 10 percent.  The sharp and deep telluric absorption features
are marginally sampled with a 2 pixel-wide slit so when the object and
telluric correction spectra are ratioed, residuals remain at the
wavelengths of these features due to a small amount of instrumental
flexure between the object and standard star positions. In order to
minimize these systematic errors, the telluric correction spectrum is
first shifted relative to the object spectrum until the noise in these
regions is minimized.  Typically these shifts are $\sim$0.1$-$0.2 pixels
for telescope movements of 10 degrees.

In principle, the flux density levels of the telluric-corrected
spectra in two adjacent orders should match exactly in the wavelength
region where they overlap; in practice we find offsets of usually less
than one percent, although occasionally as large as three percent. The
level mismatch was removed by scaling one spectrum to the level of the
other. The scale factor was determined from a section of the overlap
region where both spectra were judged to have sufficient S/N to allow
an accurate determination.  The telluric-corrected and scaled spectra
in the individual orders were then merged together to form a single,
continuous spectrum for the science object. Regions of strong
telluric absorption were then removed.  The precise wavelength
intervals removed
depended on the transparency of the atmosphere at the time of
observation but always included the $\sim$2.5-2.8~$\micron$ and
$\sim$4.2-4.6~$\micron$ regions.  For each object, the SXD and LXD
spectra were combined in a manner similar to that used to combine the
individual orders. A scale factor was determined from the overlapping
wavelength region and then used to adjust the SXD and LXD spectra to a
common level.

The next step in the reduction process was to absolutely flux calibrate
the spectra using the 2MASS photometry listed in Table \ref{tab:Sample}.
For each spectrum, we computed correction factors based on the $JHK_S$
photometry given by,

\begin{equation}
  C_X = \frac{F^{Vega}_X \times 10^{-0.4(m_X+zp_X)}}  {\int f^{\mathrm{obs}}_{\lambda}(\lambda)S_X(\lambda)d\lambda} \label{fluxcaleqn}
\end{equation}

\noindent
where $F_{X}^{\mathrm{Vega}}$ is the Vega flux, $m_X$ is the magnitude
of the object, $zp_X$ is the passband zero point, and $S_X(\lambda)$ is
the system response function in bandpass $X$, and
$f^{\mathrm{obs}}_{\lambda}$ is the flux density of the object.  We used
the Vega fluxes (5.082$\times 10^{-10}$, 2.843$\times 10^{-10}$,
1.122$\times 10^{-10}$ W m$^{-2}$), zero points ($+$0.001, $-$0.019,
$+$0.017) and the system response functions given in
\citet{2003AJ....126.1090C}.  Each spectrum was then multiplied by a
single scale factor $<$$C$$>$ given by the weighted average of the $J$-,
$H$- and $K_S$-band correction factors ($f_\lambda$ = $f^{obs}_\lambda$
$\times$ $<$$C$$>$).  The weights were given by the errors in the scale
factors, which in turn were derived from the photometric uncertainties.
This scaling has the effect of shifting the entire spectrum up or down
so that the overall absolute flux level is correct, while simultaneously
preserving the relative flux calibration of each spectral order derived
from the observations and telluric correction procedures.  For variable
stars in the sample (mostly TPAGB stars) the absolute flux calibration
is only approximate since the 2MASS photometry was obtained at an
earlier epoch than the SpeX spectroscopy.  Furthermore, not all the SXD
and LXD spectra of individual stars were obtained at the same epoch.

The spectrum of each object is then shifted to zero radial velocity. To
do this, we first selected two stars, HD~219623 (F8 V) and HD~201092 (K7
V), as representative of F-G, and K-M stars, respectively, and measured
their apparent radial velocities using the observed wavelengths of
strong, isolated atomic lines.  For HD~219623 we measured the positions
of the Pa $\epsilon$ (0.9548590 $\mu$m), Pa~$\gamma$ (1.0052128~$\mu$m),
Pa~$\beta$ (1.282159~$\mu$m), \ion{Mg}{1} (1.4881683~$\mu$m),
\ion{Mg}{1} (1.7113304~$\mu$m), and Br~$\gamma$ (2.166120~$\mu$m) lines;
for HD~201092 we used the \ion{Si}{1} (1.0588042~$\mu$m), \ion{Mg}{1}
(1.1831408~$\mu$m), \ion{Mg}{1} (1.4881683~$\mu$m), and \ion{Mg}{1}
(1.7113304~$\mu$m) lines.  The observed radial velocities for these two
stars were determined by averaging the radial velocity measurements
obtained from these lines.  The standard error on the mean radial
velocity was $\sim$3 km s$^{-1}$ for both stars. We then shifted the
spectra of these two stars in wavelength to correspond to zero radial
velocity. To determine the radial velocities of the remaining stars in
the library, we cross-correlated the 1.05$-$1.10 $\mu$m spectra of the
F-G stars (which contain isolated lines of Si, C, Mg, and Fe) against
that of HD~219623 and the 2.285$-$2.33 $\mu$m spectra of the K and M
stars (which contain the $\Delta \nu$=$+$2 CO overtone bands) against
that of HD~201092.  The peak of each cross-correlation function was
fitted with a second order polynomial to determine the radial velocity.
Based on our implementation of the cross-correlation technique
described by \citet{1979AJ.....84.1511T} as well as a comparison of
the radial velocities derived from different wavelength regions in
each spectrum, we estimate the uncertainty in our radial velocity
values to be generally less than 30 km s$^{-1}$.  The spectrum of
each object was then shifted to zero radial velocity using the radial
velocity derived from the cross correlation. It should be noted that, in
order to preserve the accuracy of our data, we did not re-sample or
re-bin the final spectra to a common wavelength scale after shifting
them, and therefore each spectrum has a unique wavelength array.  In
most cases, the velocity shifts are small: the average shift was found
to be 4.5 $\pm$ 37 km/s with a maximum of about 131 km/s. Both values
are smaller than our velocity resolution (150 km/s per resolution
element).

\begin{figure}
\centerline{\includegraphics[width=6in,angle=0]{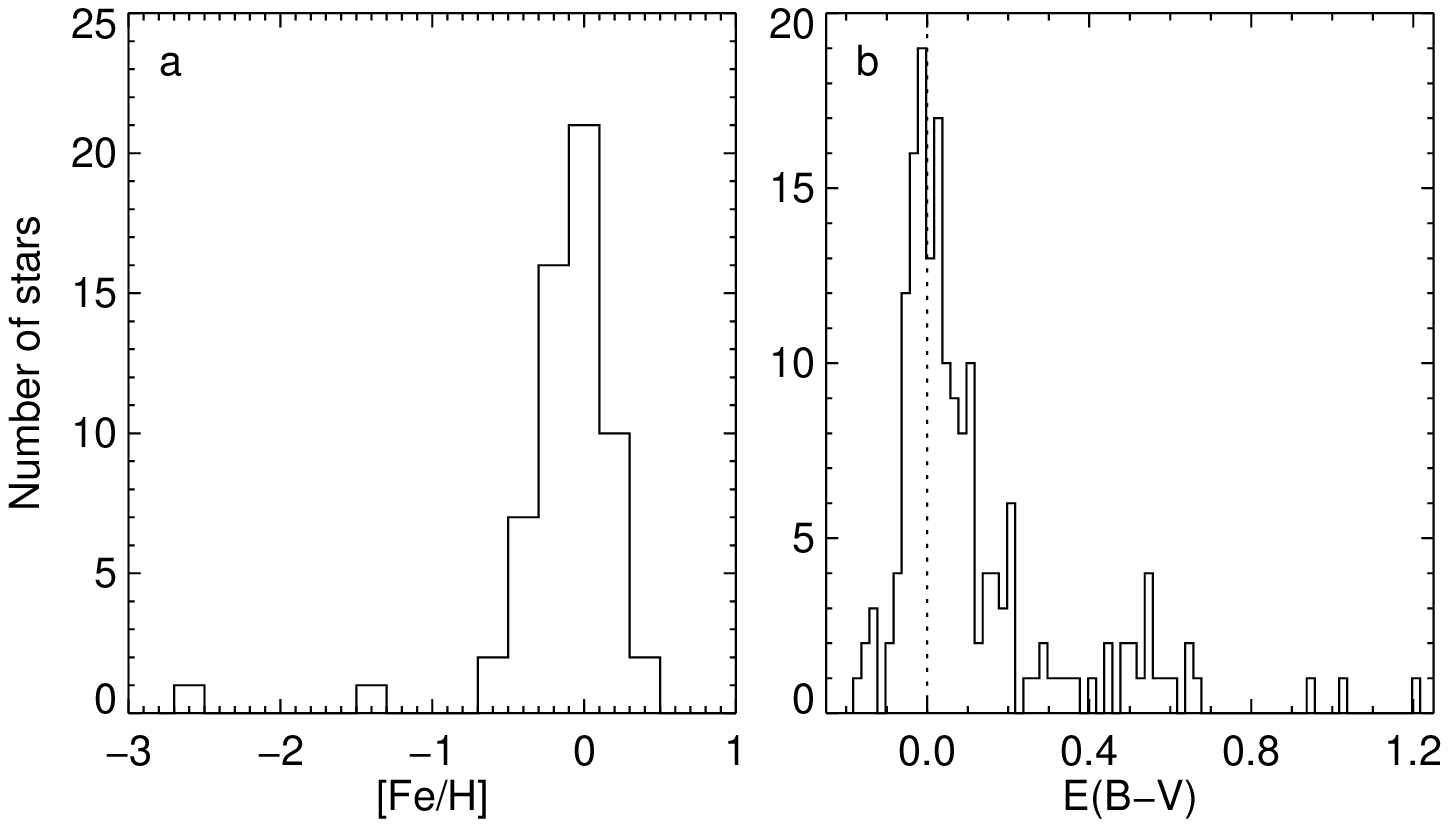}}
\caption{\label{fig:FeH}a \textit{Left:} Distribution of metallicities
  for stars in our sample with spectroscopic measurements of [Fe/H] (61
  out of 210, mostly F, G, and K stars) from
  \citet{1997A&AS..124..299C}. The mean is $-$0.1 and the dispersion 0.2
  dex, which is typical for stars in the solar neighborhood
  \citep[within 40 pc,][]{2004A&A...418..989N}.  b \textit{Right:}
  Distribution of $E(B-V)$ color excesses for the stars in our sample.
  The width of the peak at negative values indicates the uncertainty in
  the color excesses is 0.036 mag.}
\end{figure}

\begin{figure}
\centerline{\includegraphics[width=3.5in,angle=0]{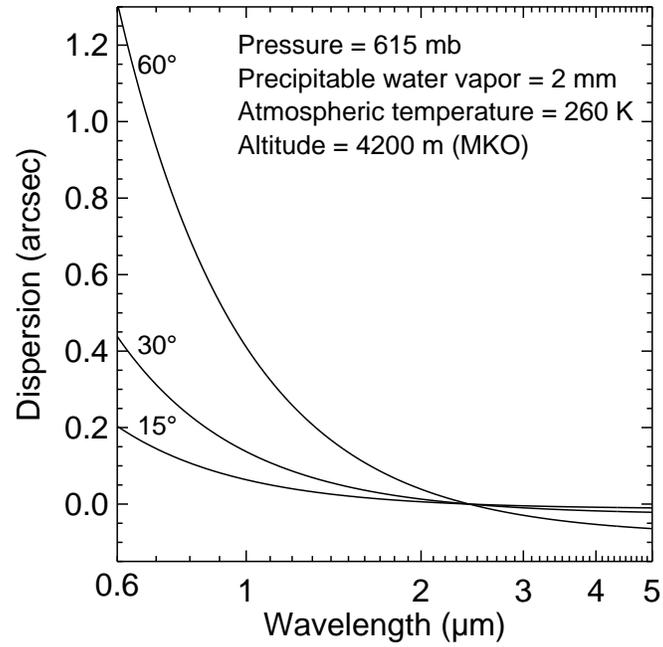}}
\caption{\label{fig:atmosdisp}Atmospheric dispersion for the summit of
  Mauna Kea and the conditions indicated, as a function of wavelength
  (relative to 2.4~$\micron$) and zenith angle. Since the magnitude of
  atmospheric dispersion in the NIR is significant compared to the slit
  width used (0$\farcs$3) observations were made at the parallactic
  angle to minimize slit losses and to measure the spectral shape more
  accurately.}
 \end{figure}

\begin{figure}
\centerline{\includegraphics[width=6in,angle=0]{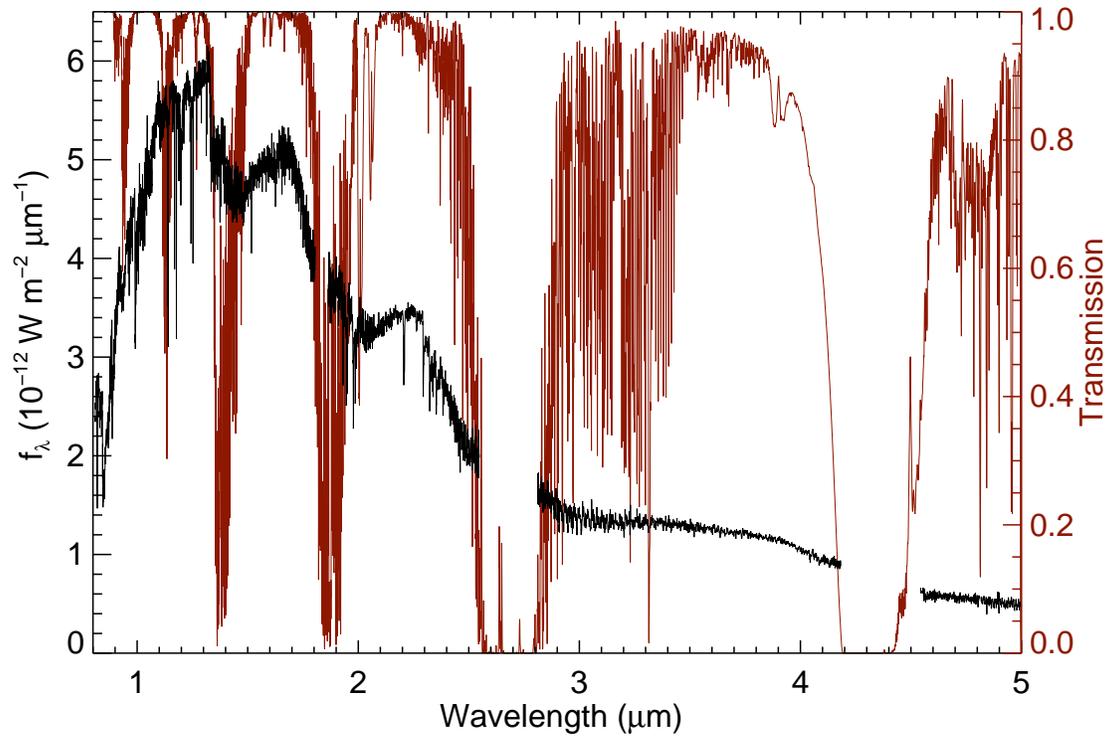}}
\caption{\label{fig:AtmosSpec}The atmospheric transmission for Mauna Kea
  (4200~m, airmass 1.15, precipitable water vapor 2~mm) overplotted with
  the spectrum of Gl~406 (M6~V).}
\end{figure}

Although the stars in our sample are generally bright and nearby, some
stars show evidence of interstellar reddening. For those applications
requiring true spectral energy distributions (e.g. EPS studies)
reddening needs to be corrected.  Consequently, as a final step in the
data reduction we corrected the spectra for reddening.  We determined
the $E(B-V)\equiv(B-V)-(B-V)_0$ color excess from the observed ($B-V$)
color and and an intrinsic $(B-V)_0$ color appropriate for its spectral
type.  The observed colors were taken from the
\citet{2006yCat.2168....0M} catalog and are on the Johnson photometric
system.  For the few stars that were not included in this catalog, we
adopted the (B-V) color given by \citet{2001KFNT...17..409K},
\citet{1992ApJS...82..351L}, and \citet{1994ApJS...93..187B}.  We
adopted the calibration of intrinsic colors as a function of spectral
type by \citet{1970A&A.....4..234F}.  For the M dwarfs, we found that
the \citet{1970A&A.....4..234F} values gave unreasonably large color
excess values for stars that are very nearby (less than 20 pc).
Furthermore, while other calibrations of intrinsic colors \citep[e.g.,
][]{SK82} agree well with that of \citet{1970A&A.....4..234F} at earlier
spectral types, they differ markedly for M stars, with the later
calibrations becoming progressively redder.  For these reasons, we
adopted the intrinsic colors given by \citet{1992ApJS...82..351L} for
the M dwarfs, which generally yield very small (or even negative) values
for $E(B-V)$ for these stars.  To derive intrinsic colors for stars with
intermediate spectral types and luminosity classes not tabulated in the
combined FitzGerald/Leggett calibration we performed a two-dimensional
surface interpolation over the intrinsic color values as a function of
spectral subtype and luminosity class.  The distribution of E(B-V)
values for the stars in our sample is shown in Figure \ref{fig:FeH}b.
As expected for the bright (and generally nearby) stars in our sample,
the distribution is peaked near 0. Fitting a Gaussian to the values of
$E(B-V) < 0$ indicates that the uncertainty on the color excesses is
about $\sigma$=0.036 mag. All color excesses determined to be less than
zero were set to 0 in the dereddening process. Furthermore, based on our
findings for the standard deviation of the distribution, we chose not to
correct any spectra with $E(B-V) < 0.108$.  The 66 dereddened stars
along with their ($B-V$), ($B-V)_0$, $E(B-V)$, and $A_V$ values are
given in Table \ref{tab:Extinction}.

The reddening-corrected spectrum $f^\mathrm{cor}_\lambda (\lambda)$ is
given by,

\begin{equation}
  f^\mathrm{cor}_\lambda (\lambda) = f_\lambda (\lambda) \times 10^{(0.4\times A_\lambda)}
\end{equation}
\noindent
where $f_\lambda (\lambda)$ is the absolutely flux calibrated spectrum
and $A_\lambda$ is the extinction law as a function of wavelength.  We
adopted the NIR law given by \citep{2007ApJ...663..320F} with $R_V$=3.0.
For this law,
\begin{equation}
A_\lambda = 1.057 \times E(B-V)\lambda^{-1.84}.
\label{eq:AV}
\end{equation}

We note that these dereddened spectra should be used with care because
the dereddening process assumes an intrinsic stellar color and a mean
Galactic extinction law that may not be accurate or appropriate in all
cases.  For example, we find $E(B-V)=0.06$ (or $A_V$=0.18) for the M6
dwarf Gl 406 which resides at a distance of only 2.4 pc
\citep{1995gcts.book.....V}.  No significant extinction is expected at
this distance.  In addition, the M giant stars and PAGB and TPAGB stars
may have local extinction due to dust formation in their cool outer
atmospheres which is difficult to separate for any interstellar
extinction that may be present.  Variable stars are also problematic due
to changes in the intrinsic color. Therefore unless otherwise noted, we
will continue to use the uncorrected spectra in the remainder of the
analysis, including the figures.  Nevertheless, the dereddened spectra
are available on the IRTF website (see footnote 2).

\begin{figure}
\centerline{\includegraphics[width=3.5in,angle=0]{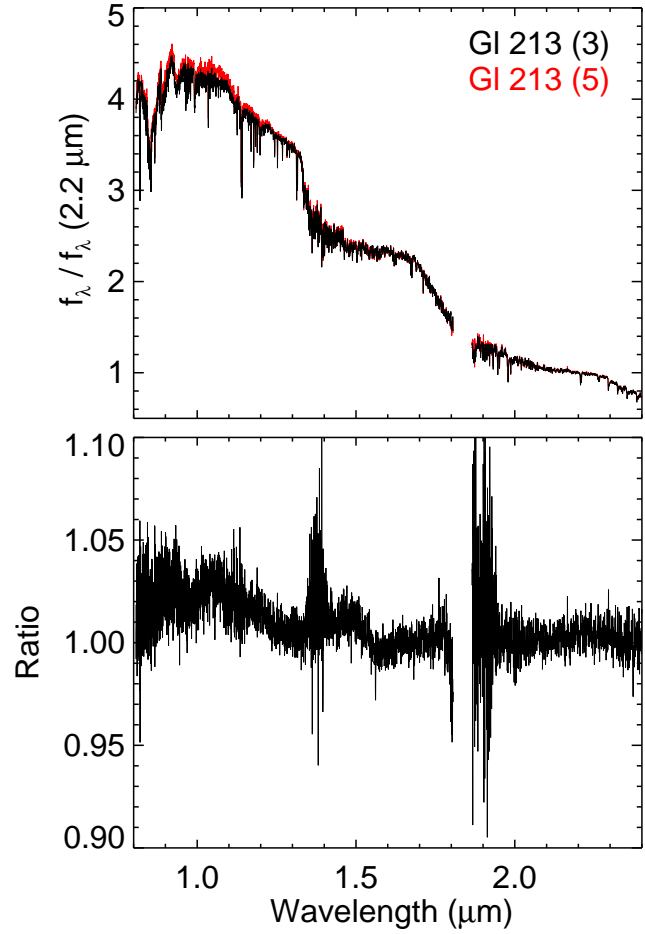}}
\caption{\label{fig:Slope}\textit{Top:} The third (black) and fifth (red)
spectra in a sequence of ten consecutive spectra of  Gl~213 (M4~V). These 
two spectra show the largest slope difference in the sequence. The  
difference is equivalent to  $J-K_{S}$=0.015. \textit{Bottom:}
  The flux ratio of the two spectra showing that the difference in slope
  is small across the range 0.8 to 2.4$\micron$. As described in
  \S\ref{Red} we can measure spectral slopes (i.e. photometric colors)
  accurate to a few percent.}
\end{figure}

\begin{figure}
\centerline{\includegraphics[width=3.5in,angle=0]{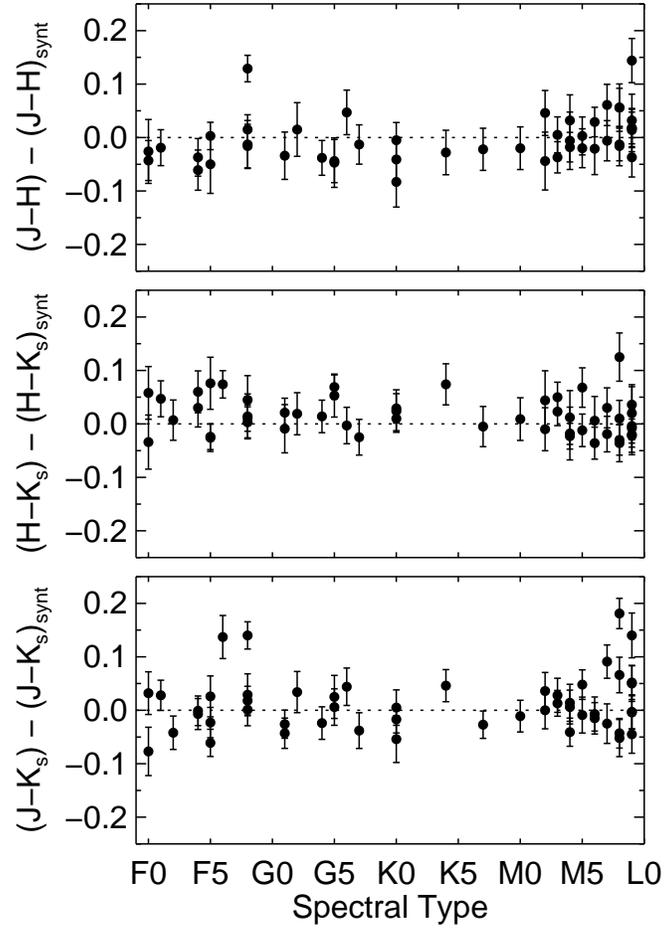}}
\caption{\label{fig:Residuals} Computed residuals (observed $-$
  synthetic) between the observed and synthesized 2MASS $J-H$, $H-K_S$,
  and $J-K_S$ colors of 53 library stars with relatively good (5\%)
  2MASS photometry.  The average residuals are $<\Delta_{J-H}> =
  0.00\pm0.04$ (RMS), $<\Delta_{H-K_S}> = 0.02\pm0.04$ (RMS), and
  $<\Delta_{J-K_S}> = 0.01\pm0.05$ (RMS). The plotted error bars show
  the 2MASS error in magnitudes. The error on the synthetic color is
  insignificant by comparison. The residuals imply that the measured
  spectral slope of these cool stars is accurate to within a few
  percent.}
 \end{figure}

We have found that the observed spectral slope from any given object
varies slightly between individual observations.  An example of this
small effect is illustrated in Figure \ref{fig:Slope}.  In order to
quantify this effect, we have computed the differences ($\Delta_{X-Y} =
(X-Y)_{\mathrm{obs}} - (X-Y)_{\mathrm{synth}}$) between the colors
derived from the 2MASS photometry and those derived from synthetic
photometry on our observed spectra.  The synthetic color for any two of
the 2MASS bandpasses is given by,


\begin{align}
X-Y & = &-2.5 \times \log \left [ \frac{\int f_\lambda S_X(\lambda) d\lambda}{F_X^\mathrm{Vega}} \right ] + \nonumber \\
    & & 2.5 \times \log \left [ \frac{\int f_\lambda S_Y(\lambda) d\lambda}{F_Y^\mathrm{Vega}} \right ] + \nonumber \\
    &  &  (zp_Y - zp_X),
\end{align}
\noindent 
where the symbols have the same meaning as in Equation 1.  The results
for the 53 stars with relatively good (5\%) 2MASS photometry (from Table
\ref{tab:Sample}) are plotted in Figure \ref{fig:Residuals}.  The
average differences are $<\Delta_{J-H}> = 0.00\pm0.04$ (RMS),
$<\Delta_{H-K_S}> = 0.02\pm0.04$, and $<\Delta_{J-K_S}> =
0.01\pm0.05$. Although the colors derived from the 2MASS photometry are
not as precise as those measured from the spectra via synthetic
photometry (better than 1\%), the photometric residuals on the sample of
53 cool stars indicate that our measurements of spectral slope are
accurate to within a few percent for F, G, K, and M spectral types.
Similar (but larger) effects have been observed by
\citet{2003SPIE.4839.1117G} while using adaptive optics with medium
resolution spectroscopy.  Their simulations demonstrate that spectral
slope variations result from changes in the fraction of light from the
object transmitted by a finite width slit as a function of
wavelength. Therefore, temporal changes in seeing, guiding, and
differential atmospheric refraction, are probably the cause of the small
variations we observe.  Although SpeX does not have an atmospheric
dispersion corrector, we minimized the effects of differential
refraction (and therefore the wavelength dependent light loss through
the slit) by moving the internal image rotator to observe at the
parallactic angle \citep[see Figure~6 in][]{2004SPIE.5492.1498R} and by
combining multiple spectra together.

As a further test of the accuracy of the spectral slopes, we have
computed synthetic $Z-Y$, $Y-J$, $J-H$, $H-K$, and $K-L'$ colors of
eight A0~V stars observed with the same instrumental setup.  The
synthetic color for any two bandpasses $X$ and $Y$ is given by,
\begin{align}
  X-Y & = & -2.5 \times \log \left [ \frac{\int \lambda f_{\lambda}(\lambda)S_X(\lambda)d\lambda}{\int \lambda f^{\mathrm{Vega}}_{\lambda}(\lambda)S_X(\lambda)d\lambda} \right ] + \nonumber \\
& & 2.5 \times \log \left [ \frac{\int \lambda f_{\lambda}(\lambda)S_Y(\lambda)d\lambda}{\int \lambda f^{\mathrm{Vega}}_{\lambda}(\lambda)S_Y(\lambda)d\lambda} \right ], 
\label{coloreqn}
\end{align}
\noindent
where $f_{\lambda}(\lambda)$ is the flux density of the object,
$f^{\mathrm{Vega}}_{\lambda}(\lambda)$ is the flux density of Vega, and
$S(\lambda)$ is the system response function for each bandpass which we
assume to be given by the product of the filter transmission and the the
atmospheric transmission.  Equation \ref{coloreqn} assumes that Vega has
zero color in all passband combinations.  We used the Wide Field Camera
\citep[WFCAM;][]{2007A&A...467..777C} $ZYJHK$ filter profiles that
include the atmospheric transmission at an airmass of 1.3 with a
precipitable water vapor content of 1.0 mm \citep{2006MNRAS.367..454H}.
The $Z$ and $Y$ bands are custom filters designed for the UKIRT Infrared
Deep Sky Survey \citep[UKIDSS; ][]{2007MNRAS.379.1599L} centered at
$\sim$0.88 and $\sim$1.03 $\mu$m while the $JHK$ bands were constructed
to the specifications of the Mauna Kea Observatories Near-Infrared
(MKO-NIR) filter set \citep{2002PASP..114..180T}.  We constructed an
$L'$ system response function using the MKO-NIR $L'$-band filter profile
and ATRAN \citep{1992nstc.rept.....L} to calculate the atmospheric
transmission under the same atmospheric conditions.  For
$f_{\lambda}^{\mathrm{Vega}} (\lambda)$ we used a Kurucz model of Vega
($T_{\mathrm{eff}}=9550$ K, $\log g = 3.950$, $v_{\mathrm{rot}} = 25$ km
s$^{-1}$, and $v_{\mathrm{turb}} = 2$ km s$^{-1}$), scaled to the flux
density at $\lambda$=5556 \AA \ given by \citet{1995AA...296..771M}.
The factors of $\lambda$ inside the integrals convert the energy flux
densities $f_{\lambda}$ to photon flux densities, which ensures that the
integrated fluxes are proportional to the observed photon count-rate
\citep[e.g.,][]{1986HiA.....7..833K, 1992A&A...264..557B}.

The mean colors of the 8 A0 V stars are $<$$Z-Y$$>$$ =-0.015\pm0.007$
(RMS), $<$$Y-J$$>$$=0.007\pm0.006$ (RMS), $<$$J-H$$>$$=0.019\pm0.005$
(RMS), $<$$H-K$$>$$=-0.005\pm0.004$, $<$$K-L'$$>$$=-0.001\pm0.010$.  
Given that we expect the mean colors to be zero, the mean colors 
of the A0V stars represent the precision in our
ability to measure spectral slope.  Taken together, the 2MASS and A0V star
photometry indicates that we can measure spectral slope to within a few
percent. Finally, slope variations introduced
due to uncertainties in the spectral types of the A0 V standard stars
are small since an uncertainty of a 0.5 subtype at A0 is equivalent to
${\delta}_{J-H}=0.003$, ${\delta}_{H-K}=0.005$, and
${\delta}_{J-L'}=0.010$, \citep{2000asqu.book.....C}.

\section{DATA AND ANALYSIS}

\subsection{\label{Spectra}The Spectra}

\begin{figure}
\centerline{\includegraphics[width=5in,angle=0]{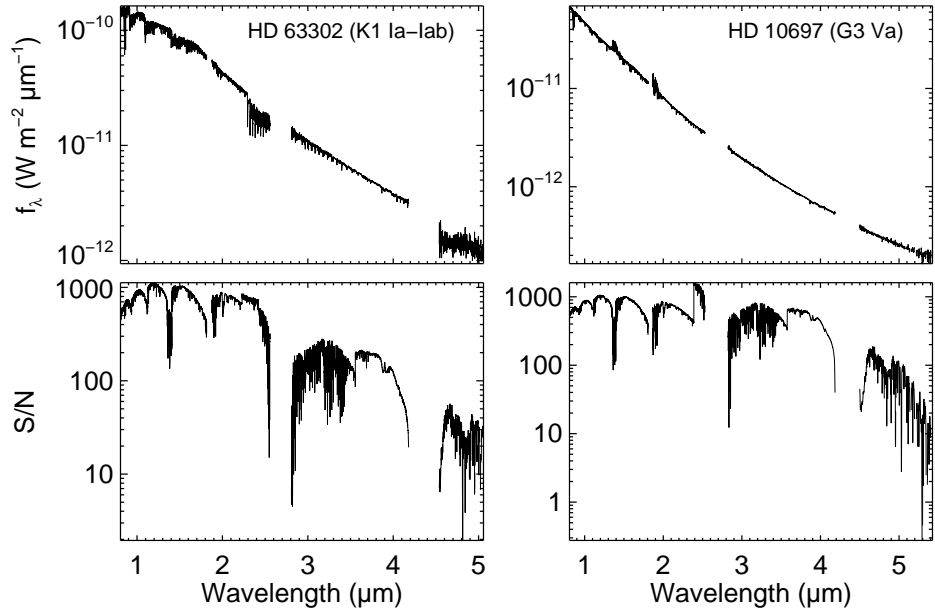}}
\caption{\label{fig:SNR} \textit{Left.} The flux and S/N spectra of
  HD~63302 (K1~Ia-Iab).  Telluric correction is excellent.
  \textit{Right.} The flux and S/N spectra of HD~10697 (G3~Va). Note the
  flux error in the spectrum at $\sim$1.9~\micron~and
  $\sim$1.4~\micron~despite the fact that the S/N$>$100. This is due to
  relatively poor telluric correction in these regions for this
  particular observation.  Therefore, care must be taken interpreting
  features in regions of high telluric contamination even if the S/N is
  high.}
 \end{figure}

Digital versions of the spectra are available at the IRTF website, which
contains a full description of the data products.  The data are
available in text or Spextool FITS format and files can be downloaded
individually or bundled together in a tar file. The files contain
wavelength, flux, and error. The errors include the photon
Poisson noise and read noise \citep{2004PASP..116..352V} for both the
object and associated telluric standard, which are then propagated
through each step of the reduction process. The file headers contain
more information (including object name, epoch, spectral type, observing
modes, 2MASS $JHK$ magnitudes, measured radial velocity, flux, and
wavelength units).  As an example, the flux and S/N spectra of HD~63302
(K1~1a-Iab) and HD~10696 (G3~V), are shown in Figure~\ref{fig:SNR}.  We
estimate that the actual S/N in our fully reduced stellar spectra is
limited to less than 1000 by systematic errors in the flat field
measurement even though the formal S/N can be greater than 1000. Also,
systematic errors in telluric correction (e.g. if the airmasses of an
object and standard star are not ideally matched), systematic
errors in the slope, as well as any errors in the spectral type of the
A0 V telluric standard are not accounted for in the formal S/N
estimate (see Figure \ref{fig:SNR}).

Representative spectra, covering 0.8 to 5.0~$\mu$m, of dwarfs, giants,
and supergiants in our sample are shown in Figures \ref{fig:FGKM_V} to
\ref{fig:FGKM_I2II}.  Given the large wavelength range and change in
flux wavelength and spectral type, it is difficult to display the
spectra at a scale that allows close examination of all of the
interesting spectral features in one plot. For display purposes we
therefore use $\lambda f_{\lambda}$ (normalized at the given wavelength)
as a function of $\log \lambda$.  The plots show the general trend of
spectral features with MK spectral type.  Feature identifications and
changes with spectral type are described in more detail in
\S\ref{Lines}, \S\ref{Mol}, and \S\ref{Features}.

It should be noted that MK spectral classification is based on
comparing an $optical$ spectrum to a set of stars (anchor points) that
define certain spectral types. Therefore it is not necessarily the
case that the trends in NIR spectral features will follow the optical
types. Also, the exact numbers for the effective temperature, surface gravity, and
composition (i.e. metallicity) are model dependent (i.e. not directly
observable) and will change as models improve. Although there are no
formal MK classification criteria for the NIR, much work has been
conducted on NIR spectral classification.  The pioneering study on
cool stars was done by \citet{1986ApJS...62..501K} who identified a
number of temperature and luminosity sensitive atomic (\ion{Na}{1},
\ion{Ca}{1}, Br~$\gamma$) and molecular (CO, H$_2$O) features in the
$K$ band. Subsequent studies developed a variety of spectral type
versus equivalent width (EW) indices for cool stars. Notable examples
amongst these are the studies by \citet{1993AA...280..536O},
\citet{1996AAS..116..239D}, \citet{1998ApJ...508..397M},
\citet{1997AJ....113.1411R}, \citet{2000AJ....120.2089F},
\citet{2003ApJ...593.1074G}, \citet{2007ApJ...671..781D},
\citet{2004ApJS..151..387I}, and \citet{2008AA...489..885M}. We do not
propose to add to these and other NIR classification schemes but
present the IRTF Spectral Library as a resource for further
investigation. Examples of EWs of several prominent features in data
as a function of spectral type and luminosity are, however, presented
in \S\ref{EW}.

As an example of the mismatches in the trends of the NIR
spectral features with MK type, Figure \ref{fig:KIII} shows a spectral
sequence of K giant stars from 0.8 to 2.45 $\mu$m based on their MK
types. From the trend in overall spectral shape and depth of the first
overtone vibration-rotation bands of CO in the $K$-band
($\sim$2.29-2.5 $\mu$m), two of the stars appear out of sequence (they
are bluer and with shallower CO absorption than expected) and should
appear earlier in the sequence by about one spectral subtype.  When
the stars are corrected for reddening (see Figure~\ref{fig:KIIIdered}
and Table~\ref{tab:Extinction}) the sequence of spectral shapes behave
as expected but the CO band depths still appear to be out of sequence.
A possible explanation is that these stars are slightly metal poor
\citep[see for example][]{2008AA...489..885M}.  However, this
explanation is inconsistent with the assigned MK spectral types, one
star being of approximately solar metallicity (HD~137759, K2~III), and
the other star being slightly metal rich (HD~114960, K3.5~IIIb CN0.5 CH0.5)
although neither star has a formal [Fe/H] measurement.  Another
possibility is the uncertainty in assigning the MK spectral type as
discussed in \S\ref{Sample_Selection}. Given these uncertainties,
spectral classification using the IRTF Spectral Library is probably
best done by comparing a given stellar spectrum of an unknown type to
an ensemble of spectra for a sequence of MK types, rather than trying
to find the closest individual spectral match. The effect of reddening
should also be considered.

Plots of the dwarf sequence F3, F5, F9, G2, G8, K1, K7, M3, and M7, for
the $I$, $Y$, $J$, $H$, $K$, and $L$ bands, respectively, are shown in
Figures \ref{fig:FGKM_VI} to \ref{fig:FGKM_VL}.  Luminosity effects at
spectral types F, G, K, and M, across the 0.8-5~$\mu$m range, are shown
in Figures \ref{fig:FI2VSeq} to \ref{fig:MI2VSeq}.  Plots showing
luminosity effects at spectral types F5 and G6 for the $I$, $Y$, $J$,
$H$, $K$, and $L$ bands, respectively, are shown in Figures
\ref{fig:FG_I2VI} to \ref{fig:FG_I2VL}; likewise luminosity effects at
spectral types K5 and M5 are shown in Figures \ref{fig:KM_I2VI} to
\ref{fig:KM_I2VL}.  A 0.8-5~$\mu$m sequence of M, S, and C giants is
plotted in Figure \ref{fig:MSC}. All these spectra are discussed in more
detail in the subsequent sections.

\subsection{\label{Lines}Atomic Line Identifications}

The S/N of the spectra is high enough that almost all of the absorption
features seen in the spectra are real and not noise.  However given that
the resolving power of the spectra is $R\sim$2000, the carriers of only
the strongest and most isolated lines can be unambiguously identified.
We therefore use the line identifications presented in the high
resolution ($R\sim$100000) spectral atlases of the solar photosphere
\citep{1993asps.book.....W,2003assi.book.....W} and Arcturus \citep[K1.5
III,][]{1995iaas.book.....H,2000vnia.book.....H}.  We selected only
those absorption lines in the atlases with depths less than 0.8 (where
the continuum has been normalized to unity) because weaker lines are
only marginally detectable in our spectra.

Tables \ref{tab:Solar Atomic Lines} and \ref{tab:Arcturus Atomic
Lines} list all atomic metal lines with depths less than 0.8 in the
solar and Arcturus atlases, respectively.  The wavelengths of the lines
are in vacuum and were taken primarily from the atlases themselves.
The revised solar atlas \citep{2003assi.book.....W} does not present
the wavelengths of the lines so we obtained them from an earlier
version of the atlas \citep{1991aass.book.....L}.  A few of the lines
in the revised atlas either lack identifications in the original atlas
or have since been identified as being carried by two lines.  We
determined the wavelengths of these lines using the National Institute
of Standards and Technology Atomic Spectra
Database\footnote{\url{http://physics.nist.gov/PhysRefData/ASD/index.html}},
the Atomic Line
List\footnote{\url{http://www.pa.uky.edu/~peter/atomic/}}, and the
\ion{Fe}{1} line list compiled by \citet{1994ApJS...94..221N}.

\subsection{\label{Mol}Molecular Band Identifications}

Molecular absorption bands are prominent in the spectra of late-type
stars.  Identifications for many of these features can be found in e.g.,
\citet{1969ARA&A...7..249S}, \citet{1990A&A...231..440B},
\citet{1989MNRAS.241..247B}, \citet{2000AAS..146..217L}.
Table~\ref{tab:MolecularBands} lists the vacuum wavelengths and
identifications for each of the band heads identified in the spectra.
Below we describe some of the absorption bands in detail.

\subsubsection{TiO} Absorption bands of TiO that arise from the $\gamma$
($A$ $^3\Phi$ $-$ $X$ $^3\Delta$), $\epsilon$ ($E$ $^3\Pi$ $-$ $X$
$^3\Delta$), $\delta$ ($ b$ $^1\Pi$ $-$ $a$ $^1\Delta$), and $\phi$ ($b$
$^1\Pi$ $-$ $d$ $^1\Sigma$) systems are conspicuous in the spectra of
late-type stars over the 0.8 to 1.5 $\mu$m wavelength range (see Figures
\ref{fig:FGKM_V}, \ref{fig:FGKM_III}, \ref{fig:FGKM_I2II},
\ref{fig:FGKM_VI}, \ref{fig:KM_I2VI}, \ref{fig:KM_I2VY},
\ref{fig:KM_I2VJ}).  The $\gamma$ and $\epsilon$ systems involve triplet
electronic states and thus exhibit triple-headed bands while the
$\delta$ and $\phi$ systems involve singlet electronic states and thus
exhibit only single-headed bands.

The $\Delta \nu=-2$ bands of the $\gamma$ system (0.82$-$0.86 $\mu$m)
are often identified in both low and high resolution spectra of
late-type stars
\citep[e.g.,][]{1991ApJS...77..417K,1998MNRAS.301.1031T,2007A&A...473..245R}.
In particular, four TiO band heads ($R_2$, $Q_2$, $R_1$, $Q_1$) are
identified between 0.849 and 0.860~$\mu$m (see Figures \ref{fig:FGKM_VI}
and \ref{fig:TiO}).  However as can be seen in both our data and the
high resolution spectra of late-type M and early-type L dwarfs,
\citep{1998MNRAS.301.1031T,2007A&A...473..245R}, two additional band
heads of similar depths exist at 0.8508 and 0.8582 $\mu$m.  Together,
the six band heads appear to form two sets of triplet band heads.  The
1$-$1 and 2$-$2 bands of the $\epsilon$ system exhibit band heads at
these wavelengths and therefore it is likely that these features arise
from the $\epsilon$ system alone or are a combination of the $\gamma$
and $\epsilon$ systems.  We also note that three of the $\Delta \nu=-2$
band head classifications listed in \citet{1957msmo.book.....G} were
marked as uncertain so we have confirmed these using the theoretical
line list of \citet{1998cpmg.conf..321S}.

It has been known for some time that absorption bands of TiO are present
in the spectra of M giant stars near 0.93 $\mu$m
\citep{1965ApJ...141..965S}.  However, to our knowledge, these bands
were not detected in the spectra of M dwarfs until the work of
\citet{2006AJ....131.1797C}. These band heads arise from the
$\Delta\nu=-1$ band of the $\epsilon$ system.  Although telluric and
intrinsic H$_2$O absorption make identifications in this region
difficult we nevertheless are able to identify the triplet band heads of
the 0$-$1 (0.9211, 0.9221, 0.9233 $\mu$m), 1$-$2 (0.9279, 0.9289, 0.9300
$\mu$m), and 2$-$3 (0.9345, 0.9356, 0.9368 $\mu$m) transitions (see
Figures \ref{fig:FGKM_VI} and \ref{fig:KM_I2VI}).  Finally, we note that
while the wavelengths of the 2$-$3 and 3$-$4 band heads measured by
\citet{1977JMoSp..64..382L} agree with our observations, they do not
match the positions of the band heads predicted by
\citet{1998cpmg.conf..321S}.

\subsubsection{VO} Absorption bands that arise from the $B$
$^4\Pi$$-$$X$ $^4\Sigma^+$~\footnote{This system is identified as the
  ``C'' system in \citet{1957msmo.book.....G},
  \citet{1978A&AS...34..409S}, and \citet{1998MNRAS.301.1031T}.} and $A$
$^4\Pi$ $-$$X$$^4\Sigma^-$ systems of VO are present in the spectra of M
stars over the 0.8 to 1.3~$\mu$m wavelength range.  We identify only the
0$-$0 ($\sim$1.06~$\mu$m) and 0$-$1 ($\sim$ 1.18~$\mu$m) bands of the
$A$ $^4\Pi$ $-$$X$$^4\Sigma^-$ system (see Figures \ref{fig:FGKM_III},
\ref{fig:FGKM_VY}, \ref{fig:KM_I2VY}, \ref{fig:MIIIwater}).  Although
there are certainly absorption features arising from the 0$-$1 band of
$B$ $^4\Pi$$-$$X$ $^4\Sigma^+$ system centered near $\sim$0.85~$\mu$m in
the spectra of late-type stars
\citep{1952ApJ...115...82K,1998MNRAS.301.1031T}, they are simply too
weak to be detected in our data.

\subsubsection{ZrO} A series of four ZrO band heads arising from the
0$-$0 band of the system $b$ $^3\Pi$$-$$a$ $^3\Delta$ of ZrO
\citep{1979ApJ...234..401P} are found in the spectra of M giant stars
and carbon stars (see Figures \ref{fig:FGKM_I2II}, \ref{fig:MI2VSeq},
\ref{fig:KM_I2VI}, \ref{fig:MSC}). Three additional band heads
are also present at 0.932060, 0.934417, and 0.935860 $\mu$m but their
corresponding transitions are unknown.  Only three of the band heads
($R_{1(c)}$, $R_{1(d)}$, and 0.932060 $\mu$m) can be conclusively
identified in the spectra given the complexity of the spectra at these
wavelengths, but we include the other band heads for completeness.
Additional ZrO band heads
\citep{1981ApJS...47..201H,1998AJ....116.2520J} in the 1.0$-$1.2 $\mu$m
wavelength range are not seen in our spectra.

\subsection{\label{Features}Variations of Spectral Features with Spectral Type and Luminosity}

In the following sections we describe the variations of spectral
features with spectral type and luminosity of the stars in our
sample. These changes are illustrated with the representative spectra
given in Figures~\ref{fig:FGKM_V} to \ref{fig:MSC} which include feature
identifications. Technically these identifications are only correct for
Arcturus and the Sun so we have identified the features over only
Arcturus and a solar analog HD 76151 (G2 V).  We caution against
assuming the same lines are present at much earlier and later spectral
types than K1.5 III, and G2 V. More complete spectral sequences are
plotted in the Appendix (Figures~\ref{fig:F_VI} to \ref{fig:M_IL}) but
without feature identifications.

Many of the features and variations in their strengths have been
previously recognized: \citet{1998AJ....116.2520J}, and
\citet{2000ApJ...535..325W} at $J$; \citet{1998ApJ...508..397M} at $H$;
\citet{1986ApJS...62..501K}, \citet{1996ApJS..107..312W}, and
\citet{1997ApJS..111..445W} at $K$; \citet{2002AJ....124.3393W}, and
\citet{2002AA...390.1033V} at $L$; \citet{1992AAS...96..593L} at
1.4-2.5~\micron~for normal stars; \citet{2000AAS..146..217L} at
0.5-2.5~\micron~for luminous cool stars; and
\citet[][0.5-2.5~\micron]{2001A&A...371.1065L}, and
\citet[][3-8~\micron]{1998A&A...340..222A} for carbon stars.

\subsubsection{F Stars}

Representative F star spectra are shown in Figures \ref{fig:FGKM_V},
\ref{fig:FGKM_III}, \ref{fig:FGKM_I2II},
\ref{fig:FGKM_VI}-\ref{fig:FI2VSeq}, and
\ref{fig:FG_I2VI}-\ref{fig:FG_I2VL}.  The NIR spectra of F stars are
dominated by the neutral hydrogen (\ion{H}{1}) absorption lines of the
Paschen (n=3), Brackett (n=4), Pfund (n=5), and Humphreys (n=6) series,
in order of increasing wavelength and decreasing strength. The Brackett
series ($H$ band) is a good luminosity indicator, smoothly decreasing in
strength from supergiants through giants to dwarfs (Figures
\ref{fig:FGKM_VH}, \ref{fig:FI2VSeq}, and \ref{fig:FG_I2VH}).  The Pfund
series ($K$ band, series limit 2.33~$\mu$m) can also be strong in early
F supergiants (Figure \ref{fig:FGKM_I2II}). Between the \ion{H}{1}
lines, spectra are dominated by features due to neutral metal species
(e.g. \ion{Si}{1} at 1.06-1.09 $\mu$m and 1.16-1.21 $\mu$m, Figures
\ref{fig:FG_I2VY} and \ref{fig:FG_I2VJ}).  The strongest feature in the
spectra is the \ion{Ca}{2} triplet at 0.86 $\mu$m (Figure
\ref{fig:FG_I2VI}). The only indication of molecular absorption is the
very weak CN feature at 1.09 $\mu$m in the latest-type F stars (Figure
\ref{fig:FGKM_I2II}).

\subsubsection{G Stars}

Representative G star spectra are shown in Figures \ref{fig:FGKM_V},
\ref{fig:FGKM_III}, \ref{fig:FGKM_I2II},
\ref{fig:FGKM_VI}-\ref{fig:FGKM_VL}, \ref{fig:GI2VSeq}, and
\ref{fig:FG_I2VI}-\ref{fig:FG_I2VL}.  \ion{H}{1} absorption weakens
significantly towards late-type G stars while the neutral metals are
stronger than seen in F stars, although the strongest feature is again
the \ion{Ca}{2} triplet at 0.86~\micron~(Figures \ref{fig:FGKM_V},
\ref{fig:FGKM_III}, and \ref{fig:FGKM_I2II}).  The strongest neutral
metal lines are those due to Mg (in the $H$ band at about 1.50~\micron,
1.58~\micron, and 1.71~\micron, Figure \ref{fig:FG_I2VH}). Molecular
absorption due to CO and CN strengthen with decreasing effective
temperature in G stars and these also provide the best luminosity
indicators. The first CO overtone bands in the $K$ band ($\sim$2.29-2.5
\micron) are strongest in supergiants and become progressively weaker
with decreasing luminosity (Figure \ref{fig:FG_I2VK}). The CN band head
at 1.09~\micron~weakens with decreasing luminosity in mid- to late-type
G stars (Figure \ref{fig:FG_I2VY}).

\subsubsection{K Stars}

Representative K star spectra are shown in Figures \ref{fig:FGKM_V},
\ref{fig:FGKM_III}, \ref{fig:FGKM_I2II},
\ref{fig:FGKM_VI}-\ref{fig:FGKM_VL}, \ref{fig:KI2VSeq}, and
\ref{fig:KM_I2VI}-\ref{fig:KM_I2VL}.  \ion{H}{1} absorption becomes very
weak in K stars and is effectively absent by late K
(e.g. Brackett~$\gamma$ in Figure
\ref{fig:FGKM_VK}, with a slight dependence on luminosity).  
Neutral metal absorption features reach a maximum
depth in the spectra of K and M stars and lines due to \ion{Al}{1} at
1.31~\micron, \ion{Mg}{1} at about 1.50~\micron~and 1.53~\micron,
\ion{Al}{1} at 1.67~\micron, and \ion{Mg}{1} at 1.71~\micron, are
particularly strong in the spectra of K~dwarfs and early-M~dwarfs
(Figures \ref{fig:FGKM_VJ} and \ref{fig:FGKM_VH}).  Lines from ionized
metals weaken with progressively later spectral types (e.g. \ion{Ca}{2}
triplet at 0.86~\micron, Figures \ref{fig:FGKM_V}, \ref{fig:FGKM_III},
and \ref{fig:FGKM_I2II}).  Molecular absorption continues to strengthen
in K~stars as effective temperature falls.  The broad $H$-band bump due
to the H$^{-}$ opacity minimum at 1.6~$\mu$m first becomes evident in
early-type K stars, and strengthens with decreasing effective
temperature (Figure \ref{fig:FGKM_III}). This feature was first observed
in the pioneering balloon observations of
\cite{1964ApJ...140..833W}. Molecular features present in the spectra of
$K$ stars are the second CO overtone vibration-rotation bands in the
$H$~band (Figure \ref{fig:KM_I2VH}), the CN band head at
1.40~\micron~(Figures \ref{fig:FGKM_III} and \ref{fig:FGKM_I2II}), OH
(1-0 and 2-1) in the $L$~band (Figure \ref{fig:KM_I2VL}), the SiO first
overtone vibration-rotation band at 4.00-4.18~\micron~(Figure
\ref{fig:KM_I2VL}), as well as the first overtone CO ($K$~band, Figure
\ref{fig:KM_I2VK}) and CN bands (band heads at 0.91~\micron~and
0.94~\micron, Figure \ref{fig:FGKM_I2II}) that are also visible in the
spectra of G~stars. Molecular features weaken with decreasing luminosity
class and provide some of the best surface gravity indicators in K
stars; the CO, CN, and SiO features are particularly sensitive to
surface gravity.  Other luminosity class indicators include \ion{Mg}{1}
at 1.49~$\mu$m and 1.71~$\mu$m, Figure \ref{fig:KM_I2VH}).  These lines
are significantly stronger in dwarfs than in giants and supergiants of
the same spectral type (see Figure~\ref{fig:EWs}).

\subsubsection{M Stars}

Representative M star spectra are shown in Figures \ref{fig:FGKM_V},
\ref{fig:FGKM_III}, \ref{fig:FGKM_I2II},
\ref{fig:FGKM_VI}-\ref{fig:FGKM_VL}, \ref{fig:MI2VSeq}, and
\ref{fig:KM_I2VI}-\ref{fig:KM_I2VL}. Molecular absorption features
dominate the spectra of M stars. The CO ($H$~band, and band heads in the
$K$~band starting at 2.3~\micron, Figure \ref{fig:FGKM_I2II}), OH (band
heads starting at ~3.4~\micron, Figures \ref{fig:KM_I2VL} and
\ref{fig:M_IL}), and SiO absorption bands (band heads starting at
4.0~\micron, Figures \ref{fig:KM_I2VL} and \ref{fig:M_IL}) are strongest
in early-type M supergiant stars (see Table \ref{tab:MolecularBands}).
TiO (several band heads starting at 0.82~\micron) and ZrO (band head at
0.93~\micron) absorption also increases from mid-M to later spectral
types in supergiants (Figures \ref{fig:MI2VSeq} and \ref{fig:M_II}).
Similar trends are seen in M giants with the addition of significant
broad H$_2$O absorption at about 1.4~\micron, 1.9~\micron, and 2.7~\micron,
starting at about M6~III (Figures \ref{fig:FGKM_III}
and \ref{fig:MIIIwater}), and with ZrO replaced by VO (band heads at
1.03~\micron~and 1.17~\micron, Figure \ref{fig:FGKM_III}) in late-type M
giants (TPAGB stars).  The $H$-band bump first seen in the spectra of
early-type K stars is strongest in mid-type M giants and supergiants
(Figures \ref{fig:FGKM_III} and \ref{fig:FGKM_I2II}).  The $H$-band
spectra of M~dwarfs are dominated by numerous FeH absorption features
\citep[Figure \ref{fig:FGKM_VH}, and Figure~7
in][]{2003ApJ...582.1066C}.  In M stars the best luminosity class
indicators are the FeH band head at 0.99~$\mu$m
\citep[Figure~\ref{fig:KM_I2VY},][]{1969PASP...81..527W}, the second CO
overtone bands in the $H$ band (strong in M supergiants and giants,
Figure \ref{fig:KM_I2VH}), the first CO overtone bands in the $K$ band
(strong in M supergiants and giants, Figure \ref{fig:KM_I2VK}), and the
first SiO overtone bands at about 4.0~\micron~(strong in M supergiants
and giants, Figure \ref{fig:KM_I2VL}).  Most neutral metal features
weaken in the late-M spectral types (e.g. \ion{Ca}{1} triplet at
2.26~\micron, Figures~\ref{fig:FGKM_VK} and \ref{fig:M_IIIK}).  The
exceptions are the alkali lines, namely the \ion{Na}{1} doublets at 0.82
\micron, 1.14 \micron, and 2.20 \micron, and the \ion{K}{1} doublets at
1.17 \micron~and 1.25 \micron~(Figures~\ref{fig:KM_I2VI},
\ref{fig:KM_I2VJ}, and \ref{fig:KM_I2VK}).  These lines are strong in
mid- to late-type M dwarfs and weak in corresponding supergiants and
giants, and are consequently excellent luminosity class or surface
gravity indicators. The \ion{Ca}{2} triplet at 0.86~\micron~is
significantly weaker in M stars relative to earlier spectral types where
it blends with TiO absorption (Figure~\ref{fig:FGKM_VI}), and is absent
by mid- to late-type M stars (slightly dependent upon luminosity,
Figures~\ref{fig:M_VI}, \ref{fig:M_IIII}, and \ref{fig:M_II}).

\subsubsection{\label{Carbon}Carbon and S Stars}

The sequence M-MS-S-SC-C is thought to be one of increasing
carbon-to-oxygen ratio as well as increasing $s$-process element
abundance during AGB evolution \citep{1979ApJ...234..538A}.  Our sample
contains one MS, four S, five C-N, one C-R, and two C-J stars (see Table
\ref{tab:Sample}).
In normal (oxygen-rich, C~$<$~O) M giants some of the oxygen is used up
to make CO but most goes into making metal oxides such as TiO. In
typical carbon stars (C~$>$~O) all the oxygen is used up in the
production of CO and the remaining carbon goes into carbon compounds
such as C$_2$, CH, CN, and C$_3$. The TiO so characteristic of M stars
is replaced by these carbon compounds. So-called S stars (C~$\sim$~O)
are intermediate between C stars and M giants. Zirconium has a stronger
affinity for oxygen than titanium, but is much less abundant, so in
normal M stars ZrO features are weak or absent.  With increasing Zr
abundance due to the $s$-process any oxygen remaining from CO formation
goes into ZrO and so in S stars ZrO predominates and TiO is weakened.

In the optical, the continuum of most carbon stars is largely obscured
by absorption features from carbon compounds.  Consequently, it is very
difficult to use the standard atomic lines to sort spectra into types
that can be calibrated in terms of effective temperature, luminosity,
and composition (i.e. a three-dimensional MK system). Nevertheless,
improved optical spectra have led to a revised MK classification scheme
for C~stars due to \citet{1993PASP..105..905K} and
\citet{1996ApJS..105..419B}.  In this scheme the notation C-R$x$,
C-J$x$, C-N$x$, C-L$x$, and C-H$x$ corresponds to different spectral
types, where increasing digit $x$ represents decreasing effective
temperature. Although the spectral types probably represent different
stellar populations, the types are defined entirely by features in the
observed spectra \citep[see Table~2,][]{1996ApJS..105..419B}.  Due to
differences in mass, original composition, and environment, not all
carbon stars are enriched in the same way.  In terms of evolutionary
status, the spectral types C-R, C-J, C-N, C-L, and C-H, are thought to
characterize red giants, giants, TPAGB stars, PAGB stars, and binary
stars undergoing mass transfer, respectively.  Further symbols can be
added to the notation to indicate luminosity class and composition.
\citep[For details of the notation see][]{1996ApJS..105..419B}.

The spectral classification scheme as developed by
\citet{1976aasc.book.....K} and revised by \citet{1979ApJ...234..538A}
for S stars is similar to that of C stars. The notation S$x$ indicates S
star effective temperature, where increasing $x$ digit represents
decreasing effective temperature. Additional symbols can be added to
indicate composition.  However, the effective temperature sequence of S
stars relative to C stars is uncertain.

Figure \ref{fig:MSC} shows a sequence of M, S, and C-N giants, all of
approximately the same effective temperature, illustrating the effects
of increasing carbon enrichment and presumed AGB evolution.  The M0~IIIb
star shows features typical of late giant stars - strong CO absorption
features in the $H$ and $K$ bands, the Ca II triplet and a TiO band head
at about 0.85~\micron, a CN band head at about 1.1~\micron, and the SiO
absorption series at about 4~\micron. The S star (S4.5~Zr~2~Ti~4) is
similar except for a strong ZrO band head at about 0.93~\micron. The two
carbon stars (both C-N~4.5) display the effects of increasing carbon
enrichment (C2~4.5 and C2~5.5 respectively). In addition to the first
overtone CO band at about 2.29~\micron, which is present in the M giant
and S star, strong CN band heads are observed at 0.9~$\mu$m, 1.1~$\mu$m,
and 1.4~\micron, together with C$_2$ absorption at about 1.2~\micron~and
1.75~\micron, and C$_2$H$_2$ and HCN features at about 3.1~\micron.  The
very cool carbon star R~Lep (HD~31996, C7,6e (N4)) shows additional
absorption features due to HCN and C$_2$H$_2$ at 1.65~\micron~and
2.5~\micron, the HCN $\upsilon_1 + \upsilon_2$ bands at 3.56~\micron,
and the broad blend of the CS first overtone and HCN $\upsilon_1 -
\upsilon_2$ bands at $\sim$3.9~\micron. The HCN, C$_2$H$_2$, and CS
features are identified by \cite{1980ApJ...235..104G} and
\cite{1998A&A...340..222A} (but are not given in
Table~\ref{tab:MolecularBands}).

\section{Example Applications of the Library}

Potential applications of the IRTF Spectral Library can take advantage
of the 0.8-5~\micron~wavelength range at $R$$\sim$2000, preserved
spectral continuum shape, and absolute flux calibration. For example,
the Library has been used to model the atmospheres of cool dwarfs
\citep{2008ApJ...678.1372C}, confirm the presence of a gapped primordial
disk around LkCa15 \citep{2008ApJ...682L.125E}, and investigate the
stellar populations and activity in the nuclei of Seyfert galaxies
\citep{2009ApJ...694.1379R}.  Technical applications include using the
spectra to design filters and to calibrate different photometric
systems.  This is made possible because the system response and telluric
effects are carefully removed from our spectra.  Of the many potential
applications, in this section we discuss just two: the measurement of
equivalent widths for spectral typing, and synthetic photometry.

\subsection{\label{EW}Equivalent Widths}

As an example of the quantitative analysis that can be carried
out with the spectra in our library, we have calculated the equivalent
widths (EWs) of several prominent features seen in the data
(\ion{Ca}{2}, \ion{Na}{1}, \ion{Al}{1}, and \ion{Mg}{1}) following the
technique described in \cite{2005ApJ...623.1115C}. The wavelength
ranges used to define the continuum and the features are given in
Table \ref{tab:EWs}.  As Figure \ref{fig:EWs} clearly demonstrates,
the \ion{Ca}{2} EW provides a fairly good discriminator of luminosity
class between spectral types F and early M; the observed ranges of
\ion{Ca}{2} EW values are seen to be remarkably narrow, particularly for
the dwarfs, with little overlap among the luminosity classes. Similarly,
the \ion{Na}{1} 2.20 $\mu$m feature increases monotonically with
spectral type (temperature) between early F and mid M and therefore
provides an approximate means of estimating a stellar spectral type,
although the uncertainty in the classification can be fairly large
($\pm$ few spectral subtypes). The remarkably large, sudden, and
monotonic increase in the \ion{Na}{1} 1.14 $\mu$m doublet EW beginning
at early M implies that this line can be used as a clear indicator of
the very latest spectral subtypes. There are many other features in NIR
spectra, in addition to what we have presented here, that can be used to
determine spectral classes
\citep[e.g.][]{1986ApJS...62..501K,1998ApJ...508..397M,
  2000ApJ...535..325W,2000AJ....120.2089F,2004ApJS..151..387I,2007A&A...468..205L,
2007ApJ...671..781D}.

\subsection{\label{CCD}Synthetic Colors}

Because our spectra are flux calibrated, and the spectra slopes are
reliable, the IRTF Spectral Library can also be used to compute
synthetic magnitudes and colors as well as transformations between
various photometric systems. Table \ref{tab:MKOColors} gives the
synthetic $Z-Y$, $Y-J$, $J-H$, $H-K$, and $K-L'$ colors of the cool
stars in the library derived using Equation \ref{coloreqn}.  Synthesized
$J-H$ versus $H-K$ color-color diagrams for our sample of cool stars are
given in Figure \ref{fig:JHHK} (0.0$<J-H<$1.2) and Figure
\ref{fig:JHHKall} ($-0.5<J-H<2.5$).  A synthesized $Y-J$ versus $J-H$
diagram of the same stars is given in Figure \ref{fig:YJJH} (all stars
except two very red OH/IR stars).  The sample of 13 L and two T dwarfs
from \cite{2005ApJ...623.1115C}, and the eight T dwarf spectral
standards from \cite{2006ApJ...637.1067B} are also included in these
figures and table. In addition, Figures \ref{fig:JHHK} and
\ref{fig:YJJH} also include color-color diagrams with the corrections
for reddening discussed in \S\ref{Red}, incorporated.

The trends in the colors of the stars as a function of spectral type in
the $J-H$ versus $H-K$ diagram are very similar to those presented by
\citet[][Figure 5]{1988PASP..100.1134B} and \citet[][Figure
2.22]{2005nlds.book.....R}, with the slight difference that our
photometry is in the NIR-MKO system. Since our photometric errors are
small (at most a few percent, see \S\ref{Red}), the
scatter in these plots is due to real differences in stellar colors
produced by variations in metallicity, reddening, etc.  The most
noticeable feature of the $JHK$ color-color diagram is the bifurcation
between M dwarfs and stars of higher luminosity classes.  Stars of all
luminosity classes initially show a steep rise in $J-H$ with later
spectral type but starting at a spectral type of $\sim$M0, the $J-H$
colors of the dwarfs become bluer while that of the M giants and
supergiants continues to become redder
\citep{1970ApJ...162..217L,1975MNRAS.171P..19G,1978ApJ...220...75F}.  The $J$ and $H$ bands
probe different layers (and thus different temperatures) in an
atmosphere because H$^-$, the dominant continuum opacity source at these
wavelengths, has a minimum at $\sim$1.6 $\mu$m. The turnover in the
$J-H$ colors of the M dwarfs is therefore due to a change in the
adiabatic temperature gradient as hydrogen is increasingly converted
into H$_2$ in the high pressure (relative to giants) atmospheres of M
dwarfs \citep{1976A&A....48..443M}.  The $J-H$ colors of the M dwarfs
continue to become bluer with the onset of H$_2$O absorption which
suppresses the $H$- and $K$-band fluxes but eventually flatten out
before becoming redder again for late-type M dwarfs and L dwarfs as the
peak of the Planck function shifts further into the NIR.

We note that the locus of late-M giants (Lb, SRb, and M variables, see
Table \ref{tab:Sample}) in the $JHK$ color-color diagram appears to turn
back down towards the location of late-type M and L dwarfs (see Figure
\ref{fig:JHHK}). Figure~\ref{fig:MIIIwater} shows the spectral behavior
of a selection of late-type M giants along the locus of decreasing $J-H$
and increasing $H-K$ color.  The most distinctive features in these
spectra are the broad H$_2$O absorption features centered at
1.4~\micron~and 1.9~\micron. H$_2$O absorption is observed no earlier
than M6~III (see also Figure~\ref{fig:FGKM_III}) and then increases in
strength with later spectral types, in agreement with the trends first
observed by \cite{1971PhDT.........1F} and \cite{1974HiA.....3..307H}.
The turnover in $J-H$ is therefore a result of increasing H$_2$O line
absorption which suppresses the $H$- and $K$-band fluxes more than the
$J$-band flux.  Models of Mira variables \citep{1989A&A...213..209B,
  1996A&A...307..481B} can reproduce the observed turnover in $J-H$.  In
these models pulsation produces extended atmospheres in which water can
form in dense cool ($\leq~10^3$~K) layers formed behind periodically
outward-running shocks.

Other intrinsically red stars include late-type supergiants, carbon and
S stars. The red colors are due to the cool continuum temperatures,
together with molecular line blanketing from CO and CN in supergiants,
and from CN and C$_{2}$ in carbon stars.

One of the observable consequences of mass loss in TPAGB stars is an
approximately linear locus in the $JHK$ color-color diagram reflecting
the effects of differential extinction and dust temperature in models of 
circumstellar shells \citep[e.g.,][]{2006AJ....132..489L}. Less dusty
Mira variables are located at the blue end of this locus at
$J-H$~$\sim$~0.8, $H-K$~$\sim$~0.6, while the more deeply embedded OH/IR
stars (optically obscured stars with 1612 MHz OH line emission) are
located at the red end. The two OH/IR stars observed in our sample are
located at $J-H$~$\sim$~2, $H-K$~$\sim$~2 (see Figure
\ref{fig:JHHKall}).

Library spectra can be used to synthesize other colors and experiment
with other photometric systems. As an example, the $Y-J$ versus $J-H$
color-color diagram shown in Figure \ref{fig:YJJH} illustrates the
advantage of using the $Y$ band in combination with the standard $JHK$
bands when trying to identify T dwarfs \citep[see
also][]{2002PASP..114..708H,2006MNRAS.367..454H} compared to using $JHK$
colors alone (see Figure \ref{fig:JHHK}). The $J-H$ color can be used as
an indicator of T dwarf spectral types and is accurate to within about
one sub-type. For example, the UKIRT Infrared Deep Sky Survey
(UKIDSS) is using $YJH$ photometry in a wide area survey for T dwarfs
and cooler objects \citep{2008MNRAS.390..304P}.  Note also the
bifurcation between M dwarfs and stars of higher luminosity which is
also seen in the $JHK$ plot. Starting at M0~V, dwarfs initially become
bluer in $J-H$ with constant $Y-J$ constant, due to the change in
adiabatic temperature gradient (Figure~\ref{fig:YJJH}). This trend ends
at about M4~V at which point dwarfs become redder in $Y-J$ and $J-H$
with decreasing effective temperature. The effect of increasing H$_2$O
absorption in these late-type dwarfs is to decrease the amount of $J-H$
reddening compared to the locus of higher luminosity stars. Late-type M
giants have similar colors to late-type dwarfs since both have H$_2$O
absorption in the NIR.

\subsection{\label{Unusual}Notes on Individual Objects}

Ten of the 212 stars in the library have spectra and/or $J-H$ versus
$H-K$ colors that are different than those expected based on their
spectral type. Five of these stars are supergiants displaying emission lines,
and with redder than
normal continua that cannot be explained by standard interstellar
reddening (reddening not in the direction of the extinction vector).
As explained
below, most of these stars are probably PAGB stars.  Three of the
unusual stars are emission-line Mira variables (TPAGB stars).  Of the
two remaining unusual stars one is an M subdwarf misclassified as an M
dwarf, and one is an F dwarf with weak emission in some metal lines.

Unusual objects are circled in the $J-H$ versus $H-K$ diagram (see
Figure~\ref{fig:JHHK}):


{\bf HD~26015} is classified as F3~V by \cite{2001AJ....121.2148G}. The
spectrum is normal up to 2.22~\micron~but at longer wavelengths some
metal lines go into weak emission (\ion{Ca}{1} doublet at 2.263 and
2.267~\micron, \ion{Mg}{1} at 2.280~\micron, \ion{Na}{1} at
2.339~\micron, and \ion{Mg}{1} at 3.867~\micron). The continuum is also
slightly bluer than normal for a spectral type of F3V. The star is
classified as variable (of unspecified type) in the GCVS and is slightly
metal rich ([Fe/H]=0.2, average from SIMBAD).

{\bf HD~179821} is classified as G4~O-Ia by
\cite{1989ApJS...71..245K}. Strong emission in the \ion{Na}{1} doublet
at 2.205 and 2.209~\micron~is seen together with very strong Pfund
series absorption longward of the series limit at about
2.33~\micron. The expected first CO overtone bands in the $K$~band are
absent. From its location in the $JHK$ color-color diagram the star has
a significant NIR excess. The star is classified as a semiregular
variable giant or supergiant (type SRd) in the GCVS. Optical HST images
of HD~179821 reveal a bright star embedded in faint extended nebulosity
\citep{2000ApJ...528..861U}. \citet{2008BaltA..17...87K} reviews
observations of HD~179821 some of which are consistent with an
intermediate-mass PAGB star, while others point to a high-mass
post-red-supergiant star.

{\bf HD~6474} is classified as G4~Ia by \cite{1989ApJS...71..245K}.  The
spectral continuum is redder than normal, and the $JHK$ colors indicate
a NIR excess. Spectral features in the $K$~band appear subdued probably
due to veiling and \ion{Si}{1} at 3.745~$\mu$m~is in emission. The star
is classified as a semiregular variable (type SRd) in the GCVS and as a
UU Her-type variable by
\citet{1993A&A...280..177Z}. \cite{2007A&A...469..799S} classify
UU~Her-type variables with NIR excess due to circumstellar dust as
probable PAGB stars.

{\bf HD~333385 (BD$+$~29$^{\circ}$~3865)} is classified as G7~Ia by
\cite{1989ApJS...71..245K}.  The star is classified as a slow irregular
variable (type L) in the GCVS.  The spectrum is clearly unusual showing
a number of metal lines in emission, particularly in the $K$-band, where
the \ion{Na}{1} doublet at 2.205 and 2.209~\micron~is strong. The first
CO overtone bands in the $K$ band appear sharper than those in other
late-type G supergiants.  \ion{Si}{1} at 3.745~$\mu$m~is in
emission. The $JHK$ colors indicate a large NIR excess.  Using
high-dispersion optical echelle spectra \cite{2000AstL...26..398K}
conclude that HD~333385 is probably a PAGB star.

{\bf HD~165782} is classified as K0~Ia by \cite{1989ApJS...71..245K}.
The spectrum shows weak absorption features in the $K$~band probably due
to veiling, and the \ion{Na}{1} doublet at 2.205 and 2.209~\micron~is in
emission.  The first CO overtone bands in the $K$ band appear sharp and
\ion{Si}{1} at 3.745~$\mu$m~is in emission.  The $JHK$ colors indicate a
NIR excess. The star is classified as a semiregular variable (type SRd)
in the GCVS.  An OH maser is reported by \cite{1998A&AS..127..185N} and
\cite{1993A&A...267..515O} classify HD~165782 as a PAGB star.

{\bf HD~212466} is classified as K2~O-Ia by \cite{1989ApJS...71..245K}.
The spectrum shows SiO in emission at 4.00, 4.04, and 4.08~\micron~and
the first CO overtone bands in the $K$ band appear sharp.  The $JHK$
colors indicate a large NIR excess. The star is classified as a
semiregular variable (type SRd) in the GCVS.  A Si emission feature at
10~$\mu$m~is cited as evidence of mass loss by
\cite{1998MNRAS.301.1083S}.

{\bf Gl~299} is classified as M4~V by
\cite{1994AJ....108.1437H}. However, its location below the locus of M
dwarfs in the $J-H$ versus $H-K$ plot is more consistent with an M
subdwarf \citep[][Figure A3]{1988PASP..100.1134B}. This is confirmed by
its spectrum which shows weak CO overtone absorption at 2.29~\micron, a
weak \ion{Na}{1} doublet at 2.205 and 2.209~\micron, and weak \ion{K}{1}
at 1.516~\micron, compared to a normal M4~V
\citep[e.g.][]{2006AJ....131.1797C}.

{\bf HD~14386 (Mira)} is classified as M5e-M9e~III and as a Mira
variable (type M) in the GCVS.  This archetypal variable star was
observed on three occasions. On 2003 January 14 the emission line due to
Pa$\beta$ (1.28~\micron, EW $-$0.4~\AA) was detected, on 2003 September
20 emission lines due to Pa$\gamma$ (1.094~\micron, EW $-$0.9~\AA) and
Pa$\beta$ (1.282~\micron, EW~$-$0.7~\AA) were detected, and on 2003
November 6 the emission line due to \ion{Na}{1} (1.269~\micron,
EW~$-$1.0~\AA) was detected.  (Non-detections of these lines are roughly
EW~$>-$0.2~\AA).  Mira-type variables can emit in a variety of metal and
hydrogen lines, probably originating in atmospheric shock waves
resulting from pulsation \citep[e.g.][]{2001A&A...369.1027R}. The lines
are known to come and go depending upon cycle and phase \citep[e.g.][in
the NIR]{2000AAS..146..217L}.

{\bf BRI ~B2339$-$0447} is classified as M7-8~III by
\cite{1997AJ....113.1421K} and as a Mira variable (type M) in the GCVS.
The spectrum shows Pa$\gamma$ (1.0944~\micron, E.W.~1.2~\AA), Pa$\beta$
(1.282~\micron, EW~0.8~\AA), and Br$\gamma$ (2.166~\micron,
E.W.~0.5~\AA) in emission. Pa$\delta$ (1.005~\micron) in emission was
also detected but blended with a TiO band head.

{\bf IRAS~1403$-$1042} is classified as M8-9~III by
\cite{1997AJ....113.1421K} and as a Mira variable (type M) in the GCVS.
The spectrum shows Pa$\gamma$ (1.094~\micron, E.W.~1.9~\AA) and
Pa$\beta$ (1.282~\micron, E.W.~4.8~\AA) in emission (see Figure
\ref{fig:JHHK}). Pa$\delta$ (1.005~\micron) in emission was also
detected but blended with a TiO band head.

\section{Summary}

We have constructed a medium resolution (R$\sim$2000) near-infrared
(0.8-5 \micron) spectral library of 210 cool stars.  The stars all have
well established MK classifications and have near-solar
metallicities. The sample covers F, G, K, and M stars with luminosity
classes between I and V, and includes some AGB, carbon, and S
stars. Sample selection, data reduction, and data calibration are
carefully described. The continuum shape of the spectra are measured to
an accuracy of a few percent, and the spectra are absolutely flux calibrated
using 2MASS photometry. Synthesized color-color diagrams are constructed
from the spectra and their use demonstrated.  Spectral features are
described and detailed lists of atomic and molecular features are
tabulated.  Several unusual stars in the sample are identified and
described.  The library is available in digital form from the IRTF
website.

\acknowledgments

Observations for The IRTF Spectral Library required a significant
investment in telescope time and the authors would like to thank IRTF
Director, Alan Tokunaga, and the IRTF TAC for their support. Observing
was ably assisted telescope operators Bill Golisch, Dave Griep, and Paul
Sears. We would also like to thank the IRTF day crew and engineering
staff for excellent maintenance of the telescope and instrumentation.
Natasha F\"{o}rster-Schreiber helped us with sample selection during the
early stages of the project.  J.T.R. would like to thank Katelyn Allers
for providing observations of five late-type M giants.
M.C.C. acknowledges financial support from the IRTF and by the National
Aeronautics and Space Administration through the {\em Spitzer Space
  Telescope} Fellowship Program through a contract issued by the Jet
Propulsion Laboratory, California Institute of Technology, under a
contract with NASA.  W.D.V. and M.C.C acknowledge financial support from
IRTF during several trips to Honolulu.  This publication makes use of
data from the Two Micron All Sky Survey, which is a joint project of the
University of Massachusetts and the Infrared Processing and Analysis
Center and is funded by NASA and the National Science Foundation.  This
research has made use of the SIMBAD database, operated at CDS,
Strasbourg, France; NASA's Astrophysics Data System Bibliographic
Services; and the SpeX Prism Spectral Libraries, maintained by Adam
Burgasser \footnote{\url{http://www.browndwarfs.org/spexprism}}.  We
thank the anonymous referee for a thorough reading of the manuscript and
insightful suggestions for improving it.

\noindent
{\em Facilities:} IRTF (SpeX)

\bibliographystyle{apj}
\bibliography{IRTFSL}

\newcommand\snorm{\mbox{1.2 }}
\newcommand\znorm{\mbox{0.9 }}
\newcommand\ynorm{\mbox{1.0 }}
\newcommand\jnorm{\mbox{1.2 }}
\newcommand\hnorm{\mbox{1.6 }}
\newcommand\knorm{\mbox{2.2 }}
\newcommand\lnorm{\mbox{3.6 }}

\newcommand\srange{\mbox{0.81$-$5.0 }}
\newcommand\zrange{\mbox{0.815$-$0.95 }}
\newcommand\yrange{\mbox{0.95$-$1.10 }}
\newcommand\jrange{\mbox{1.15$-$1.34 }}
\newcommand\hrange{\mbox{1.48$-$1.78 }}
\newcommand\krange{\mbox{1.92$-$2.47 }}
\newcommand\lrange{\mbox{3.3$-$4.1 }}

\clearpage

\clearpage

\clearpage

\clearpage

\clearpage

\clearpage

\clearpage

\begin{sidewaysfigure}
\centerline{\includegraphics[width=6in,angle=90]{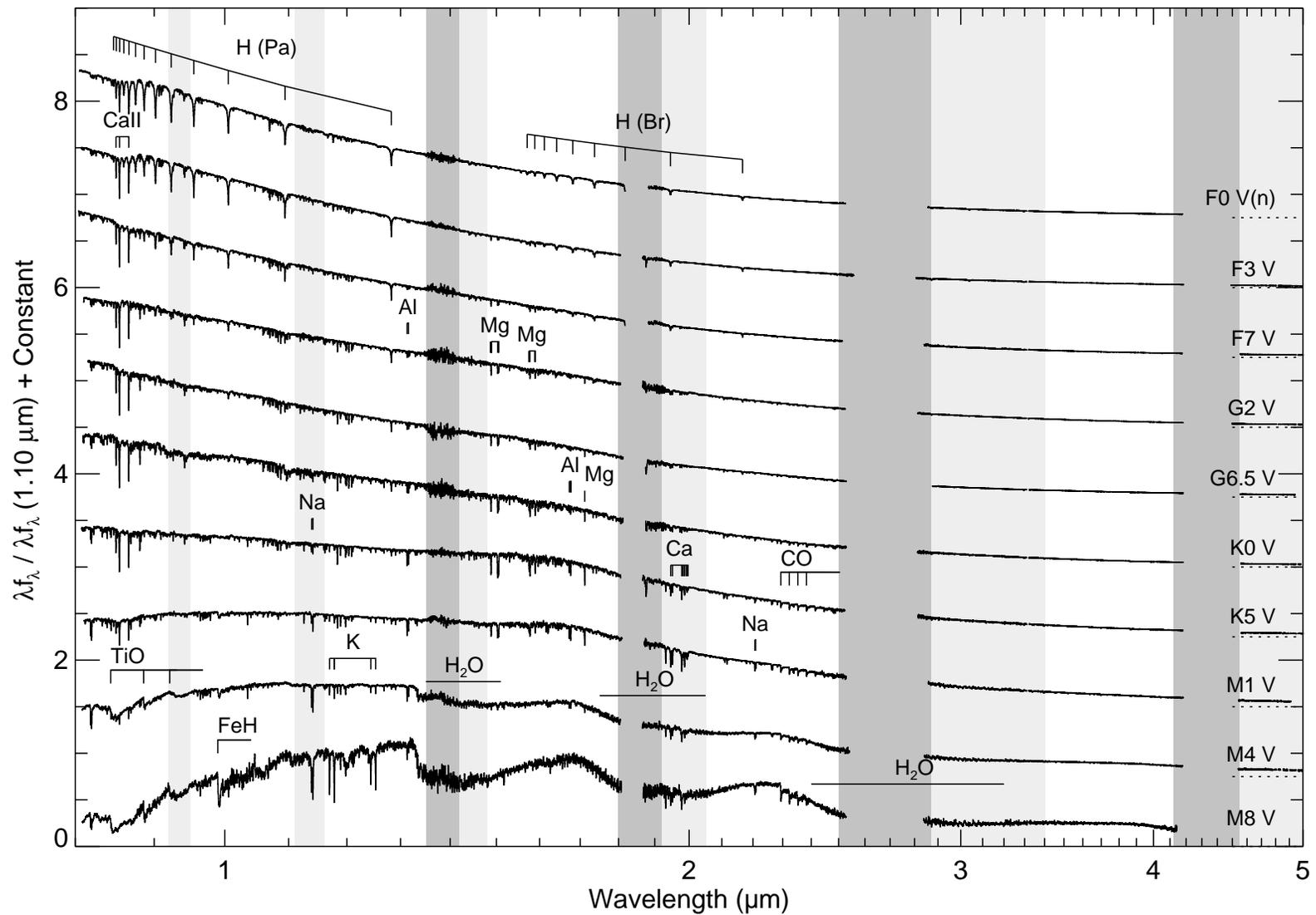}}
\caption{\label{fig:FGKM_V}Dwarf sequence from 0.8$-$5 $\mu$m.  The
  spectra are of HD~108159 (F0~V(n)), HD~26015 (F3~V), HD~126660 (F7~V),
  HD~76151 (G2~V), HD~115617 (G6.5~V), HD~145675 (K0~V), HD~36003
  (K5~V), HD~42581 (M1~V), Gl~213 (M4~V), and Gl~752B (M8~V).  The
  spectra have been normalized to unity at 1.10~$\mu$m and offset by
  constants (dotted lines).  Regions of strong (transmission $<$ 20\%)
  telluric absorption are shown in dark grey while regions of moderate
  (transmission $<$ 80\%) telluric absorption are shown in light grey.}
 \end{sidewaysfigure}

\clearpage

\begin{sidewaysfigure}
\centerline{\includegraphics[width=6in,angle=90]{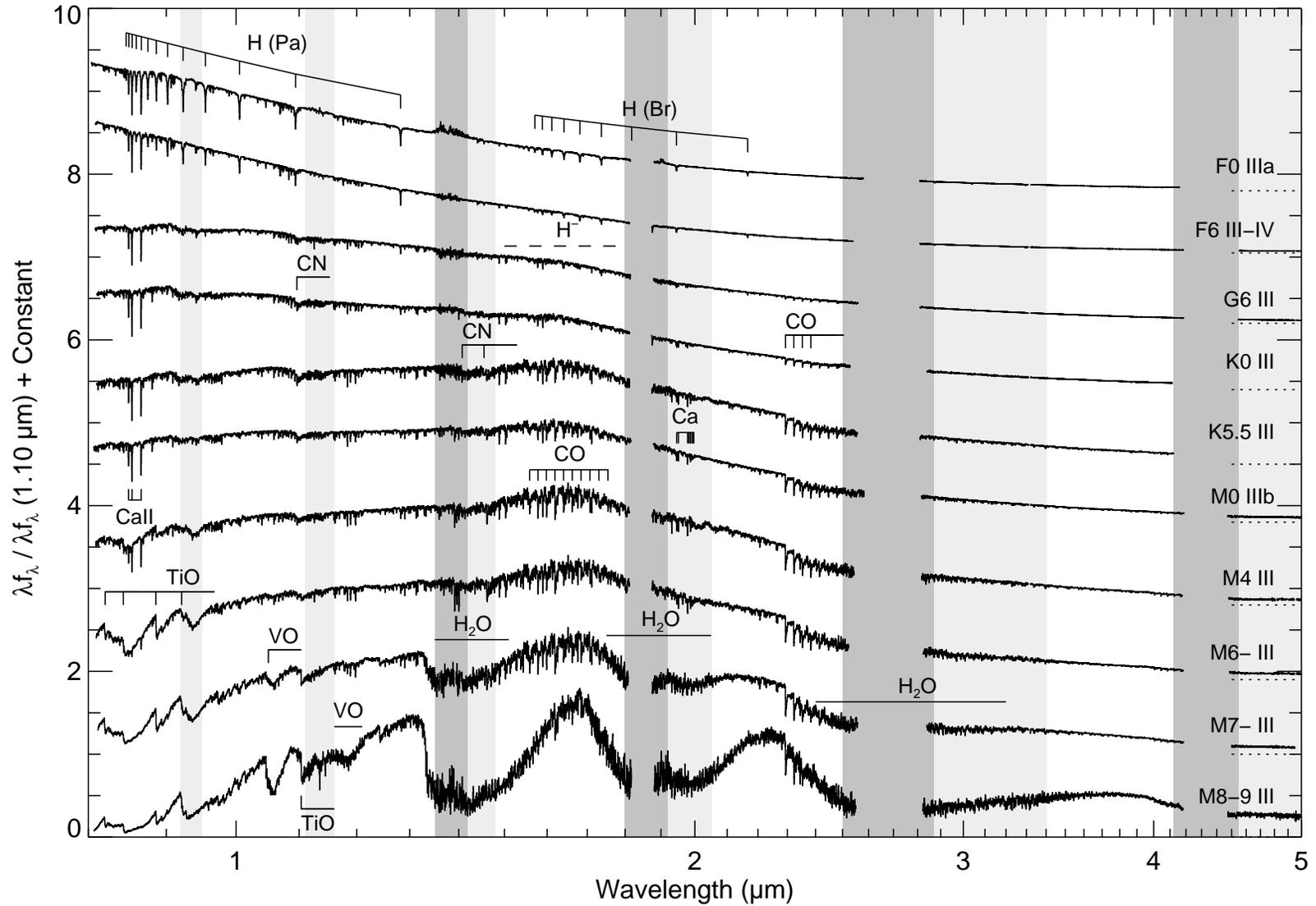}}
\caption{\label{fig:FGKM_III}Giant sequence from 0.8$-$5 $\mu$m.  The
  spectra are of HD~89025 (F0~IIIa), HD~160365 (F6~III-IV), HD~27277
  (G6~III), HD~100006 (K0~III), HD~120477 (K5.5~III), HD~213893
  (M0~IIIb), HD~4408 (M4~III), HD~18191 (M6-~III), HD~108849 (M7-~III),
  and IRAS~21284$-$0747 (M8-9~III).  The spectra have been normalized to
  unity at 1.10 $\mu$m and offset by constants (dotted lines).  Regions
  of strong (transmission $<$ 20\%) telluric absorption are shown in
  dark grey while regions of moderate (transmission $<$ 80\%) telluric
  absorption are shown in light grey.}
 \end{sidewaysfigure}

\clearpage

\begin{sidewaysfigure}
\centerline{\includegraphics[width=6in,angle=90]{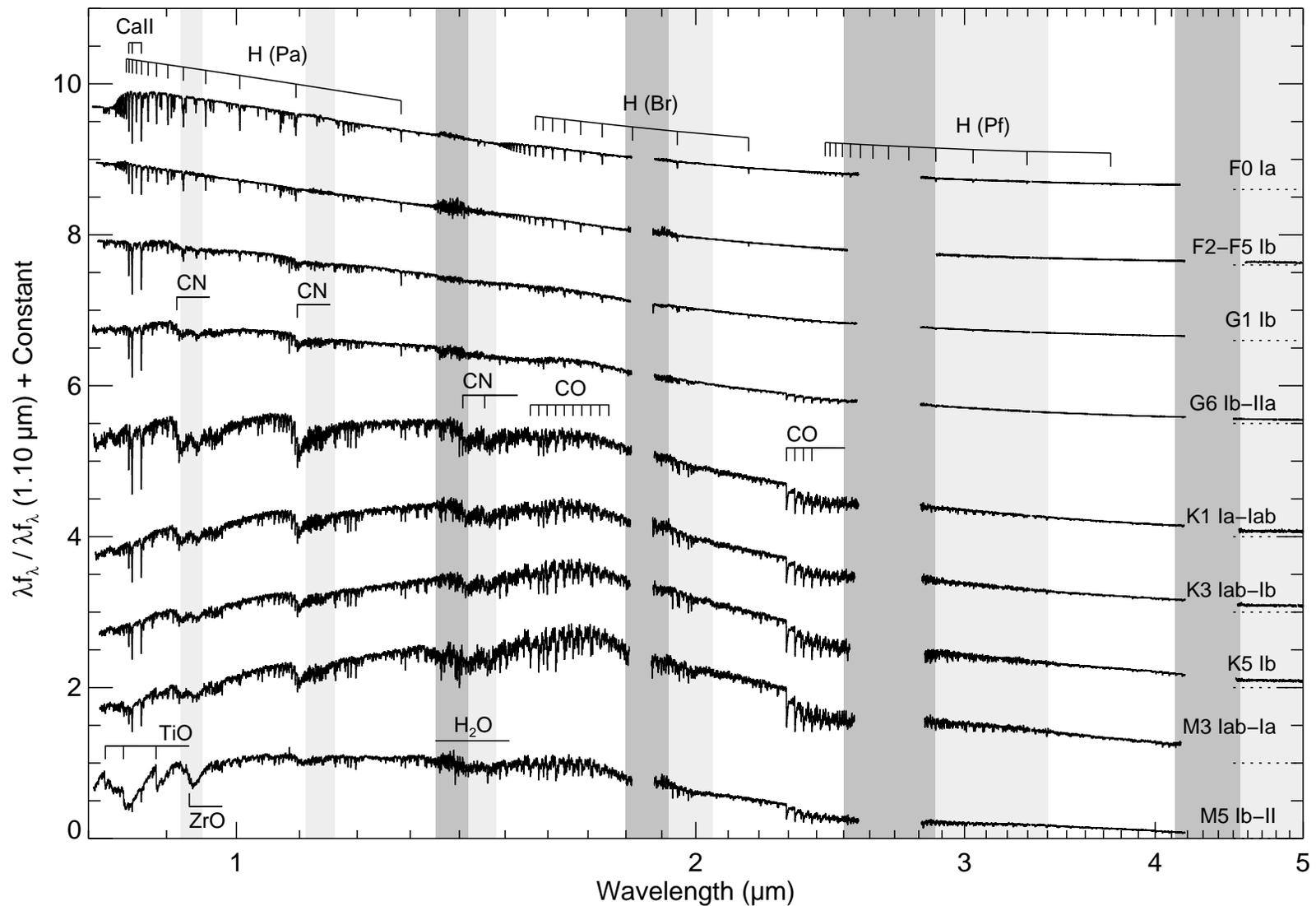}}
\caption{\label{fig:FGKM_I2II}Supergiant sequence from 0.8$-$5.0 $\mu$m.
  The spectra are of HD~7927 (F0~Ia), BD$+$38~2803 (F2-F5~Ib), HD~74395
  (G1~Ib), HD~202314 (G6~Ib-IIa~Ca1~Ba0.5), HD~63302 (K1~Ia-Iab),
  HD~187238 (K3~Iab-Ib), HD~216946 (K5~Ib Ca1 Ba0.5), CD$-$31~4916
  (M3~Iab-Ia), and HD~156014 (M5~Ib-II).  The spectra have been
  normalized to unity at 1.10 $\mu$m and offset by constants (dotted
  lines).  Regions of strong (transmission $<$ 20\%) telluric absorption
  are shown in dark grey while regions of moderate (transmission $<$
  80\%) telluric absorption are shown in light grey.}
 \end{sidewaysfigure}

\clearpage

\begin{sidewaysfigure}
\centerline{\includegraphics[width=6in,angle=90]{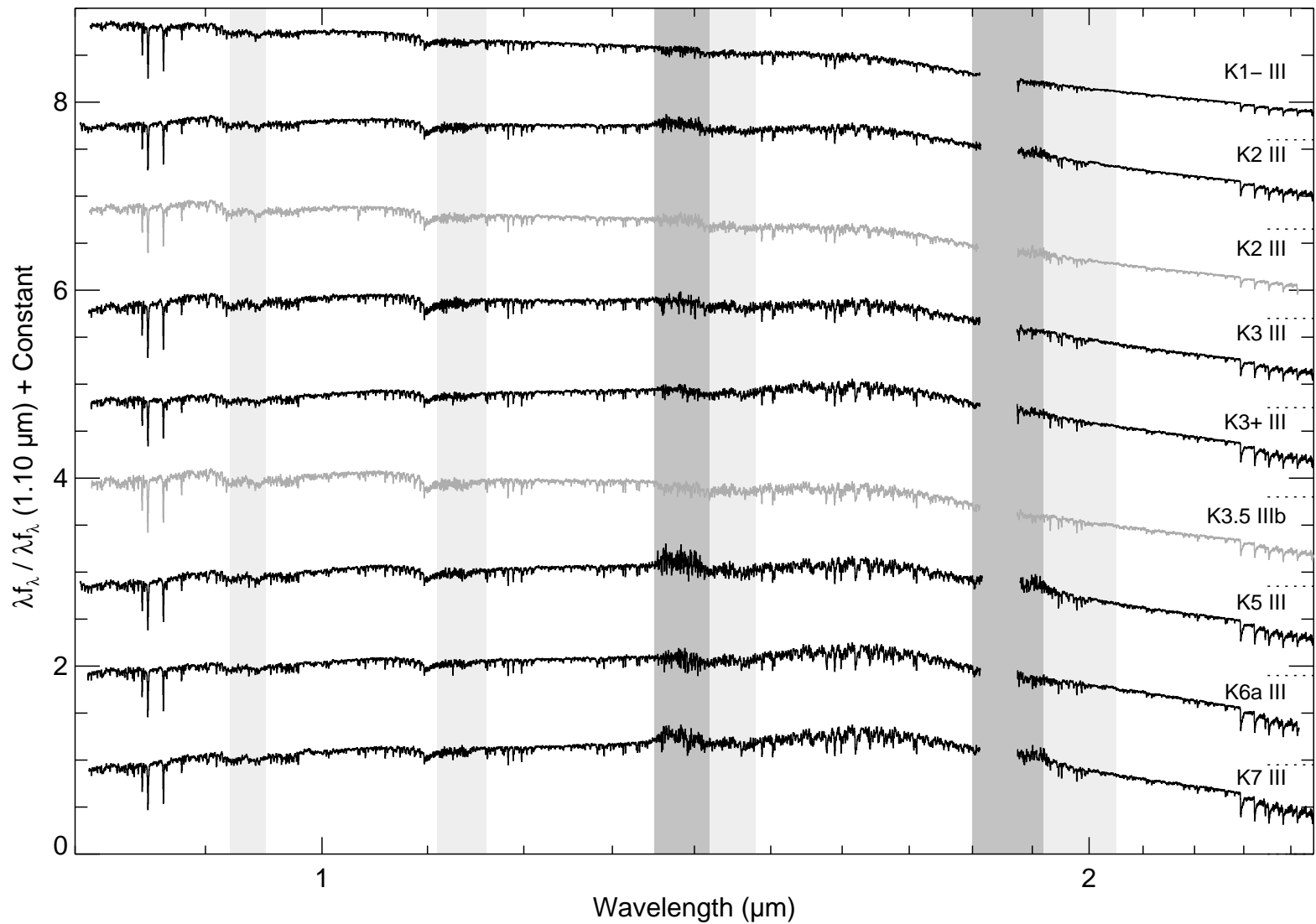}}
\caption{\label{fig:KIII}K giant sequence from 0.8$-$2.45 $\mu$m.  The
  stars are HD~36124 (K1-~III~Fe-0.5), HD~132935 (K2~III), HD~137759
  (K2~III), HD~221246 (K3~III), HD~99998 (K3+~III~Fe$-$0.5), HD~114960
  (K3.5~IIIb~CN0.5~CH0.5), HD~181596 (K5~III), HD~3346 (K6~IIIa), and
  HD~194193 (K7~III). Stars forming a smooth spectral sequence are
  plotted in black, while the two stars (HD~137759 and HD~114960) which
  appear slightly out of sequence (see \S\ref{Spectra}) are plotted in
  grey.}
\end{sidewaysfigure}

\clearpage

\begin{sidewaysfigure}
\centerline{\includegraphics[width=6in,angle=90]{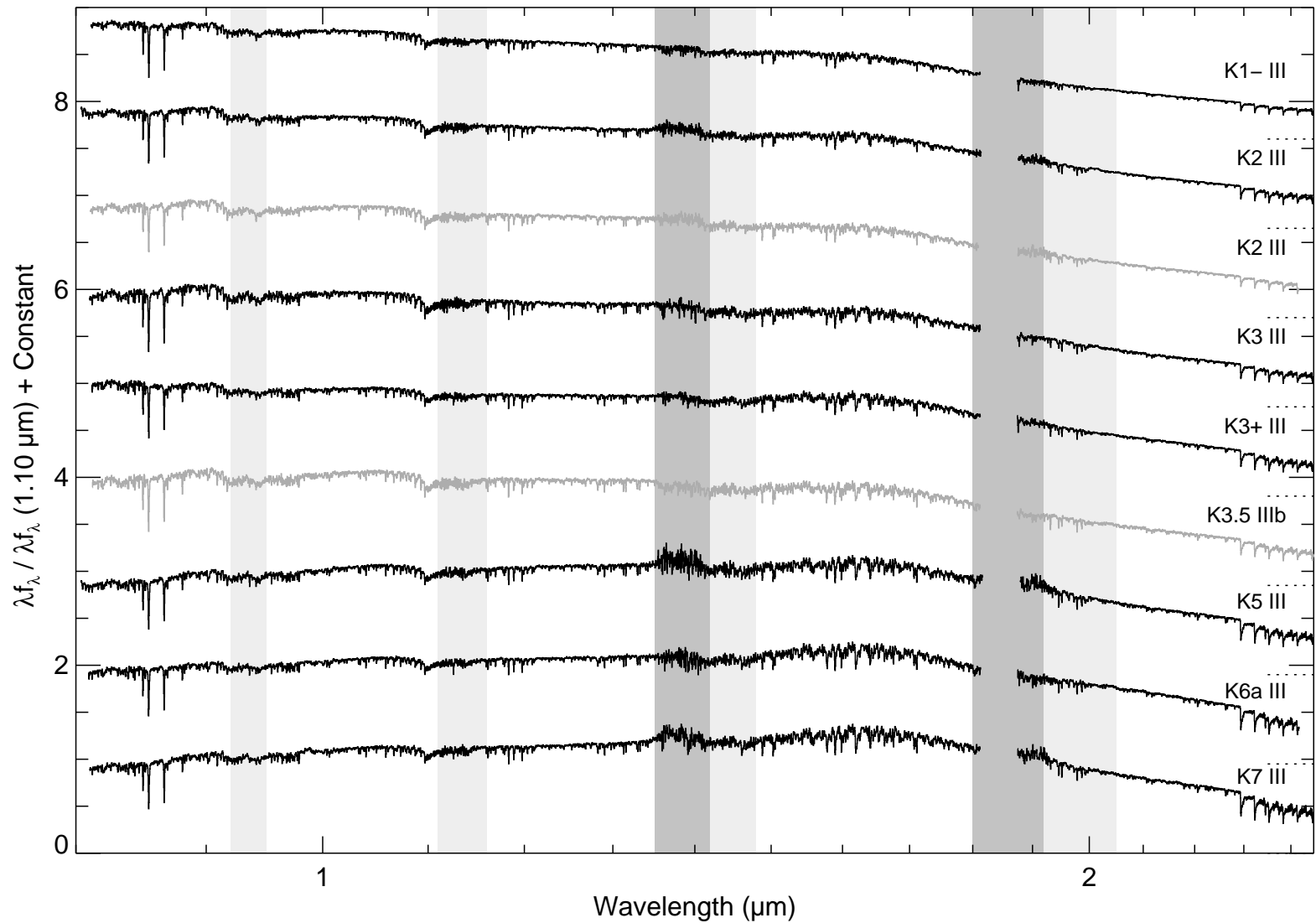}}
\caption{\label{fig:KIIIdered}Same as Figure~\ref{fig:KIII} except
  that the dereddened spectra are plotted. The spectral continuum shapes
  now behave as expected but the CO band depths (wavelengths
  $\geq$2.29$\micron$) of the two stars HD~137759 and HD~114960 (plotted
  in grey) still appear slightly out of sequence (see \S\ref{Spectra}).}
\end{sidewaysfigure}

\clearpage

\begin{sidewaysfigure}
\centerline{\includegraphics[width=6in,angle=90]{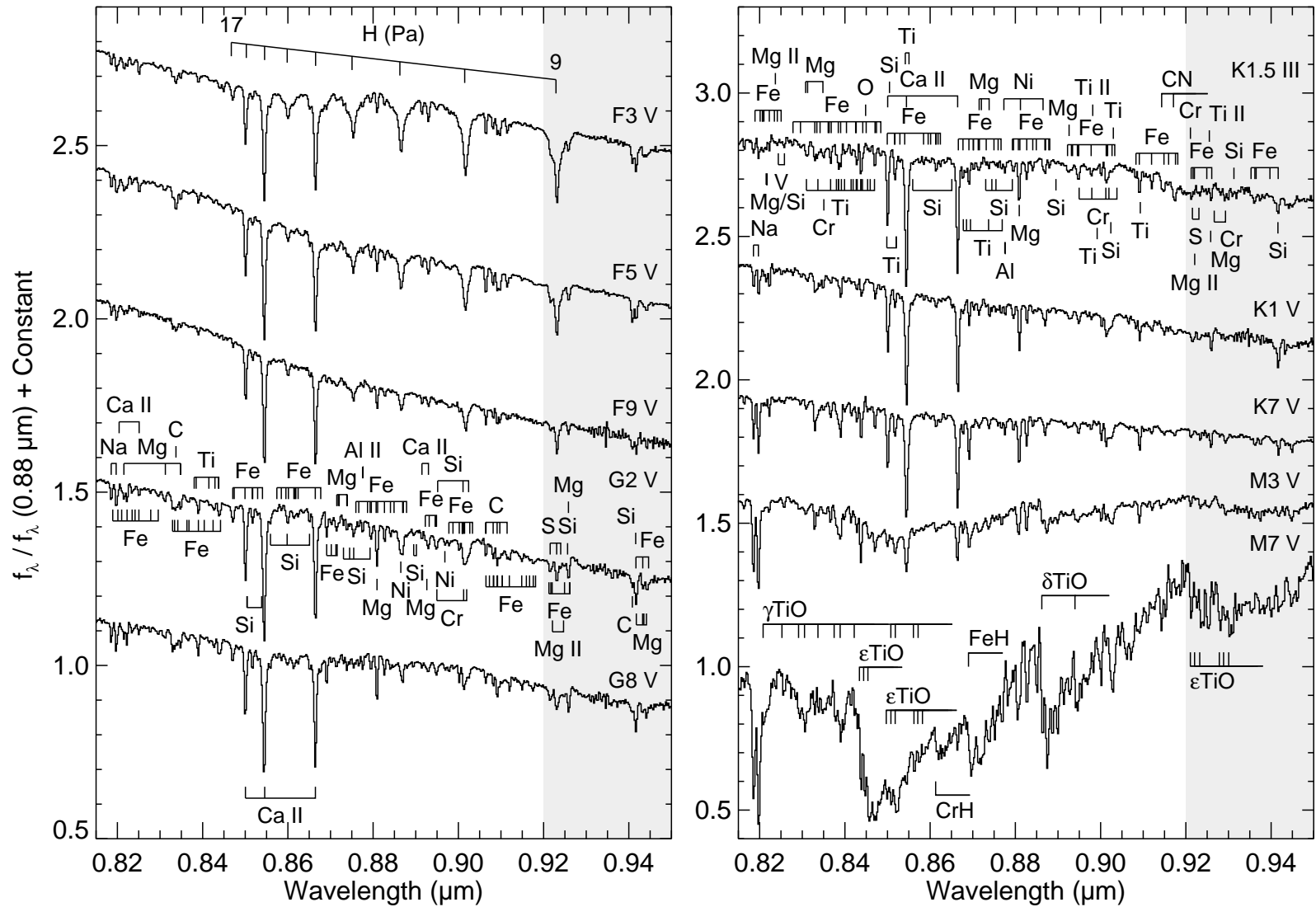}}
\caption{\label{fig:FGKM_VI} A sequence of F, G, K, and M dwarf stars
  plotted over the $I$ band (0.82$-$0.95~ $\mu$m).  The spectra are of
  HD~26015 (F3~V), HD~27524 (F5~V), HD~165908 (F9~V metal weak),
  HD~76151 (G2~V), HD~101501 (G8~V), HD~10476 (K1~V), HD~237903 (K7~V),
  Gl~388 (M3~V), and Gl~644C (vB~8) (M7~V).  The K1.5~III comparison
  star is Arcturus (HD~124897).  The spectra have been normalized to
  unity at 0.88~$\mu$m and offset by constants.}
\end{sidewaysfigure}

\clearpage

\begin{sidewaysfigure}
\centerline{\includegraphics[width=6in,angle=90]{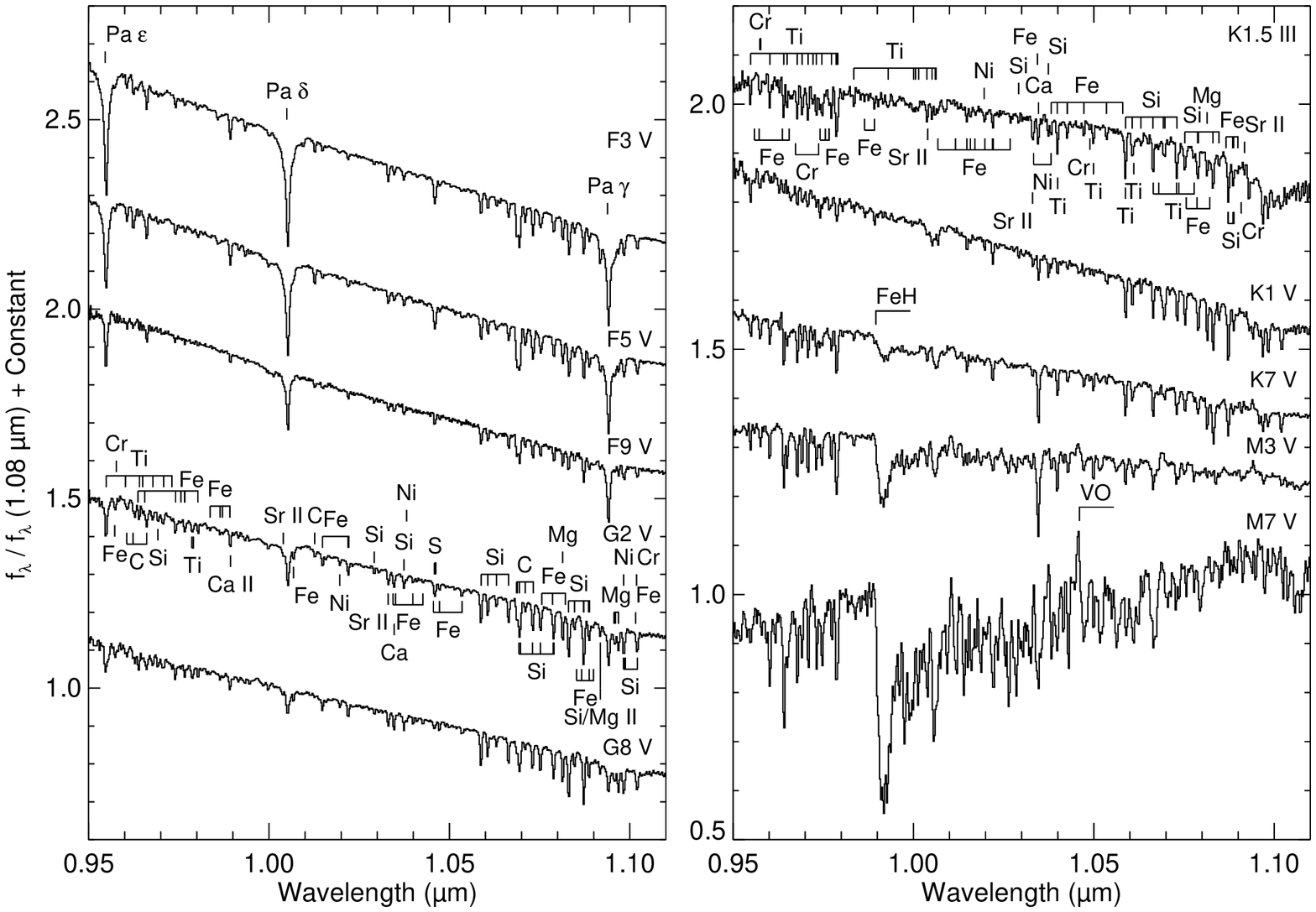}}
\caption{\label{fig:FGKM_VY} Same as Figure \ref{fig:FGKM_VI} except
  over the $Y$ band (0.95$-$1.10 $\mu$m).  The spectra have been
  normalized to unity at 1.08 $\mu$m and offset by constants.}
\end{sidewaysfigure}

\clearpage

\begin{sidewaysfigure}
\centerline{\includegraphics[width=6in,angle=90]{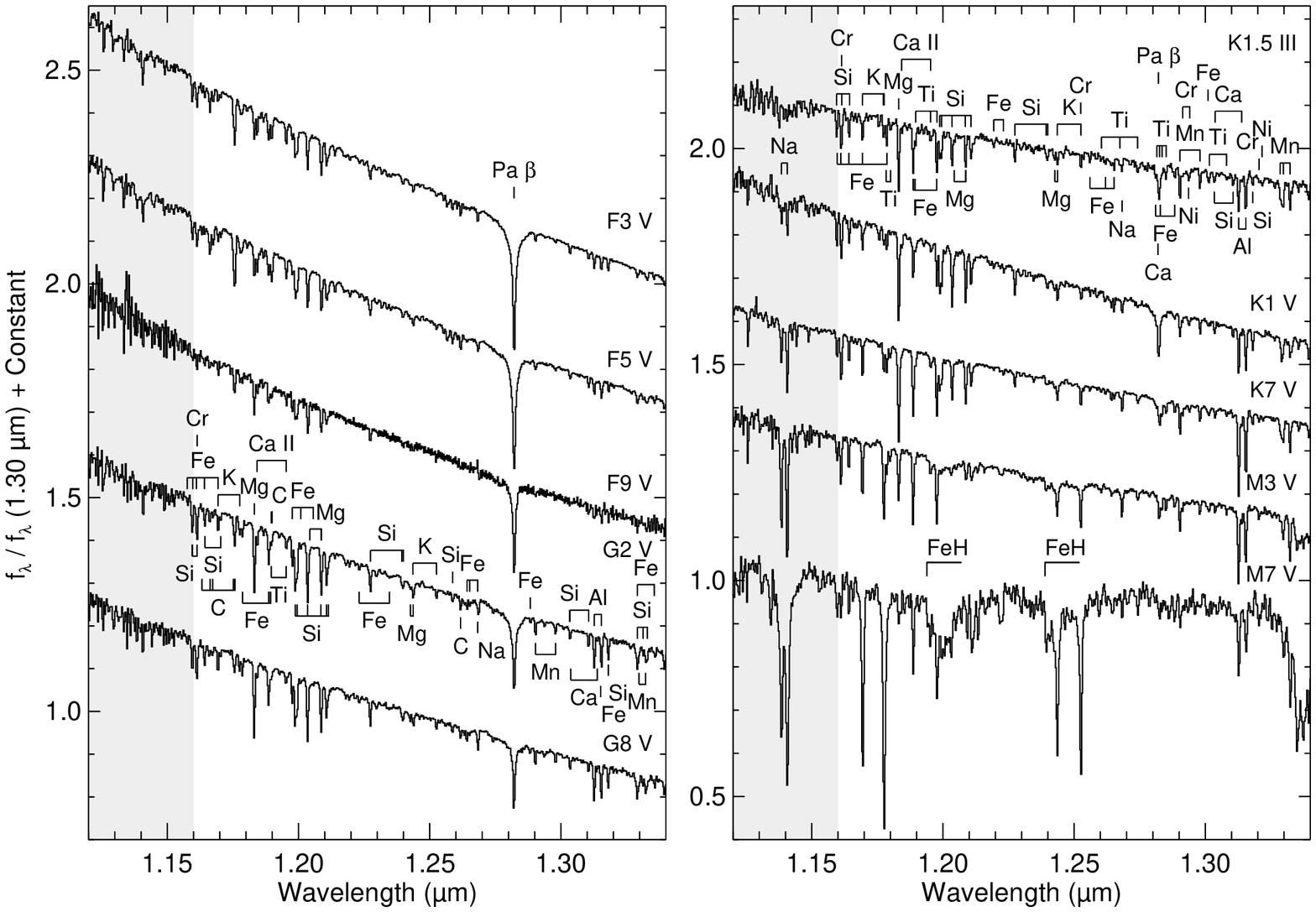}}
\caption{\label{fig:FGKM_VJ} Same as Figure \ref{fig:FGKM_VI} except
  over the $J$ band (1.12$-$1.34 $\mu$m).  The spectra have been
  normalized to unity at 1.30 $\mu$m and offset by constants.}
 \end{sidewaysfigure}

\clearpage

\begin{sidewaysfigure}
\centerline{\includegraphics[width=6in,angle=90]{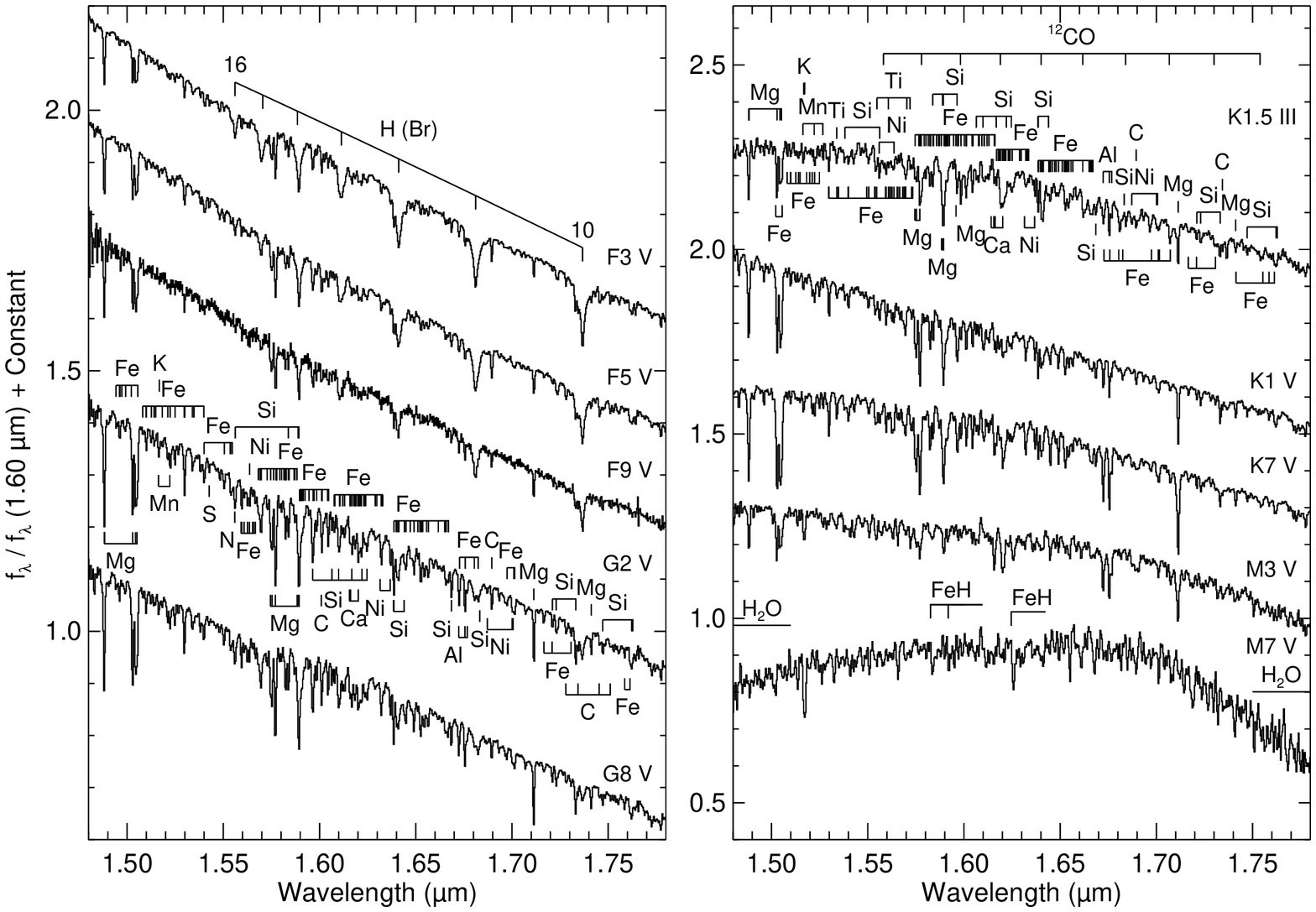}}
\caption{\label{fig:FGKM_VH} Same as Figure \ref{fig:FGKM_VI} except
  over the $H$ band (1.48$-$1.78 $\mu$m).  The spectra have been
  normalized to unity at 1.60 $\mu$m and offset by constants.}
 \end{sidewaysfigure}

\clearpage

\begin{sidewaysfigure}
\centerline{\includegraphics[width=6in,angle=90]{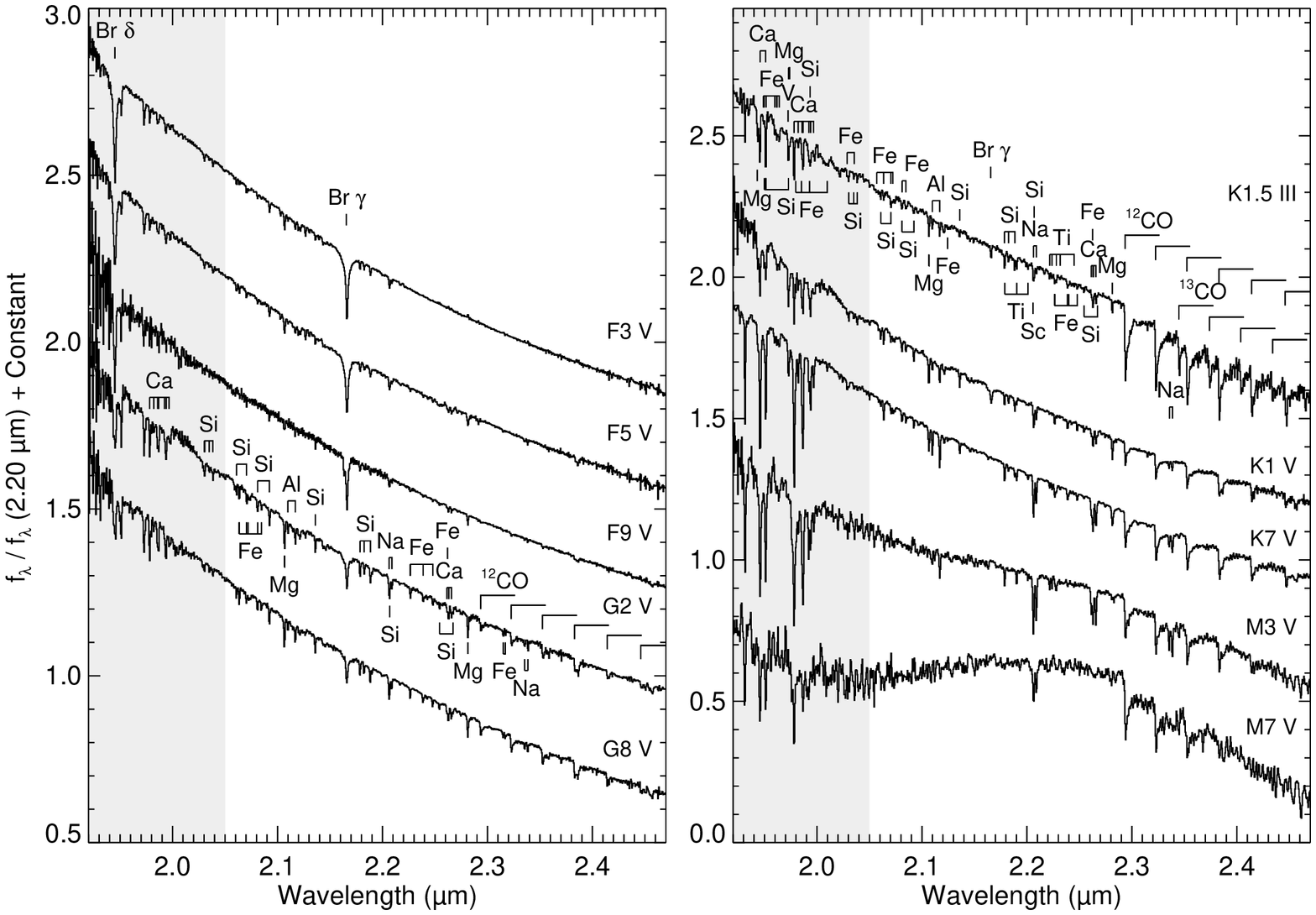}}
\caption{\label{fig:FGKM_VK} Same as Figure \ref{fig:FGKM_VI} except
  over the $K$ band (1.92$-$2.5 $\mu$m).  The spectra have been
  normalized to unity at 2.20 $\mu$m and offset by constants.}
 \end{sidewaysfigure}

\clearpage

\begin{sidewaysfigure}
\centerline{\includegraphics[width=6in,angle=90]{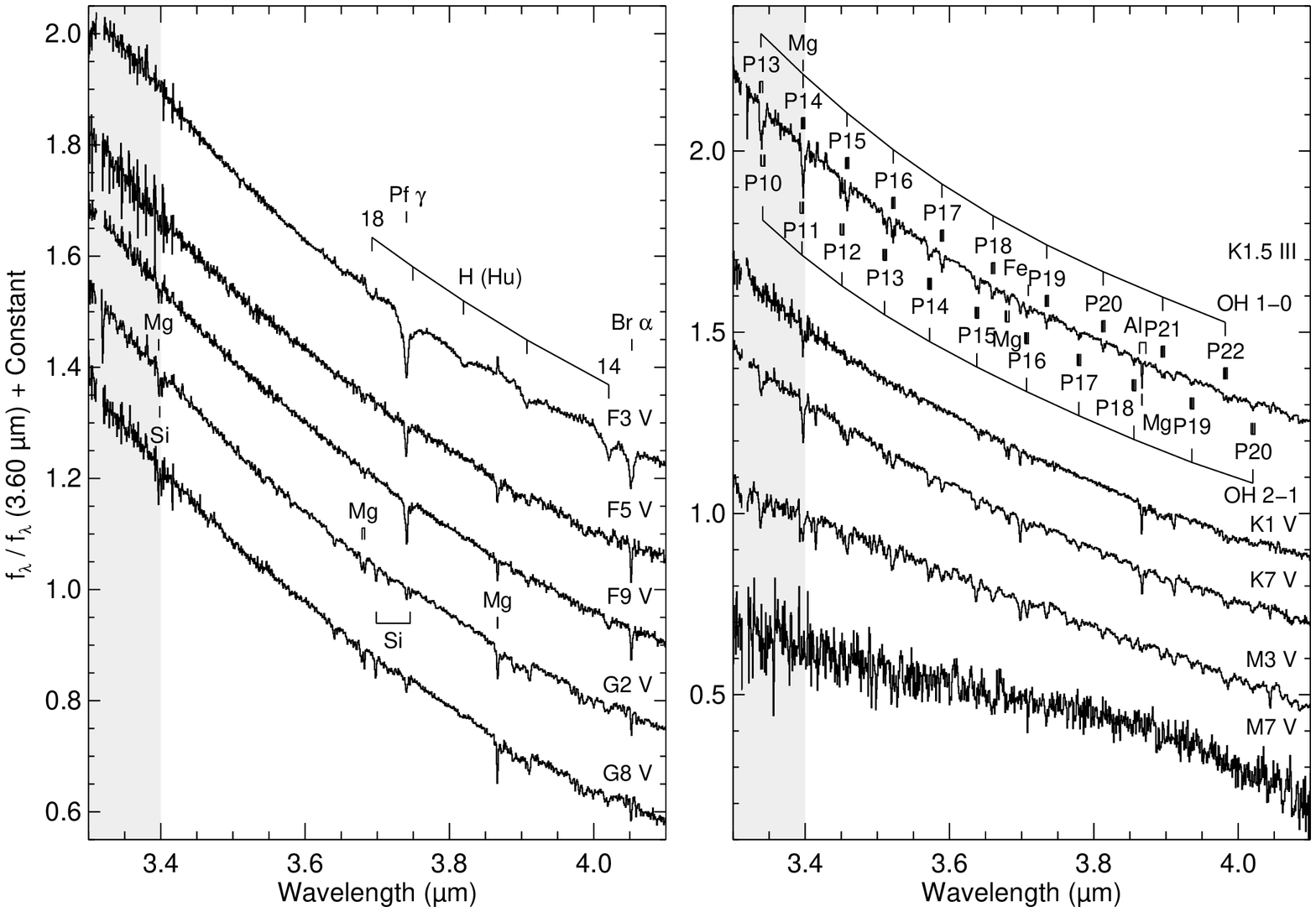}}
\caption{\label{fig:FGKM_VL} Same as Figure \ref{fig:FGKM_VI} except
  over the $L$ band (3.3$-$4.2 $\mu$m).  The spectra have been
  normalized to unity at 3.60 $\mu$m and offset by constants.}
 \end{sidewaysfigure}

\clearpage

\begin{sidewaysfigure}
\centerline{\includegraphics[width=6in,angle=90]{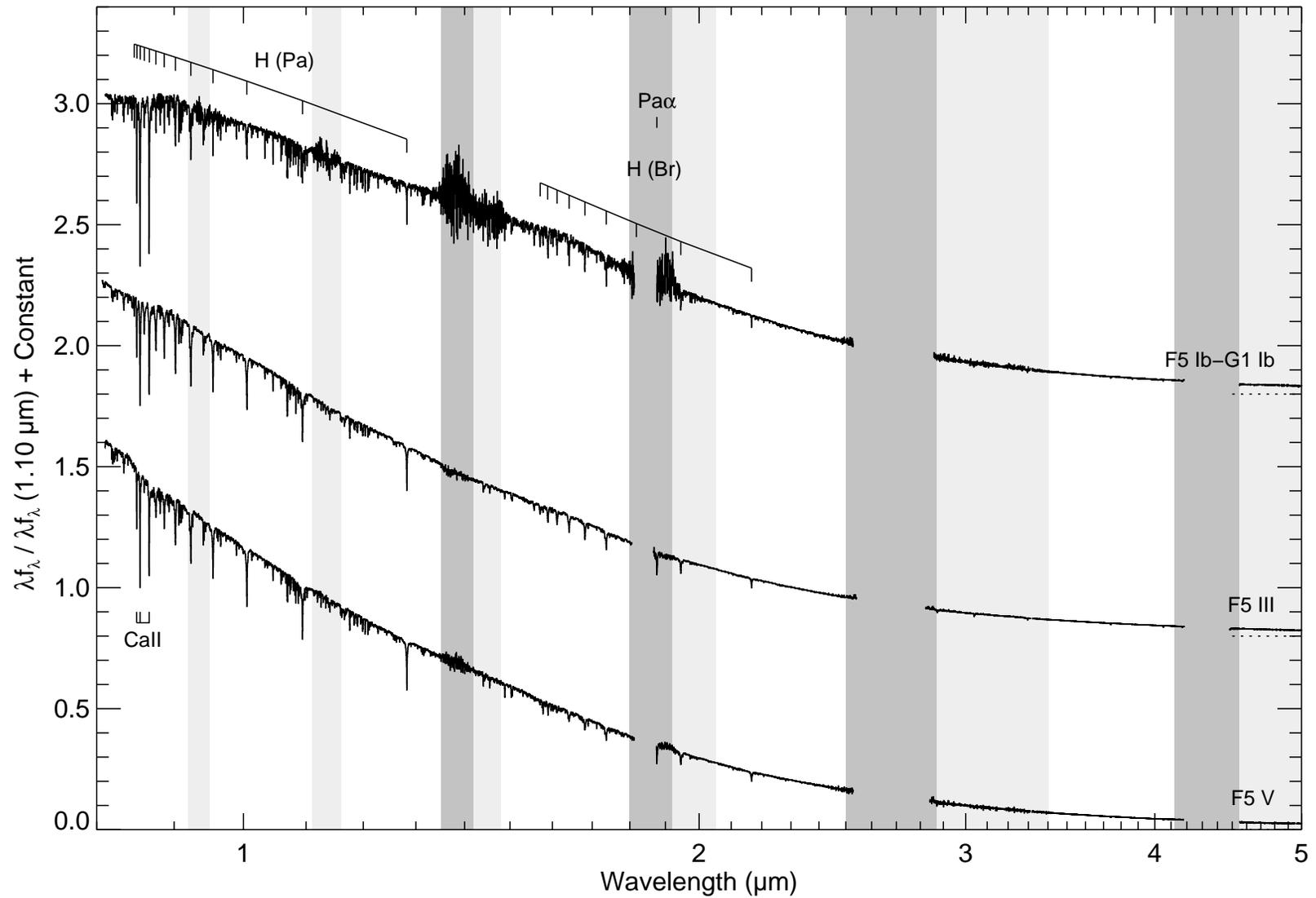}}
\caption{\label{fig:FI2VSeq} Luminosity effects at spectral type F.  The
  spectra are of HD~27524 (F5~V), HD~17918 (F5~III), and HD~213306
  (F5~Ib$-$G1~Ib) and have been normalized to unity at 1.10 $\mu$m and
  offset by constants (dotted lines).  Regions of strong (transmission
  $<$ 20\%) telluric absorption are shown in dark grey while regions of
  moderate (transmission $<$ 80\%) telluric absorption are shown in
  light grey.}
 \end{sidewaysfigure}

\clearpage

\begin{sidewaysfigure}
\centerline{\includegraphics[width=6in,angle=90]{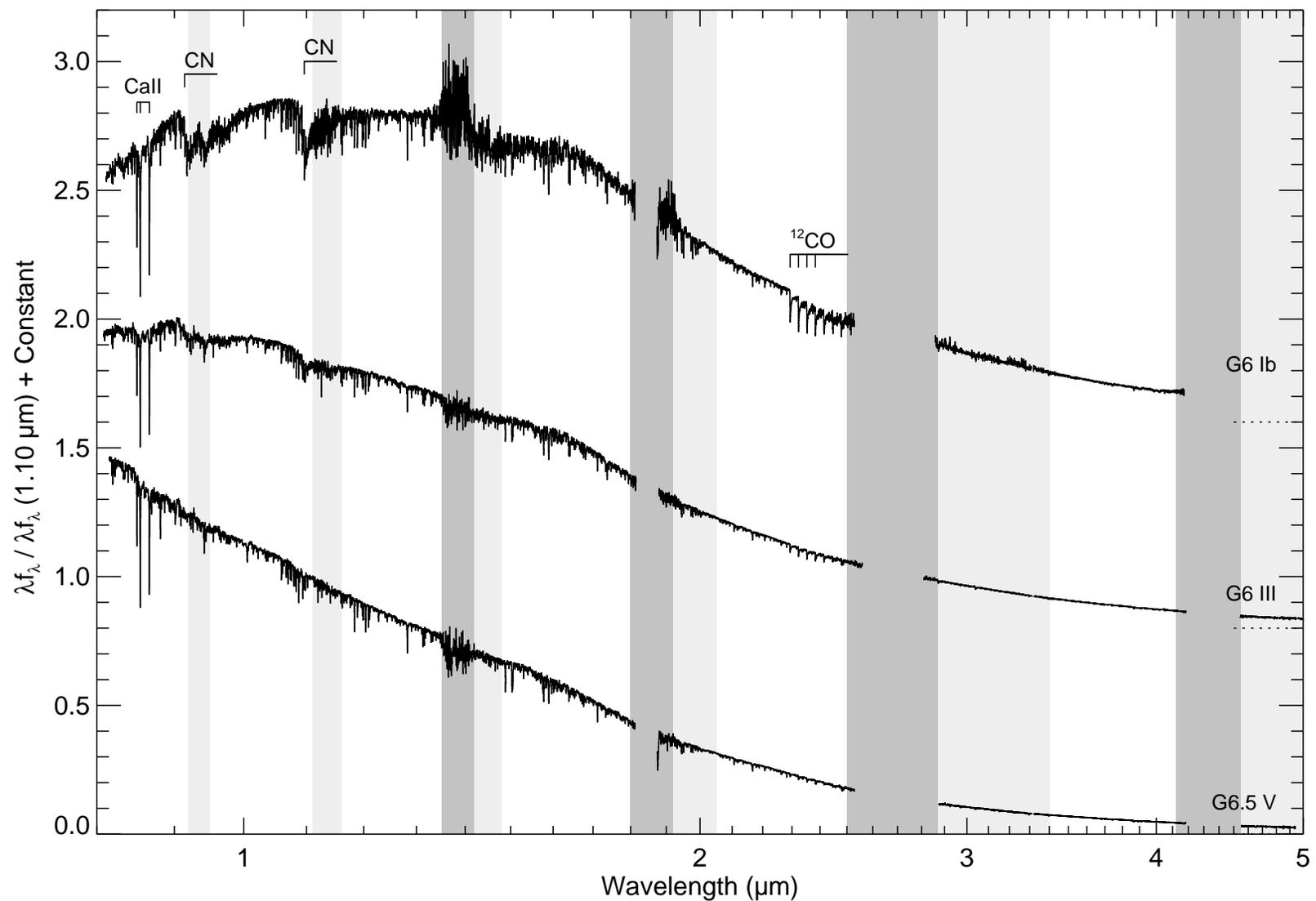}}
\caption{\label{fig:GI2VSeq} Luminosity effects at spectral type G.  The
  spectra are of HD~115617 (G6.5~V), HD~27277 (G6~III), and HD~161664
  (G6~Ib H$\delta$ 1) and have been normalized to unity at 1.10 $\mu$m
  and offset by constants (dotted lines).  Regions of strong
  (transmission $<$ 20\%) telluric absorption are shown in dark grey
  while regions of moderate (transmission $<$ 80\%) telluric absorption
  are shown in light grey.}
 \end{sidewaysfigure}

\clearpage

\begin{sidewaysfigure}
\centerline{\includegraphics[width=6in,angle=90]{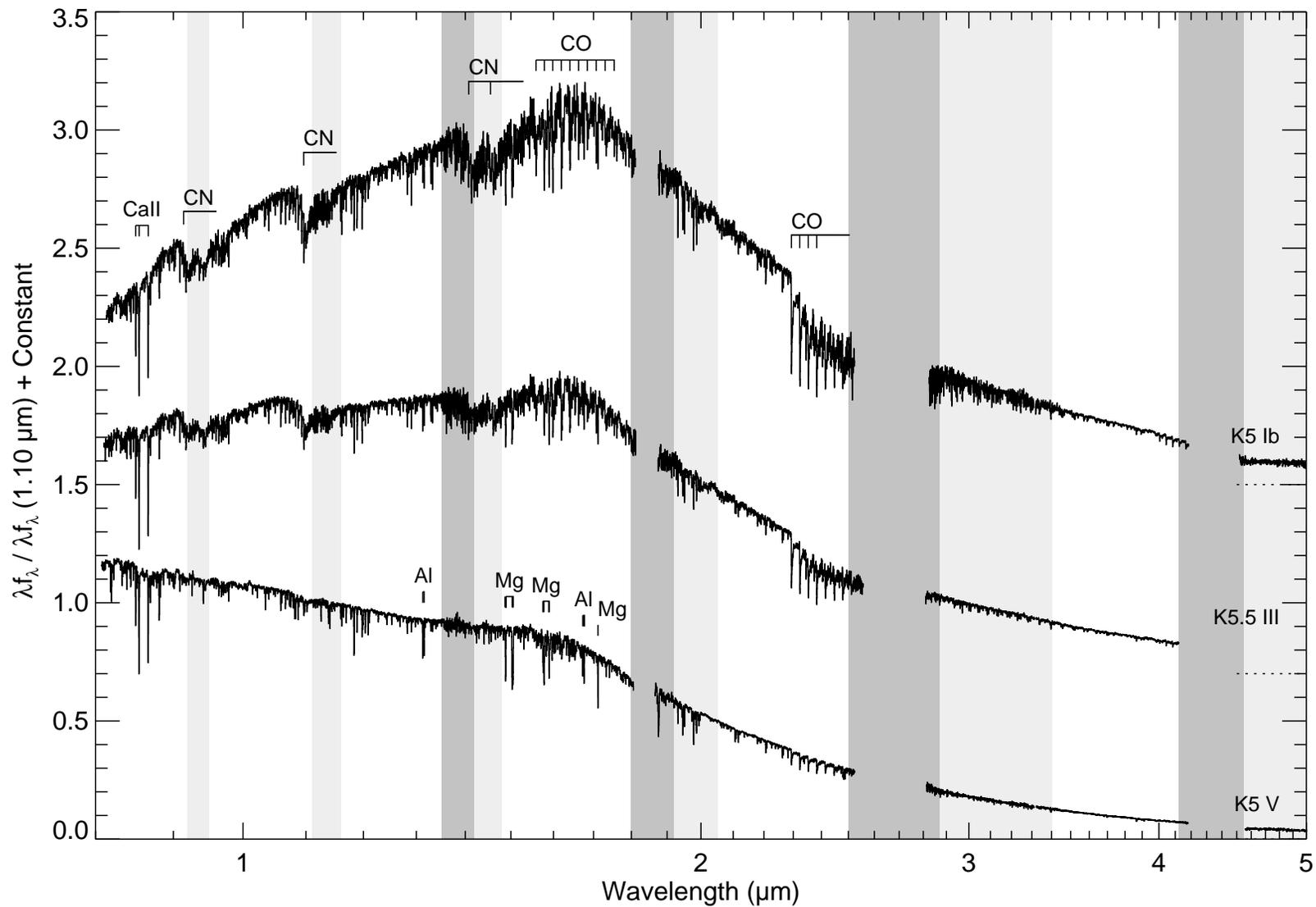}}
\caption{\label{fig:KI2VSeq} Luminosity effects at spectral type K.  The
  spectra are of HD~36003 (K5~V), HD~120477 (K5.5~III), and HD~216946
  (K5~Ib) and have been normalized to unity at 1.10 $\mu$m and offset by
  constants (dotted lines).  Regions of strong (transmission $<$ 20\%)
  telluric absorption are shown in dark grey while regions of moderate
  (transmission $<$ 80\%) telluric absorption are shown in light grey.}
 \end{sidewaysfigure}

\clearpage

\begin{sidewaysfigure}
\centerline{\includegraphics[width=6in,angle=90]{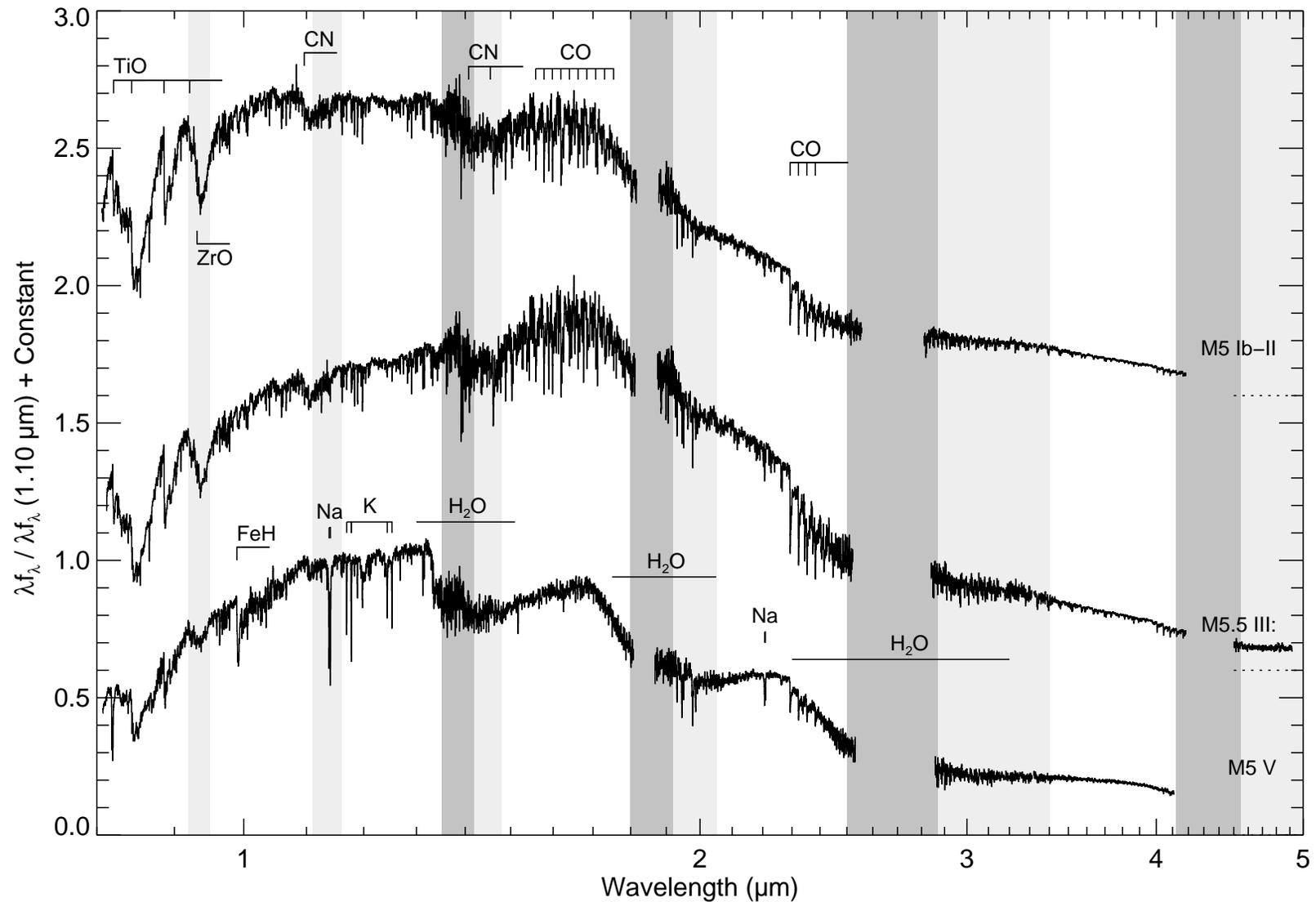}}
\caption{\label{fig:MI2VSeq} Luminosity effects at spectral type M.  The
  spectra are of Gl~866ABC (M5~V), HD~94705 (M5.5~III:), and HD~156014
  (M5~Ib-II) and have been normalized to unity at 1.10 $\mu$m and offset
  by constants (dotted lines).  Regions of strong (transmission $<$
  20\%) telluric absorption are shown in dark grey while regions of
  moderate (transmission $<$ 80\%) telluric absorption are shown in
  light grey.}
\end{sidewaysfigure}

\clearpage

\begin{sidewaysfigure}
\centerline{\includegraphics[height=8in,angle=90]{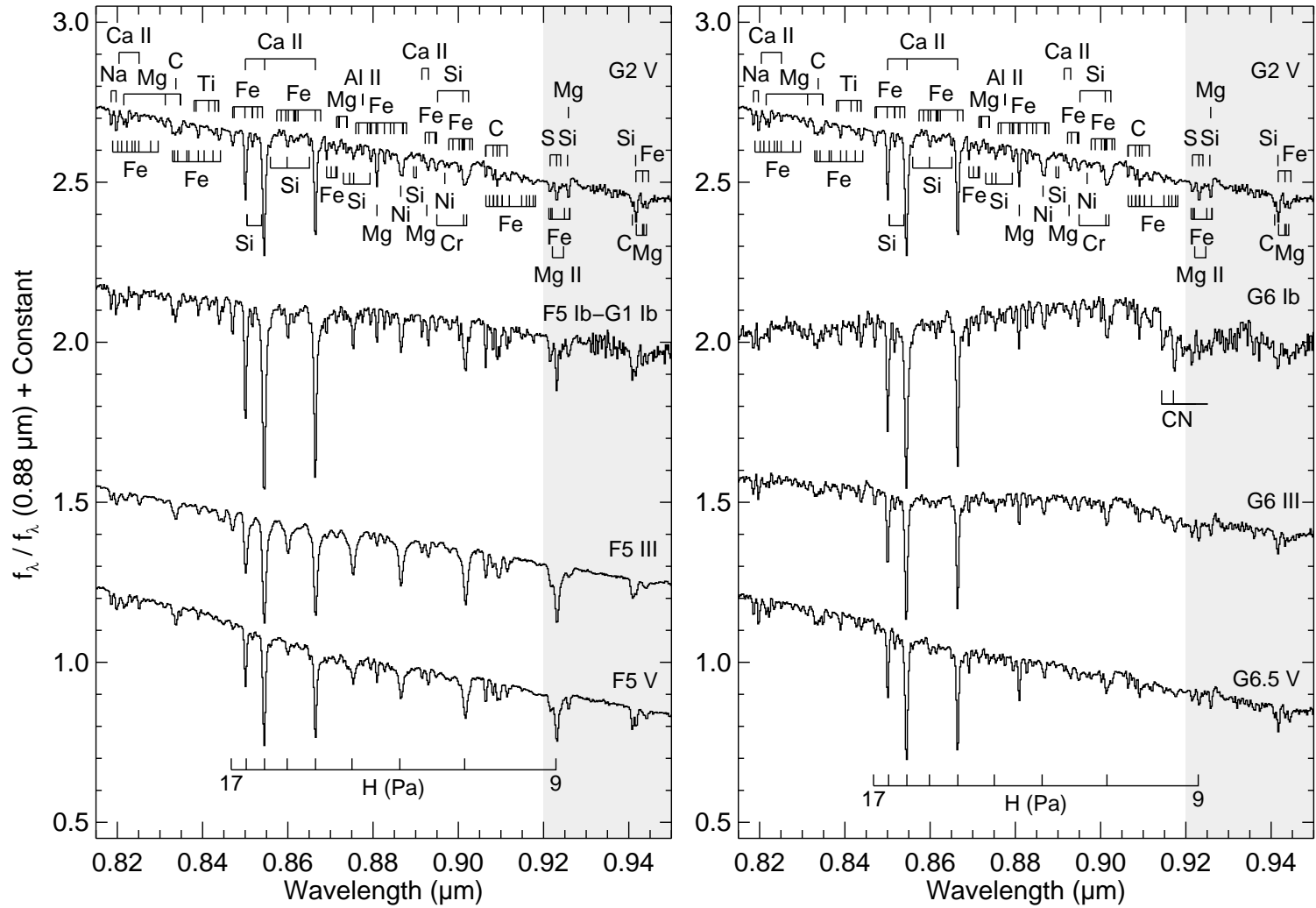}}
\caption{\label{fig:FG_I2VI}: Luminosity effects at spectral types F5
  (\textit{left}) and G6 (\textit{right}) plotted over the $I$ band
  (0.82$-$0.95 $\mu$m).  The spectra are of HD~213306 (F5~Ib-G1~Ib),
  HD~17918 (F5~III), HD~27524 (F5~V), HD~161664 (G6~Ib~H$_{\delta}$1),
  HD~27277 (G6~III), and HD~115617 (G6.5~V).  The G2~V comparison star
  is HD~76151.  The spectra have been normalized to unity at 0.88~$\mu$m
  and offset by constants.}
\end{sidewaysfigure}

\clearpage

\begin{sidewaysfigure}
\centerline{\includegraphics[height=8in,angle=90]{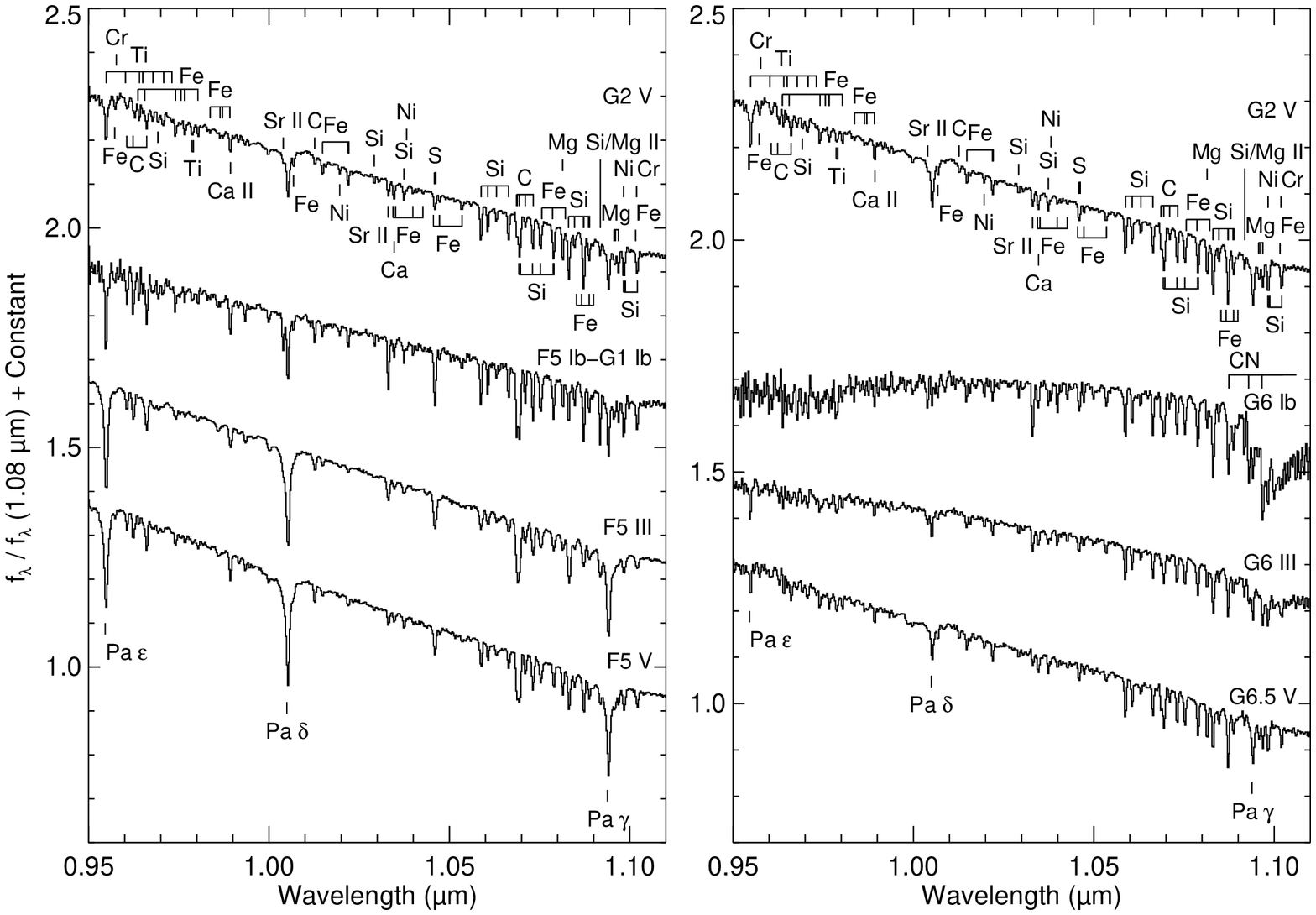}}
\caption{\label{fig:FG_I2VY}: Same as Figure \ref{fig:FG_I2VI} except
  over the $Y$ band.  The spectra have normalized to unity at 1.08
  $\mu$m and offset by constants.}
 \end{sidewaysfigure}

\clearpage

\begin{sidewaysfigure}
\centerline{\includegraphics[height=8in,angle=90]{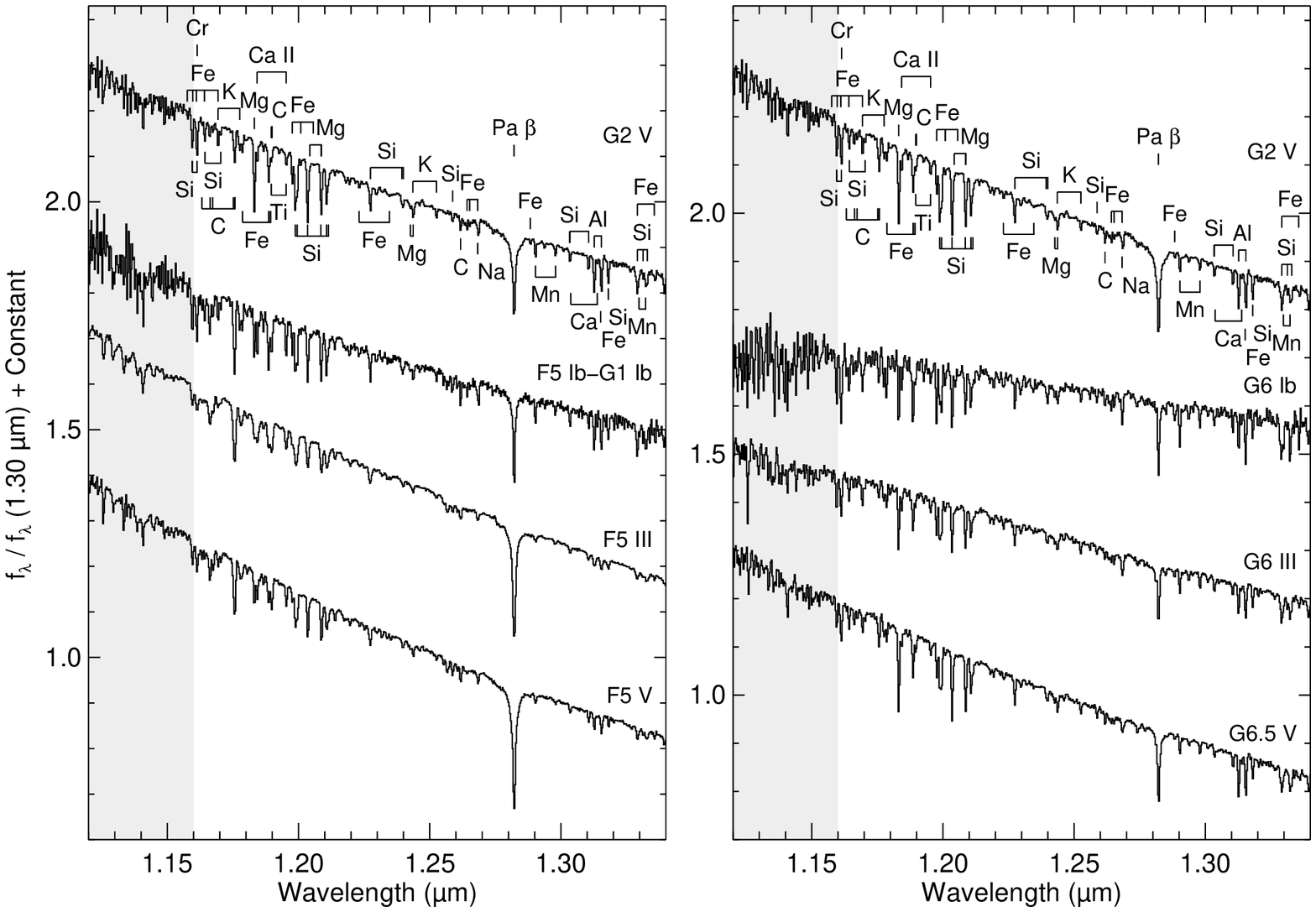}}
\caption{\label{fig:FG_I2VJ}: Same as Figure \ref{fig:FG_I2VI} except
  over the $J$ band.  The spectra have normalized to unity at 1.30
  $\mu$m and offset by constants.}
\end{sidewaysfigure}

\clearpage

\begin{sidewaysfigure}
\centerline{\includegraphics[height=8in,angle=90]{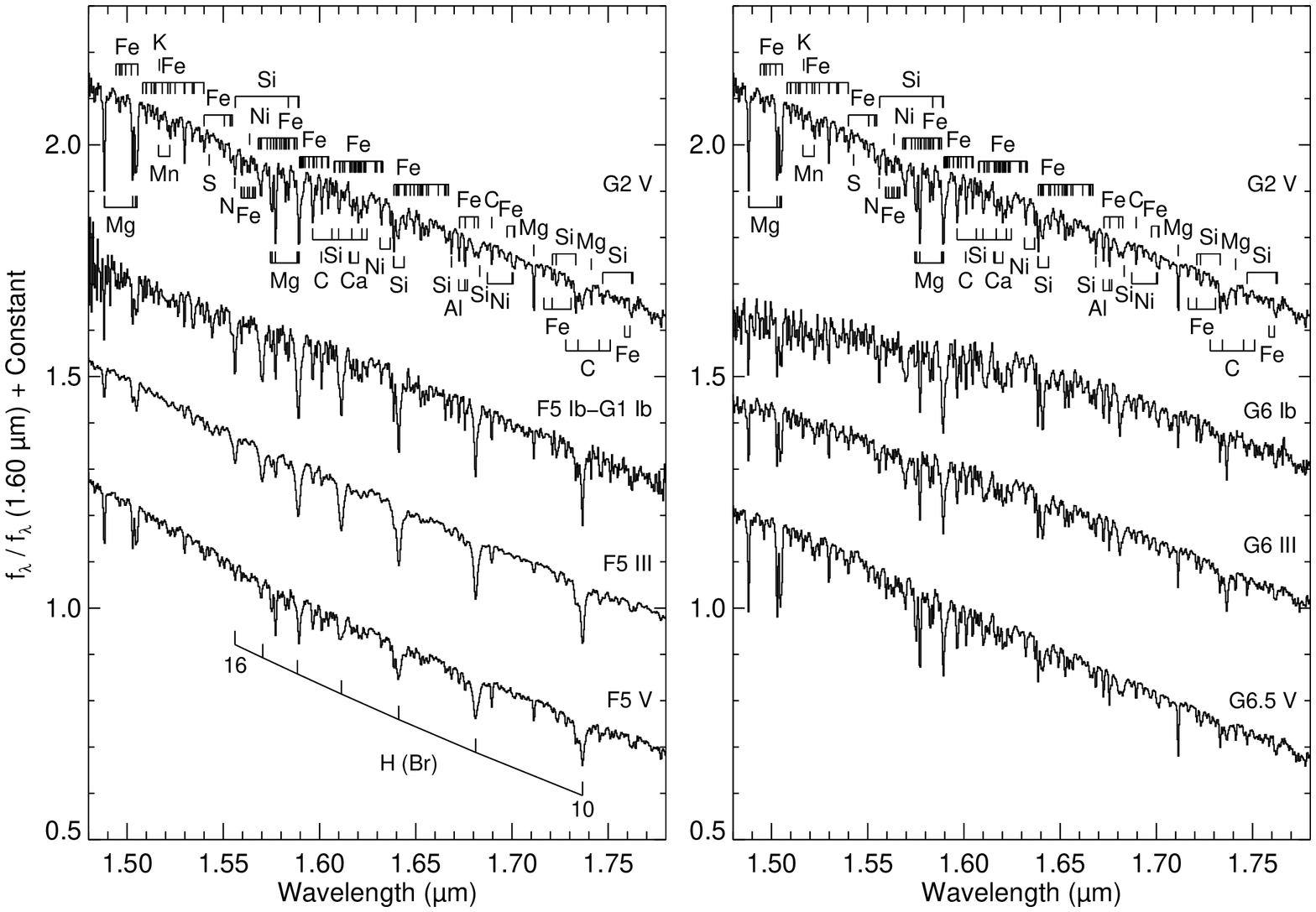}}
\caption{\label{fig:FG_I2VH}: Same as Figure \ref{fig:FG_I2VI} except
  over the $H$ band.  The spectra have normalized to unity at 1.60
  $\mu$m and offset by constants.}
\end{sidewaysfigure}

\clearpage

\begin{sidewaysfigure}
\centerline{\includegraphics[height=8in,angle=90]{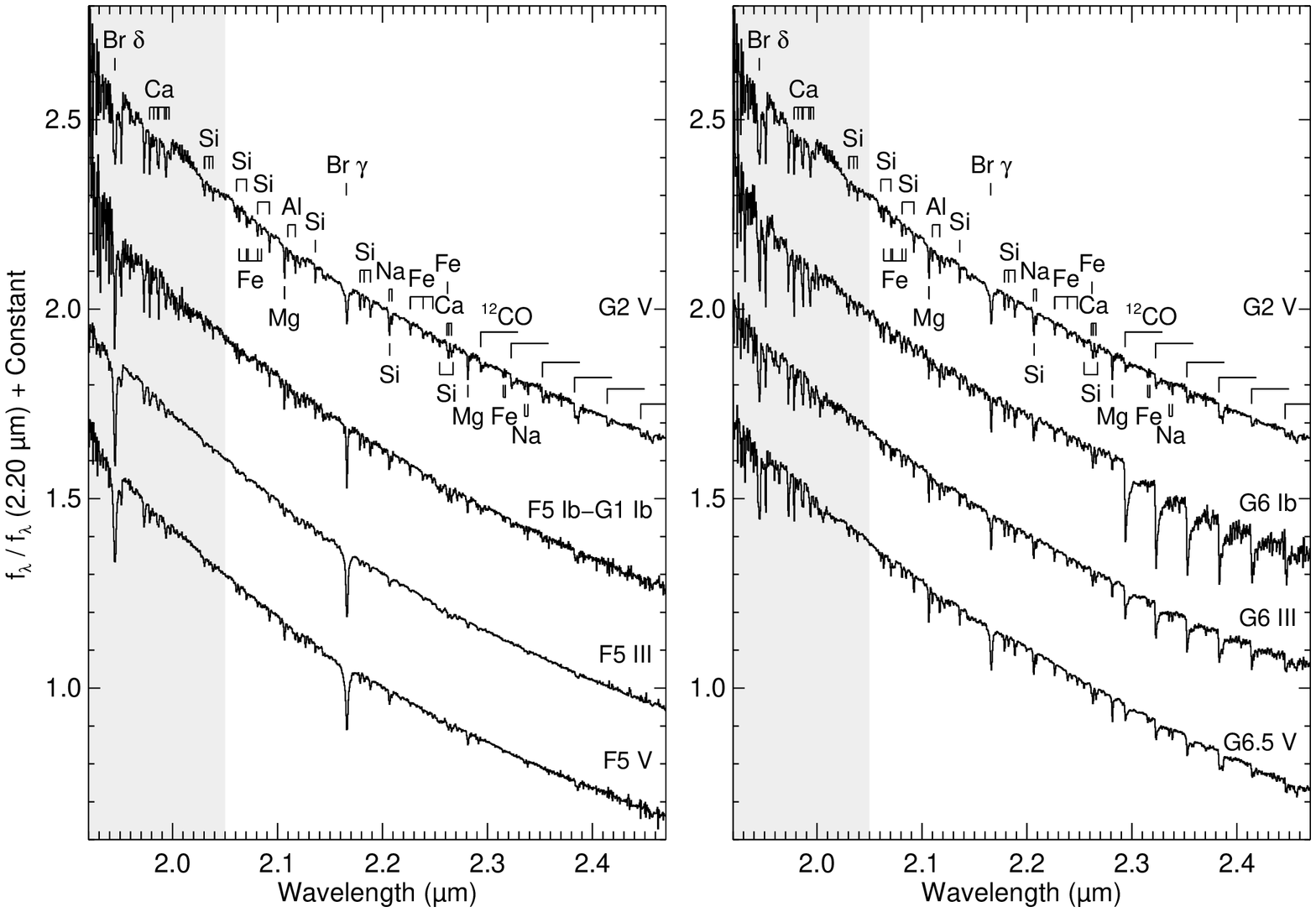}}
\caption{\label{fig:FG_I2VK}: Same as Figure \ref{fig:FG_I2VI} except
  over the $K$ band.  The spectra have normalized to unity at 2.20
  $\mu$m and offset by constants.}
\end{sidewaysfigure}

\clearpage

\begin{sidewaysfigure}
\centerline{\includegraphics[height=8in,angle=90]{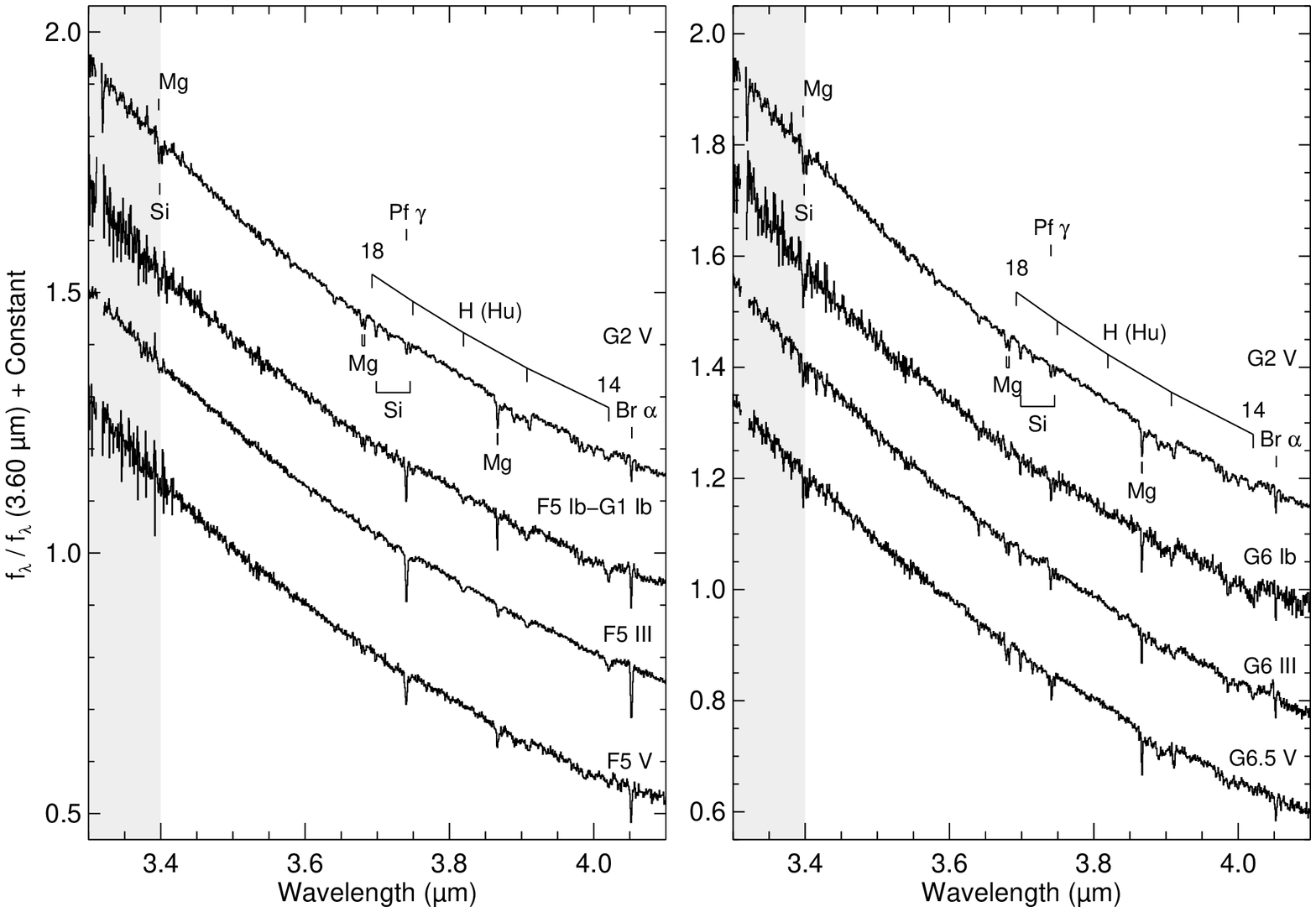}}
\caption{\label{fig:FG_I2VL}: Same as Figure \ref{fig:FG_I2VI} except
  over the $L$ band.  The spectra have normalized to unity at 3.60
  $\mu$m and offset by constants.}
\end{sidewaysfigure}

\clearpage

\begin{sidewaysfigure}
\centerline{\includegraphics[height=8in,angle=90]{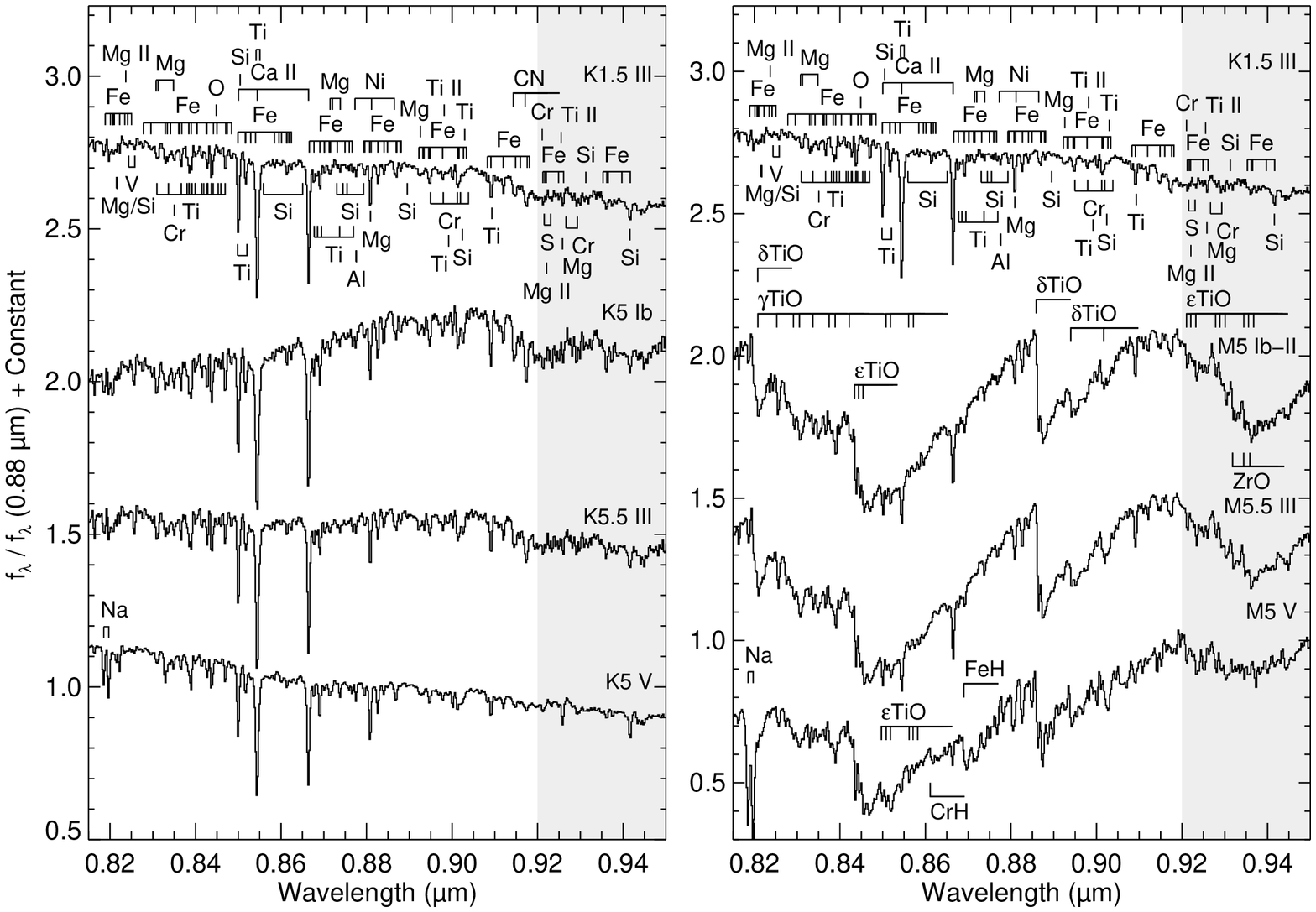}}
\caption{\label{fig:KM_I2VI}: Luminosity effects at spectral types K5
  (\textit{left}) and M5 (\textit{right}) plotted over the $I$ band
  (0.82$-$0.95 $\mu$m).  The spectra are of HD~216946 (K5~Ib), HD~120477
  (K5.5~III), HD~36003 (K5~V), HD~156014 (M5~Ib-II), HD~94705
  (M5.5~III:), and Gl~866ABC (M5~V).  The K1.5~III comparison star is
  Arcturus (HD~124897).  The spectra have been normalized to unity at
  0.88~$\mu$m and offset by constants.}
\end{sidewaysfigure}

\clearpage

\begin{sidewaysfigure}
\centerline{\includegraphics[height=8in,angle=90]{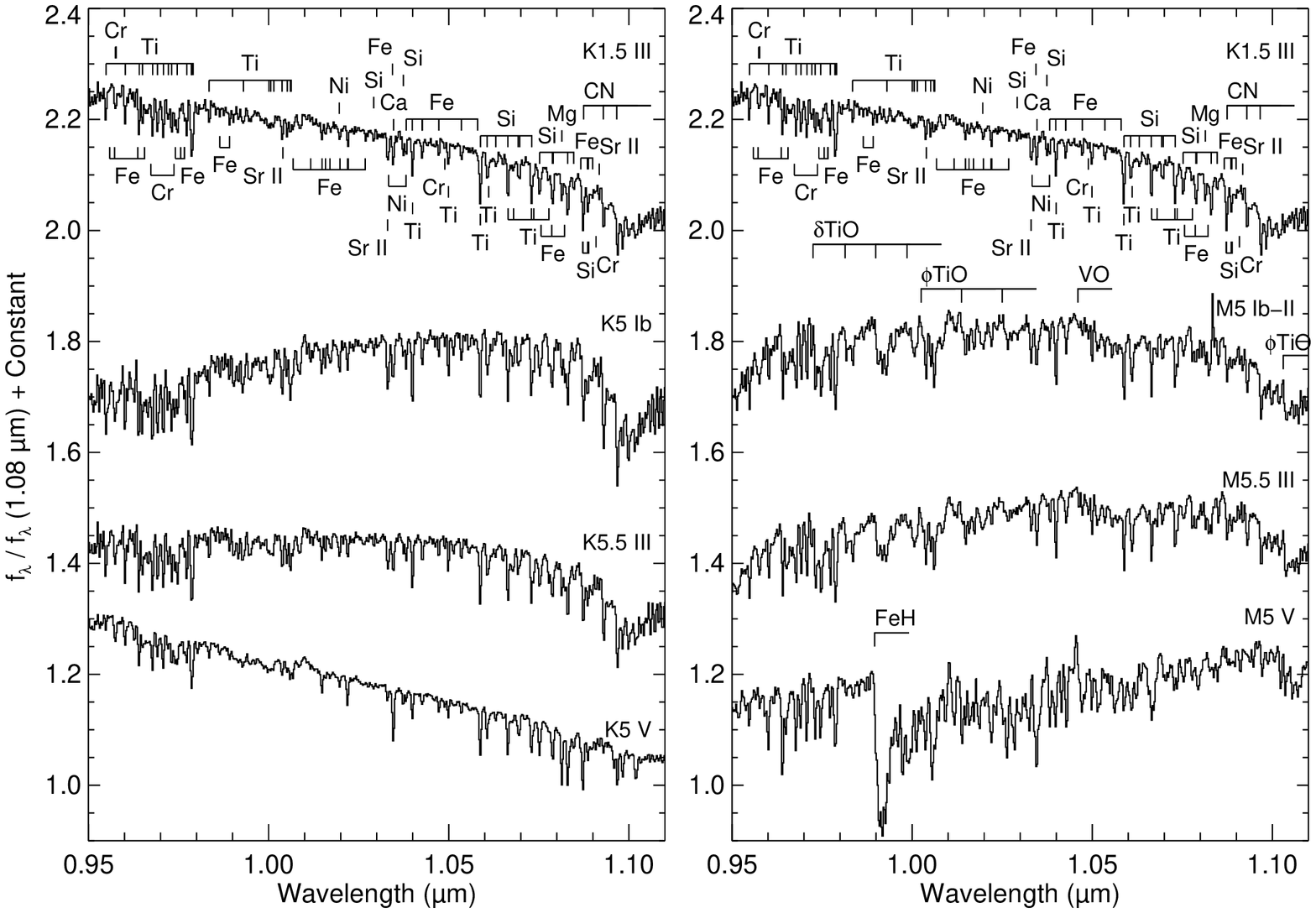}}
\caption{\label{fig:KM_I2VY}: Same as Figure \ref{fig:KM_I2VI} except
  over the $Y$ band.  The spectra have normalized to unity at 1.08
  $\mu$m and offset by constants.}
\end{sidewaysfigure}

\clearpage

\begin{sidewaysfigure}
\centerline{\includegraphics[height=8in,angle=90]{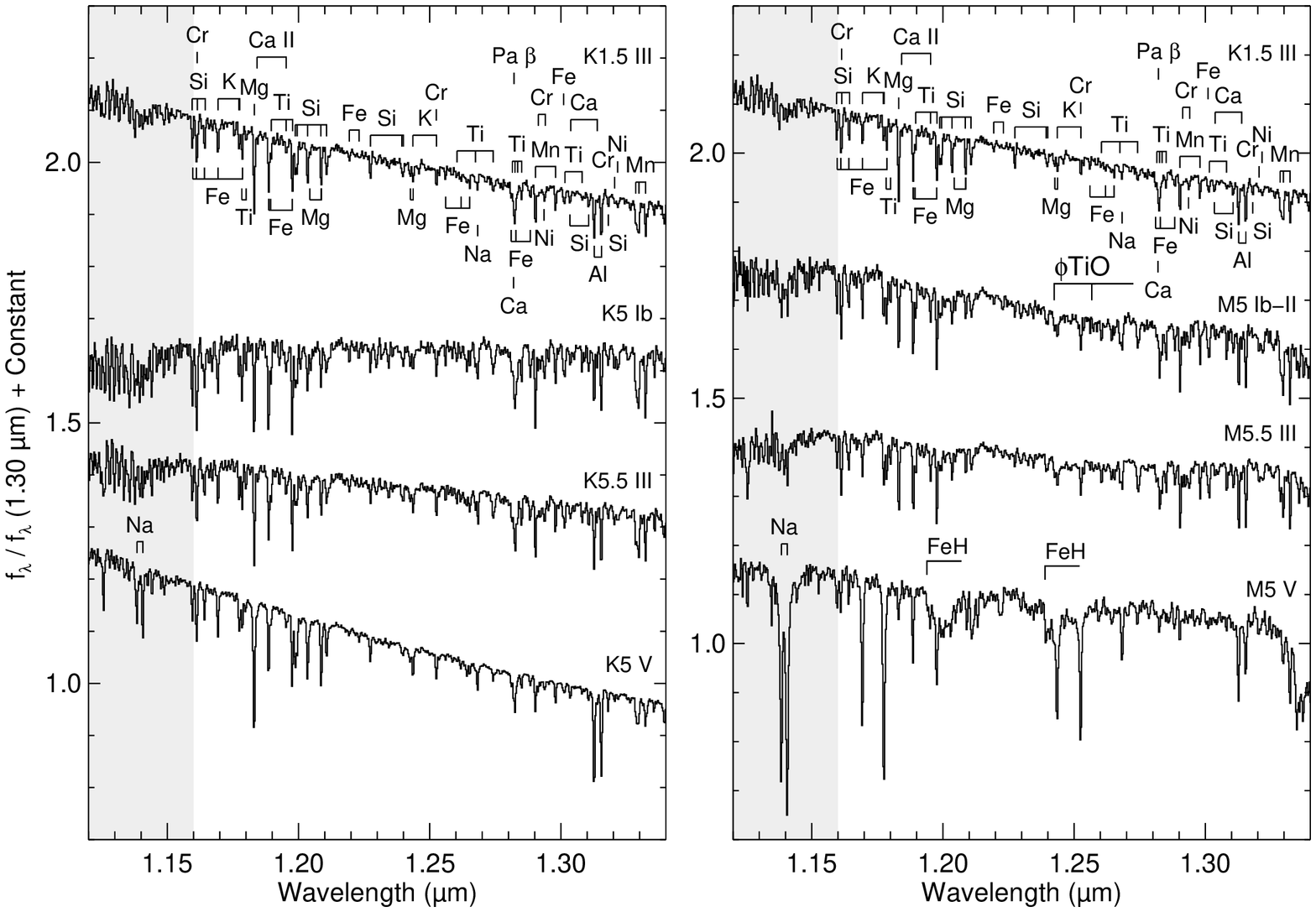}}
\caption{\label{fig:KM_I2VJ}: Same as Figure Same as Figure
  \ref{fig:KM_I2VI} except over the $J$ band.  The spectra have
  normalized to unity at 1.30 $\mu$m and offset by constants.}
\end{sidewaysfigure}

\clearpage

\begin{sidewaysfigure}
\centerline{\includegraphics[height=8in,angle=90]{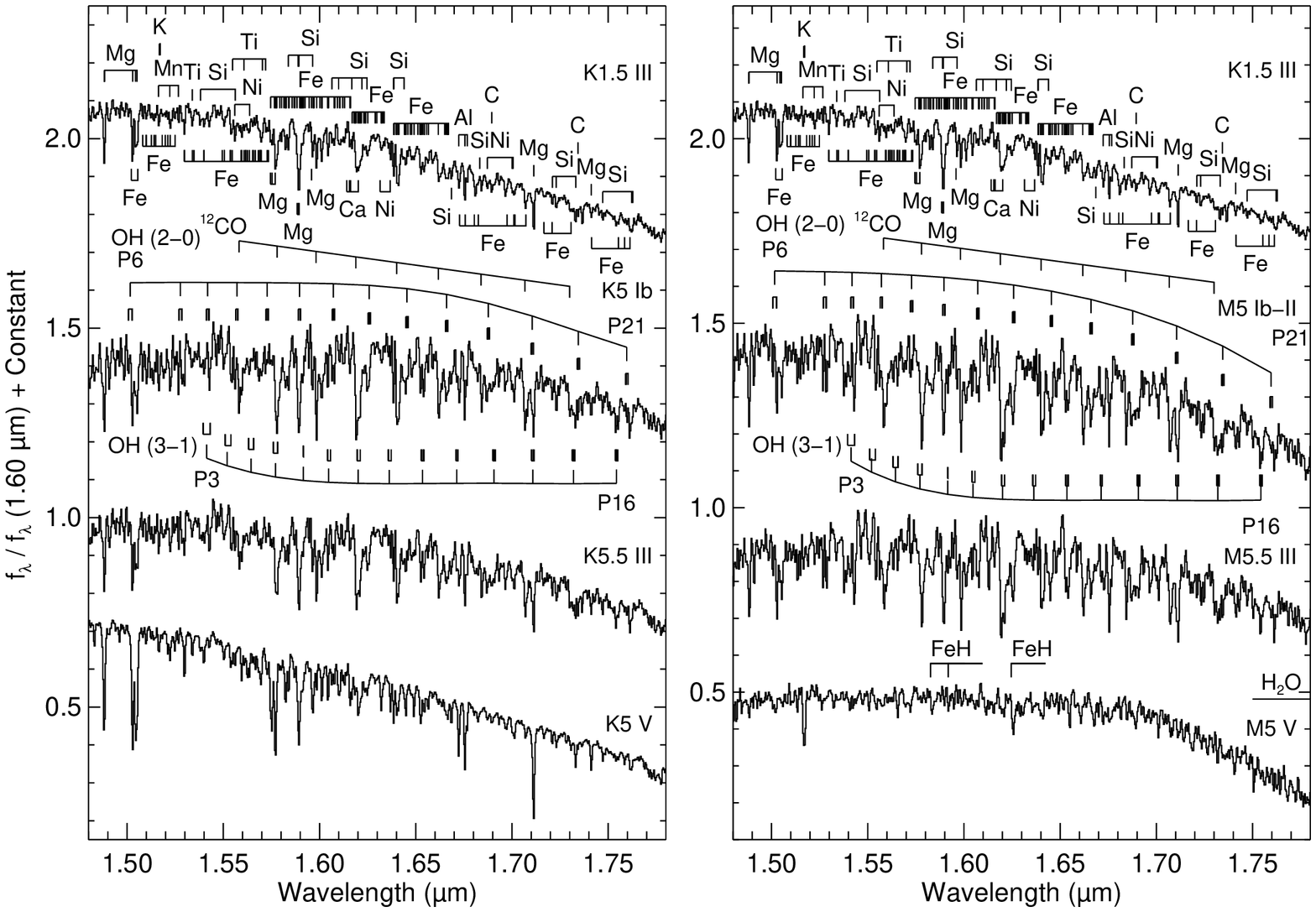}}
\caption{\label{fig:KM_I2VH}: Same as Figure \ref{fig:KM_I2VI} except
  over the $H$ band.  The spectra have normalized to unity at 1.60
  $\mu$m and offset by constants.}
\end{sidewaysfigure}

\clearpage

\begin{sidewaysfigure}
\centerline{\includegraphics[height=8in,angle=90]{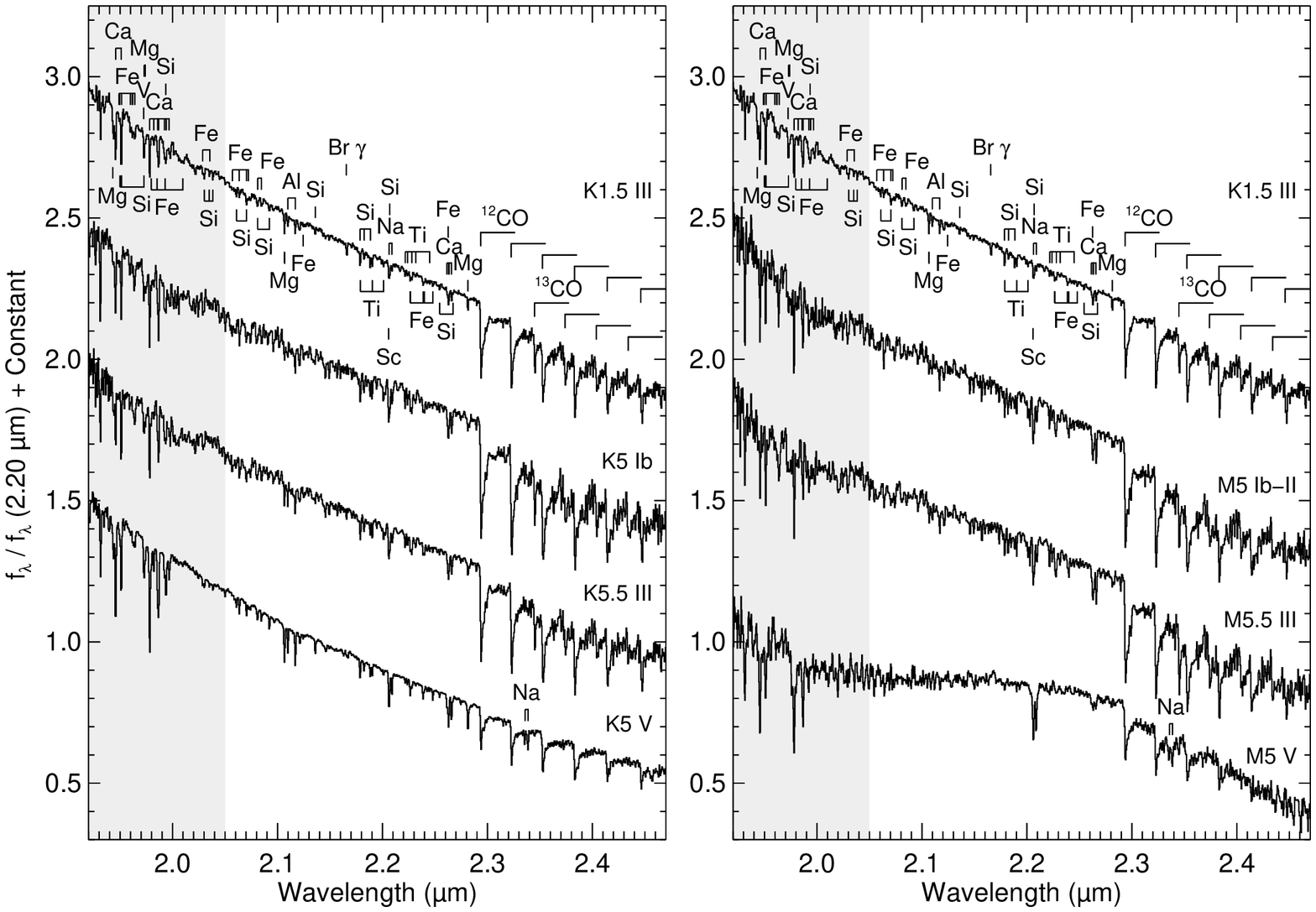}}
\caption{\label{fig:KM_I2VK}: Same as Figure \ref{fig:KM_I2VI} except
  over the $K$ band.  The spectra have normalized to unity at 2.20
  $\mu$m and offset by constants.}
\end{sidewaysfigure}

\clearpage

\begin{sidewaysfigure}
\centerline{\includegraphics[height=8in,angle=90]{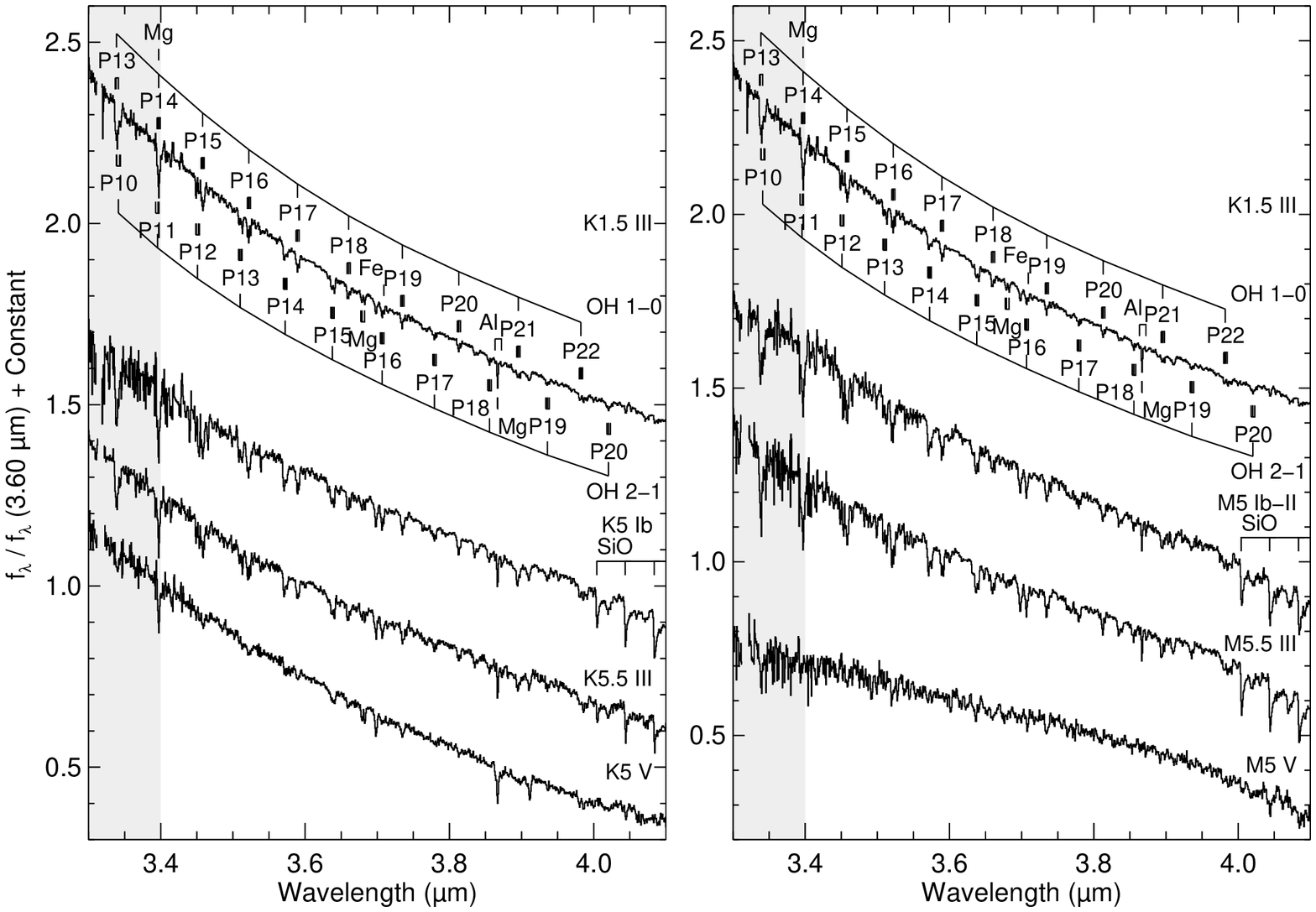}}
\caption{\label{fig:KM_I2VL}: Same as Figure \ref{fig:KM_I2VI} except
  over the $L$ band.  The spectra have normalized to unity at 3.60
  $\mu$m and offset by constants.}
\end{sidewaysfigure}

\clearpage

\begin{sidewaysfigure}
\centerline{\includegraphics[height=8in,angle=90]{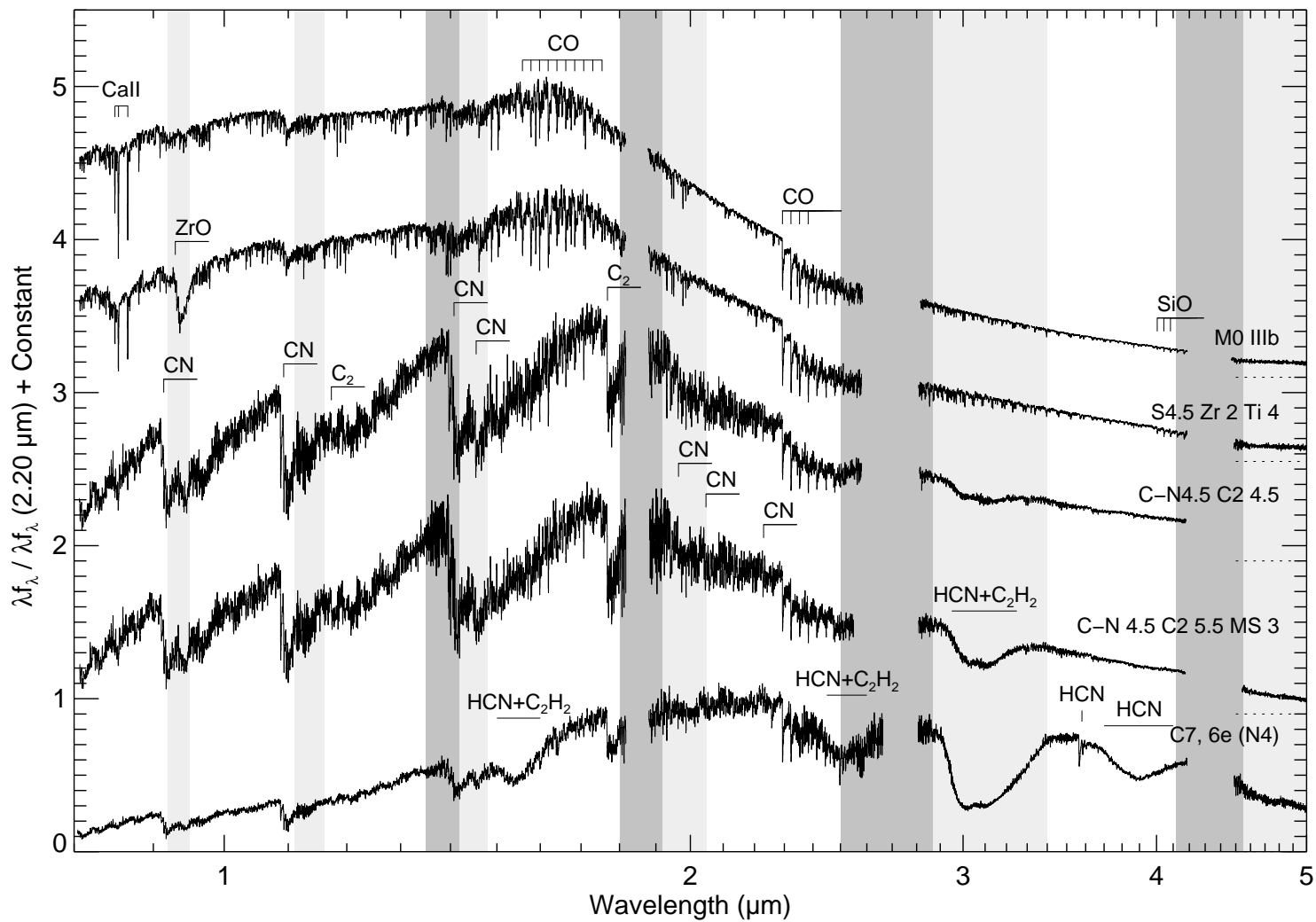}}
\caption{\label{fig:MSC}A sequence of M, S, and C giants of
  approximately the same effective temperature, illustrating the effects
  of increasing carbon abundance over oxygen during AGB evolution.  The
  spectra are of HD~213893 (M0~IIIb), HD~64332 (S4.5~Zr~2~Ti~4), HD~92055
  (C-N4.5~C2~4.5), and HD~76221 (C-N~4.5~C2~5.5~MS~3), and have been
  normalized to unity at 1.08~$\mu$m and offset by constants (dotted
  lines).  Also plotted is the very cool carbon star R~Lep (HD~31996,
  C7,6e~(N4)).  Regions of strong (transmission $<$ 20\%) telluic
  absorption are shown in dark grey while regions of moderate
  (transmission $<$ 80\%) telluric absorption are shown in light grey.}
\end{sidewaysfigure}

\clearpage

\begin{figure}
\centerline{\includegraphics[width=6in,angle=0]{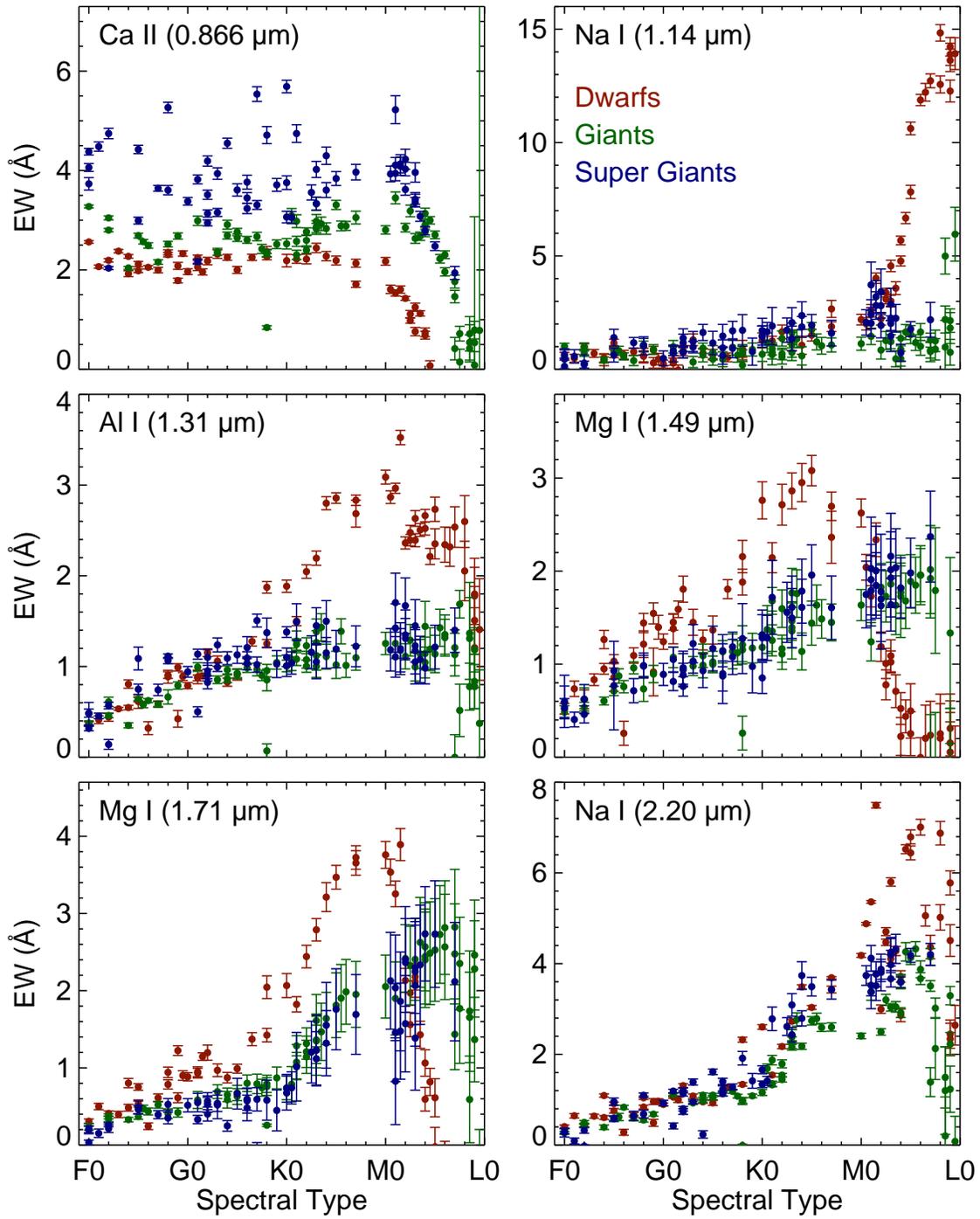}}
\caption{\label{fig:EWs}The equivalent widths of prominent atomic
  absorption lines as a function of spectral type.  The \ion{Ca}{2} EW
  is a good luminosity class indicator. The \ion{Na}{1} 2.20~\micron~EW
  provides a good indication of spectral subtype, while the 
  \ion{Na}{1} 1.14~\micron~doublet EW can be used as a clear indicator of the
  very latest spectral subtypes.}
\end{figure}

\clearpage

\begin{figure}
\centerline{\includegraphics[width=3.5in,angle=0]{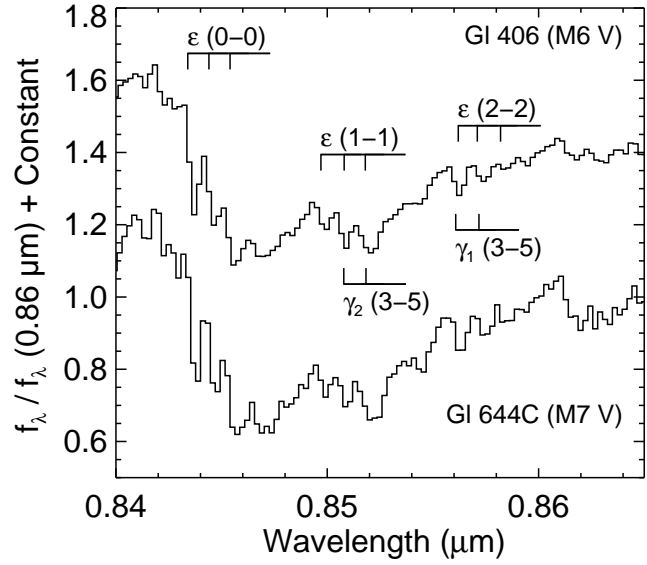}}
\caption{\label{fig:TiO} The spectrum of Gl 406 (M6 V) and Gl 644C (vB
  8, M7 V) centered on the $\Delta\nu$~=~0 $\epsilon$ system of TiO.
  The 0-0, 1-1, and 2-2 bands are indicated.  Also indicated are the
  $\Delta\nu$~=~-2 band heads of the $\gamma$ system.  The set of band
  heads at $\sim$0.853~$\mu$m and $\sim$0.858~$\mu$m are most likely a
  combination of both the $\epsilon$ and $\gamma$ systems even though
  only the $\gamma$ heads are indicated in the literature.}
\end{figure}

\clearpage

\begin{figure}
\centerline{\includegraphics[width=6.5in,angle=90]{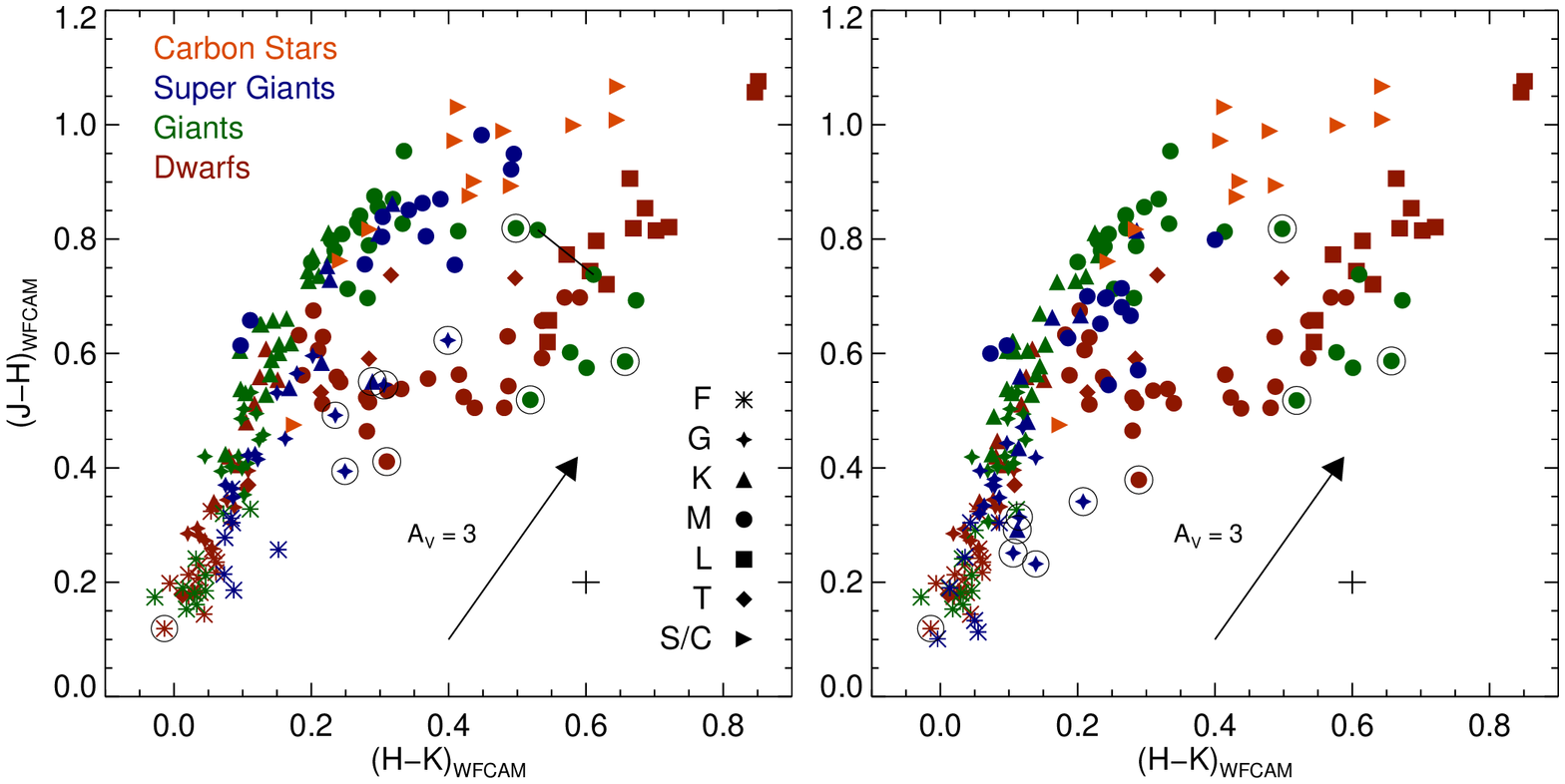}}
\vspace{-1.5in}
\caption{\label{fig:JHHK}\textit{Left:} Synthesized WFCAM $J-H$ versus
  $H-K$ diagram for our sample of cool stars (see
  Table~\ref{tab:MKOColors}), plus additional L and T~dwarfs from
  \citet{2005ApJ...623.1115C}, and T dwarfs from
  \citet{2006ApJ...637.1067B}. Several objects with extreme colors fall
  outside the plotted range and are plotted in Figure~\ref{fig:JHHKall}.
  Note the bifurcation between dwarfs and stars of higher luminosity
  beyond spectral type M0~V. Stars with odd spectra are circled (see
  \S\ref{Unusual}). HD~69243 (M6e-M9e~III, Mira variable) was observed
  at two epochs and the change in color is indicated (see also Figure
  \ref{fig:MIIIwater}).  The $+$ sign indicates the accuracy of the
  photometry, and the arrow indicates the direction and magnitude of the
  reddening vector for A$_V$=3.  \textit{Right:} The same plot but now
  with the correction for reddening included (the 77 dereddened stars
  are given in Table~\ref{tab:Extinction}).  Note how most M supergiant
  stars move closer to the main sequence.}
\end{figure}

\clearpage

\begin{figure}
\centerline{\includegraphics[width=3.5in,angle=0]{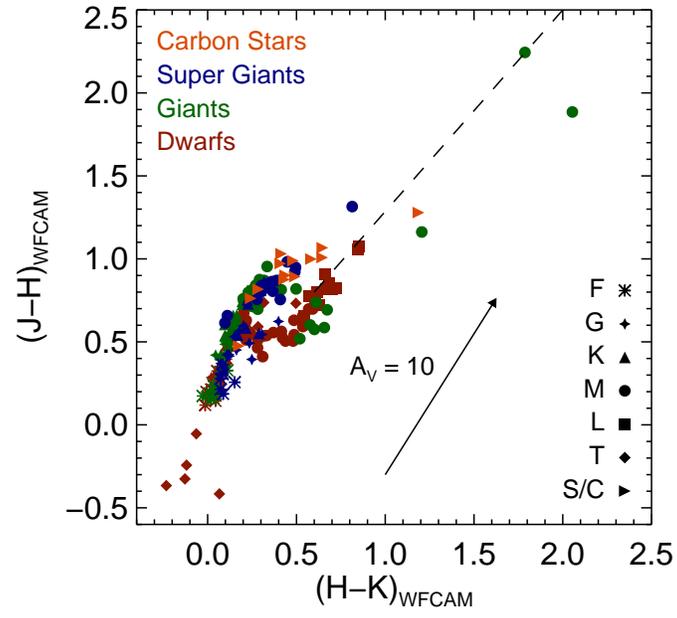}}
\caption{\label{fig:JHHKall} Same as Figure~\ref{fig:JHHK} except all
  stars are plotted. The additional stars include some very red Mira and
  OH/IR stars, plus several blue T~dwarfs. The colors of the Mira and
  OH/IR stars follow the linear locus (\textit{dashed line}) expected
  from models of circumstellar dust shells in TPAGB
  stars \citep[e.g.,][]{2006AJ....132..489L}.}
\end{figure}

\clearpage

\begin{sidewaysfigure}
\centerline{\includegraphics[width=6in,angle=90]{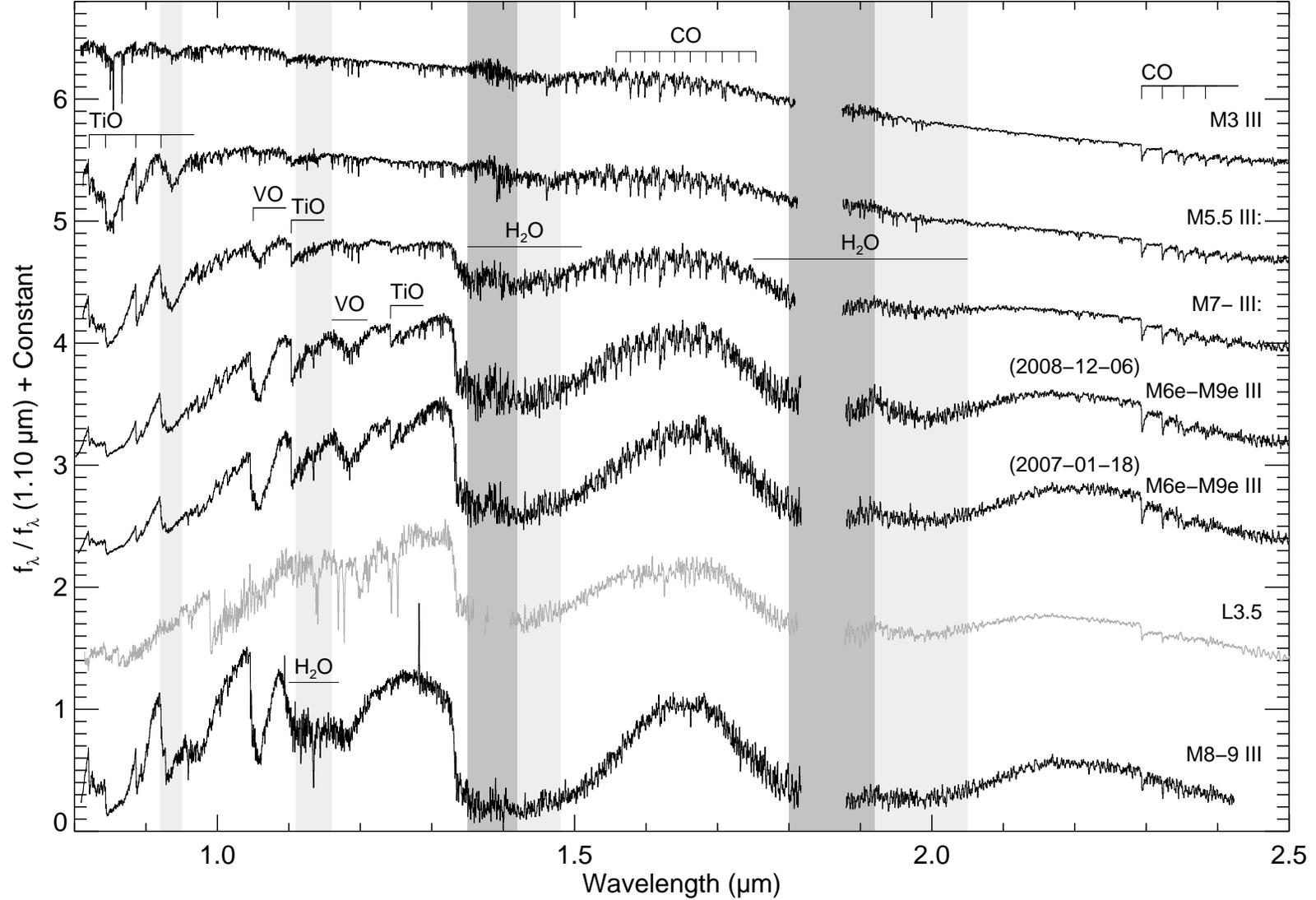}}
\caption{\label{fig:MIIIwater} The spectral behavior of M giant stars
  along the locus of decreasing $J-H$ and increasing $H-K$ in
  Figure~\ref{fig:JHHK} (plotted from the top).  Strong water absorption
  at 1.4\micron~and 1.9\micron~reduces the flux emitted at $H$ relative
  to $J$ and $K$. As a result, the locus of M giant stars in the $J-H$
  versus $H-K$ diagram appears to turn down toward late-M and early-L
  dwarfs (see \S\ref{CCD} and Figure~\ref{fig:JHHK}). The spectra
  plotted (black) are HD~39045 (M3~III), HD~94705 (M5.5~III:), HD~108849
  (M7-~III:), HD~69243 (M6e-M9e~III) on 2008 Dec 06, HD~69243 (M6e-M9e~III) on
  2007 Jan 18, and IRAS~14303-1042 (M8-9~III). Also plotted is the
  spectrum (grey) of 2MASS J00361617$+$1821104 \citep[L3.5,
  ][]{2005ApJ...623.1115C} which has very similar $JHK$ colors to
  HD~69243 (M6e-M9e~III).}
\end{sidewaysfigure}

\clearpage

\begin{figure}
\centerline{\includegraphics[width=6.5in,angle=90]{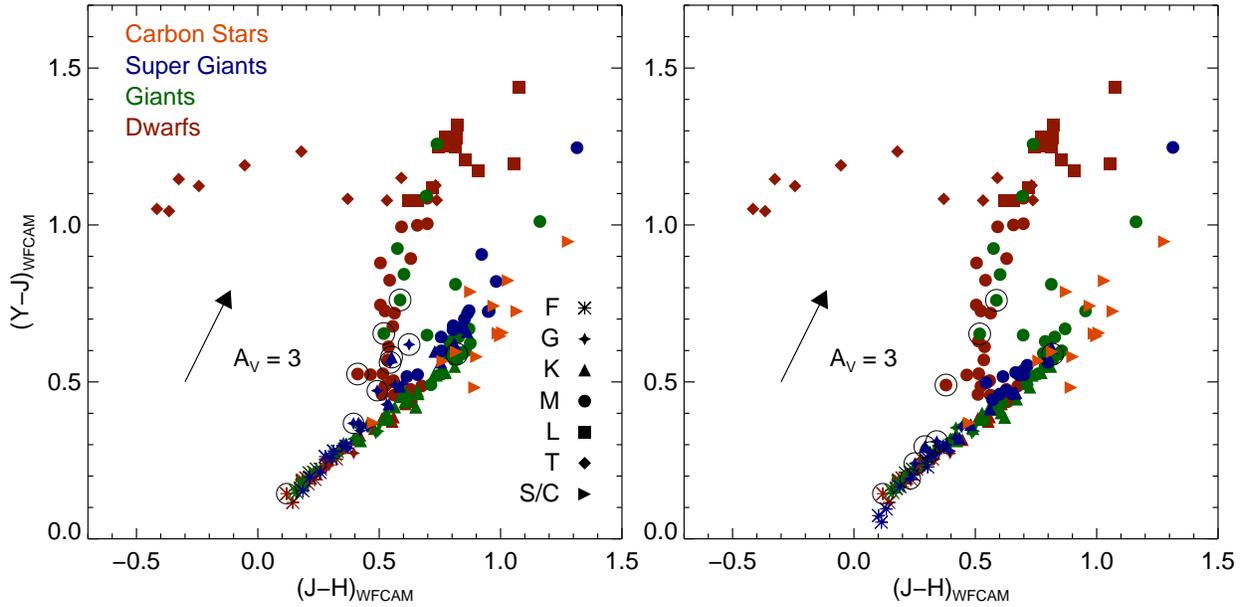}}
\vspace{-1.5in}
\caption{\label{fig:YJJH}\textit{Left:} Synthesized WFCAM $Y-J$ versus
  $J-H$ diagram for our sample of cool stars (see
  Table~\ref{tab:MKOColors}), plus additional L and T~dwarfs from
  \citet{2005ApJ...623.1115C}, and T dwarfs from
  \citet{2006ApJ...637.1067B}.  The plot illustrates the advantage of
  using a $Y$ filter when trying to identify T~dwarfs compared to using
  $JHK$ colors (see Figure~\ref{fig:JHHK}), and is an example of using
  the library to experiment with other photometric systems. Note the
  bifurcation between dwarfs and stars of higher luminosity beyond
  spectral type M0~V.  Several late-type M giants have similar colors to
  late-type dwarfs since both have water absorption in the
  NIR. \textit{Right:} The same plot but now with the correction for
  reddening included (the 77 dereddened stars are given in
  Table~\ref{tab:Extinction}).  Note how most M supergiant stars move
  closer to the main sequence.}
\end{figure}

\clearpage

\begin{deluxetable}{lccccc}
\tablecolumns{6}
\tabletypesize{\scriptsize} 
\tablewidth{0pc}
\tablecaption{\label{tab:History}Near-IR Spectral Libraries}
\tablehead{
\colhead{Spectral Library} & 
\colhead{$\lambda$} & 
\colhead{Spectral} & 
\colhead{Luminosity} & 
\colhead{Number of} & 
\colhead{Resolving} \\
\colhead{Reference} & 
\colhead{($\micron$)} & 
\colhead{Type} & 
\colhead{Class} & 
\colhead{Stars} & 
\colhead{Power}} 
\startdata
 
\citet{1970AJ.....75..785J} & 1.2$-$2.5 & A$-$M &I$-$V & 32 & 550  \\
\citet{1986ApJS...62..501K} & 2.0$-$2.5 & F$-$M &I$-$V & 26 & 2500$-$3100 \\
\citet{1986ApJS...62..373L} & 1.5$-$2.5 & C &          & 30 & 75,000 \\
\citet{1988ApJ...324.1071M}& 1.0$-$2.5 & Be, Ae& I    & 13 & 500 \\
\citet{1990PNAOJ...1..259T}& 1.5$-$2.5 & C &           & 33& 2000\\
\citet{1991ApJ...378..742T}& 0.45$-$2.45 & M   & III   & 32& 1000\\
\citet{1992AAS...96..593L} & 1.4$-$2.5 & O$-$M &I$-$V & 56 & 550 \\
\citet{1993AA...280..536O} & 1.5$-$1.7 & G$-$M &I$-$V & 40 & 1500 \\
\enddata
\end{deluxetable}

\clearpage
\LongTables
\begin{landscape}
\begin{deluxetable*}{llllllccr@{ $\pm$ }lr@{ $\pm$ }lr@{ $\pm$ }l}
\tablecolumns{14}
\setlength{\tabcolsep}{0.04in}
\tabletypesize{\scriptsize} 
\tablewidth{0pc}
\tablecaption{\label{tab:Sample}The Sample}
\tablehead{
\colhead{Object} & 
\colhead{HR} & 
\colhead{Other} & 
\colhead{Spectral} & 
\colhead{Ref.} & 
\colhead{Variability\tablenotemark{a}} & 
\colhead{$V$\tablenotemark{b}} & 
\colhead{$B-V$\tablenotemark{b}} & 
\multicolumn{2}{c}{{$J$}} &
\multicolumn{2}{c}{{$H$}} &
\multicolumn{2}{c}{{$K_s$}} \\
\colhead{} & 
\colhead{Number} & 
\colhead{Name} & 
\colhead{Type} & 
\colhead{} & 
\colhead{Type} & 
\colhead{(mag)} & 
\colhead{(mag)} & 
\multicolumn{2}{c}{{(mag)}} &
\multicolumn{2}{c}{{(mag)}} &
\multicolumn{2}{c}{{(mag)}} \\
\colhead{(1)} & 
\colhead{(2)} & 
\colhead{(3)} & 
\colhead{(4)} & 
\colhead{(5)} & 
\colhead{(6)} & 
\colhead{(7)} & 
\colhead{(8)} & 
\multicolumn{2}{c}{{(9)}} &
\multicolumn{2}{c}{{(10)}} &
\multicolumn{2}{c}{{(11)}}}
 
\startdata
 
HD 7927 & HR 382 & $\phi$ Cas & F0 Ia & 1 & var & 4.987 & 0.680 & 3.751 & 0.296 & 3.537 & 0.248 & 3.186 & 0.268 \\
HD 135153 & HR 5660 & i Lup & F0 Ib$-$II & 1 &  & 4.904 & 0.376 & 4.258 & 0.252 & 4.007 & 0.194 & 3.991 & 0.276 \\
HD 6130 & HR 292 &  & F0 II & 1 &  & 5.919 & 0.491 & 4.991 & 0.274 & 4.690 & 0.254 & 4.406 & 0.020 \\
HD 89025 & HR 4013 & $\zeta$ Leo & F0 IIIa & 1 & var & 3.435 & 0.307 & 2.700 & 0.232 & 2.628 & 0.208 & 2.631 & 0.280 \\
HD 13174 & HR 623 & 14 Ari & F0 III$-$IVn & 2 &  & 4.989 & 0.335 & 4.441 & 0.308 & 4.357 & 0.240 & 4.062 & 0.036 \\
HD 27397 & HR 1351 & h Tau & F0 IV & 1 & DSCTC & 5.593 & 0.283 & 5.002 & 0.037 & 4.927 & 0.047 & 4.853 & 0.015 \\
HD 108519 &  &  & F0 V(n) & 1 &  & 7.705 & 0.273 & 7.093 & 0.021 & 6.998 & 0.031 & 6.965 & 0.040 \\
HD 173638 & HR 7055 &  & F1 II & 1 & *I & 5.710 & 0.595 & 4.213 & 0.262 & 3.908 & 0.228 & 3.840 & 0.298 \\
\enddata
 
\tablerefs{(1) \citet{2001AJ....121.2148G}; (2) \citet{1989ApJS...69..301G}; (3) \cite{1951ApJ...113...60R}; (4) \citet{1955ApJS....2..195R}; (5) \citet{1981ApJS...45..437A}; (6) \citet{1976PASP...88...95C}; (7) \citet{1960ApJ...131..330K}; (8) \citet{1973ARAA..11...29M}; (9) \citet{1976AJ.....81..225M}; (10) \citet{2000webKN}; (11) \citet{1989ApJS...71..245K}; (12) \citet{1978BICDS..15..121J}; (13) \citet{1989BICDS..36...27G}; (14) \citet{1967AJ.....72.1334C}; (15) \citet{2001KFNT...17..409K}; (16) \citet{1972AJ.....77..486U}; (17) \citet{1991ApJS...77..417K}; (18) \citet{1994AJ....108.1437H}; (19) \citet{1998GCVS4.C......0K}; (20) \citet{1997AJ....113.1421K}; (21) \citet{1995AJ....109..797K}; (22) \citet{2001A&A...366L..13D}; (23) \citet{1999ApJ...519..802K}; (24) \citet{1979ApJ...234..538A}}
 
\tablenotetext{a}{Variability type from the General Catalog of Variable Stars \citep{1998GCVS4.C......0K}.  The variability type is described in \url{http://www.sai.msu.su/groups/cluster/gcvs/gcvs/iii/vartype.txt}.}
\tablenotetext{b}{$V$ and $B-V$ values are primarily from \citet{2006yCat.2168....0M} and are augmented with values from \citet{2001KFNT...17..409K}, \citet{1994ApJS...93..187B} and \citet{1992ApJS...82..351L}.}
\end{deluxetable*}

\clearpage
\end{landscape}

\begin{deluxetable}{ccccccc}
\tablecolumns{7}
\tabletypesize{\scriptsize}
\tablewidth{0pc}
\tablecaption{\label{tab:Composition}Spectral Composition of Library}
\tablehead{
\colhead{Spectral} &
\multicolumn{6}{c} {Luminosity Class} \\
\colhead{Type} &
\colhead{0} &
\colhead{I} &
\colhead{II} &
\colhead{III} &
\colhead{IV} &
\colhead {V}}

\startdata
 F   & 0 &  7 & 4 &  10 & 5 & 15 \\
 G   & 1 & 12 & 6 & 13 & 3 & 11 \\
 K   & 1 &  7 & 5 & 19 & 2 &  8 \\
 M   & 0 & 14 & 2 & 27 & 0 & 25 \\
 S   & $\cdots$  & $\cdots$   & $\cdots$  &  5 & $\cdots$  & $\cdots$   \\
 C   & $\cdots$  & $\cdots$  &  $\cdots$  &  8 & $\cdots$  &  $\cdots$  \\
\enddata
\end{deluxetable}

\clearpage
\LongTables
\begin{landscape}
\begin{deluxetable}{llclccll}
\tablecolumns{8}
\tabletypesize{\scriptsize}
\tablewidth{0pc}
\tablecaption{\label{tab:ObsLog}Log of SpeX Observations}
\tablehead{
\colhead{Object} & 
\colhead{Spectral} & 
\colhead{UT Date} & 
\colhead{Spectroscopy} & 
\colhead{$R$} & 
\colhead{Exp. Time} & 
\colhead{A0 V} & 
\colhead {Sky} \\
\colhead{} & 
\colhead{Type} &
\colhead{} & 
\colhead{Mode} &
\colhead{} & 
\colhead{(sec)} & 
\colhead{Standard} &
\colhead{Conditions} \\
\colhead{(1)} & 
\colhead{(2)} & 
\colhead{(3)} & 
\colhead{(4)} & 
\colhead{(5)} & 
\colhead{(6)} &
\colhead{(7)} &
\colhead {(8)}}

\startdata
HD 7927 & F0 Ia & 2005 Oct 06 & SXD & 2000 & \phn120 & HD 12365 & Clear\\ 
 &  & 2005 Oct 06 & LXD1.9 & 2500 & \phn150 & HD 12365 & Clear\\ 
HD 135153 & F0 Ib$-$II & 2007 Jun 24 & SXD & 2000 & \phn\phn96 & HD 146624 & Clear\\ 
 &  & 2007 Jun 24 & LXD2.1 & 2500 & \phn300 & HD 146624 & Clear\\ 
HD 6130 & F0 II & 2001 Sep 01 & SXD & 2000 & \phn100 & HD 12365 & Thin Cirrus\\ 
 &  & 2001 Sep 01 & LXD1.9 & 2500 & \phn200 & HD 12365 & Thin Cirrus\\ 
HD 89025 & F0 IIIa & 2007 Apr 25 & SXD & 2000 & \phn100 & HD 88960 & Clear\\ 
 &  & 2007 Apr 25 & LXD1.9 & 2500 & \phn200 & HD 88960 & Clear\\
\enddata
\end{deluxetable}

\clearpage
\end{landscape}

\begin{deluxetable}{llcccc}
\tablecolumns{6}
\tabletypesize{\scriptsize} 
\tablewidth{0pc}
\tablecaption{\label{tab:Extinction}Stars with $E(B-V)$ $>$ 0.075.}
\tablehead{
\colhead{Object} & 
\colhead{Spectral} & 
\colhead{$(B-V)$} &
\colhead{$(B-V)_0$} &
\colhead{$E(B-V)$} &
\colhead{$A_V$} \\
\colhead{} & 
\colhead{Type} & 
\colhead{(mag)} &
\colhead{(mag)} &
\colhead{(mag)} &
\colhead{(mag)} \\
\colhead{(1)} & 
\colhead{(2)} & 
\colhead{(3)} & 
\colhead{(4)} & 
\colhead{(5)} & 
\colhead{(6)}}
 
\startdata
 
HD 6130 & F0 II &  0.491  &  0.200  & 0.291 & 0.9 \\
HD 7927 & F0 Ia &  0.680  &  0.150  & 0.530 & 1.6 \\
HD 135153 & F0 Ib$-$II &  0.376  &  0.180  & 0.196 & 0.6 \\
HD 173638 & F1 II &  0.595  &  0.220  & 0.375 & 1.1 \\
HD 182835 & F2 Ib &  0.593  &  0.180  & 0.413 & 1.2 \\
HD 16232 & F4 V &  0.510  &  0.420  & 0.090 & 0.3 \\
HD 213306 & F5 Ib $-$ G1 Ib &  0.480  &  0.260  & 0.220 & 0.7 \\
HD 160365 & F6 III$-$IV &  0.556  &  0.460  & 0.096 & 0.3 \\
\enddata
\end{deluxetable}

\clearpage

\begin{deluxetable}{llllllll}
\tablecolumns{11}
\tabletypesize{\scriptsize} 
\tablewidth{0pc}
\tablecaption{\label{tab:Solar Atomic Lines}Strong Metal Lines in the Solar Spectrum}
\tablehead{
\colhead{Vacuum} & 
\colhead{Element} & 
\colhead{Vacuum} & 
\colhead{Element} & 
\colhead{Vacuum} & 
\colhead{Element} & 
\colhead{Vacuum} & 
\colhead{Element} \\ 
\colhead{Wavelength ($\mu$m)} & 
\colhead{} & 
\colhead{Wavelength ($\mu$m)} & 
\colhead{} & 
\colhead{Wavelength ($\mu$m)} & 
\colhead{} & 
\colhead{Wavelength ($\mu$m)} & 
\colhead{}}
 
\startdata
 
0.8185503 & \ion{ Na }{1} & 0.8189049 & \ion{ Fe }{1} & 0.8197077 & \ion{ Na }{1} & 0.8201178 & \ion{ Fe }{1} \\
0.8203977 & \ion{ Ca }{2} & 0.8209999 & \ion{ Fe }{1} & 0.8215293 & \ion{ Mg }{1} & 0.8222636 & \ion{ Fe }{1} \\
0.8234587 & \ion{ Fe }{1} & 0.8241394 & \ion{ Fe }{1} & 0.8250417 & \ion{ Fe }{1} & 0.8251063 & \ion{ Ca }{2} \\
0.8278187 & \ion{ Fe }{1} & 0.8295802 & \ion{ Fe }{1} & 0.8312545 & \ion{ Mg }{1} & 0.8329342 & \ion{ Fe }{1} \\
0.8334208 & \ion{ Fe }{1} & 0.8337440 & \ion{ C }{1} & 0.8341703 & \ion{ Fe }{1} & 0.8348416 & \ion{ Mg }{1} \\
0.8363099 & \ion{ Fe }{1} & 0.8367942 & \ion{ Fe }{1} & 0.8380165 & \ion{ Ti }{1} & 0.8384830 & \ion{ Ti }{1} \\
0.8390078 & \ion{ Fe }{1} & 0.8403729 & \ion{ Fe }{1} & 0.8414668 & \ion{ Ti }{1} & 0.8426453 & \ion{ Fe }{1} \\
0.8428818 & \ion{ Ti }{1} & 0.8437274 & \ion{ Ti }{1} & 0.8437971 & \ion{ Ti }{1} & 0.8441896 & \ion{ Fe }{1} \\
\enddata
\end{deluxetable}

\clearpage

\begin{deluxetable}{llllllll}
\tablecolumns{11}
\tabletypesize{\scriptsize} 
\tablewidth{0pc}
\tablecaption{\label{tab:Arcturus Atomic Lines}Strong Metal Lines in the Arcturus Spectrum}
\tablehead{
\colhead{Vacuum} & 
\colhead{Element} & 
\colhead{Vacuum} & 
\colhead{Element} & 
\colhead{Vacuum} & 
\colhead{Element} & 
\colhead{Vacuum} & 
\colhead{Element} \\ 
\colhead{Wavelength ($\mu$m)} & 
\colhead{} & 
\colhead{Wavelength ($\mu$m)} & 
\colhead{} & 
\colhead{Wavelength ($\mu$m)} & 
\colhead{} & 
\colhead{Wavelength ($\mu$m)} & 
\colhead{}}
 
\startdata
 
0.8027028 & \ion{Ti}{1} & 0.8030128 & \ion{Fe}{1} & 0.8030499 & \ion{Fe}{1} & 0.8048243 & \ion{Fe}{1} \\
0.8049814 & \ion{Fe}{1} & 0.8056425 & \ion{Mg}{1} & 0.8070439 & \ion{Ti}{1} & 0.8072310 & \ion{Zr}{1} \\
0.8074360 & \ion{Fe}{1} & 0.8077351 & \ion{Fe}{1} & 0.8082752 & \ion{Fe}{1} & 0.8087374 & \ion{Fe}{1} \\
0.8094836 & \ion{Cu}{1} & 0.8095446 & \ion{Si}{1} & 0.8095706 & \ion{V}{1} & 0.8096137 & \ion{Co}{1} \\
0.8099087 & \ion{Fe}{1} & 0.8100928 & \ion{Mg}{1} & 0.8110530 & \ion{Fe}{1} & 0.8114371 & \ion{Fe}{1} \\
0.8118973 & \ion{V}{1} & 0.8135217 & \ion{Zr}{1} & 0.8146790 & \ion{V}{1} & 0.8163294 & \ion{V}{1} \\
0.8185480 & \ion{Na}{1} & 0.8189032 & \ion{Fe}{1} & 0.8197064 & \ion{Na}{1} & 0.8201155 & \ion{Fe}{1} \\
0.8206336 & \ion{Fe}{1} & 0.8207177 & \ion{Fe}{1} & 0.8209977 & \ion{Fe}{1} & 0.8215279 & \ion{Mg}{1} \\
\enddata
\end{deluxetable}

\clearpage

\begin{deluxetable*}{lccc}
\tablecolumns{11}
\tabletypesize{\scriptsize} 
\tablewidth{0pc}
\tablecaption{\label{tab:EWs}Equivalent Width Limit Definitions}
\tablehead{
\colhead{Feature} & 
\colhead{Feature Limits} & 
\colhead{First Continuum Level} & 
\colhead{Second Continuum Level} \\ 
\colhead{} & 
\colhead{($\mu$m)} & 
\colhead{Limits ($\mu$m)} &
\colhead{Limits ($\mu$m)}}
 
\startdata
 
\ion{Ca}{2} (0.866 $\mu$m) & 0.860$-$0.875  & 0.862$-$0.864  & 0.870$-$0.873 \\
\ion{Na}{1} (1.14 $\mu$m)  & 1.120$-$1.160  & 1.125$-$1.130  & 1.150$-$1.160 \\
\ion{Al}{1} (1.313 $\mu$m) & 1.300$-$1.330  & 1.305$-$1.309  & 1.320$-$1.325 \\
\ion{Mg}{1} (1.485 $\mu$m) & 1.475$-$1.4975 & 1.4775$-$1.485 & 1.491$-$1.497 \\
\ion{Mg}{1} (1.711 $\mu$m) & 1.695$-$1.726  & 1.702$-$1.708  & 1.715$-$1.720 \\
\ion{Na}{1} (2.206 $\mu$m) & 2.185$-$2.230  & 2.192$-$2.198  & 2.213$-$2.220 \\

\enddata
\end{deluxetable*}

\clearpage

\LongTables
\begin{landscape}
\begin{deluxetable}{llr@{ $\pm$ }lr@{ $\pm$ }lr@{ $\pm$ }lr@{ $\pm$ }lr@{ $\pm$ }lr@{ $\pm$ }l}
\tablecolumns{14}
\setlength{\tabcolsep}{0.04in}
\tabletypesize{\scriptsize} 
\tablewidth{0pc}
\tablecaption{\label{tab:EW Values}EWs}
\tablehead{
\colhead{Object} & 
\colhead{Spectral} & 
\multicolumn{2}{c}{\ion{Ca}{2}} &
\multicolumn{2}{c}{\ion{Na}{1}} &
\multicolumn{2}{c}{\ion{Al}{1}} &
\multicolumn{2}{c}{\ion{Mg}{1}} &
\multicolumn{2}{c}{\ion{Mg}{1}} &
\multicolumn{2}{c}{\ion{Na}{1}} \\
\colhead{} & 
\colhead{Type} & 
\multicolumn{2}{c}{{(0.866 $\mu$m)}} &
\multicolumn{2}{c}{{(1.14 $\mu$m)}} &
\multicolumn{2}{c}{{(1.313 $\mu$m)}} &
\multicolumn{2}{c}{{(1.485 $\mu$m)}} &
\multicolumn{2}{c}{{(1.711 $\mu$m)}} &
\multicolumn{2}{c}{{(2.206 $\mu$m)}}}
 
\startdata
 
HD 7927 & F0 Ia &  3.73 &  0.12 &  0.13 &  0.36 &  0.32 &  0.03 &  0.59 &  0.18 &  0.03 &  0.03 & $-$0.12 &  0.04 \\
HD 135153 & F0 Ib$-$II &  4.38 &  0.07 &  0.45 &  0.41 &  0.48 &  0.12 &  0.60 &  0.28 &  0.20 &  0.05 &  0.28 &  0.10 \\
HD 6130 & F0 II &  4.06 &  0.08 &  0.66 &  0.25 &  0.49 &  0.05 &  0.53 &  0.10 &  0.20 &  0.04 &  0.31 &  0.05 \\
HD 89025 & F0 IIIa &  3.28 &  0.03 &  1.01 &  0.14 &  0.38 &  0.03 &  0.48 &  0.07 &  0.24 &  0.04 &  0.27 &  0.03 \\
HD 13174 & F0 III$-$IVn &  3.05 &  0.04 &  0.99 &  0.17 &  0.61 &  0.05 &  0.62 &  0.12 &  0.37 &  0.03 &  0.50 &  0.04 \\
HD 27397 & F0 IV &  2.76 &  0.04 &  0.45 &  0.23 &  0.41 &  0.05 &  0.63 &  0.09 &  0.32 &  0.03 &  0.50 &  0.05 \\
HD 108519 & F0 V(n) &  2.56 &  0.03 &  0.58 &  0.17 &  0.37 &  0.03 &  0.52 &  0.06 &  0.31 &  0.03 &  0.41 &  0.04 \\
HD 173638 & F1 II &  4.48 &  0.09 &  0.57 &  0.30 &  0.46 &  0.03 &  0.41 &  0.08 &  0.15 &  0.05 &  0.09 &  0.06 \\
\enddata
 
\end{deluxetable}

\clearpage
\end{landscape}

\clearpage

\begin{deluxetable}{lll}  
\tablecolumns{14}
\tabletypesize{\scriptsize} 
\tablewidth{0pc}
\tablecaption{\label{tab:MolecularBands}Molecular Features in FGKM Spectra}
\tablehead{
\colhead{Wavelength ($\mu$m)} &
\colhead{Transition} &
\colhead{Ref.} \\
\colhead{(1)} &
\colhead{(2)} &
\colhead{(3)}} 

\startdata
\multicolumn{3}{c}{TiO} \\
\hline
0.82009172 $R$ head       & 1$-$0 band of $\delta$ ($b$ $^1$$\Pi$ $-$ $a$ $^1$$\Delta$)                                        & 1 \\
0.82081 $R_3$ head        & 0$-$2 band of $\gamma$ ($A$ $^3\Phi$$-$ $X$ $^3\Delta$)                                            & 2 \\
0.82523 $R_2$ head        & 0$-$2 band of $\gamma$ ($A$ $^3\Phi$$-$ $X$ $^3\Delta$)                                            & 2 \\
0.82720357 $R$ head       & 2$-$1 band of $\delta$ ($b$ $^1$$\Pi$ $-$ $a$ $^1$$\Delta$)                                        & 1 \\
0.82913 $R_3$ head        & 1$-$3 band of $\gamma$ ($A$ $^3\Phi$$-$ $X$ $^3\Delta$)                                            & 2 \\
0.83052 $R_1$ head        & 0$-$2 band of $\gamma$ ($A$ $^3\Phi$$-$ $X$ $^3\Delta$)                                            & 2 \\
0.83368 $R_2$ head        & 1$-$3 band of $\gamma$ ($A$ $^3\Phi$$-$ $X$ $^3\Delta$)                                            & 2 \\
\enddata        

\tablerefs{(1) \citet{1998ApJ...498..851V}; (2) \citet{1957msmo.book.....G}; (3) \citet{1977JMoSp..64..382L}; (4) \citet{1950ApJ...111..314P}; (5) \citet{1973ApJ...180..845L}; (6) \citet{1980ApJS...42..241G}; (7) \citet{1965ims..book.....P}; (8) \citet{1976ims..book.....P}; (9) \citet{1948ApJ...108..163H}; (10) \citet{1971Brock}; (11) \citet{1987ApJS...65..721P}; (12) \citet{2001ApJ...559..424W}; (13) \citet{1967ApJS...14..171A}; (14) \citet{1979ApJ...234..401P}; (15) \cite{lagerqvist65}; (16) \cite{1982JMoSp..92..391C}; (17) \citet{1963ApJ...137...61B}; (18) \citet{1994ApJS...95..535G}; (19) \citet{1995iaas.book.....H}; (20) \cite{1974PASP...86..806B}}

\end{deluxetable}

\clearpage

\begin{deluxetable}{llccccc}
\tablecolumns{7}
\tabletypesize{\scriptsize} 
\tablewidth{0pc}
\tablecaption{\label{tab:MKOColors}Synthetic WFCAM Colors}
\tablehead{
\colhead{Object} & 
\colhead{Spectral} & 
\colhead{$Z-Y$} &
\colhead{$Y-J$} &
\colhead{$J-H$} &
\colhead{$H-K$} &
\colhead{$K-L'$} \\
\colhead{} & 
\colhead{Type} & 
\colhead{(mag)} &
\colhead{(mag)} &
\colhead{(mag)} &
\colhead{(mag)} &
\colhead{(mag)} \\
\colhead{(1)} & 
\colhead{(2)} & 
\colhead{(3)} & 
\colhead{(4)} & 
\colhead{(5)} & 
\colhead{(6)} & 
\colhead{(7)}}
 
\startdata
 
HD 7927 & F0 Ia & $+$0.161 & $+$0.211 & $+$0.257 & $+$0.152 & $+$0.208 \\
HD 135153 & F0 Ib$-$II & $+$0.113 & $+$0.153 & $+$0.186 & $+$0.087 & $+$0.044 \\
HD 6130 & F0 II & $+$0.147 & $+$0.193 & $+$0.225 & $+$0.090 & $+$0.128 \\
HD 89025 & F0 IIIa & $+$0.052 & $+$0.147 & $+$0.161 & $+$0.033 & $+$0.058 \\
HD 13174 & F0 III$-$IVn & $+$0.066 & $+$0.155 & $+$0.153 & $+$0.018 & $+$0.053 \\
HD 27397 & F0 IV & $+$0.058 & $+$0.115 & $+$0.107 & $-$0.011 & $-$0.018 \\
HD 108519 & F0 V(n) & $+$0.043 & $+$0.116 & $+$0.144 & $+$0.044 & $+$0.018 \\
HD 173638 & F1 II & $+$0.187 & $+$0.200 & $+$0.222 & $+$0.083 & $+$0.088 \\
\enddata
\end{deluxetable}

\clearpage

\appendix

\section{Spectral Sequences}
\begin{figure}[h]
\centerline{\includegraphics[width=6in,angle=0]{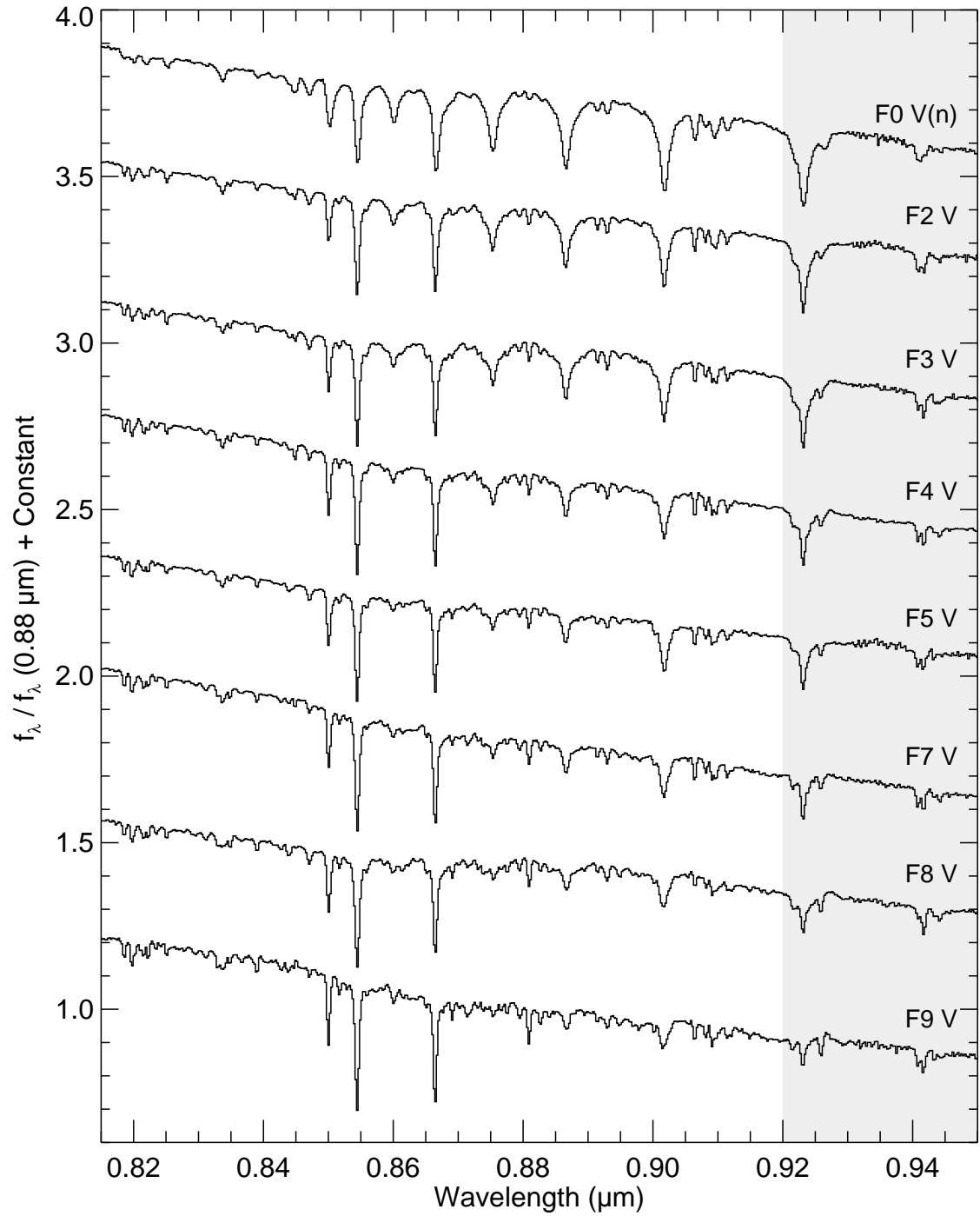}}
\caption{\label{fig:F_VI} A sequence of F dwarf stars plotted over the
  $I$ band (0.82$-$0.95~$\mu$m).  The spectra are of HD~108519 F0~V(n),
  HD 113139 (F2~V), HD~26015 (F3~V), HD~87822 (F4~V), HD~218804 (F5~V),
  HD~126660 (F7~V), HD~27393 (F8~V), and HD~176051 (F9~V).  The spectra
  have been normalized to unity at 0.88~$\mu$m and offset by constants.}
\end{figure}

\begin{figure}
\centerline{\includegraphics[width=6.0in,angle=0]{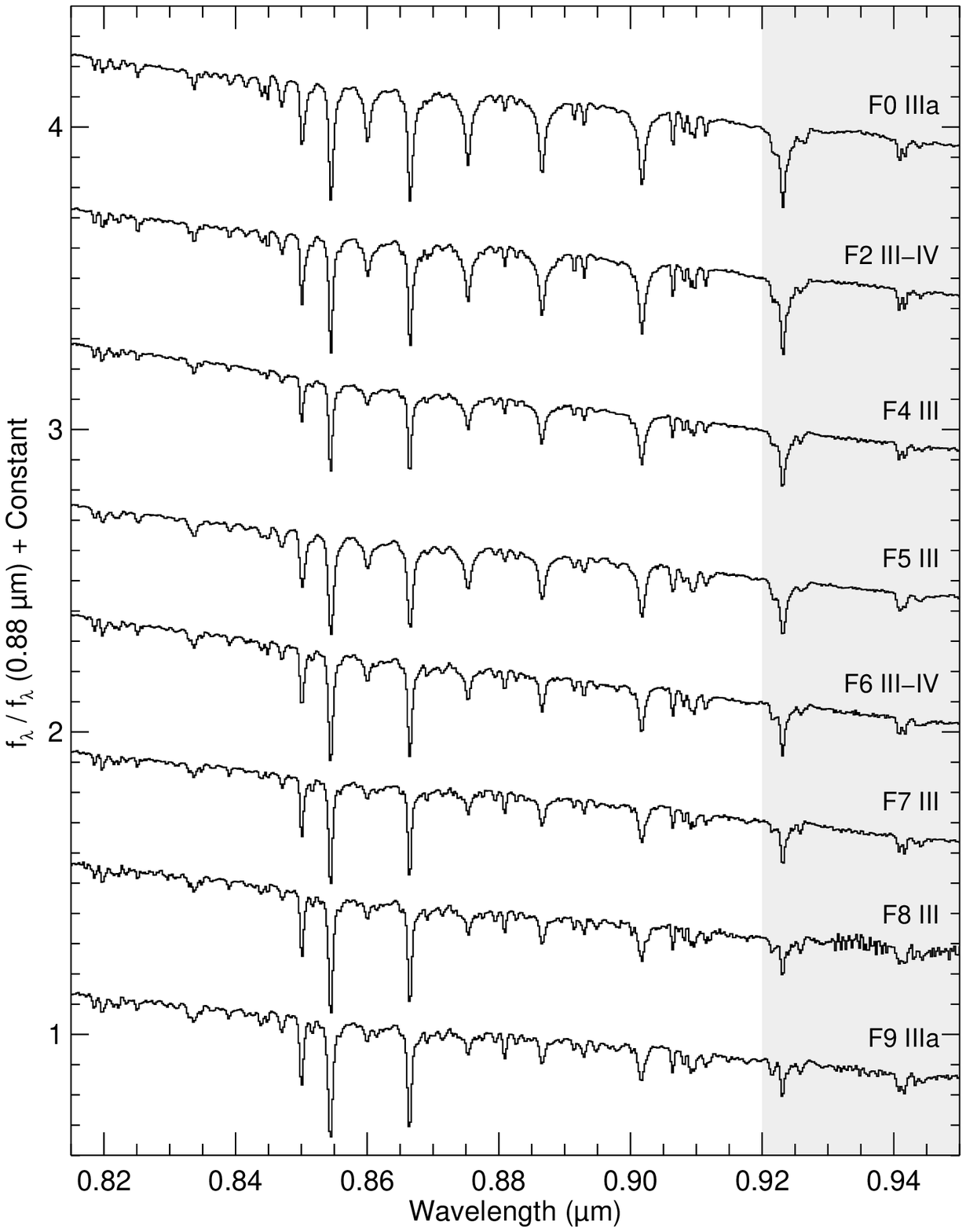}}
\caption{\label{fig:F_IIII} A sequence of F giant stars plotted over the
  $I$ band (0.82$-$0.95~$\mu$m).  The spectra are of HD~89025 (F0~IIIa),
  HD~40535 (F2~III-IV), HD~21770 (F4~III), HD~17918 (F5~III), HD~160365
  (F6~III-IV), HD~124850 (F7~III), HD~220657 (F8~III), and HD~6903
  (F9~IIIa).  The spectra have been normalized to unity at 0.88~$\mu$m
  and offset by constants.}
\end{figure}

\begin{figure}
\centerline{\includegraphics[width=6.0in,angle=0]{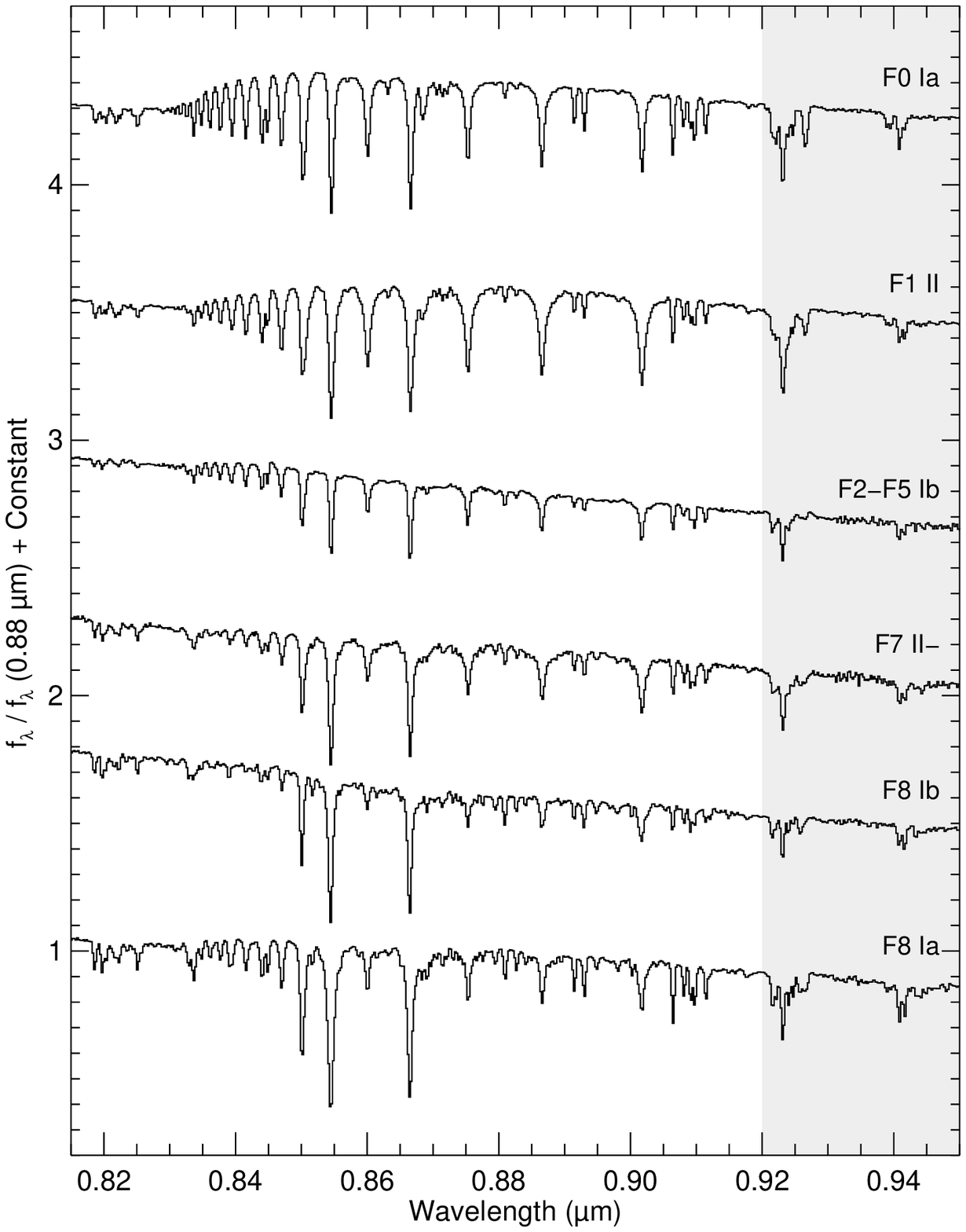}}
\caption{\label{fig:F_II} A sequence of F supergiant stars plotted over
  the $I$ band (0.82$-$0.95~$\mu$m).  The spectra are of HD~7927
  (F0~Ia), HD~173638 (F1~II), BD~+38~2803 (F2-F5~Ib), HD~201078 (F7 II$-$),
  HD~51956 (F8~Ib) and HD~190323 (F8~Ia).  The spectra have been
  normalized to unity at 0.88~$\mu$m and offset by constants.}
\end{figure}

\begin{figure}
\centerline{\includegraphics[width=6.0in,angle=0]{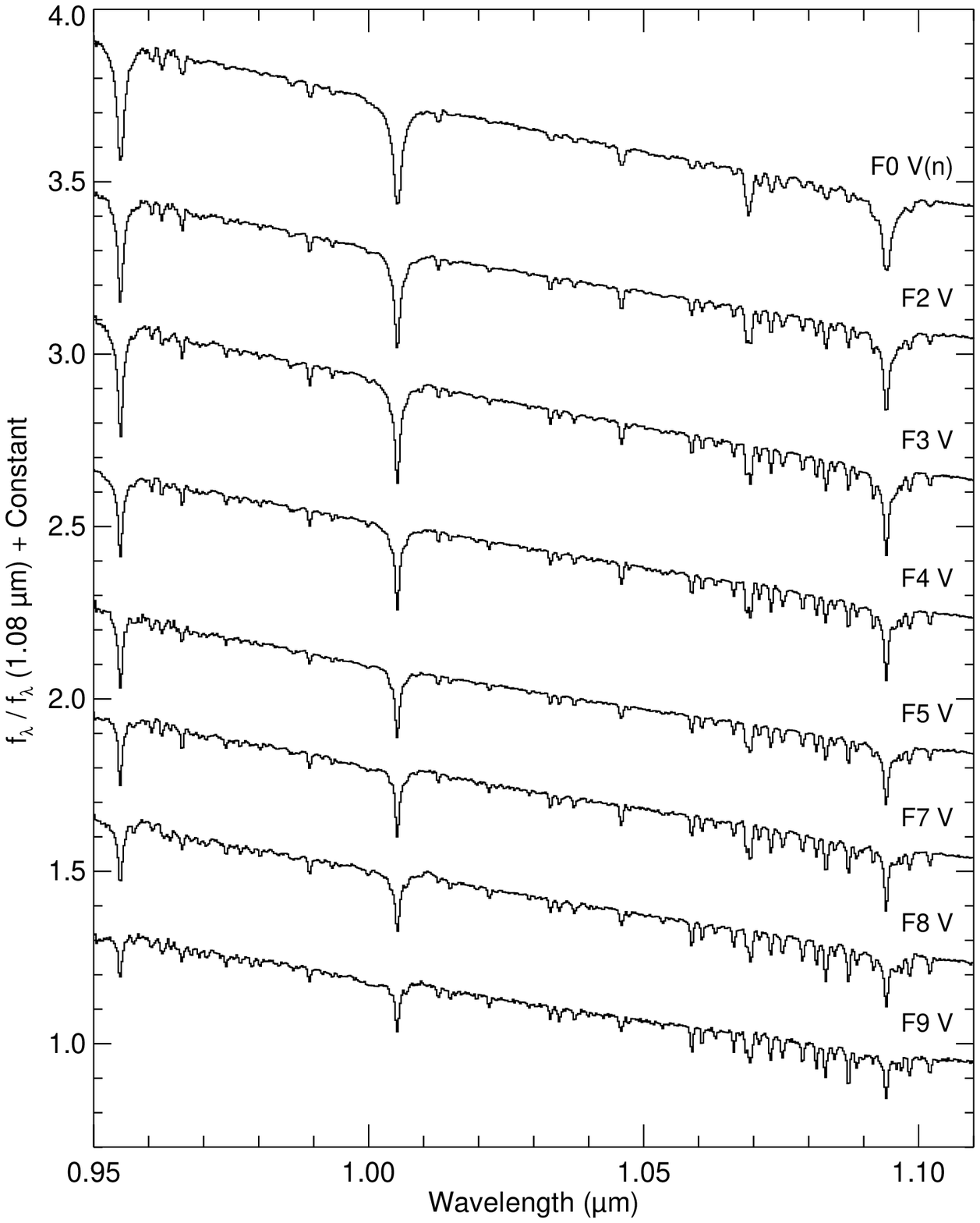}}
\caption{\label{fig:F_VY} A sequence of F dwarf stars plotted over the
  $Y$ band (0.95$-$1.10~$\mu$m).  The spectra are of HD~108519 F0~V(n),
  HD 113139 (F2~V), HD~26015 (F3~V), HD~87822 (F4~V), HD~218804 (F5~V),
  HD~126660 (F7~V), HD~27393 (F8~V), and HD~176051 (F9~V).  The spectra
  have been normalized to unity at 1.08~$\mu$m and offset by constants.}
\end{figure}

\begin{figure}
\centerline{\includegraphics[width=6.0in,angle=0]{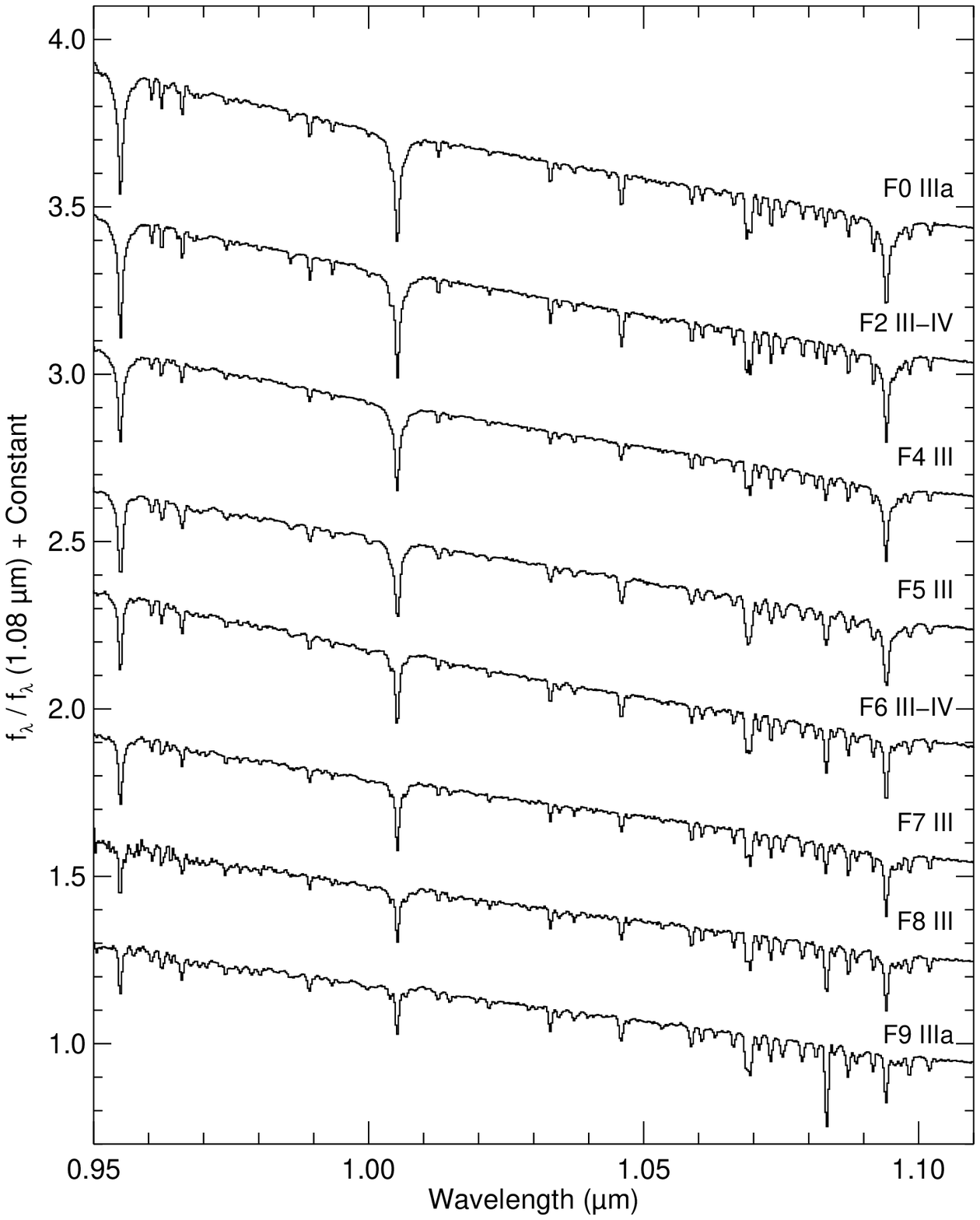}}
\caption{\label{fig:F_IIIY} A sequence of F giant stars plotted over the
  $Y$ band (0.95$-$1.10~$\mu$m).  The spectra are of HD~89025 (F0~IIIa),
  HD~40535 (F2~III-IV), HD~21770 (F4~III), HD~17918 (F5~III), HD~160365
  (F6~III-IV), HD~124850 (F7~III), HD~220657 (F8~III), and HD~6903
  (F9~IIIa).  The spectra have been normalized to unity at 1.08~$\mu$m
  and offset by constants.}
\end{figure}

\begin{figure}
\centerline{\includegraphics[width=6.0in,angle=0]{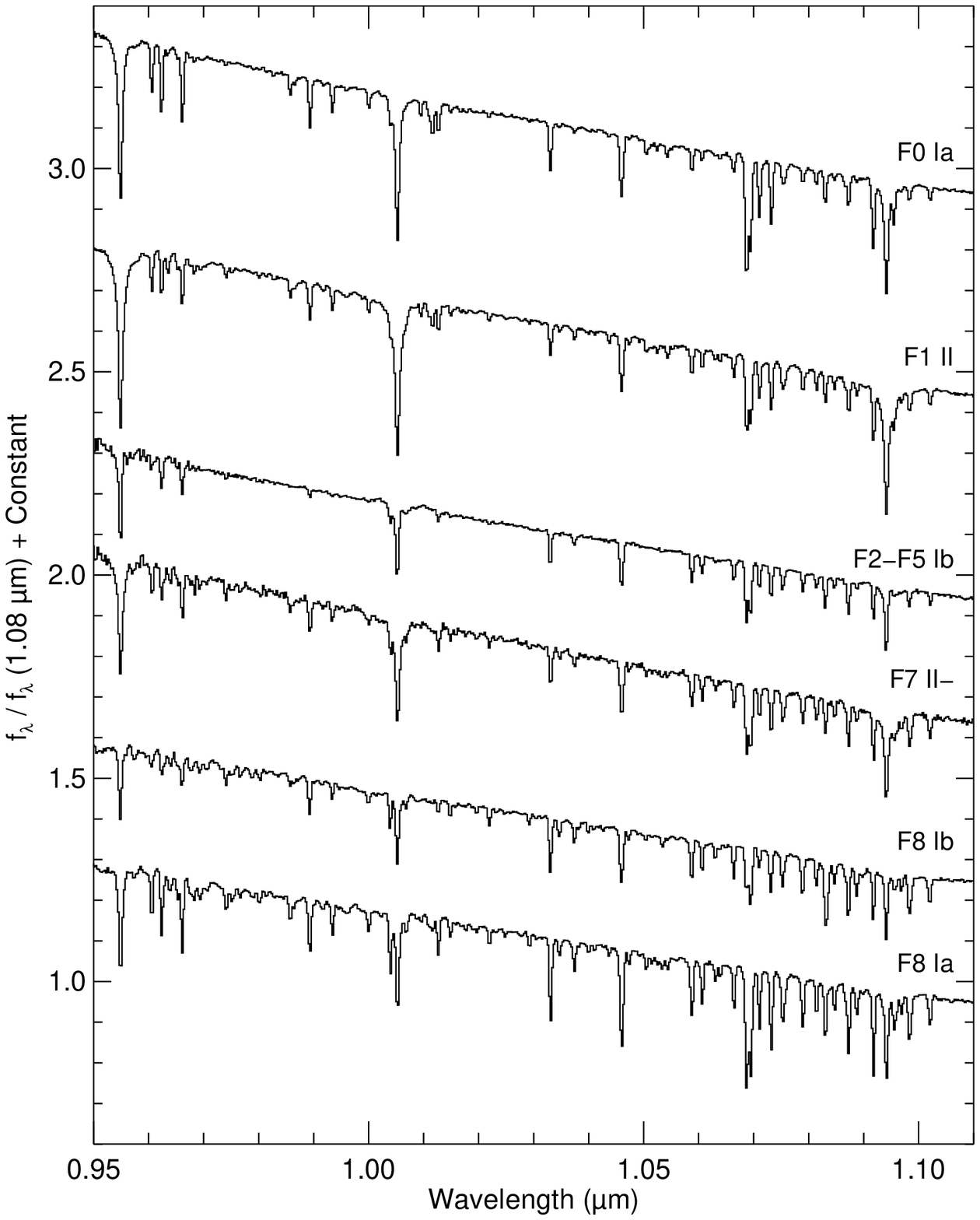}}
\caption{\label{fig:F_IY} A sequence of F supergiant stars plotted over
  the $Y$ band (0.95$-$1.10~$\mu$m).  The spectra are of HD~7927
  (F0~Ia), HD~173638 (F1~II), BD~+38~2803 (F2-F5~Ib), HD~51956 (F6~IbII),
  HD~201078 (F7 II-), and HD~190323 (F8~Ia).  The spectra have been
  normalized to unity at 1.08~$\mu$m and offset by constants.}
\end{figure}

\begin{figure}
\centerline{\includegraphics[width=6.0in,angle=0]{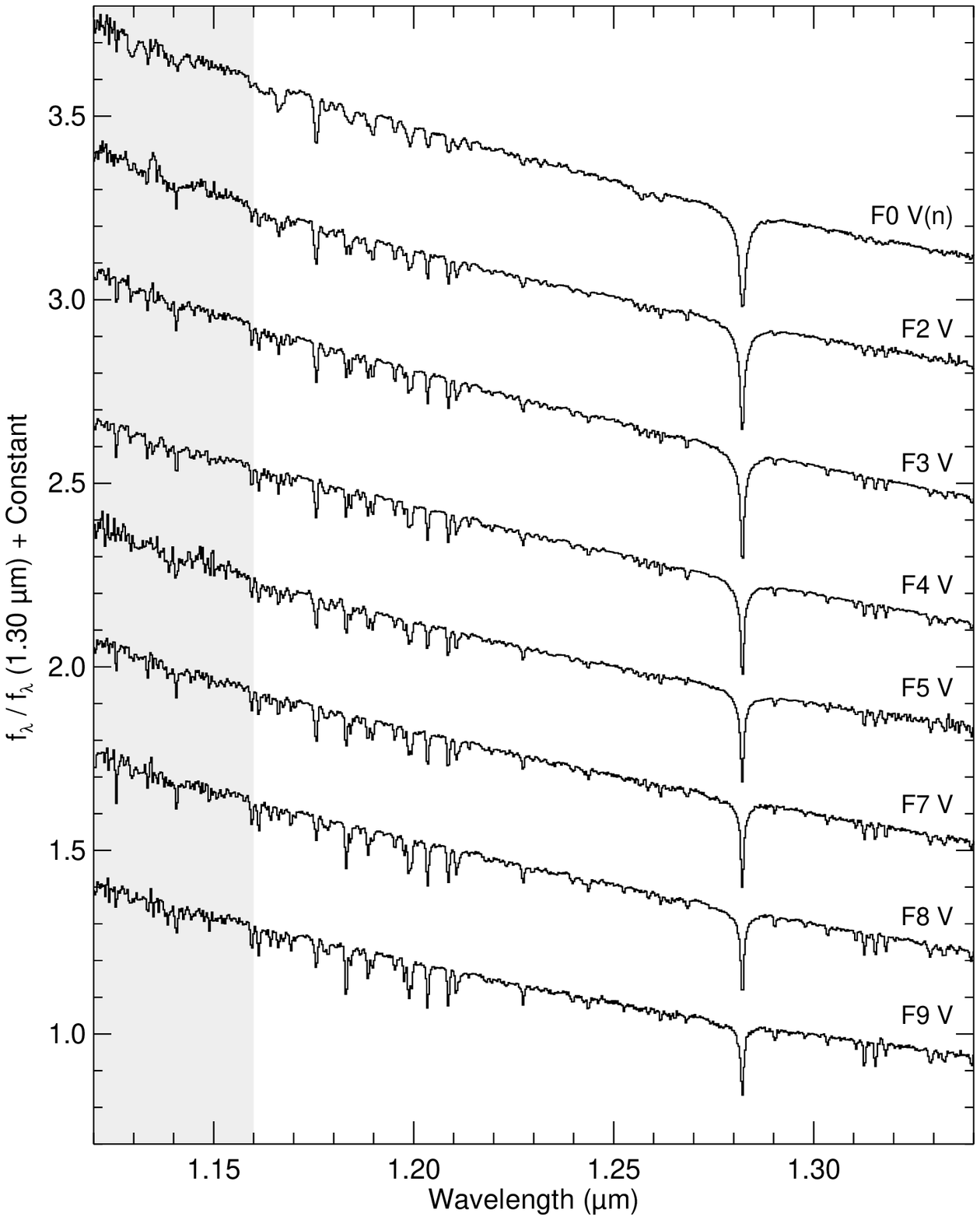}}
\caption{\label{fig:F_VJ} A sequence of F dwarf stars plotted over the
  $J$ band (1.12$-$1.34~$\mu$m).  The spectra are of HD~108519 F0~V(n),
  HD 113139 (F2~V), HD~26015 (F3~V), HD~87822 (F4~V), HD~218804 (F5~V),
  HD~126660 (F7~V), HD~27393 (F8~V), and HD~176051 (F9~V).  The spectra
  have been normalized to unity at 1.30~$\mu$m and offset by constants.}
\end{figure}

\begin{figure}
\centerline{\includegraphics[width=6.0in,angle=0]{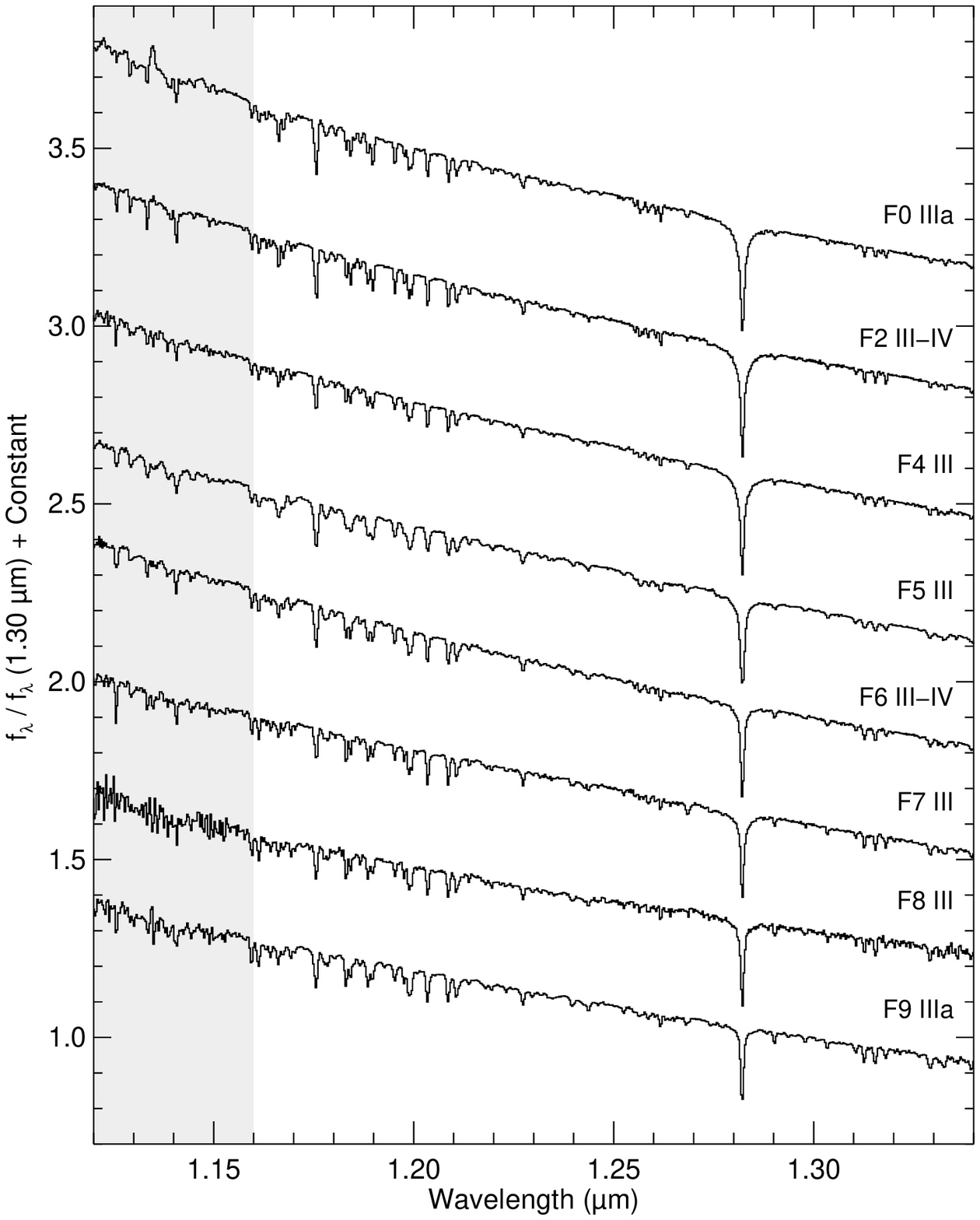}}
\caption{\label{fig:F_IIIJ} A sequence of F giant stars plotted over the
  $J$ band (1.12$-$1.34~$\mu$m).  The spectra are of HD~89025 (F0~IIIa),
  HD~40535 (F2~III-IV), HD~21770 (F4~III), HD~17918 (F5~III), HD~160365
  (F6~III-IV), HD~124850 (F7~III), HD~220657 (F8~III), HD~6903 (F9~IIIa).  The spectra
  have been normalized to unity at 1.30~$\mu$m and offset by constants.}
\end{figure}

\begin{figure}
\centerline{\includegraphics[width=6.0in,angle=0]{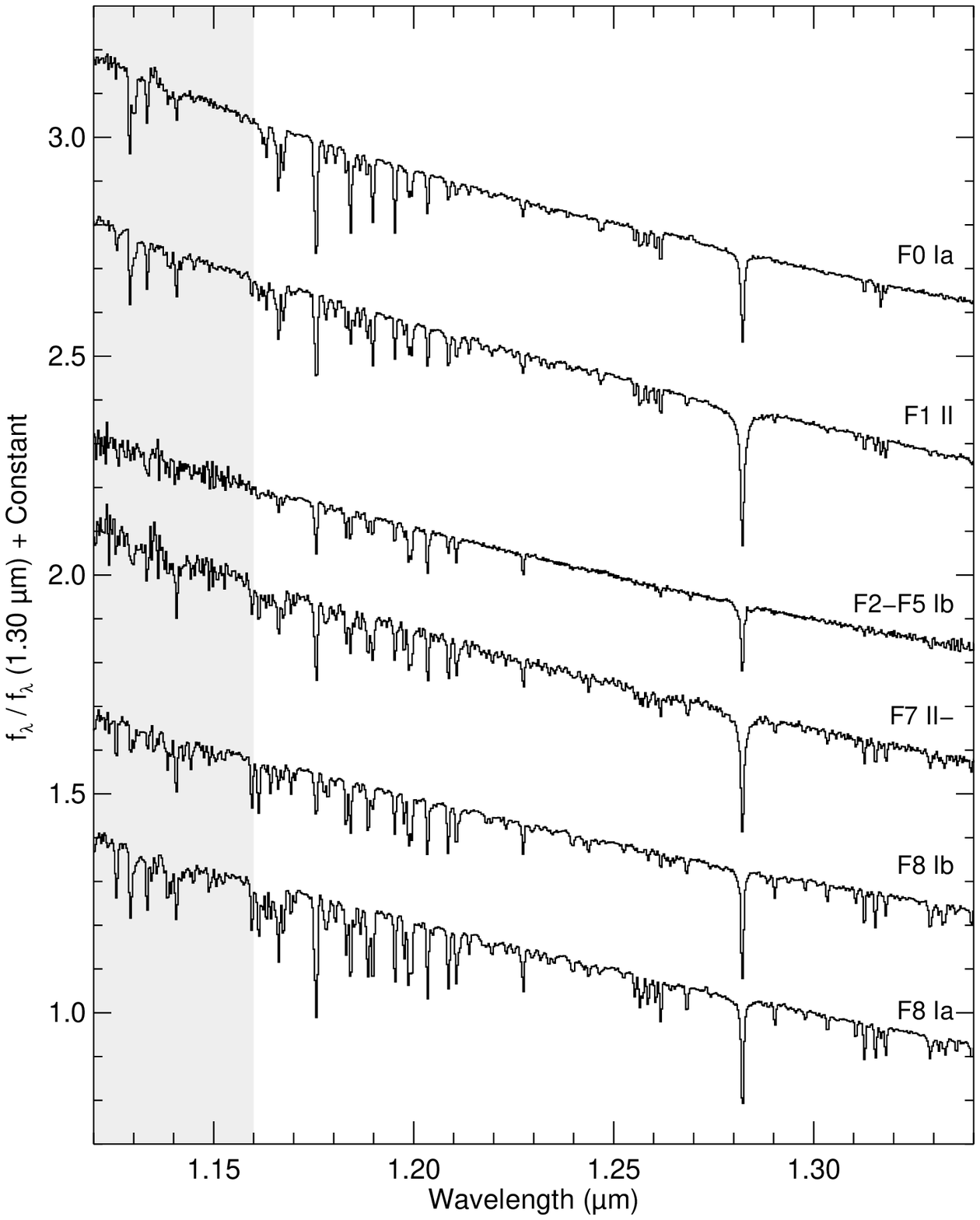}}
\caption{\label{fig:F_IJ} A sequence of F supergiant stars plotted over
  the $J$ band (1.12$-$1.34~$\mu$m).  The spectra are of HD~7927
  (F0~Ia), HD~173638 (F1~II), BD~+38~2803 (F2-F5~Ib), HD~201078 (F7 II$-$),
  HD~51956 (F8~Ib) and HD~190323 (F8~Ia).  The spectra have been
  normalized to unity at 1.30~$\mu$m and offset by constants.}
\end{figure}

\begin{figure}
\centerline{\includegraphics[width=6.0in,angle=0]{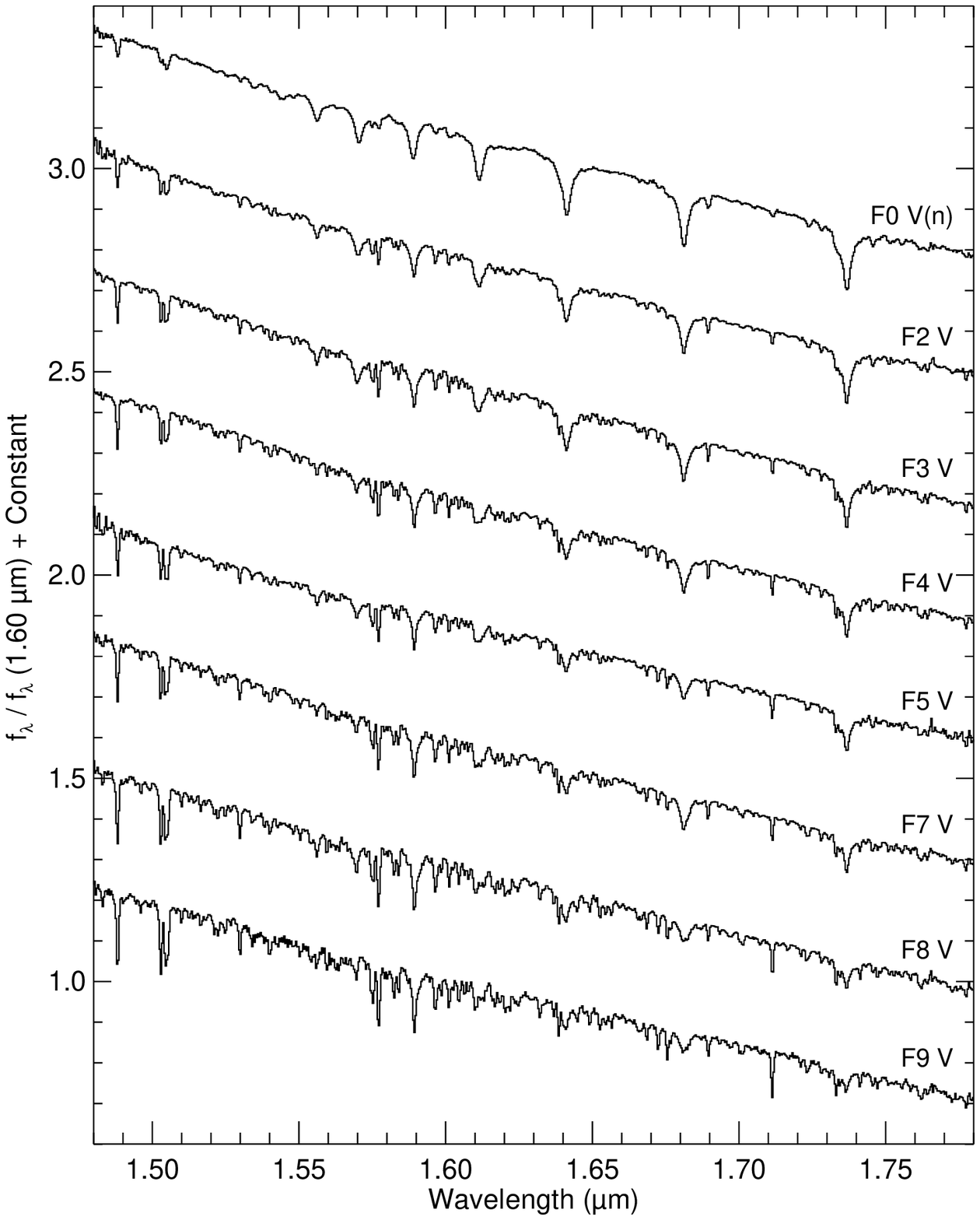}}
\caption{\label{fig:F_VH} A sequence of F dwarf stars plotted over the
  $H$ band (1.78$-$1.78~$\mu$m).  The spectra are of HD~108519 F0~V(n),
  HD 113139 (F2~V), HD~26015 (F3~V), HD~87822 (F4~V), HD~218804 (F5~V),
  HD~126660 (F7~V), HD~27393 (F8~V), and HD~176051 (F9~V).  The spectra
  have been normalized to unity at 1.60~$\mu$m and offset by constants.}
\end{figure}

\begin{figure}
\centerline{\includegraphics[width=6.0in,angle=0]{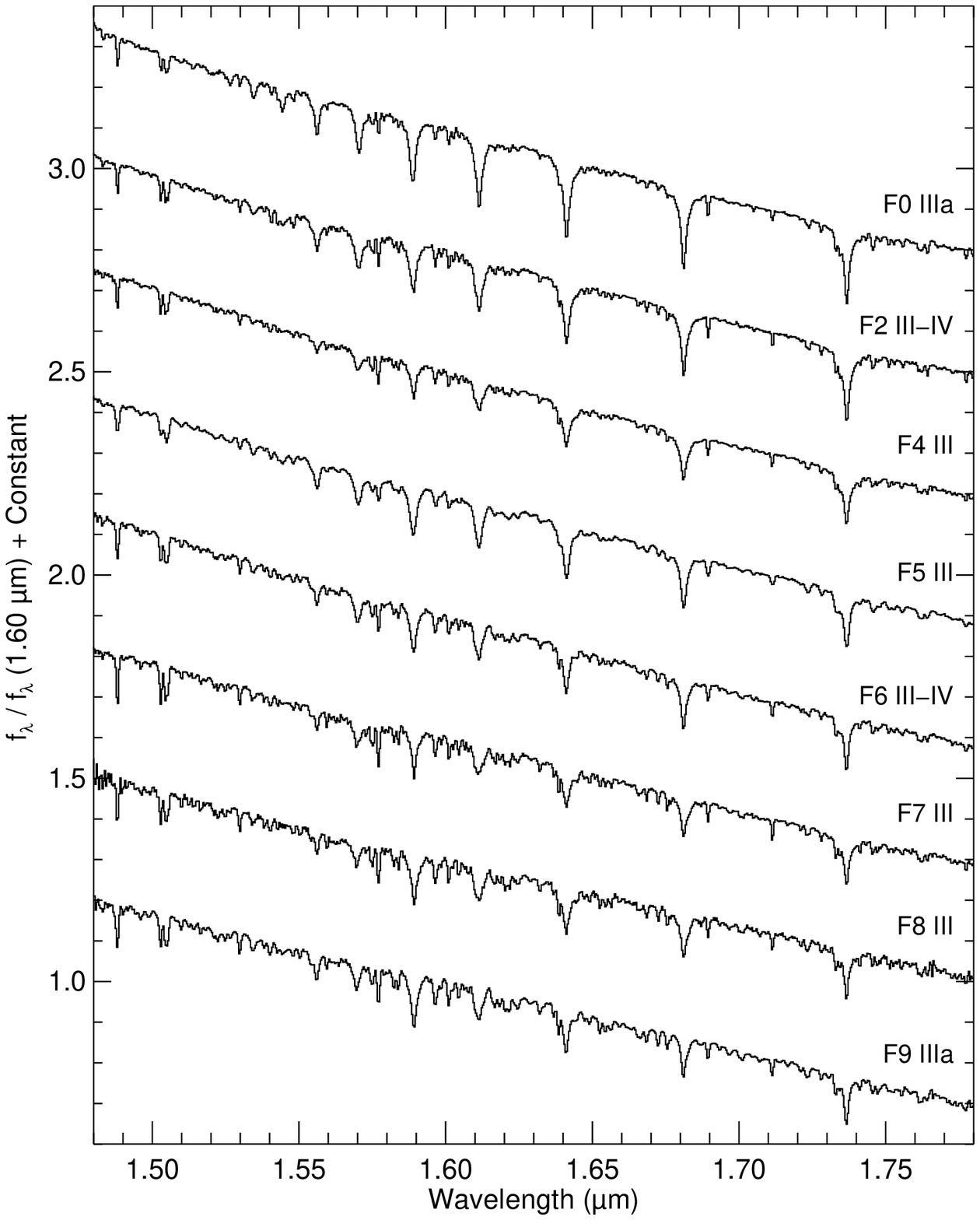}}
\caption{\label{fig:F_IIIH} A sequence of F giant stars plotted over the
  $H$ band (1.48$-$1.78~$\mu$m).  The spectra are of HD~89025 (F0~IIIa),
  HD~40535 (F2~III-IV), HD~21770 (F4~III), HD~17918 (F5~III), HD~160365
  (F6~III-IV), HD~124850 (F7~III), HD~220657 (F8~III), HD~6903
  (F9~IIIa).  The spectra have been normalized to unity at 1.60~$\mu$m
  and offset by constants.}
\end{figure}

\begin{figure}
\centerline{\includegraphics[width=6.0in,angle=0]{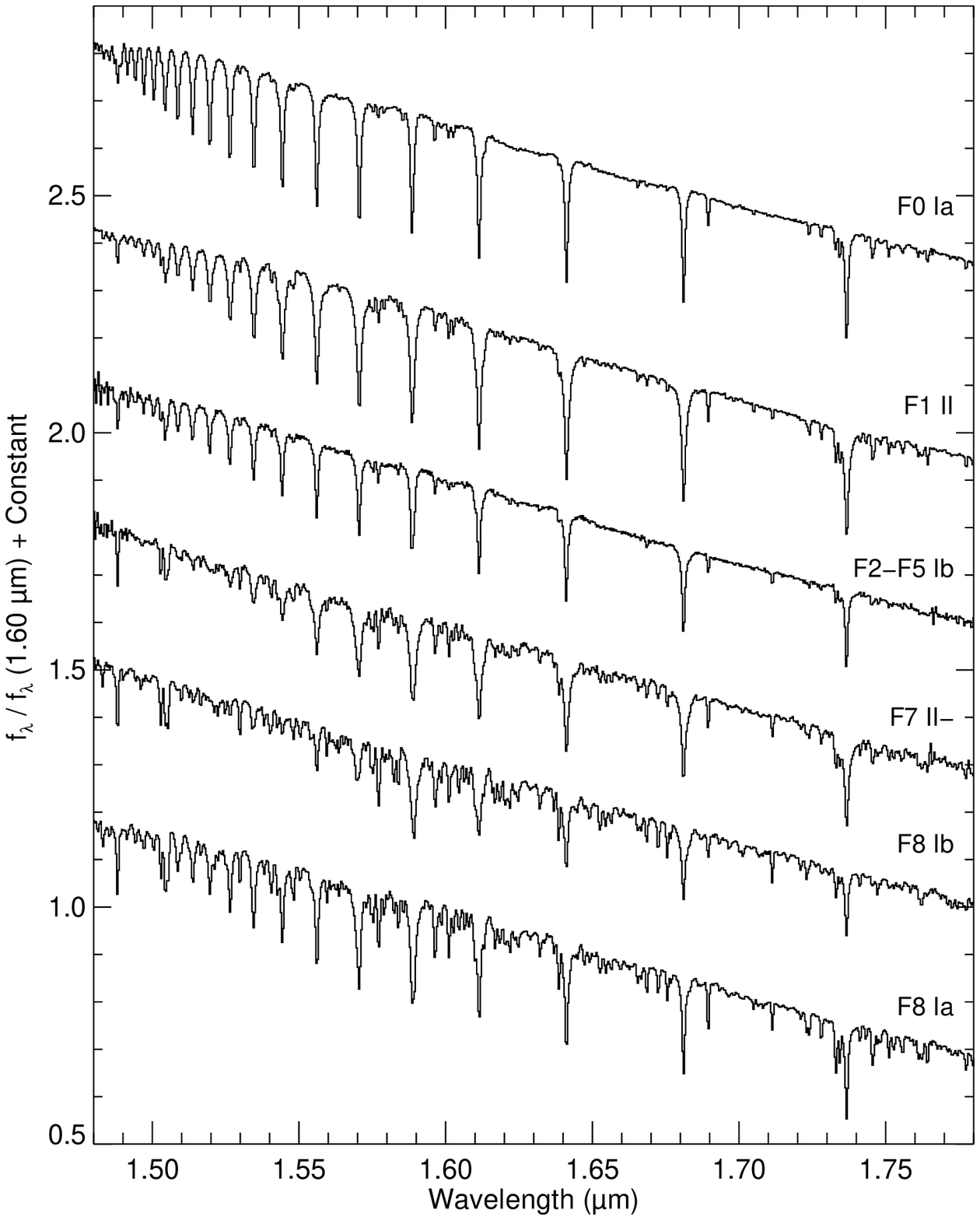}}
\caption{\label{fig:F_IH} A sequence of F supergiant stars plotted over
  the $H$ band (1.48$-$1.78~$\mu$m).  The spectra are of HD~7927
  (F0~Ia), HD~173638 (F1~II), BD~+38~2803 (F2-F5~Ib), HD~51956 (F6~IbII),
  HD~201078 (F7 II-), and HD~190323 (F8~Ia).  The spectra have been
  normalized to unity at 1.60~$\mu$m and offset by constants.}
\end{figure}

\begin{figure}
\centerline{\includegraphics[width=6.0in,angle=0]{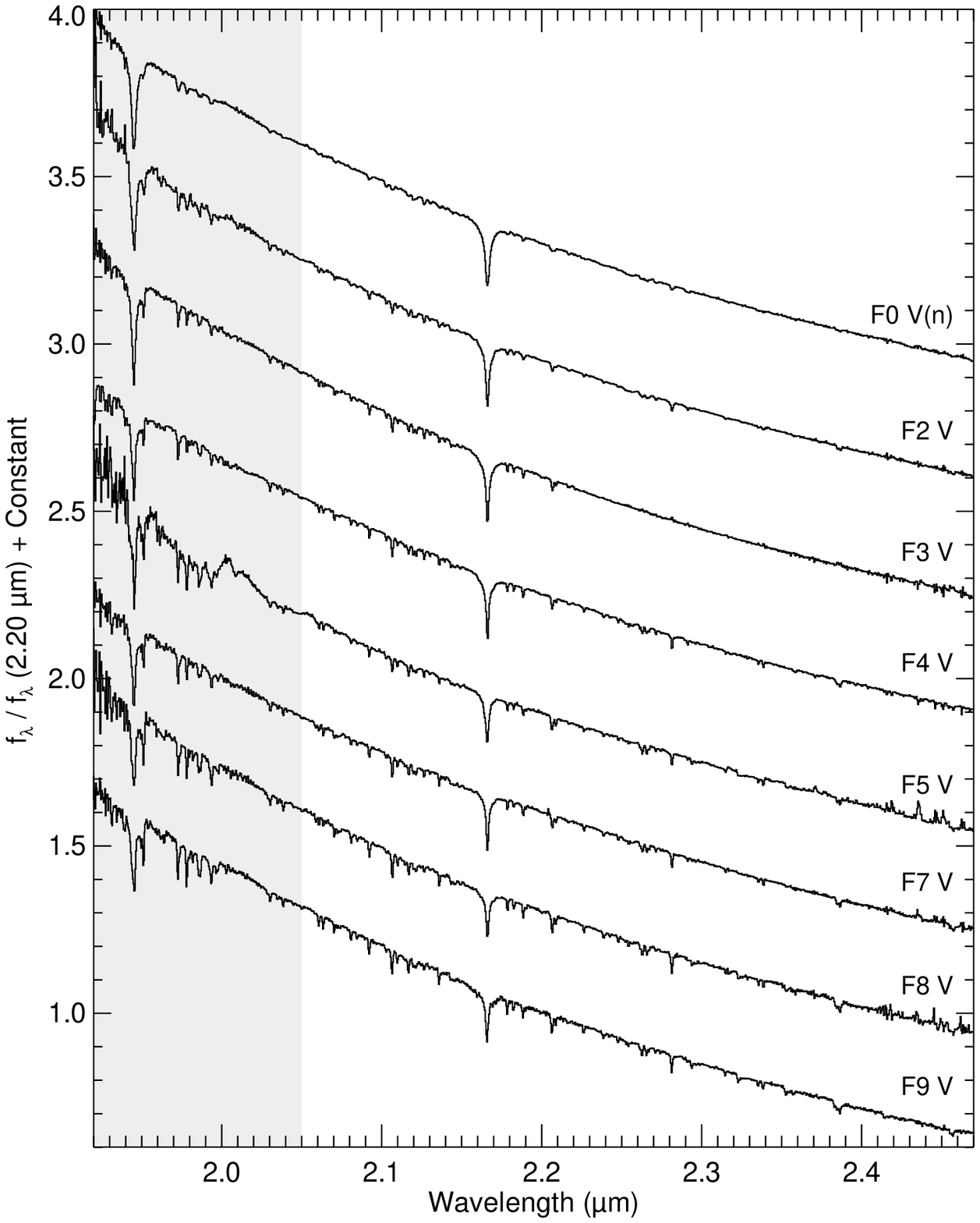}}
\caption{\label{fig:F_VK} A sequence of F dwarf stars plotted over the
  $K$ band (1.92$-$2.5~$\mu$m).  The spectra are of HD~108519 F0~V(n),
  HD 113139 (F2~V), HD~26015 (F3~V), HD~87822 (F4~V), HD~218804 (F5~V),
  HD~126660 (F7~V), HD~27393 (F8~V), and HD~176051 (F9~V).  The spectra
  have been normalized to unity at 2.20~$\mu$m and offset by constants.}
\end{figure}

\begin{figure}
\centerline{\includegraphics[width=6.0in,angle=0]{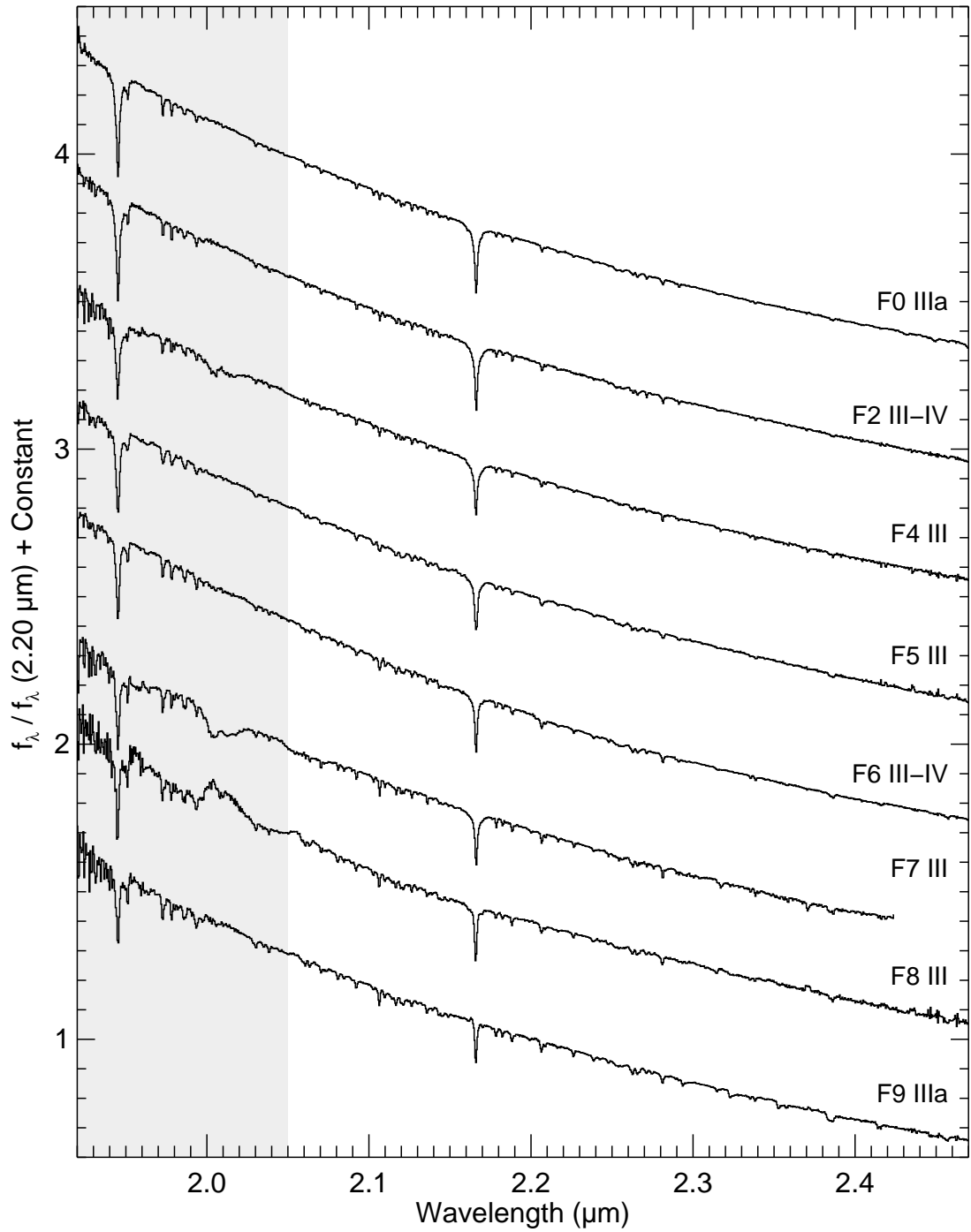}}
\caption{\label{fig:F_IIIK} A sequence of F giant stars plotted over the
  $K$ band (1.92$-$2.5~$\mu$m).  The spectra are of HD~89025 (F0~IIIa),
  HD~40535 (F2~III-IV), HD~21770 (F4~III), HD~17918 (F5~III), HD~160365
  (F6~III-IV), HD~124850 (F7~III), HD~220657 (F8~III), HD~6903 (F9~IIIa).  The spectra
  have been normalized to unity at 2.20~$\mu$m and offset by constants.
  Note the relatively poor correction of telluric CO$_2$ at
  $\sim$2.02~\micron~in the F7~III and F8~III stars (see \S\ref{Obs}).}
\end{figure}

\begin{figure}
\centerline{\includegraphics[width=6.0in,angle=0]{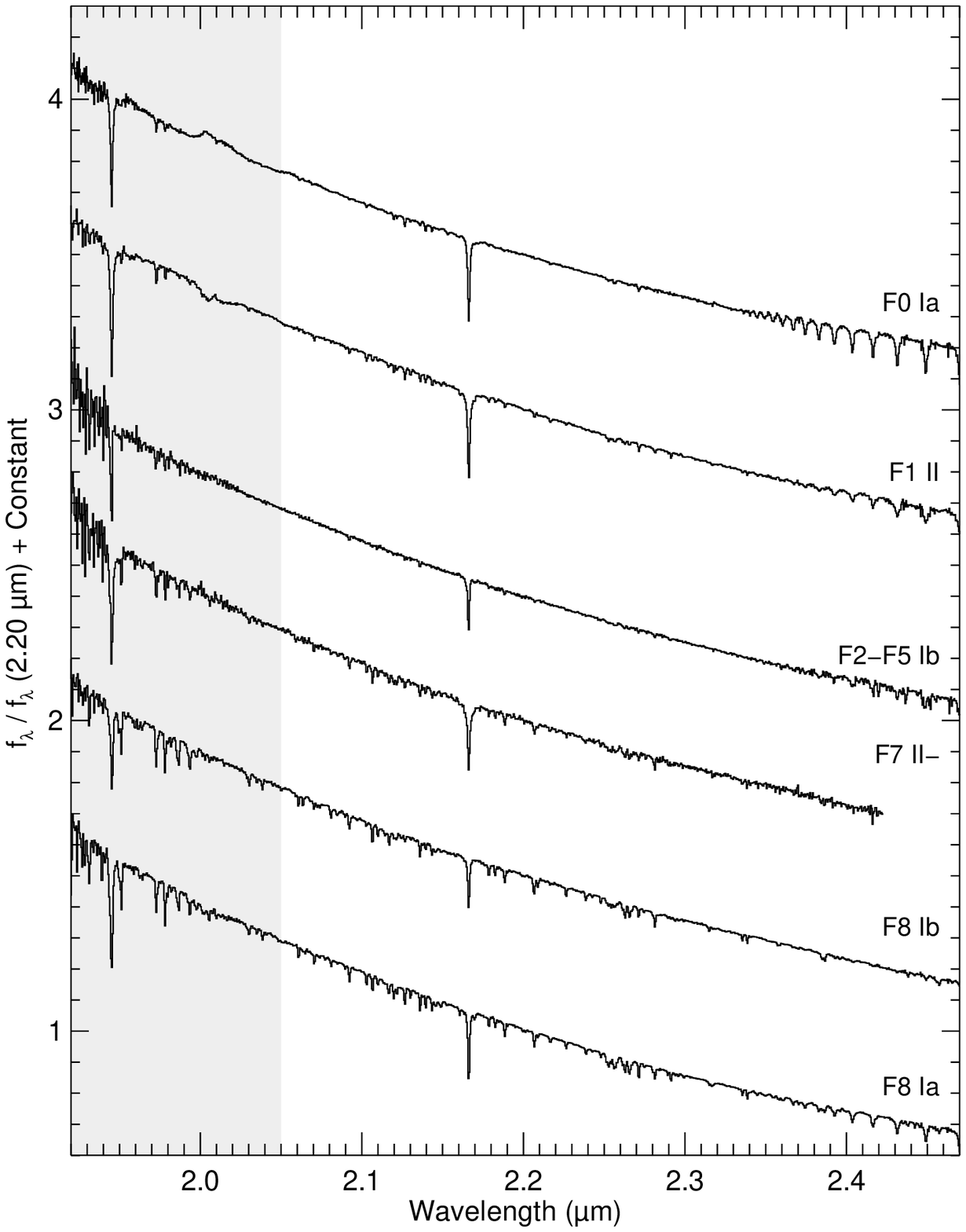}}
\caption{\label{fig:F_IK} A sequence of F supergiant stars plotted over
  the $K$ band (1.92$-$2.5~$\mu$m).  The spectra are of HD~7927 (F0~Ia),
  HD~173638 (F1~II), BD~+38~2803 (F2-F5~Ib), HD~51956 (F6~IbII), HD~201078
  (F7 II-), and HD~190323 (F8~Ia).  The spectra have been normalized to
  unity at 2.20~$\mu$m and offset by constants.}
\end{figure}

\begin{figure}
\centerline{\includegraphics[width=6.0in,angle=0]{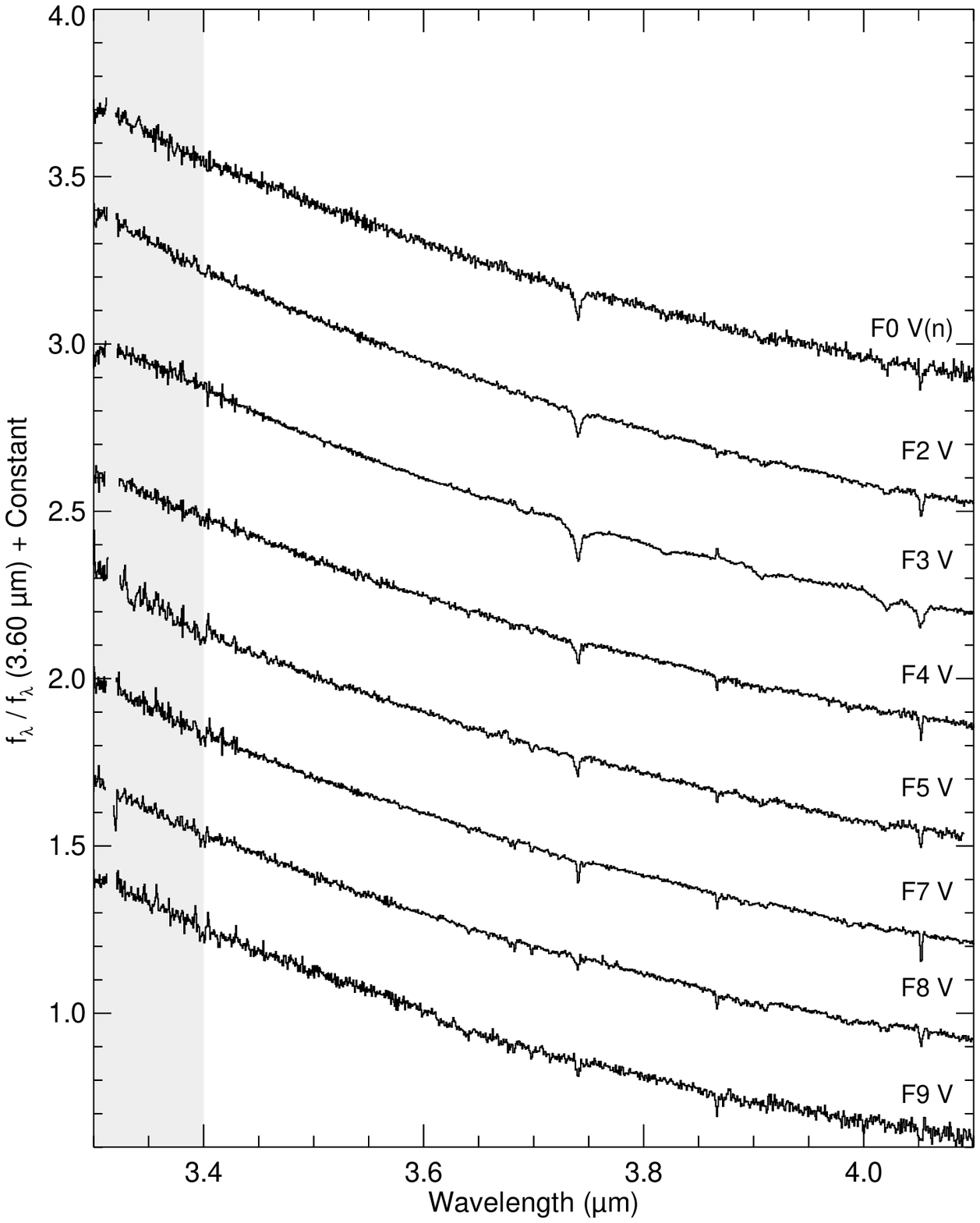}}
\caption{\label{fig:F_VL} A sequence of F dwarf stars plotted over the
  $L'$ band (3.3$-$4.1~$\mu$m).  The spectra are of HD~108519 F0~V(n),
  HD 113139 (F2~V), HD~26015 (F3~V), HD~87822 (F4~V), HD~218804 (F5~V),
  HD~126660 (F7~V), HD~27393 (F8~V), and HD~176051 (F9~V).  The spectra
  have been normalized to unity at 3.6~$\mu$m and offset by constants.}
\end{figure}

\begin{figure}
\centerline{\includegraphics[width=6.0in,angle=0]{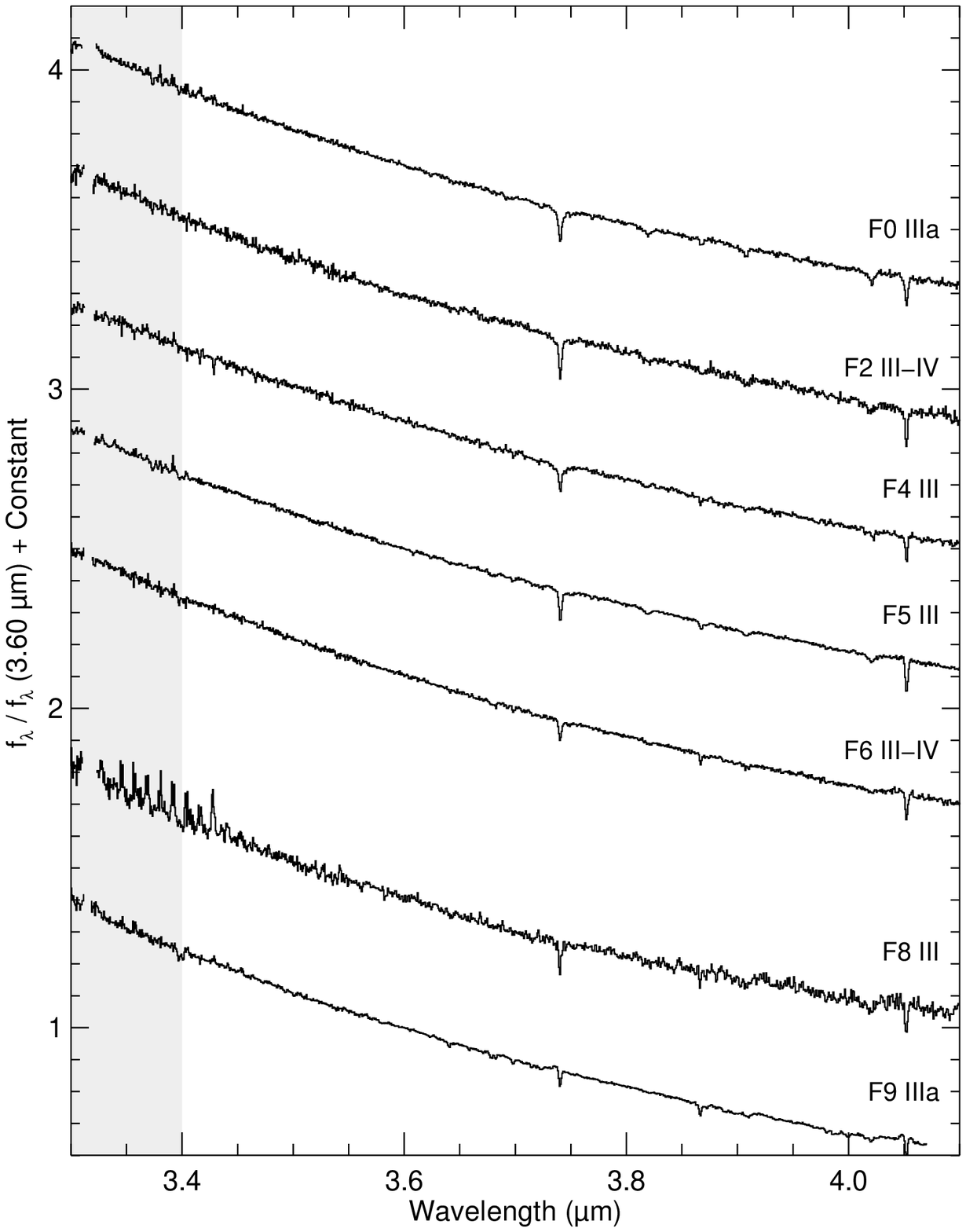}}
\caption{\label{fig:F_IIIL} A sequence of F giant stars plotted over the
  $L'$ band (3.3$-$4.1~$\mu$m).  The spectra are of HD~89025 (F0~IIIa),
  HD~40535 (F2~III-IV), HD~21770 (F4~III), HD~17918 (F5~III), HD~160365
  (F6~III-IV), HD~124850 (F7~III), HD~220657 (F8~III), and HD~6903 (F9~IIIa).  The spectra
  have been normalized to unity at 3.6~$\mu$m and offset by constants.}
\end{figure}

\begin{figure}
\centerline{\includegraphics[width=6.0in,angle=0]{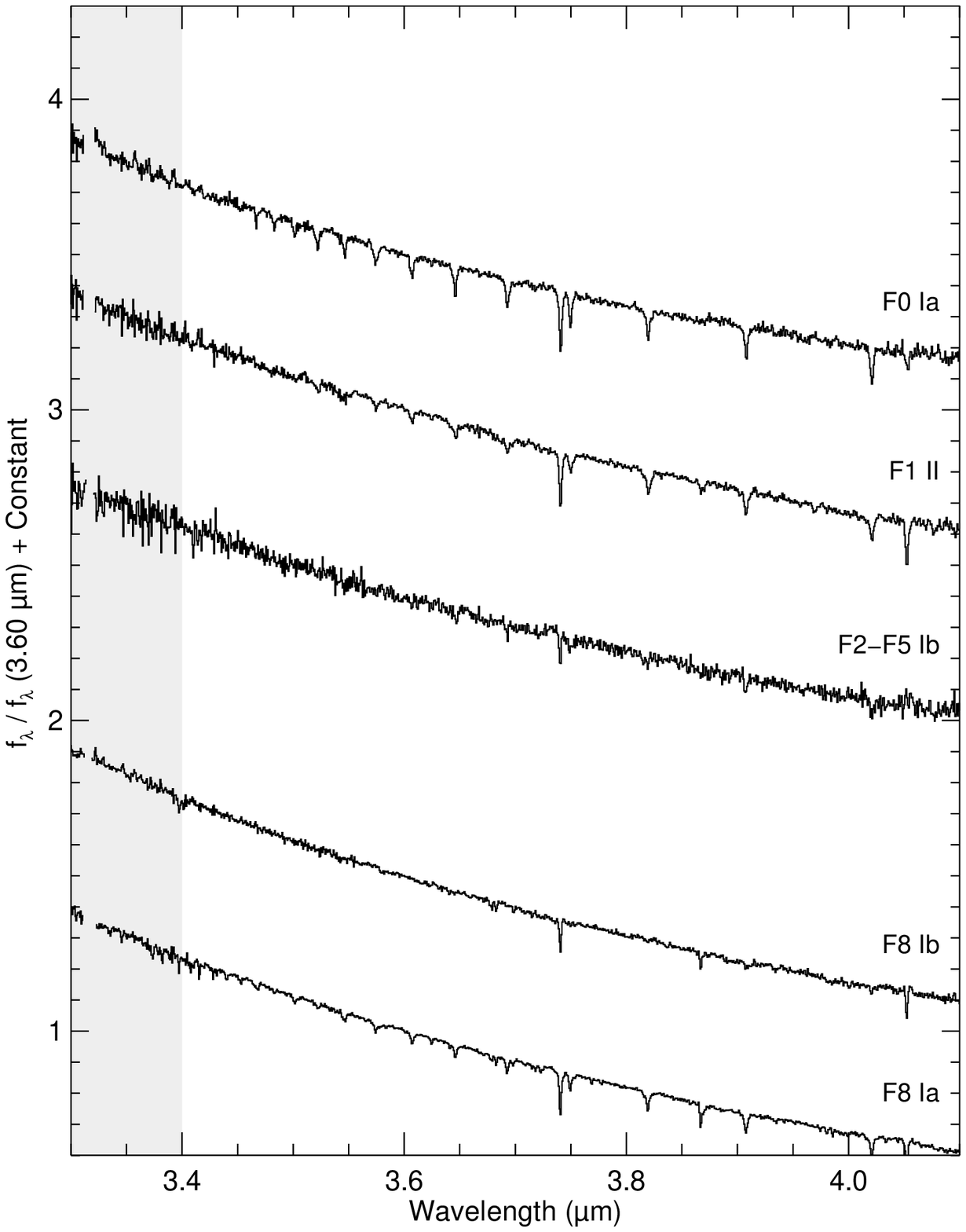}}
\caption{\label{fig:F_IL} A sequence of F supergiant stars plotted over
  the $L'$ band (1.92$-$2.5~$\mu$m).  The spectra are of HD~7927
  (F0~Ia), HD~173638 (F1~II), BD~+38~2803 (F2-F5~Ib), HD~51956 (F6~IbII),
  HD~201078 (F7 II-), and HD~190323 (F8~Ia).  The spectra have been
  normalized to unity at 3.6~$\mu$m and offset by constants.}
\end{figure}

\begin{figure}
\centerline{\includegraphics[width=6.0in,angle=0]{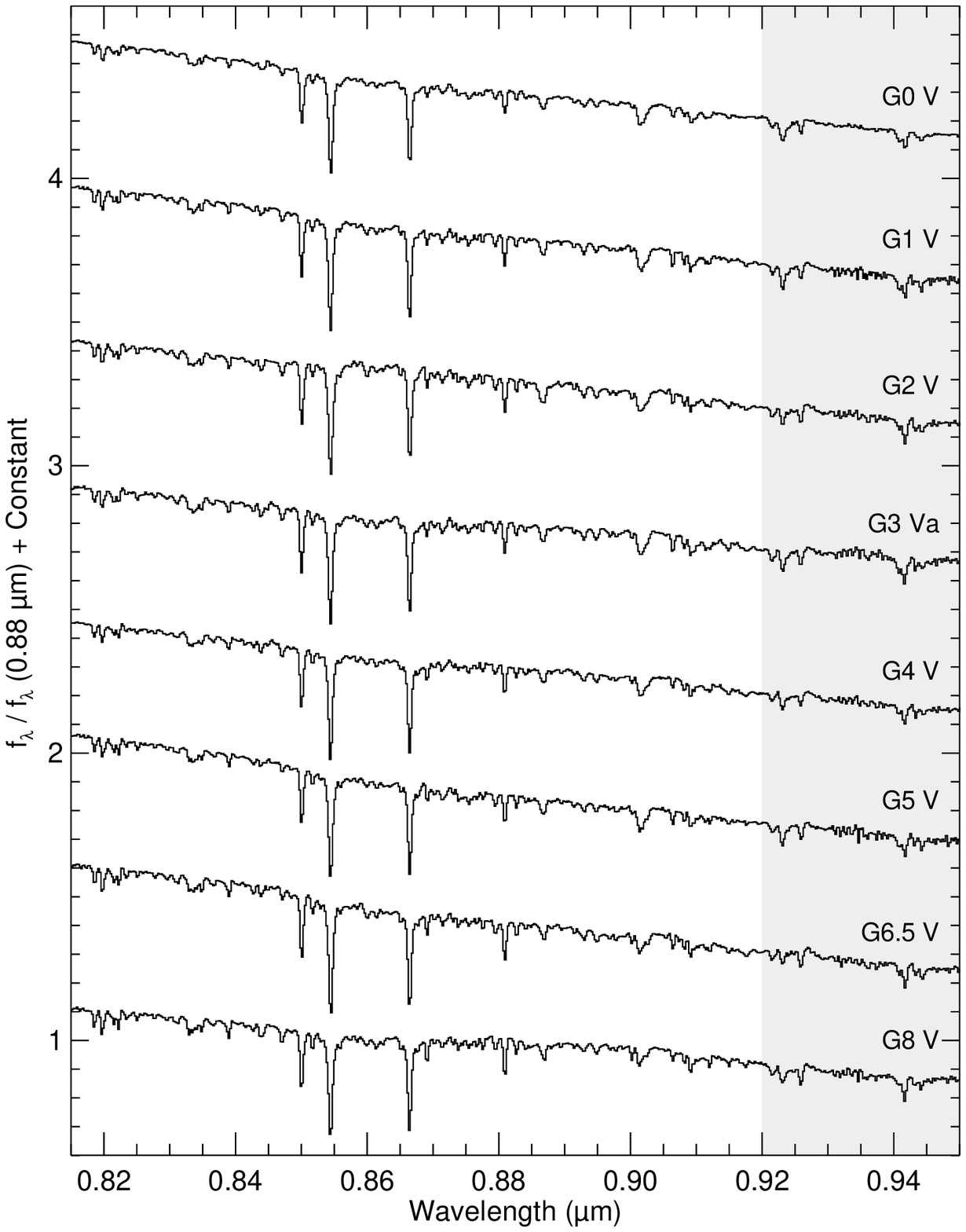}}
\caption{\label{fig:G_VI} A sequence of G dwarf stars plotted over the
  $I$ band (0.82$-$0.95~$\mu$m).  The spectra are of HD~109358 (G0~V),
  HD~10307 (G1~V), HD~76151 (G2~V), HD~10697 (G3~Va), HD~214850 (G4~V),
  HD~165185 (G5~V), HD~115617 (G6.5~V), and HD~101501 (G8~V).  The
  spectra have been normalized to unity at 0.88~$\mu$m and offset by
  constants.}
\end{figure}

\begin{figure}
\centerline{\includegraphics[width=6.0in,angle=0]{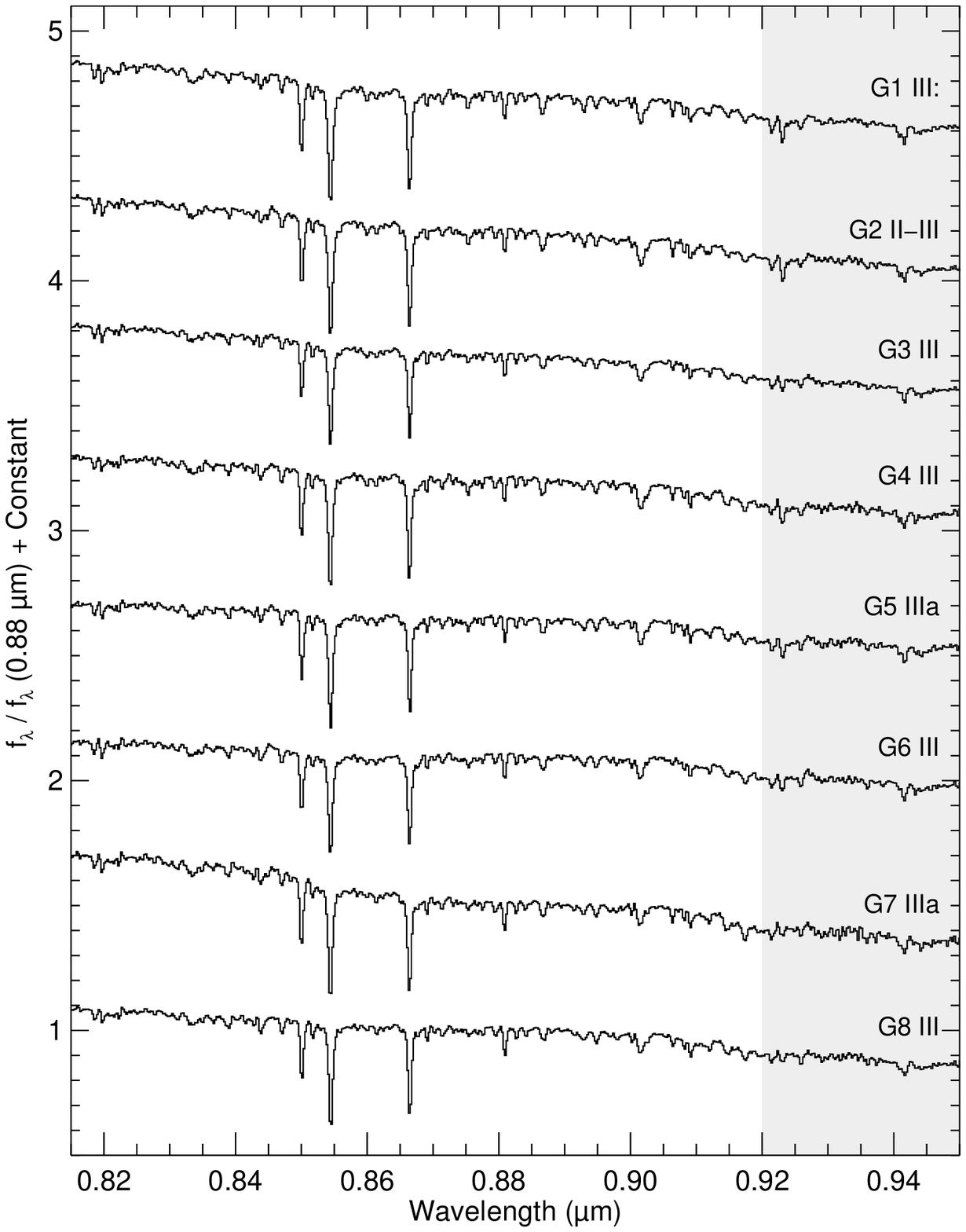}}
\caption{\label{fig:G_IIII} A sequence of G giant stars plotted over the
  $I$ band (0.82$-$0.95~$\mu$m).  The spectra are of
  HD~21018 (G1~III:CH-1:), HD~219477 (G2~II-III), HD~88639 (G3~IIIb~Fe-1),
  HD~108477 (G4~III), HD~193896 (G5~IIIa), HD~27277 (G6~III), HD~182694
  (G7~IIIa), and HD~135722 (G8~III~Fe-1).  The spectra have been
  normalized to unity at 0.88~$\mu$m and offset by constants.}
\end{figure}

\begin{figure}
\centerline{\includegraphics[width=6.0in,angle=0]{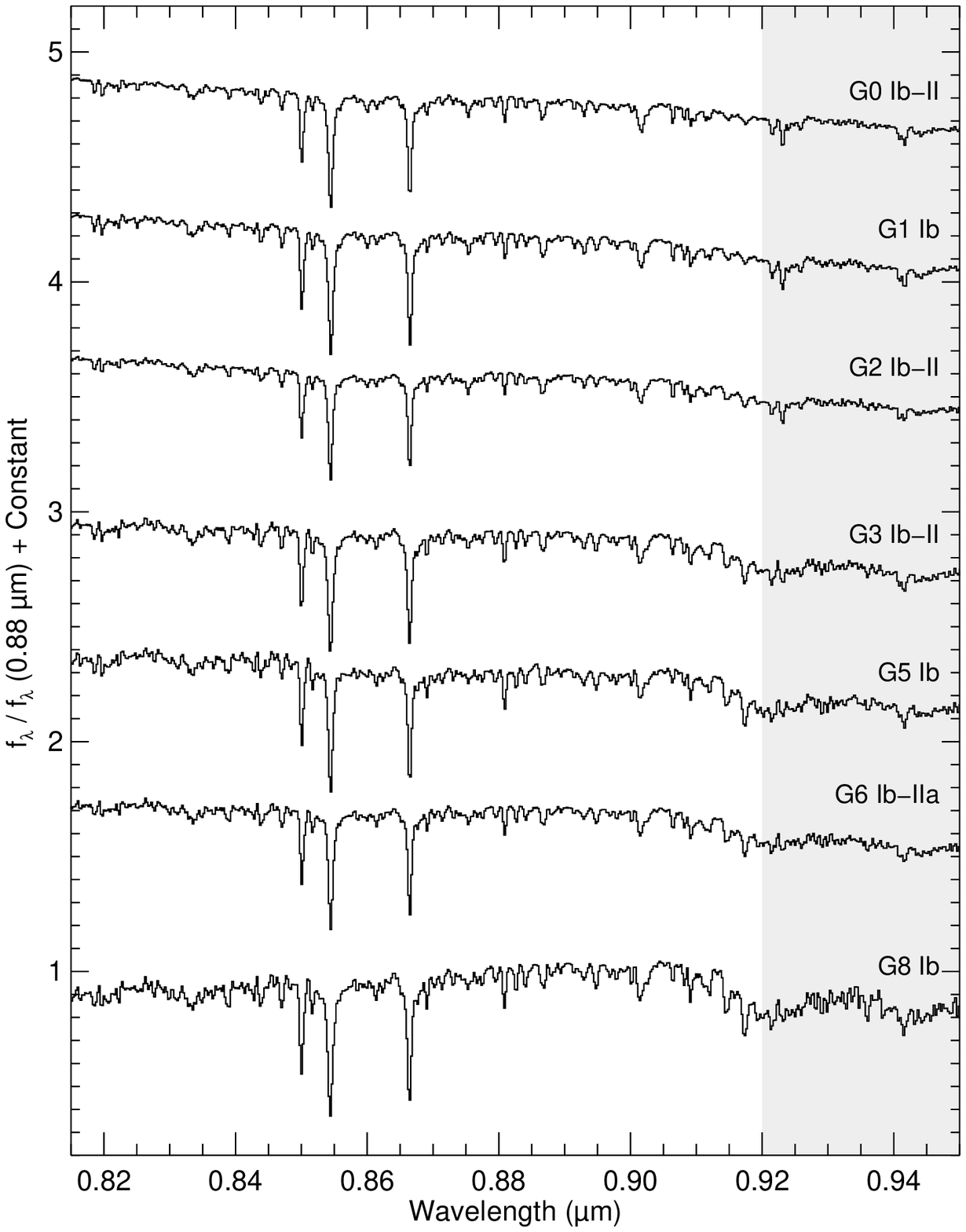}}
\caption{\label{fig:G_II} A sequence of G supergiant stars plotted over
  the $I$ band (0.82$-$0.95~$\mu$m).  The spectra are of HD~185018
  (G0~Ib-II), HD~74395 (G1~Ib), HD~3421 (G2~Ib-II), HD~192713
  (G3~Ib-II~Wk~H\&K~comp?), HD~190113 (G5~Ib), HD~202314 (G6~Ib-IIa Ca1~B0.5), and
  HD~208606 (G8~Ib).  The spectra have been normalized to unity at
  0.88~$\mu$m and offset by constants.}
\end{figure}

\begin{figure}
\centerline{\includegraphics[width=6.0in,angle=0]{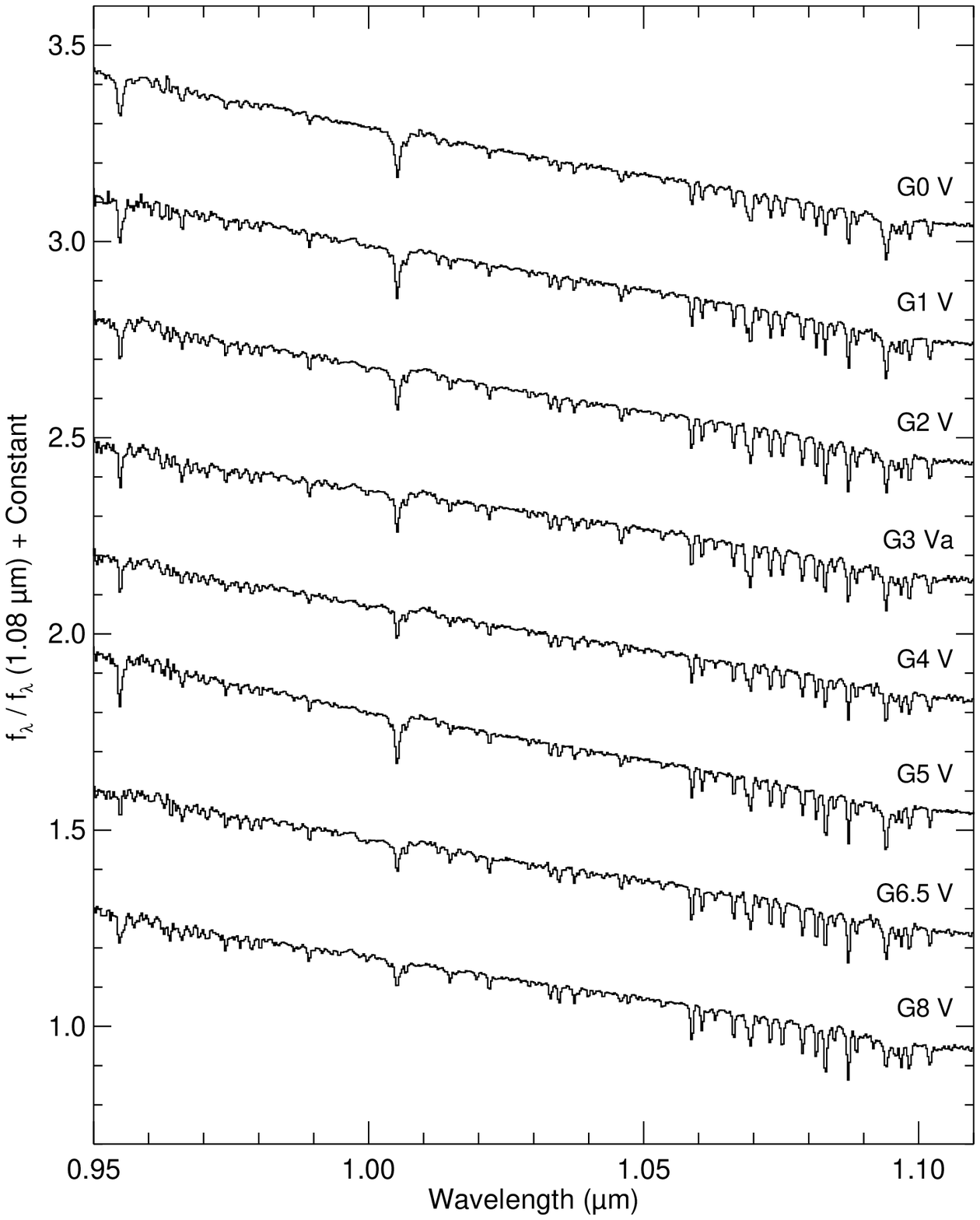}}
\caption{\label{fig:G_VY} A sequence of G dwarf stars plotted over the
  $Y$ band (0.95$-$1.10~$\mu$m).  The spectra are of HD~109358 (G0~V),
  HD~10307 (G1~V), HD~76151 (G2~V), HD~10697 (G3~Va), HD~214850 (G4~V),
  HD~165185 (G5~V), HD~115617 (G6.5~V), and HD~101501 (G8~V).  The
  spectra have been normalized to unity at 1.08~$\mu$m and offset by
  constants.}
\end{figure}

\begin{figure}
\centerline{\includegraphics[width=6.0in,angle=0]{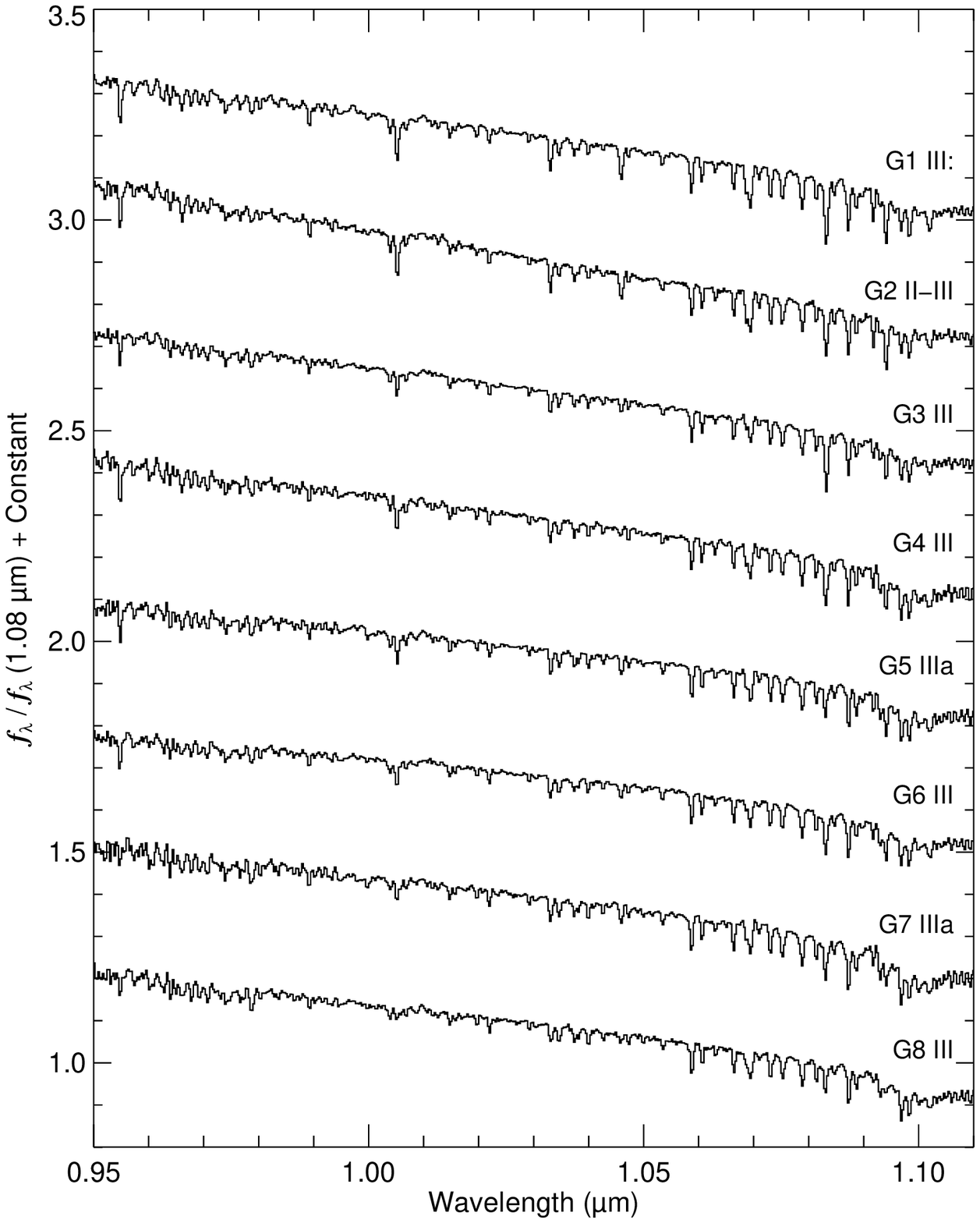}}
\caption{\label{fig:G_IIIY} A sequence of G giant stars plotted over the
  $Y$ band (0.95$-$1.10~$\mu$m).  The spectra are of
  HD~21018 (G1~III:CH-1:), HD~219477 (G2~II-III), HD~88639 (G3b~III~Fe-1),
  HD~108477 (G4~III), HD~193896 (G5~IIIa), HD~27277 (G6~III), HD~182694
  (G7~IIIa), and HD~135722 (G8~III~Fe-1).  The spectra have been
  normalized to unity at 1.08~$\mu$m and offset by constants.}
\end{figure}

\begin{figure}
\centerline{\includegraphics[width=6.0in,angle=0]{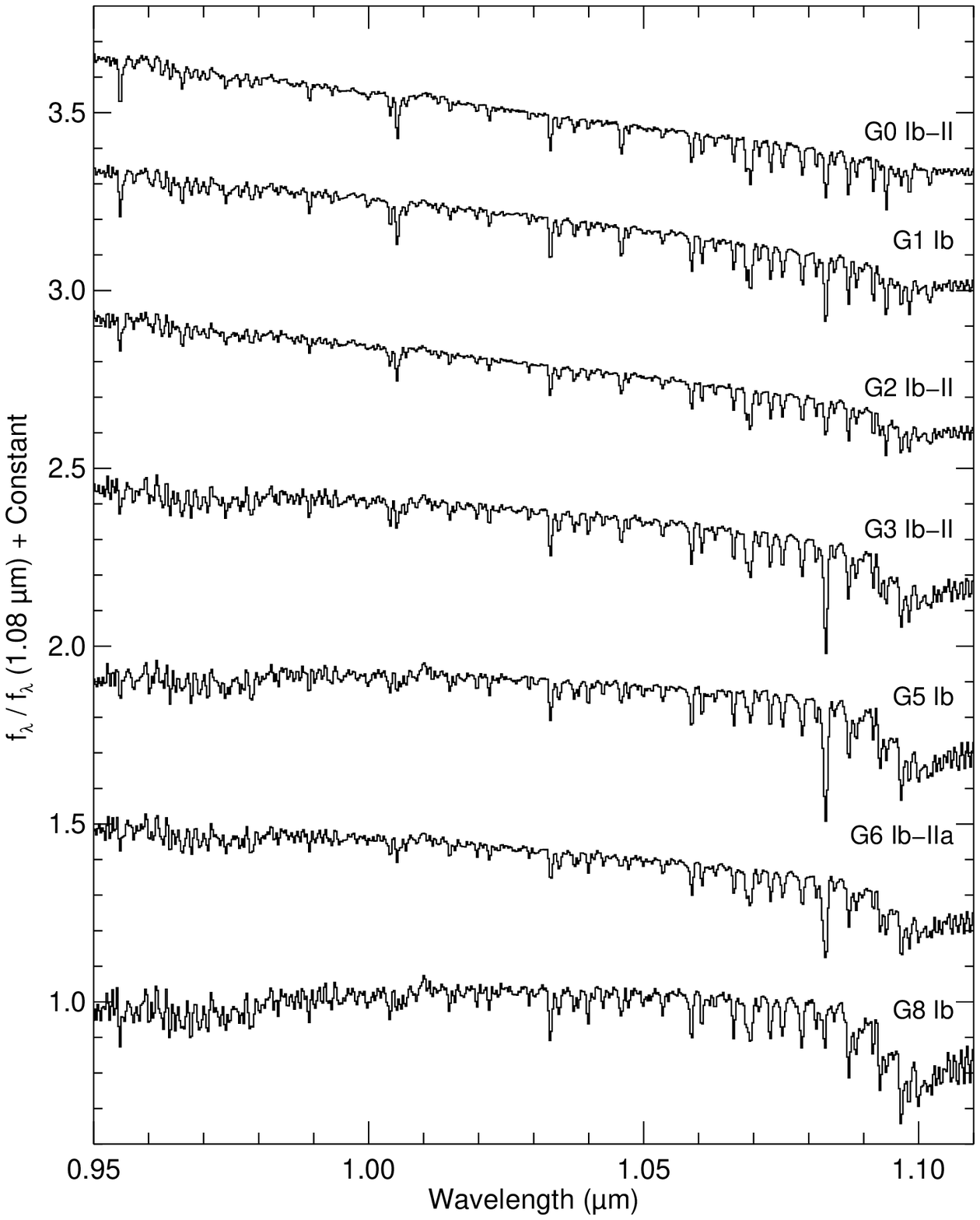}}
\caption{\label{fig:G_IY} A sequence of G supergiant stars plotted over
  the $Y$ band (0.95$-$1.10~$\mu$m).  The spectra are of HD~185018
  (G0~Ib-II), HD~74395 (G1~Ib), HD~3421 (G2~Ib-II), HD~192713
  (G3~Ib-II~Wk~H\&K~comp?), HD~190113 (G5~Ib), HD~202314 (G6~Ib-IIa Ca1~B0.5), and
  HD~208606 (G8~Ib).  The spectra have been normalized to unity at
  1.08~$\mu$m and offset by constants.}
\end{figure}

\begin{figure}
\centerline{\includegraphics[width=6.0in,angle=0]{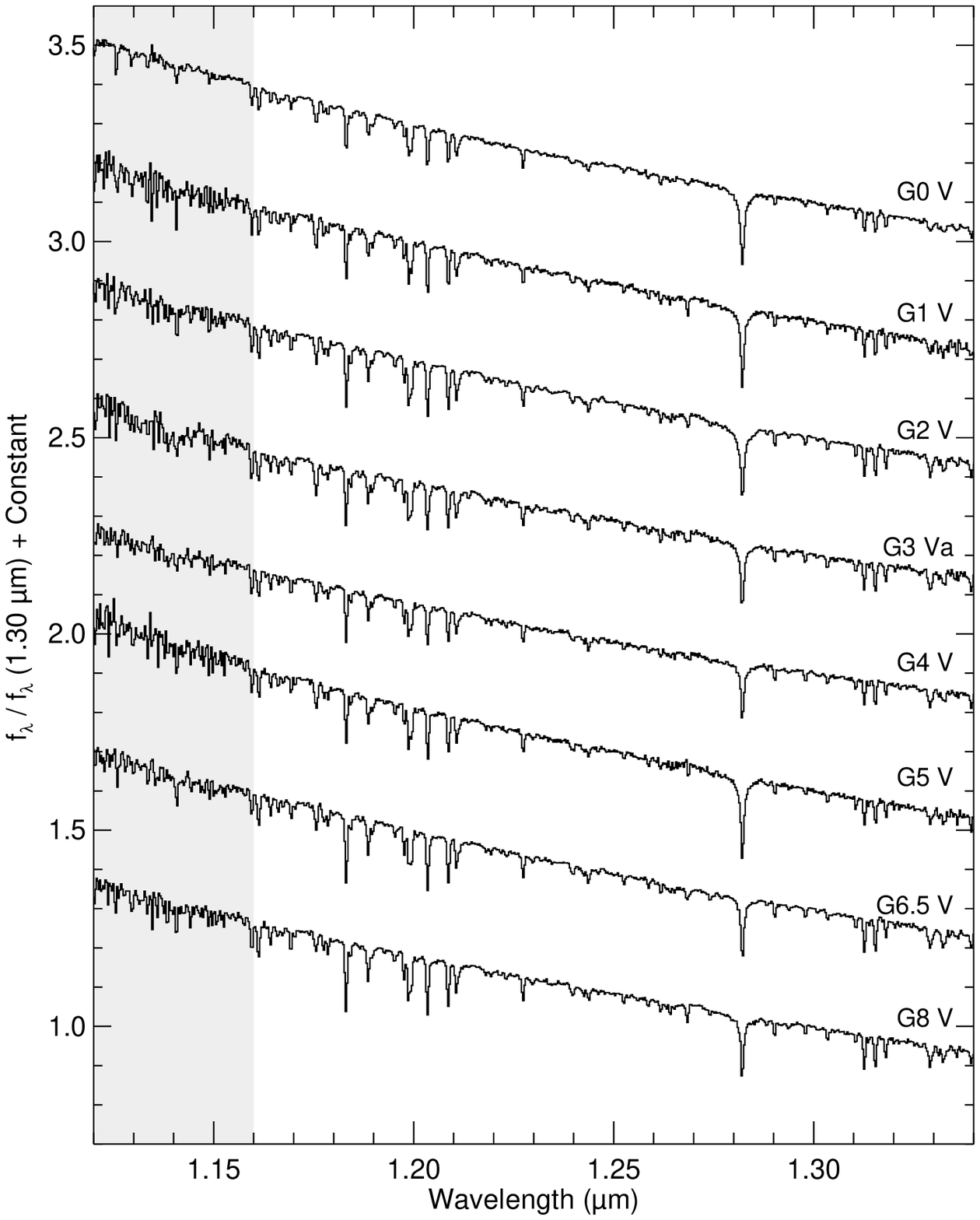}}
\caption{\label{fig:G_VJ} A sequence of G dwarf stars plotted over the
  $J$ band (1.12$-$1.34~$\mu$m).  The spectra are of HD~109358 (G0~V),
  HD~10307 (G1~V), HD~76151 (G2~V), HD~10697 (G3~Va), HD~214850 (G4~V),
  HD~165185 (G5~V), HD~115617 (G6.5~V), and HD~101501 (G8~V).  The
  spectra have been normalized to unity at 1.30~$\mu$m and offset by
  constants.}
\end{figure}

\begin{figure}
\centerline{\includegraphics[width=6.0in,angle=0]{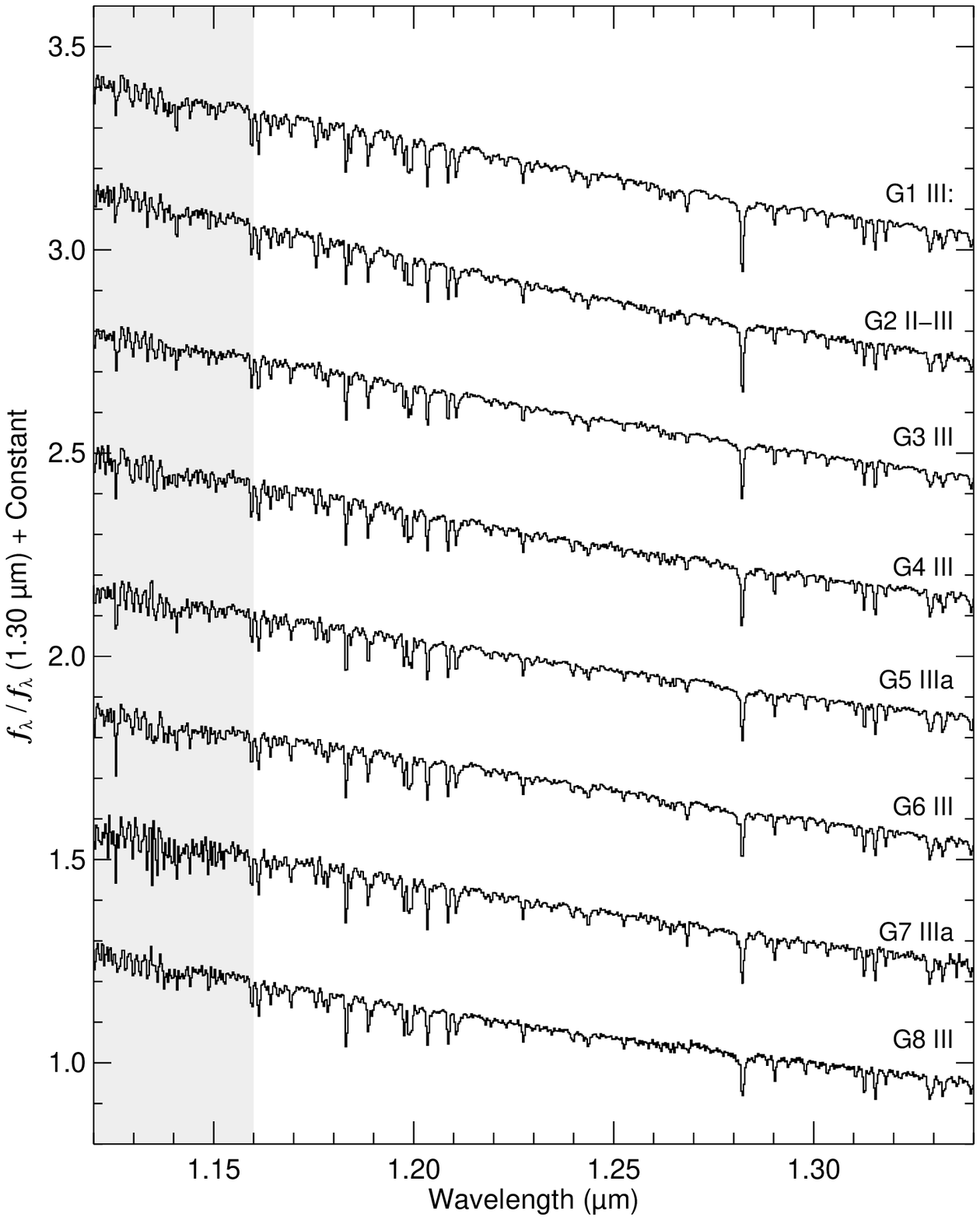}}
\caption{\label{fig:G_IIIJ} A sequence of G giant stars plotted over the
  $J$ band (1.12$-$1.34~$\mu$m).  The spectra are of
  HD~21018 (G1~III:CH-1:), HD~219477 (G2~II-III), HD~88639 (G3b~III~Fe-1),
  HD~108477 (G4~III), HD~193896 (G5~IIIa), HD~27277 (G6~III), HD~182694
  (G7~IIIa), and HD~135722 (G8~III~Fe-1).  The spectra have been
  normalized to unity at 1.30~$\mu$m and offset by constants.}
\end{figure}

\begin{figure}
\centerline{\includegraphics[width=6.0in,angle=0]{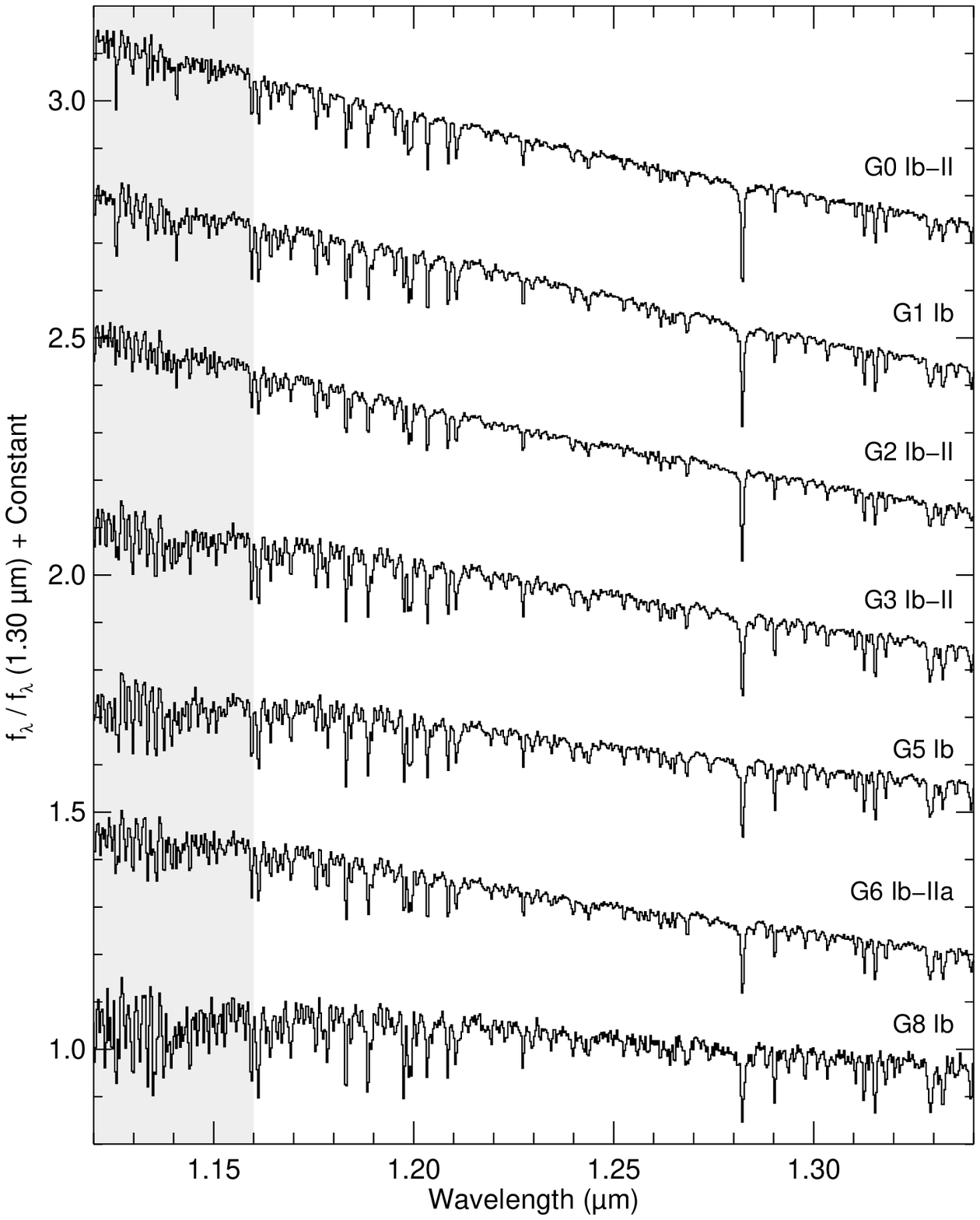}}
\caption{\label{fig:G_IJ} A sequence of G supergiant stars plotted over
  the $J$ band (1.12$-$1.34~$\mu$m).  The spectra are of HD~185018
  (G0~Ib-II), HD~74395 (G1~Ib), HD~3421 (G2~Ib-II), HD~192713
  (G3~Ib-II~Wk~H\&K~comp?), HD~190113 (G5~Ib), HD~202314 (G6~Ib-IIa
  Ca1~B0.5), and HD~208606 (G8~Ib).  The spectra have been normalized to
  unity at 1.30~$\mu$m and offset by constants.}
\end{figure}

\begin{figure}
\centerline{\includegraphics[width=6.0in,angle=0]{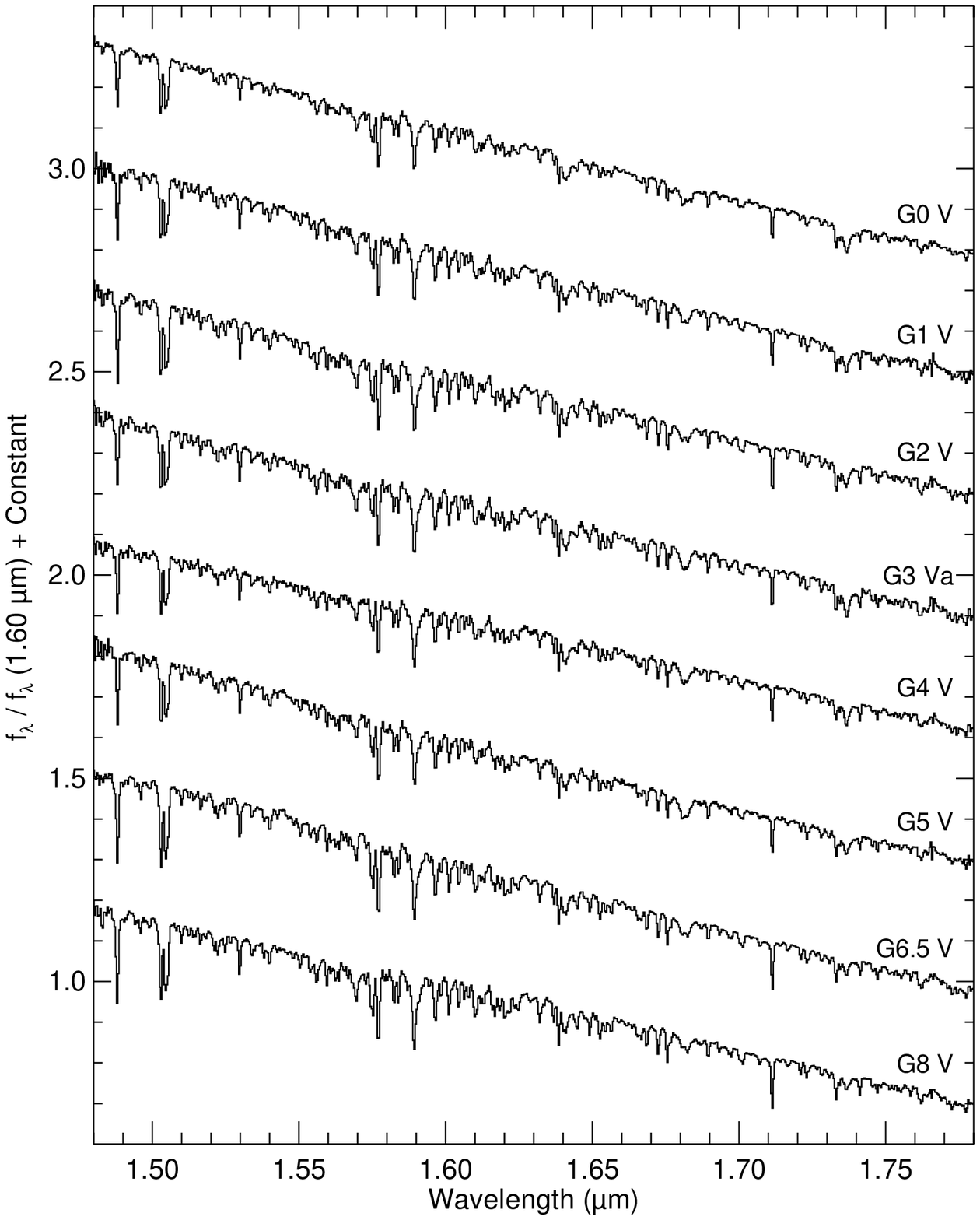}}
\caption{\label{fig:G_VH} A sequence of G dwarf stars plotted over the
  $H$ band (1.48$-$1.78~$\mu$m).  The spectra are of HD~109358 (G0~V),
  HD~10307 (G1~V), HD~76151 (G2~V), HD~10697 (G3~Va), HD~214850 (G4~V),
  HD~165185 (G5~V), HD~115617 (G6.5~V), and HD~101501 (G8~V).  The
  spectra have been normalized to unity at 1.60~$\mu$m and offset by
  constants.}
\end{figure}

\begin{figure}
\centerline{\includegraphics[width=6.0in,angle=0]{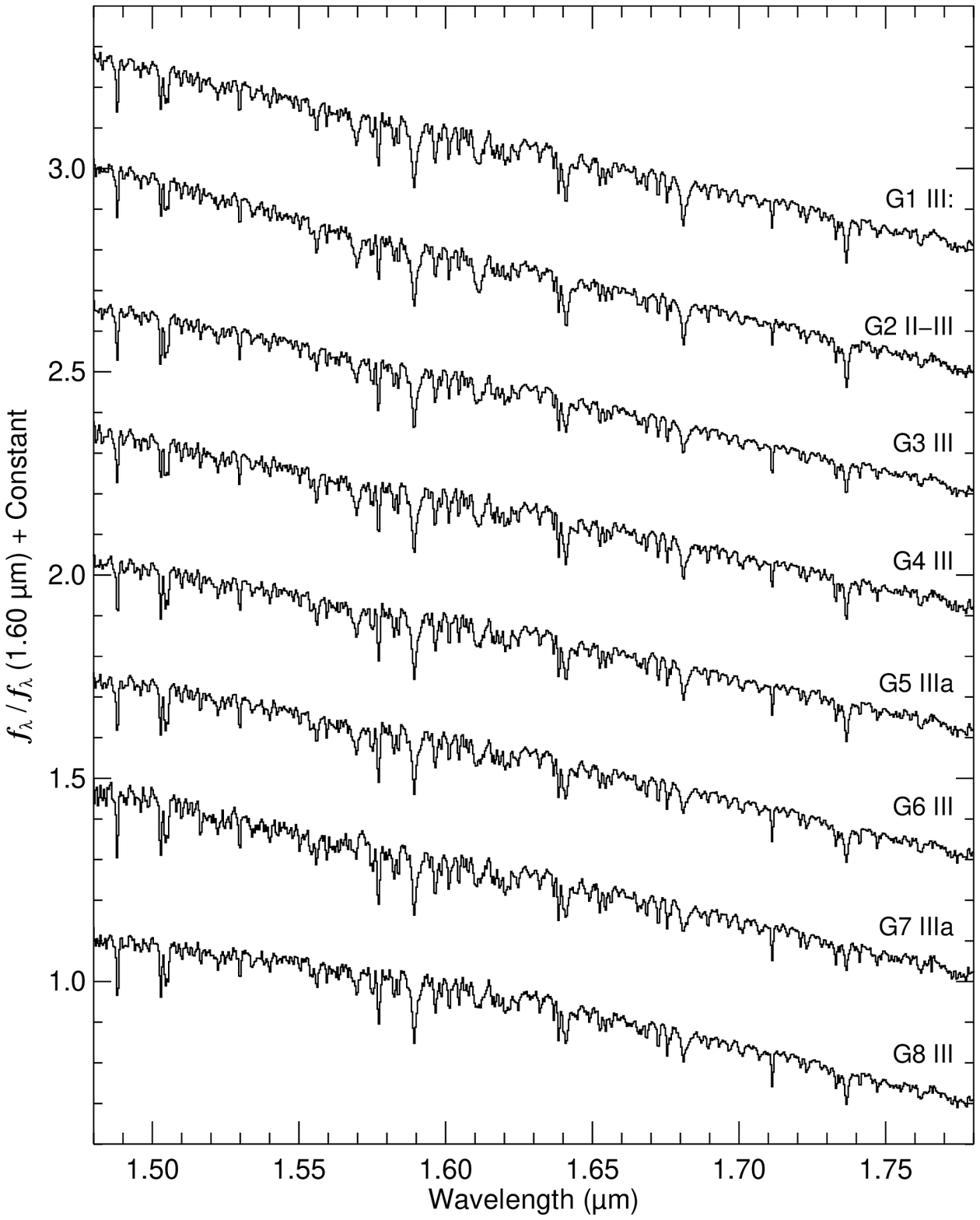}}
\caption{\label{fig:G_IIIH} A sequence of G giant stars plotted over the
  $H$ band (1.48$-$1.78~$\mu$m).  The spectra are of
  HD~21018 (G1~III:CH-1:), HD~219477 (G2~II-III), HD~88639 (G3b~III~Fe-1),
  HD~108477 (G4~III), HD~193896 (G5~IIIa), HD~27277 (G6~III), HD~182694
  (G7~IIIa), and HD~135722 (G8~III~Fe-1).  The spectra have been
  normalized to unity at 1.60~$\mu$m and offset by constants.}
\end{figure}

\begin{figure}
\centerline{\includegraphics[width=6.0in,angle=0]{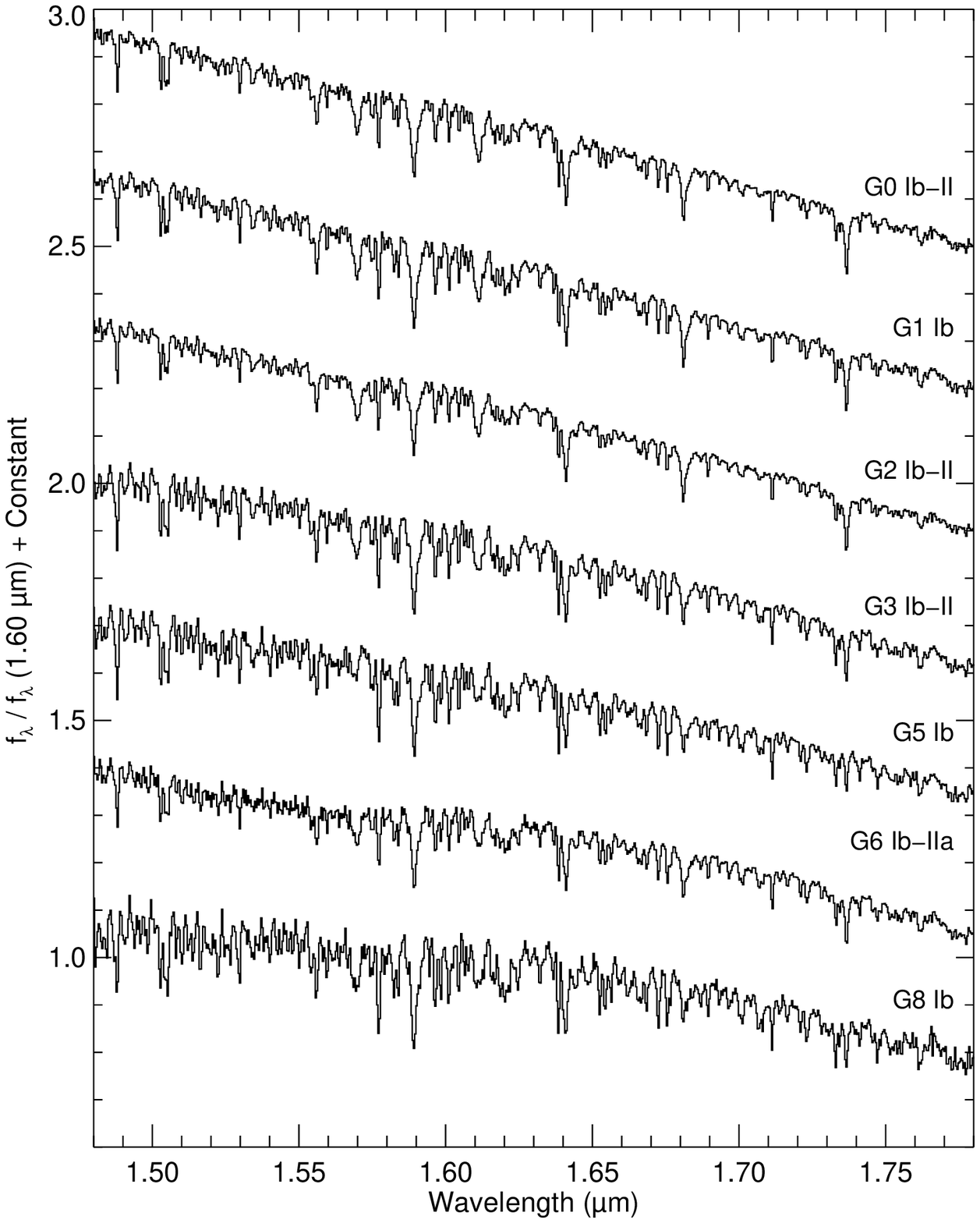}}
\caption{\label{fig:G_IH} A sequence of G supergiant stars plotted over
  the $H$ band (1.48$-$1.78~$\mu$m).  The spectra are of HD~185018
  (G0~Ib-II), HD~74395 (G1~Ib), HD~3421 (G2~Ib-II), HD~192713
  (G3~Ib-II~Wk~H\&K~comp?), HD~190113 (G5~Ib), HD~202314 (G6~Ib-IIa Ca1~B0.5), and
  HD~208606 (G8~Ib).  The spectra have been normalized to unity at
  1.60~$\mu$m and offset by constants.}
\end{figure}

\begin{figure}
\centerline{\includegraphics[width=6.0in,angle=0]{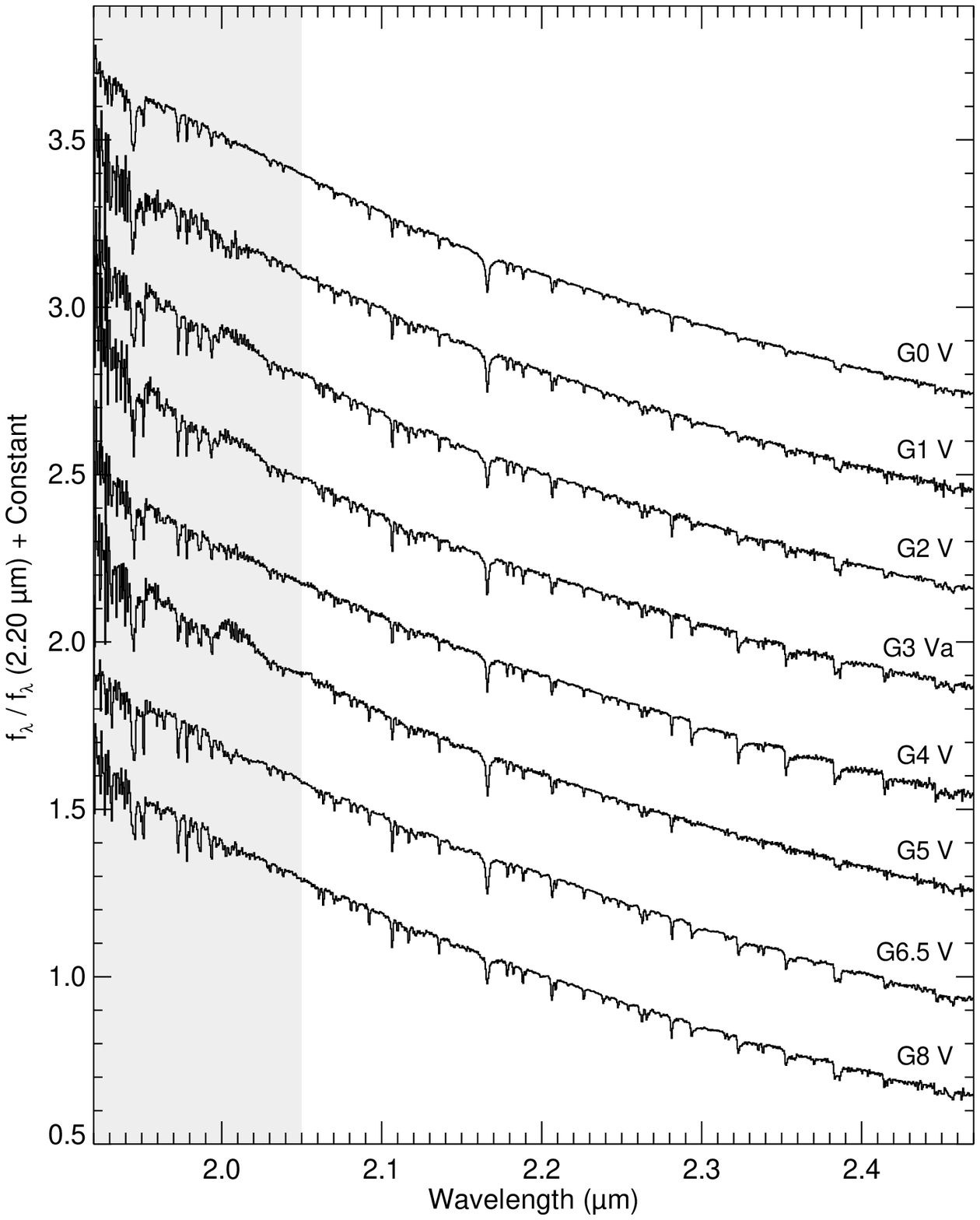}}
\caption{\label{fig:G_VK} A sequence of G dwarf stars plotted over the
  $K$ band (1.92$-$2.58~$\mu$m).  The spectra are of HD~109358 (G0~V),
  HD~10307 (G1~V), HD~76151 (G2~V), HD~10697 (G3~Va), HD~214850 (G4~V),
  HD~165185 (G5~V), HD~115617 (G6.5~V), and HD~101501 (G8~V).  The
  spectra have been normalized to unity at 2.20~$\mu$m and offset by
  constants.}
\end{figure}

\begin{figure}
\centerline{\includegraphics[width=6.0in,angle=0]{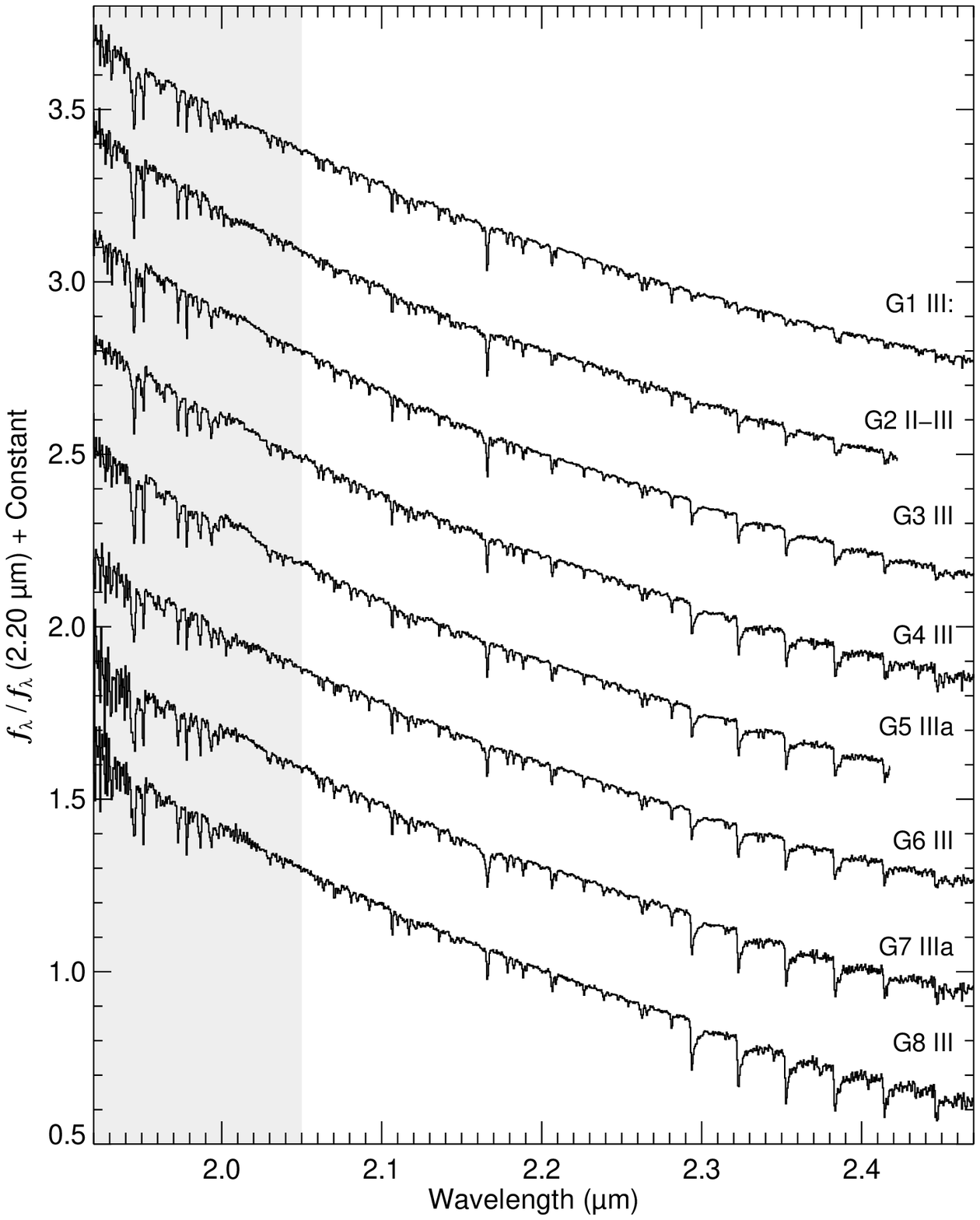}}
\caption{\label{fig:G_IIIK} A sequence of G giant stars plotted over the
  $K$ band (1.92$-$2.5~$\mu$m).  The spectra are of
  HD~21018 (G1~III:CH-1:), HD~219477 (G2~II-III), HD~88639 (G3b~III~Fe-1),
  HD~108477 (G4~III), HD~193896 (G5~IIIa), HD~27277 (G6~III), HD~182694
  (G7~IIIa), and HD~135722 (G8~III~Fe-1).  The spectra have been
  normalized to unity at 2.20~$\mu$m and offset by constants.}
\end{figure}

\begin{figure}
\centerline{\includegraphics[width=6.0in,angle=0]{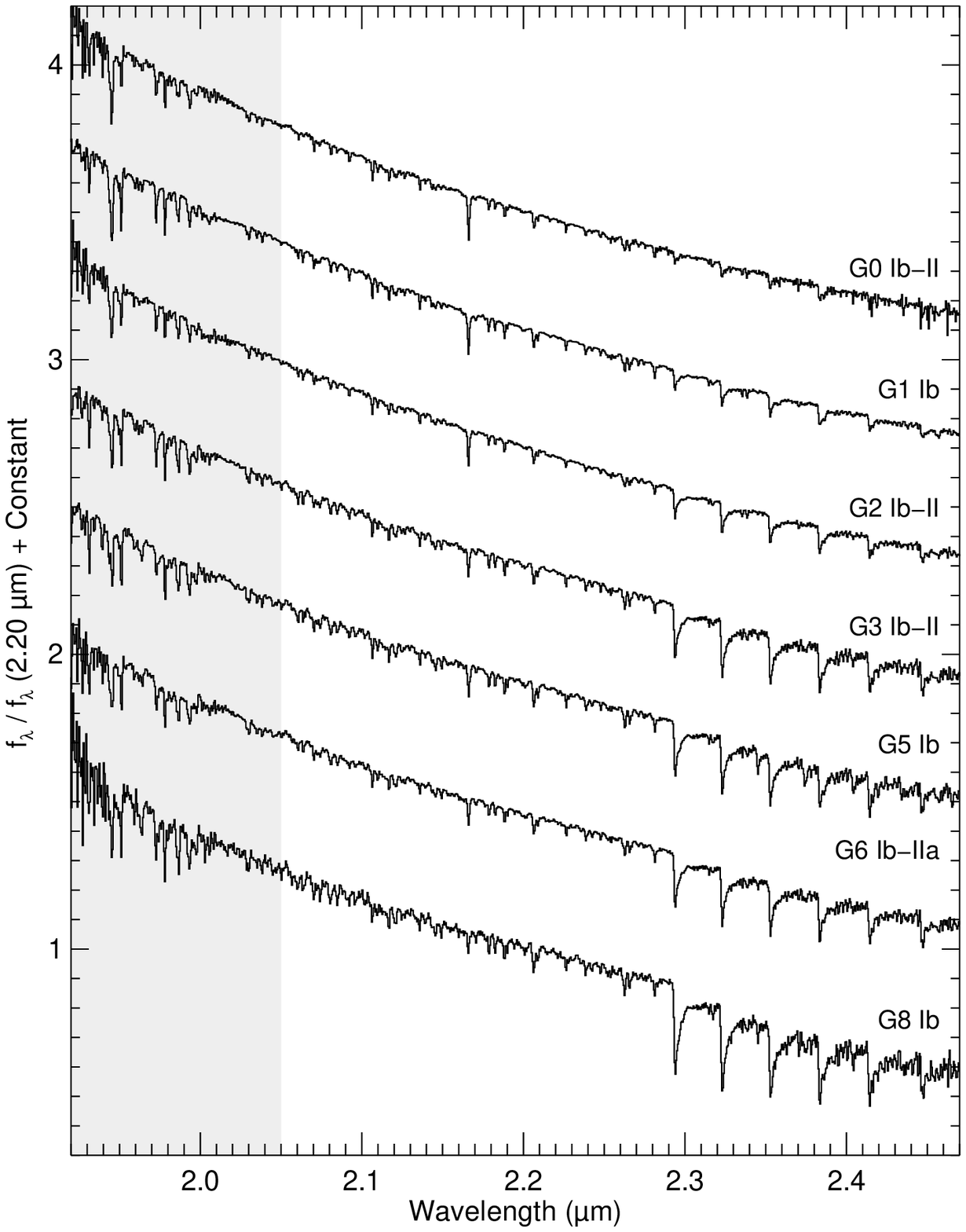}}
\caption{\label{fig:G_IK} A sequence of G supergiant stars plotted over
  the $K$ band (1.92$-$2.5~$\mu$m).  The spectra are of HD~185018
  (G0~Ib-II), HD~74395 (G1~Ib), HD~3421 (G2~Ib-II), HD~192713
  (G3~Ib-II~Wk~H\&K~comp?), HD~190113 (G5~Ib), HD~202314 (G6~Ib-IIa Ca1~B0.5), and
  HD~208606 (G8~Ib).  The spectra have been normalized to unity at
  2.20~$\mu$m and offset by constants.}
\end{figure}

\begin{figure}
\centerline{\includegraphics[width=6.0in,angle=0]{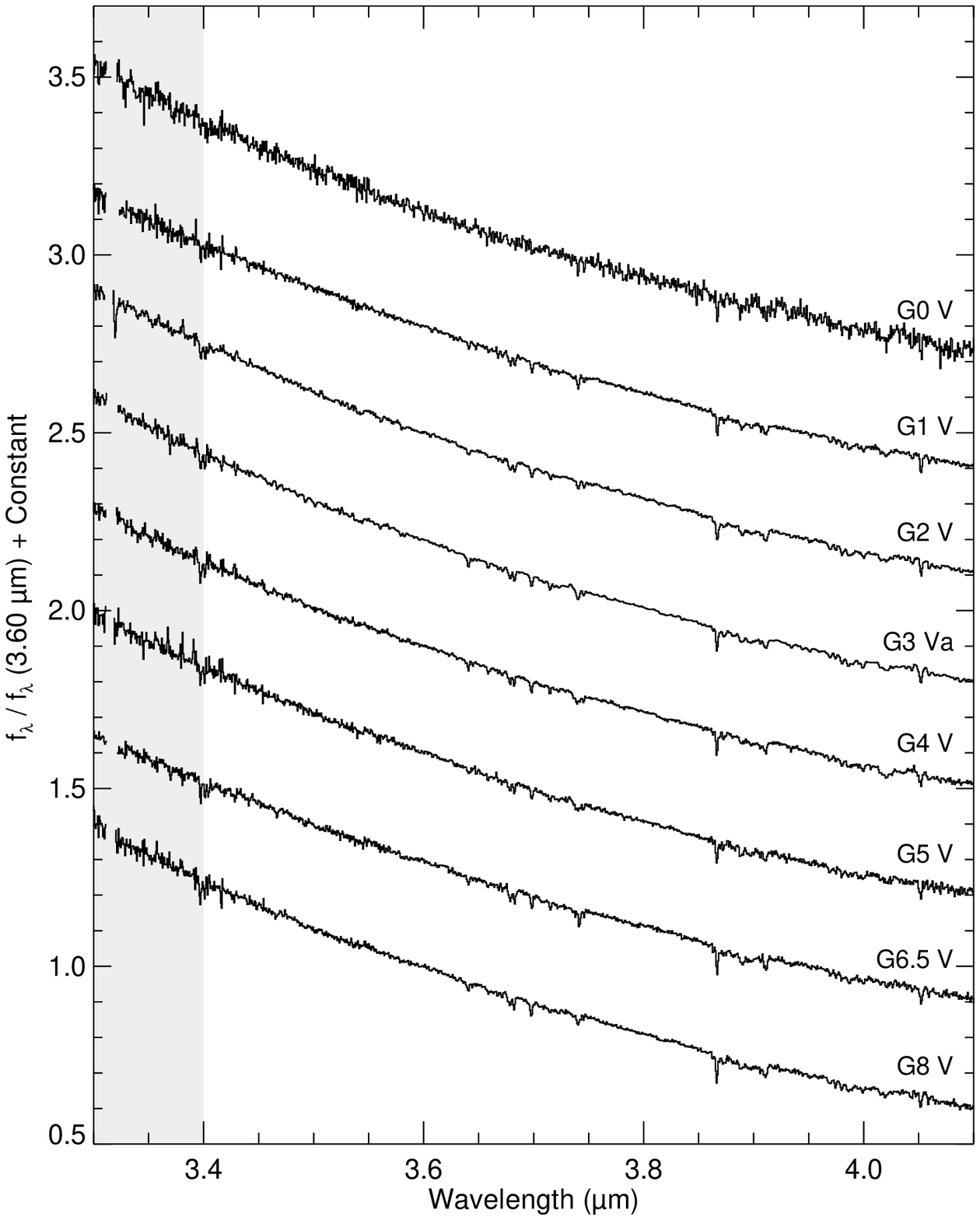}}
\caption{\label{fig:G_VL} A sequence of G dwarf stars plotted over the
  $L'$ band (3.3$-$4.1~$\mu$m).  The spectra are of HD~109358 (G0~V),
  HD~10307 (G1~V), HD~76151 (G2~V), HD~10697 (G3~Va), HD~214850 (G4~V),
  HD~165185 (G5~V), HD~115617 (G6.5~V), and HD~101501 (G8~V).  The
  spectra have been normalized to unity at 3.6~$\mu$m and offset by
  constants.}
\end{figure}

\begin{figure}
\centerline{\includegraphics[width=6.0in,angle=0]{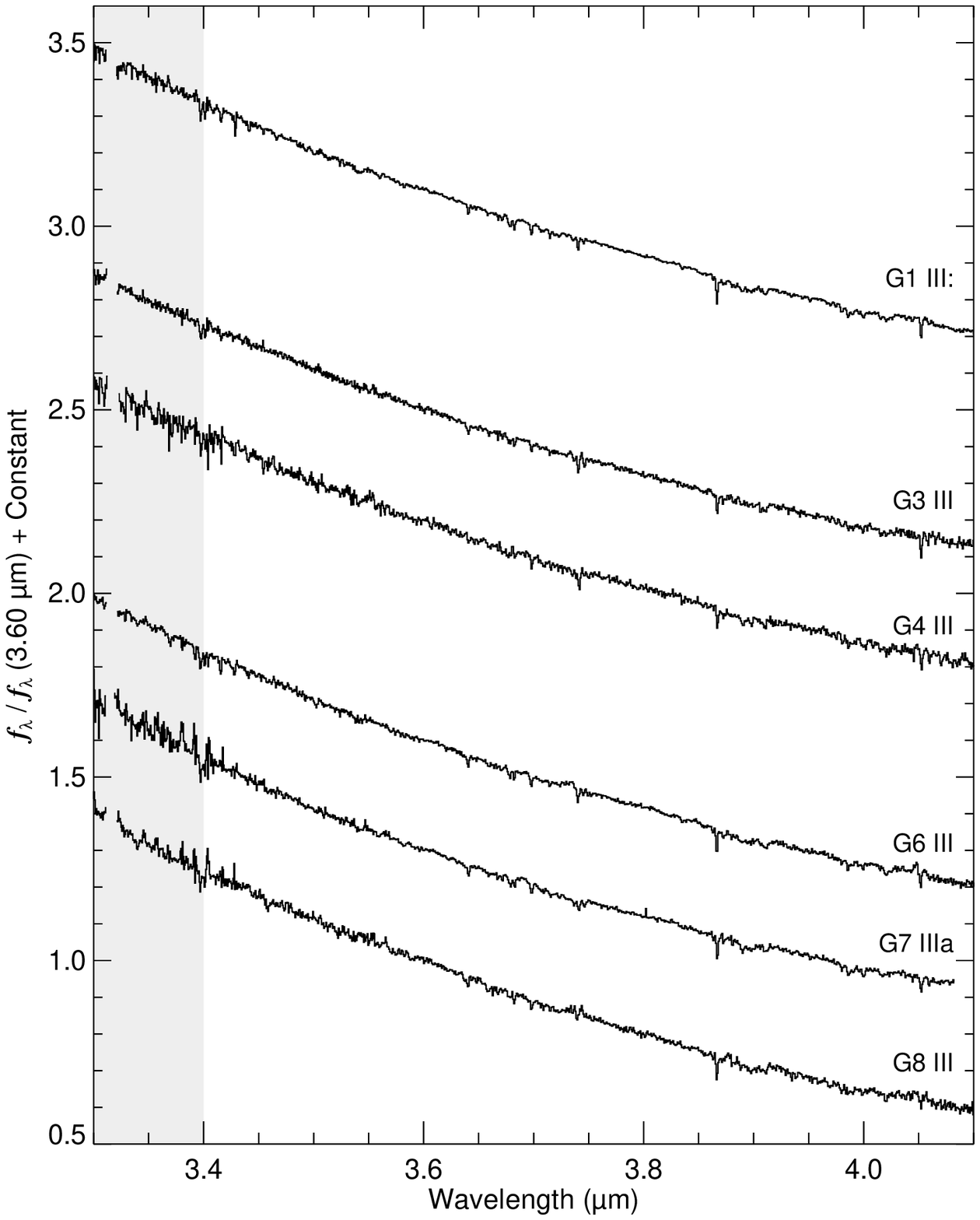}}
\caption{\label{fig:G_IIIL} A sequence of G giant stars plotted over the
  $L'$ band (3.6$-$4.1~$\mu$m).  The spectra are of
  HD~21018 (G1~III:CH-1:), HD~219477 (G2~II-III), HD~88639 (G3b~III~Fe-1),
  HD~108477 (G4~III), HD~193896 (G5~IIIa), HD~27277 (G6~III), HD~182694
  (G7~IIIa), and HD~135722 (G8~III~Fe-1).  The spectra have been
  normalized to unity at 3.6~$\mu$m and offset by constants.}
\end{figure}

\begin{figure}
\centerline{\includegraphics[width=6.0in,angle=0]{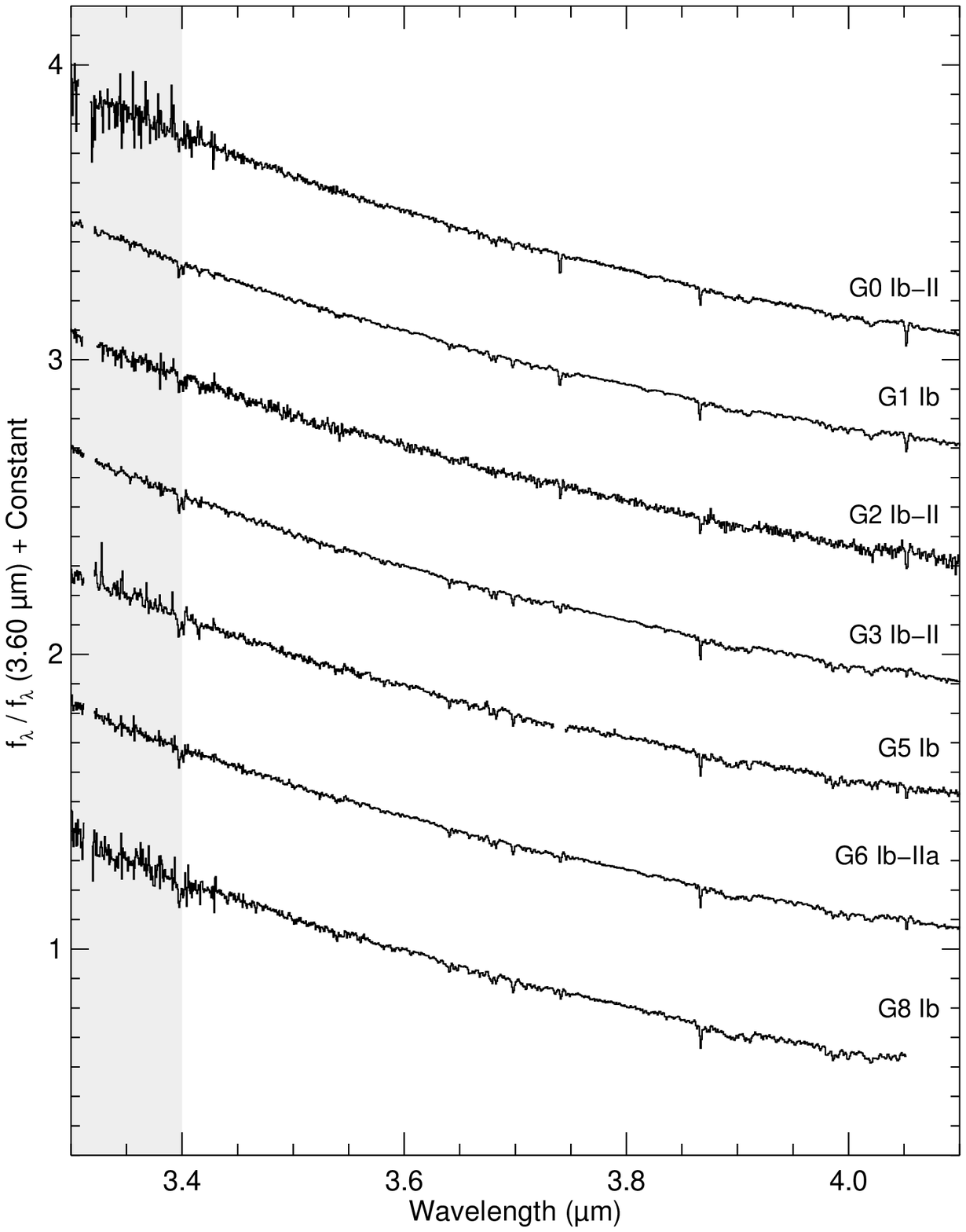}}
\caption{\label{fig:G_IL} A sequence of G supergiant stars plotted over
  the $L'$ band (3.6$-$4.1~$\mu$m).  The spectra are of HD~185018
  (G0~Ib-II), HD~74395 (G1~Ib), HD~3421 (G2~Ib-II), HD~192713
  (G3~Ib-II~Wk~H\&K~comp?), HD~190113 (G5~Ib), HD~202314 (G6~Ib-IIa Ca1~B0.5), and
  HD~208606 (G8~Ib).  The spectra have been normalized to unity at
  3.6~$\mu$m and offset by constants.}
\end{figure}

\begin{figure}
\centerline{\includegraphics[width=6.0in,angle=0]{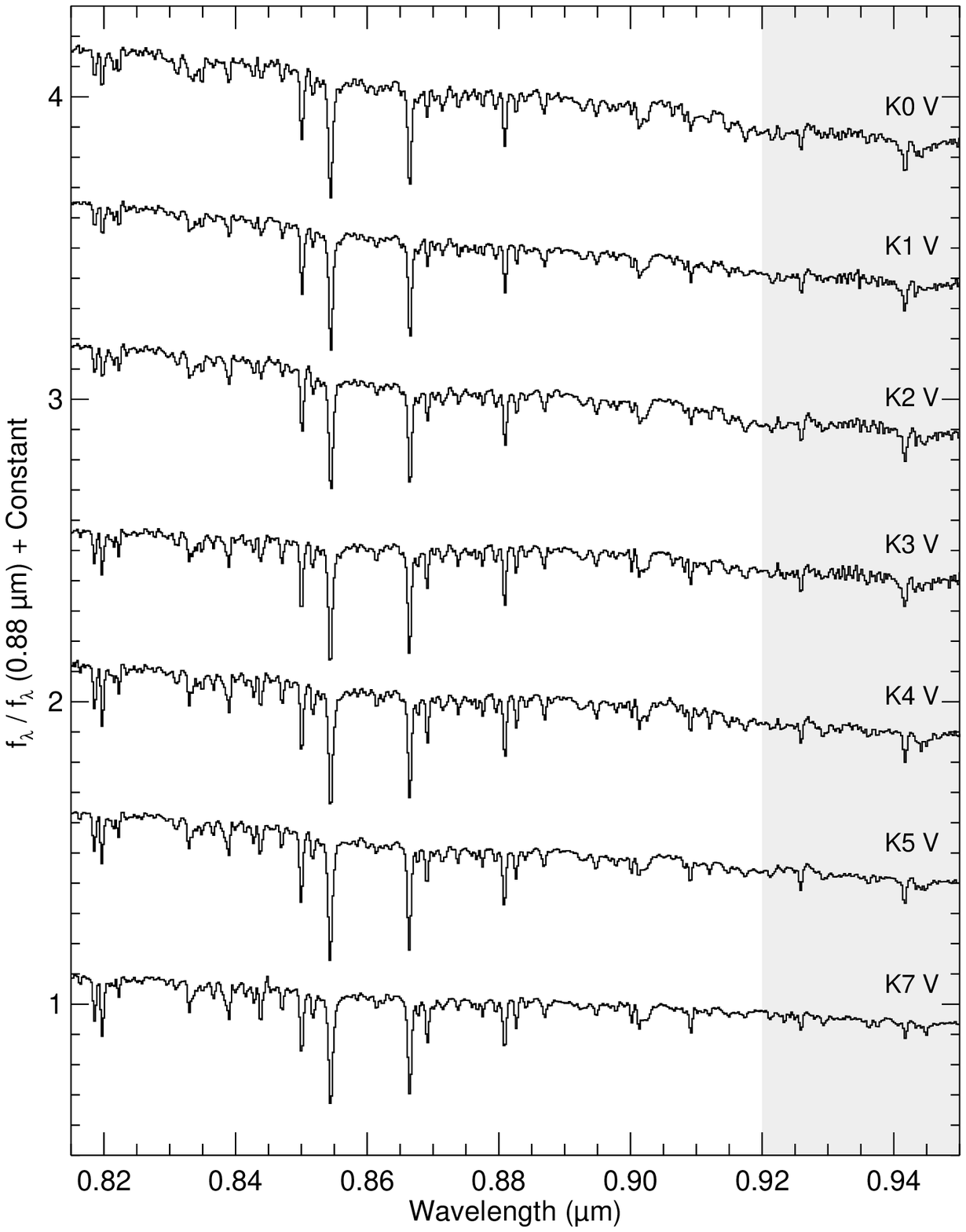}}
\caption{\label{fig:K_VI} A sequence of K dwarf stars plotted over the
  $I$ band (0.82$-$0.95~$\mu$m).  The spectra are of HD~145675 (K0~V),
  HD~10476 (K1~V), HD~3765 (K2~V), HD~219134 (K3~V), HD~45977 (K4~V),
  HD~36003 (K5~V), and HD~237903 (K7~V).  The spectra have been
  normalized to unity at 0.88~$\mu$m and offset by constants.}
\end{figure}

\begin{figure}
\centerline{\includegraphics[width=6.0in,angle=0]{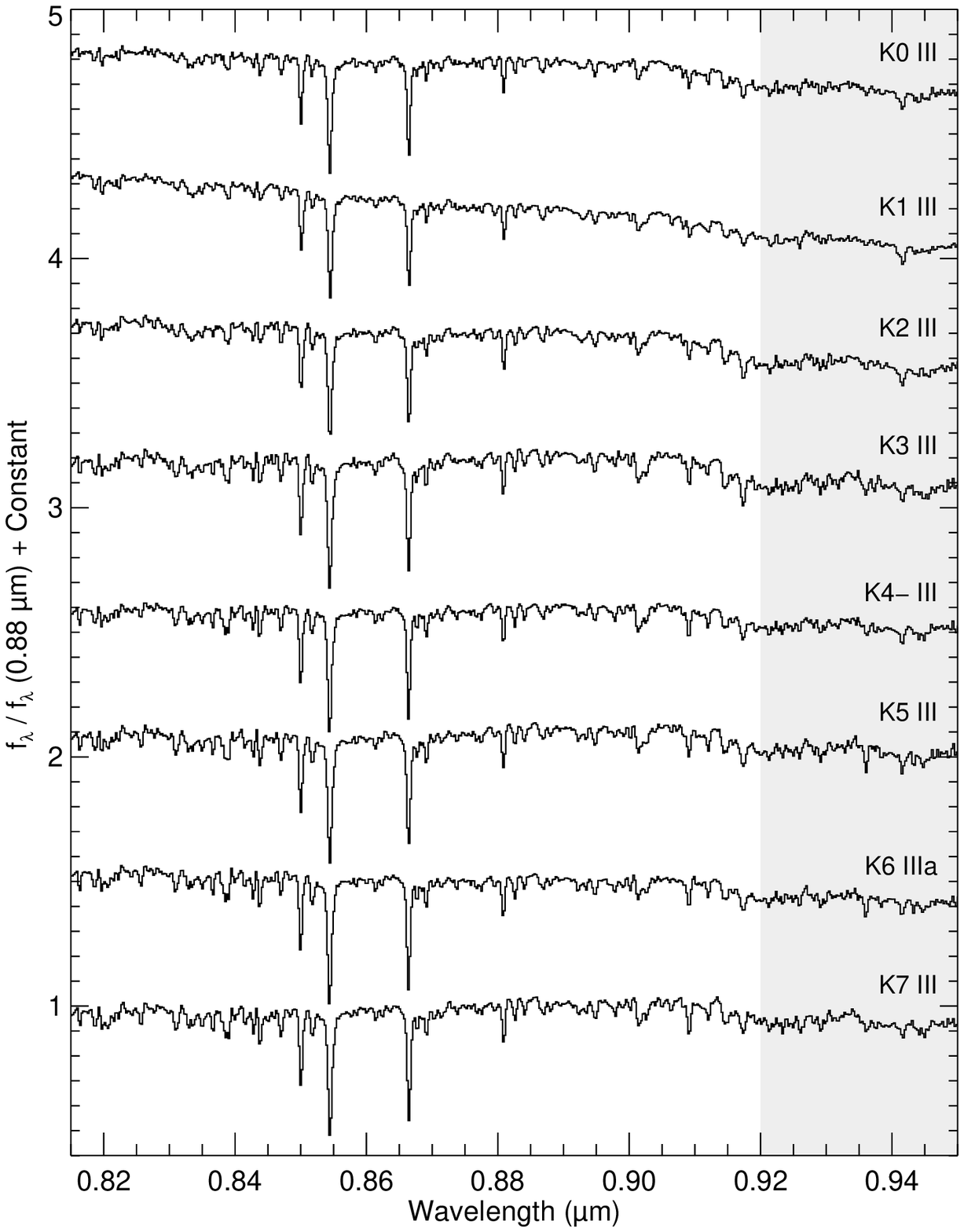}}
\caption{\label{fig:K_IIII} A sequence of K giant stars plotted over the
  $I$ band (0.82$-$0.95~$\mu$m).  The spectra are of HD~100006 (K0~
  III), HD~25975 (K1~III), HD~137759 (K2~III), HD~221246 (K3~III),
  HD~207991 (K4-~III), HD~181596 (K5~III), HD~3346 (K6~IIIa), and
  HD~194193 (K7~III).  The spectra have been normalized to unity at
  0.88~$\mu$m and offset by constants.}
\end{figure}

\begin{figure}
\centerline{\includegraphics[width=6.0in,angle=0]{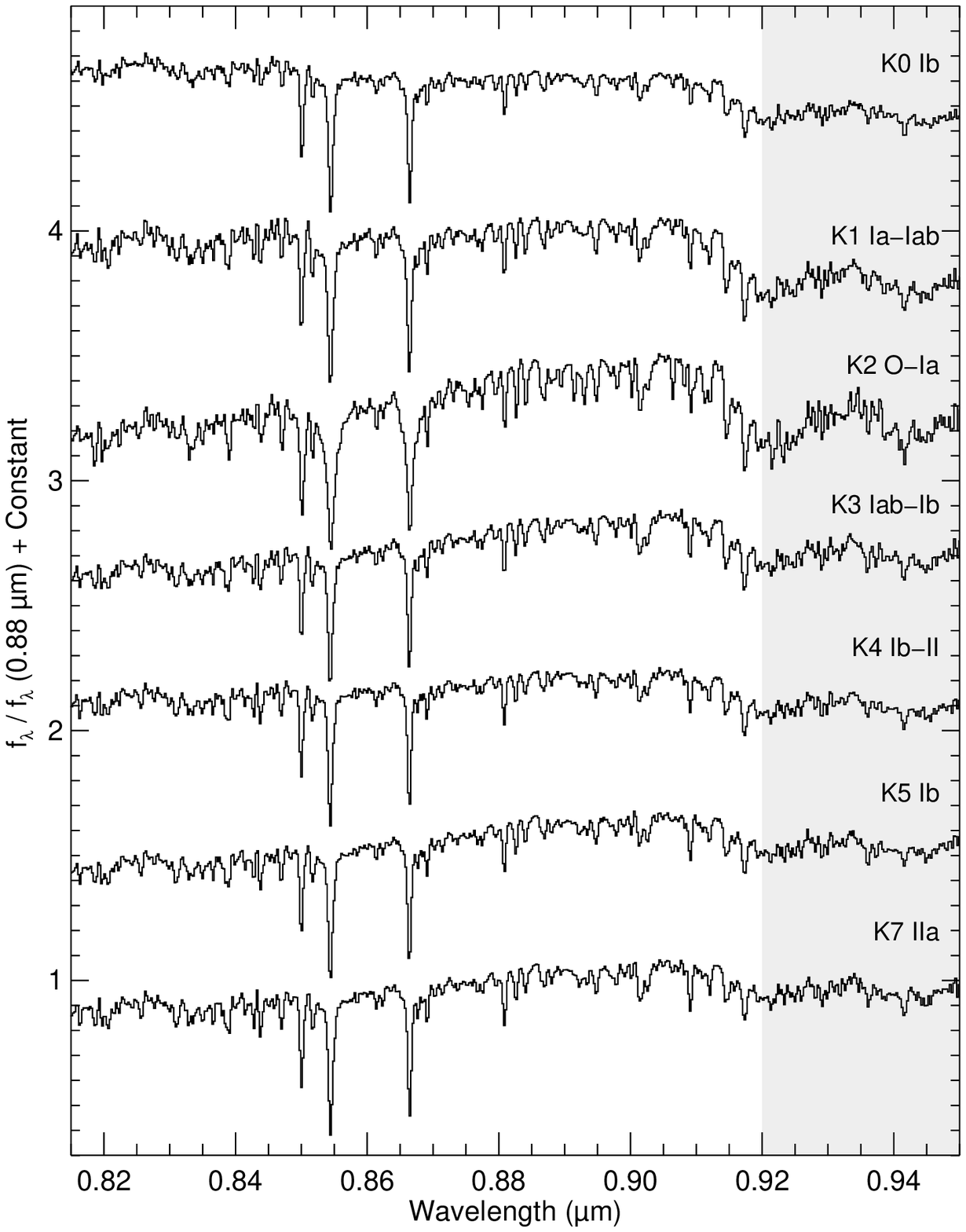}}
\caption{\label{fig:K_II} A sequence of K supergiant stars plotted over
  the $I$ band (0.82$-$0.95~$\mu$m).  The spectra are of HD~44391
  (K0~Ib), HD~63302 (K1~Ia-Iab), HD~212466 (K2~O-Ia), HD~187238
  (K3~Iab-Ib), HD~201065 (K4~Ib-II), HD~216946 (K5~Ib), and HD~181475
  (K7~IIa).  The spectra have been normalized to unity at 0.88~$\mu$m
  and offset by constants.}
\end{figure}

\begin{figure}
\centerline{\includegraphics[width=6.0in,angle=0]{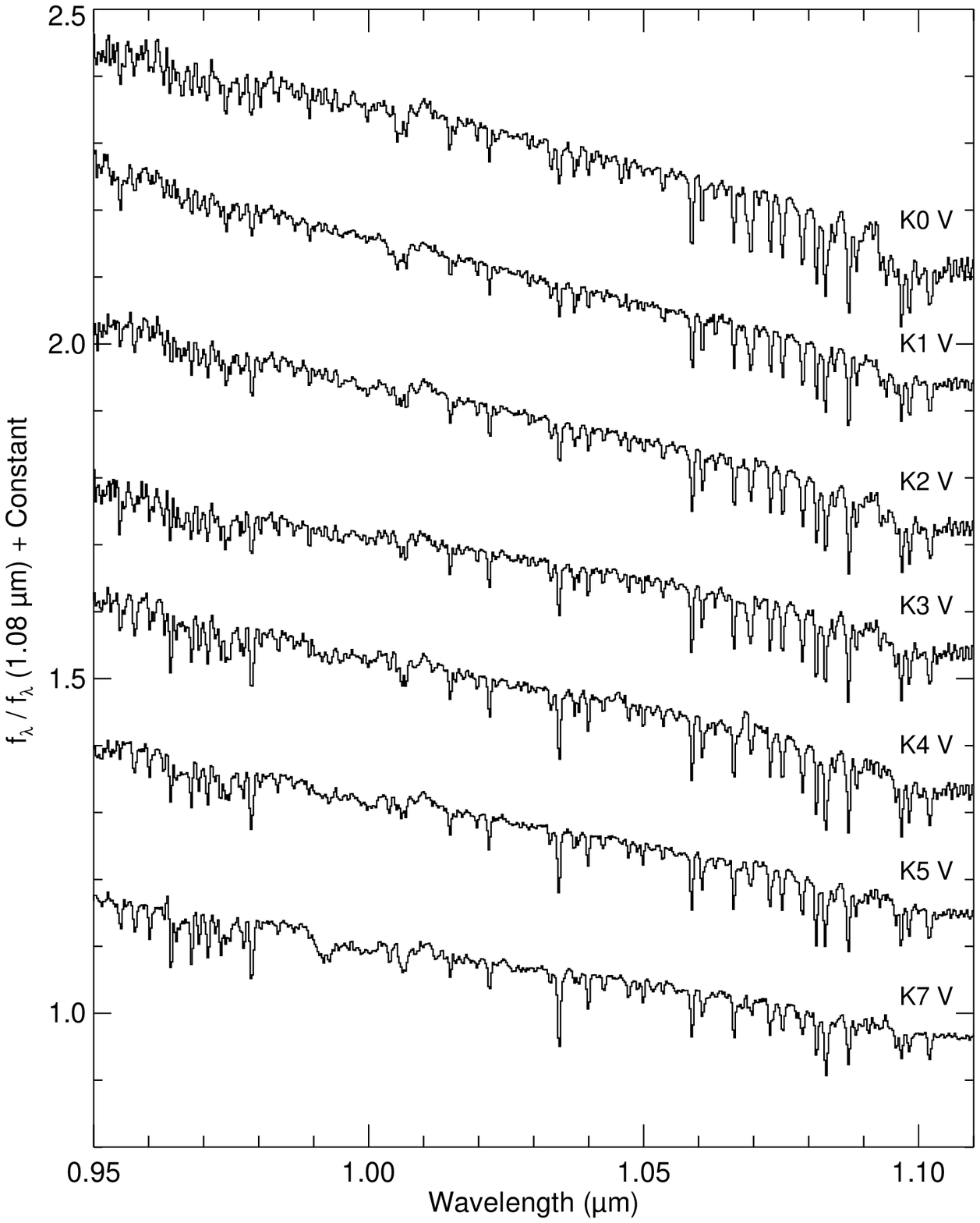}}
\caption{\label{fig:K_VY} A sequence of K dwarf stars plotted over the
  $Y$ band (0.95$-$0.10~$\mu$m).  The spectra are of HD~145675 (K0~V),
  HD~10476 (K1~V), HD~3765 (K2~V), HD~219134 (K3~V), HD~45977 (K4~V),
  HD~36003 (K5~V), and HD~237903 (K7~V).  The spectra have been
  normalized to unity at 1.08~$\mu$m and offset by constants.}
\end{figure}

\begin{figure}
\centerline{\includegraphics[width=6.0in,angle=0]{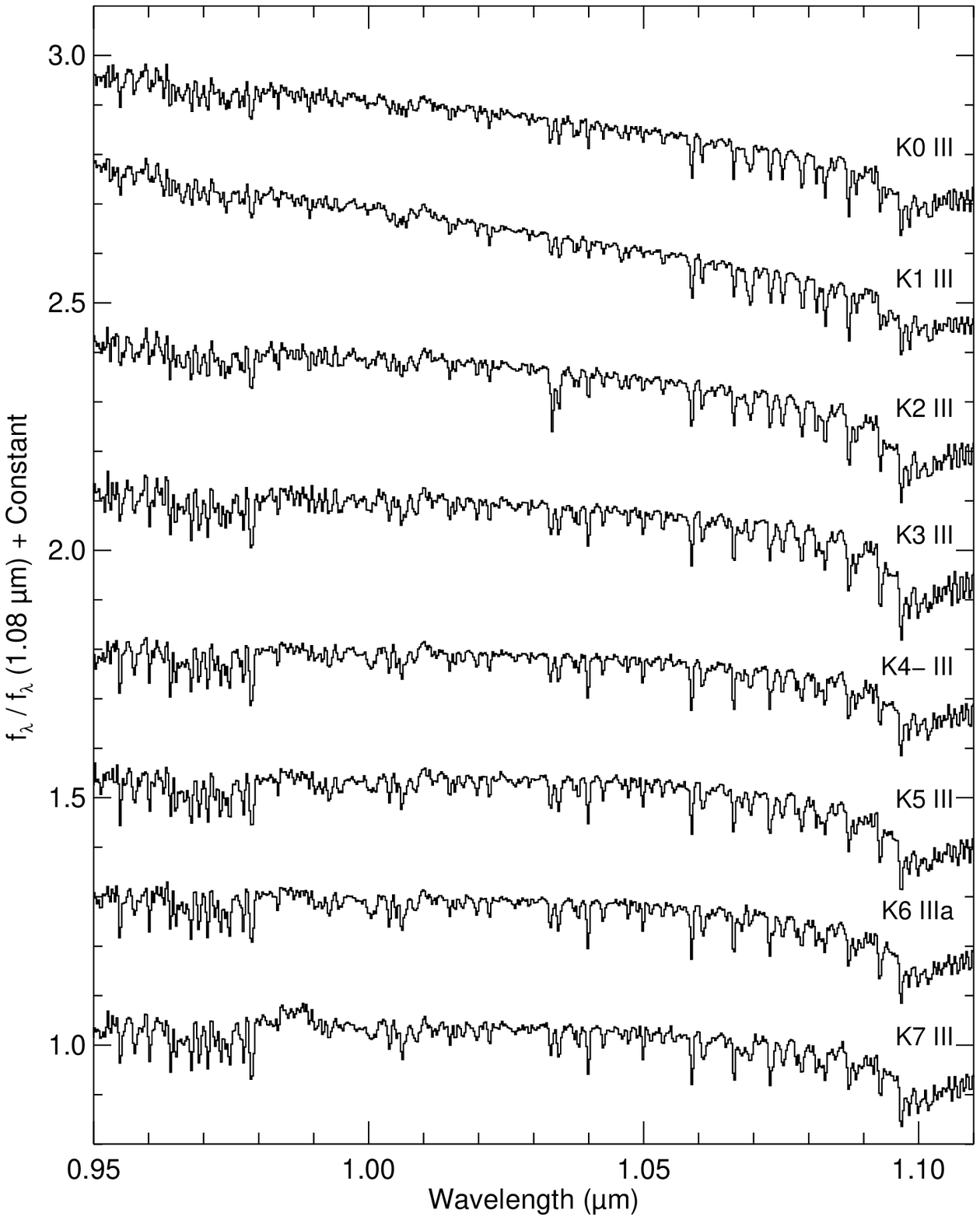}}
\caption{\label{fig:K_IIIY} A sequence of K giant stars plotted over the
  $Y$ band (0.95$-$1.10~$\mu$m).  The spectra are of HD~100006 (K0~
  III), HD~25975 (K1~III), HD~137759 (K2~III), HD~221246 (K3~III),
  HD~207991 (K4-~III), HD~181596 (K5~III), HD~3346 (K6~IIIa), and
  HD~194193 (K7~III).  The spectra have been normalized to unity at
  1.08~$\mu$m and offset by constants.}
\end{figure}

\begin{figure}
\centerline{\includegraphics[width=6.0in,angle=0]{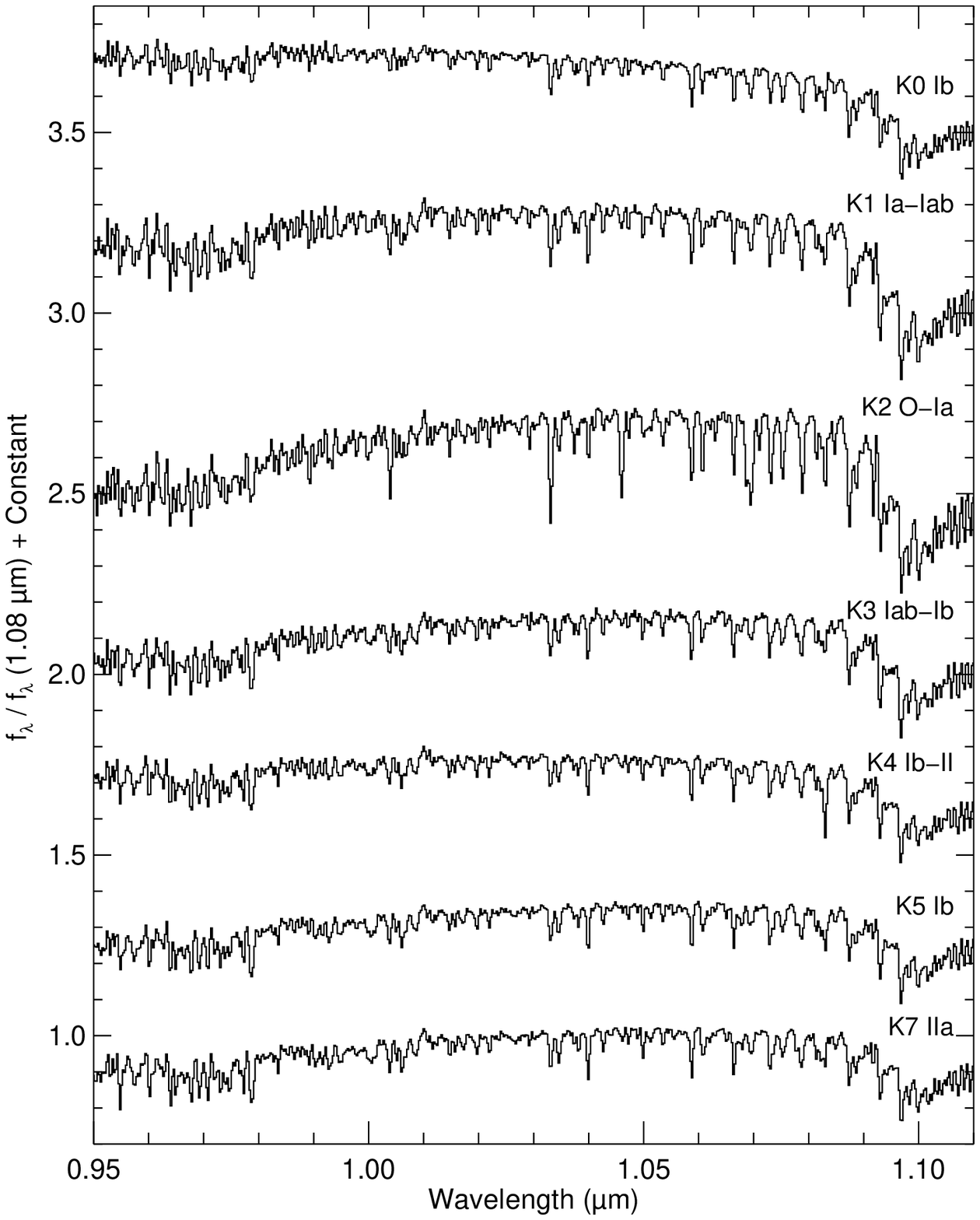}}
\caption{\label{fig:K_IY} A sequence of K supergiant stars plotted over
  the $Y$ band (0.95$-$1.10~$\mu$m).  The spectra are of HD~44391
  (K0~Ib), HD~63302 (K1~Ia-Iab), HD~212466 (K2~O-Ia), HD~187238
  (K3~Iab-Ib), HD~201065 (K4~Ib-II), HD~216946 (K5~Ib), and HD~181475
  (K7~IIa).  The spectra have been normalized to unity at 1.08~$\mu$m
  and offset by constants.}
\end{figure}

\begin{figure}
\centerline{\includegraphics[width=6.0in,angle=0]{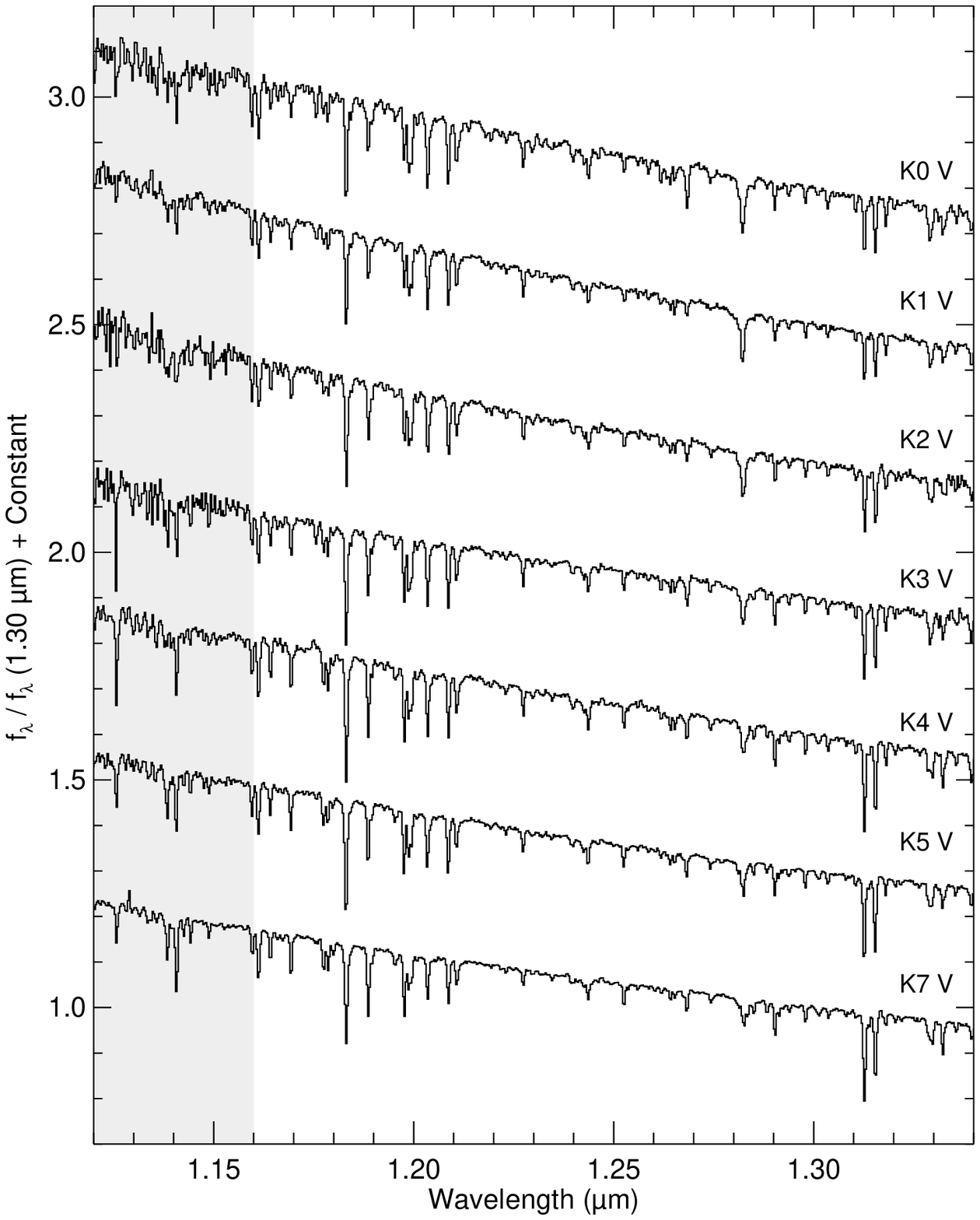}}
\caption{\label{fig:K_VJ} A sequence of K dwarf stars plotted over the
  $J$ band (1.12$-$1.34~$\mu$m).  The spectra are of HD~145675 (K0~V),
  HD~10476 (K1~V), HD~3765 (K2~V), HD~219134 (K3~V), HD~45977 (K4~V),
  HD~36003 (K5~V), and HD~237903 (K7~V).  The spectra have been
  normalized to unity at 1.30~$\mu$m and offset by constants.}
\end{figure}

\begin{figure}
\centerline{\includegraphics[width=6.0in,angle=0]{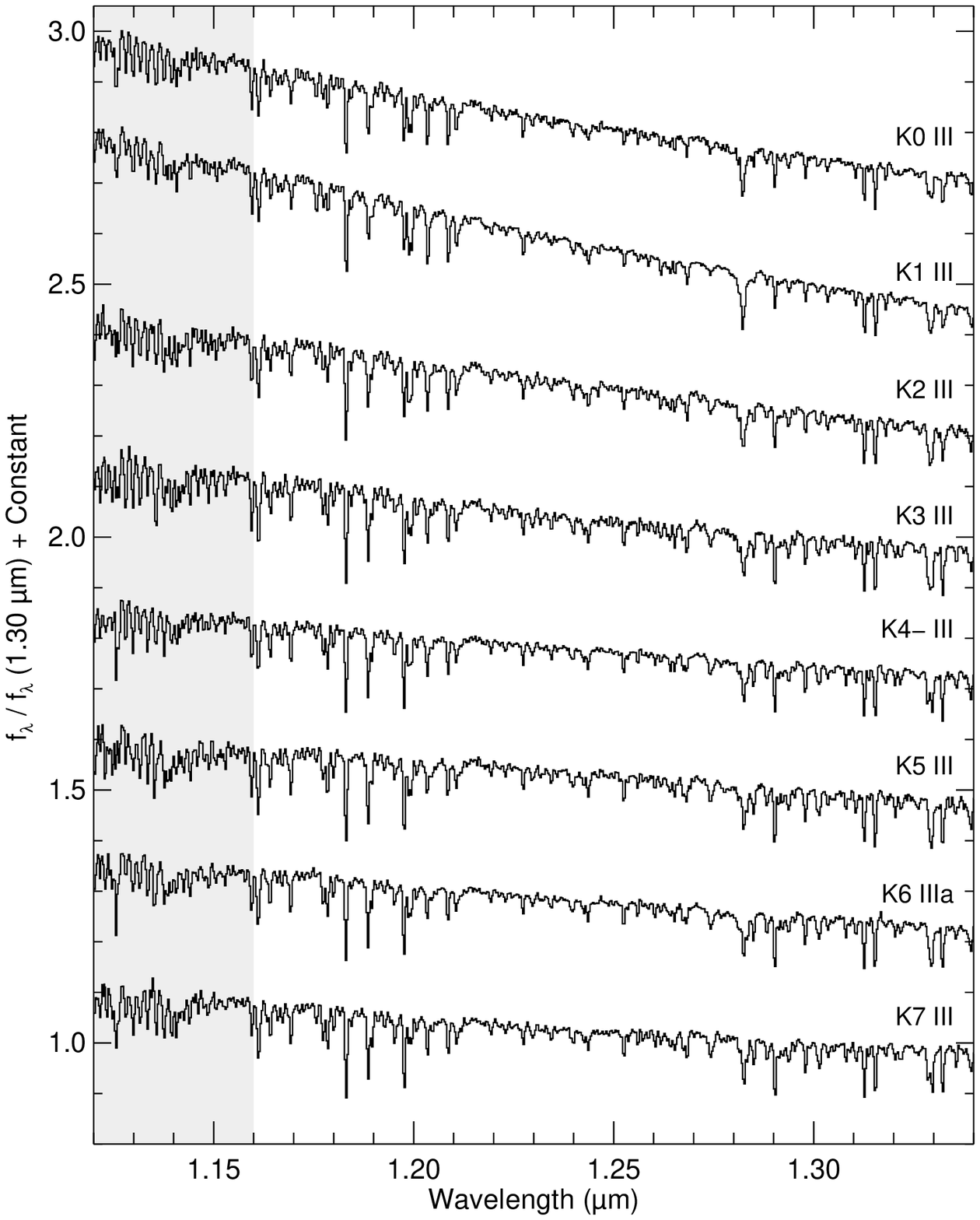}}
\caption{\label{fig:K_IIIJ} A sequence of K giant stars plotted over the
  $J$ band (1.12$-$1.34~$\mu$m).  The spectra are of HD~100006 (K0~
  III), HD~25975 (K1~III), HD~137759 (K2~III), HD~221246 (K3~III),
  HD~207991 (K4-~III), HD~181596 (K5~III), HD~3346 (K6~IIIa), and
  HD~194193 (K7~III).  The spectra have been normalized to unity at
  1.30~$\mu$m and offset by constants.}
\end{figure}

\begin{figure}
\centerline{\includegraphics[width=6.0in,angle=0]{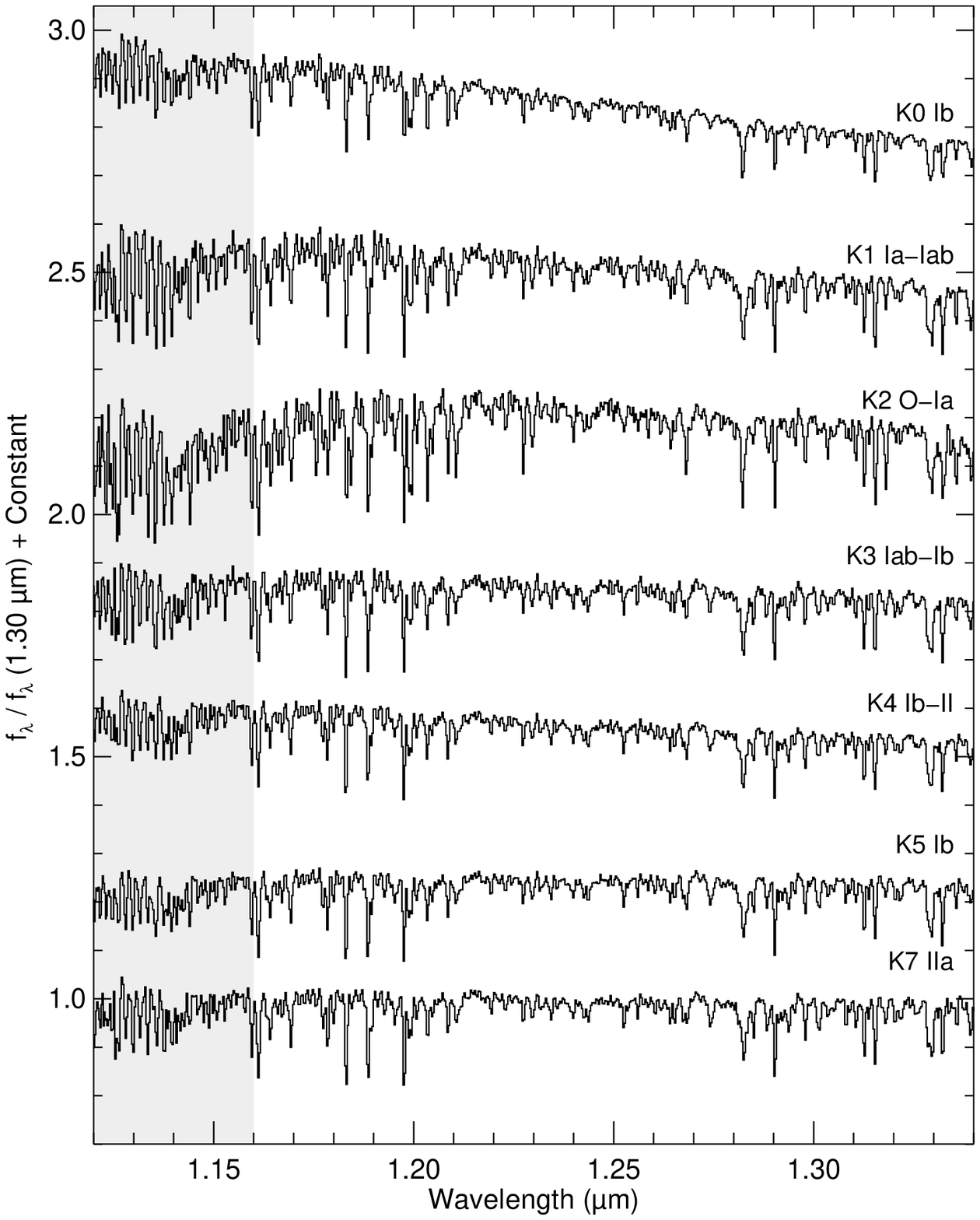}}
\caption{\label{fig:K_IJ} A sequence of K supergiant stars plotted over
  the $J$ band (1.12$-$1.34~$\mu$m).  The spectra are of HD~44391
  (K0~Ib), HD~63302 (K1~Ia-Iab), HD~212466 (K2~O-Ia), HD~187238
  (K3~Iab-Ib), HD~201065 (K4~Ib-II), HD~216946 (K5~Ib), and HD~181475
  (K7~IIa).  The spectra have been normalized to unity at 1.30~$\mu$m
  and offset by constants.}
\end{figure}

\begin{figure}
\centerline{\includegraphics[width=6.0in,angle=0]{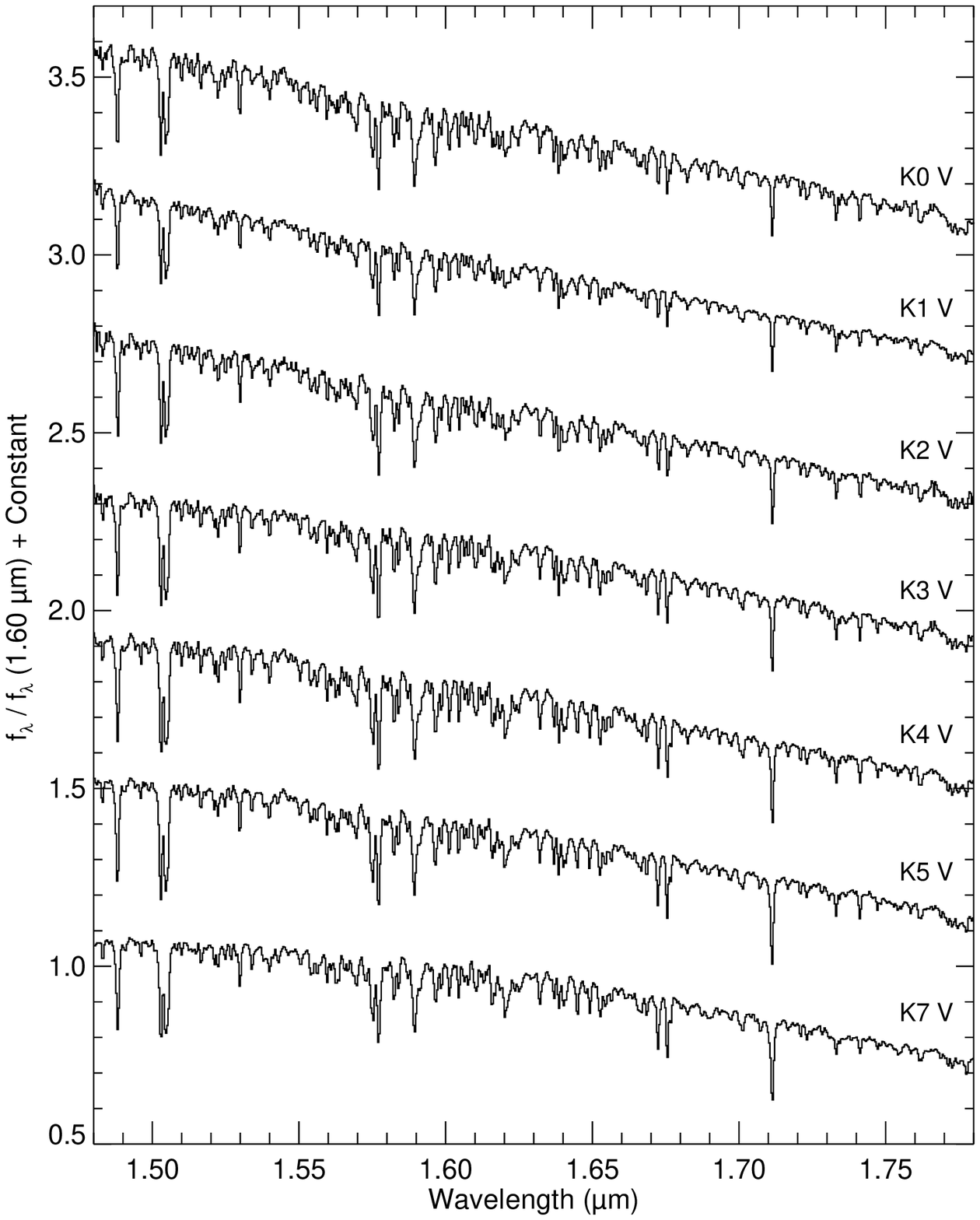}}
\caption{\label{fig:K_VH} A sequence of K dwarf stars plotted over the
  $H$ band (1.48$-$1.78~$\mu$m).  The spectra are of HD~145675 (K0~V),
  HD~10476 (K1~V), HD~3765 (K2~V), HD~219134 (K3~V), HD~45977 (K4~V),
  HD~36003 (K5~V), and HD~237903 (K7~V).  The spectra have been
  normalized to unity at 1.60~$\mu$m and offset by constants.}
\end{figure}

\begin{figure}
\centerline{\includegraphics[width=6.0in,angle=0]{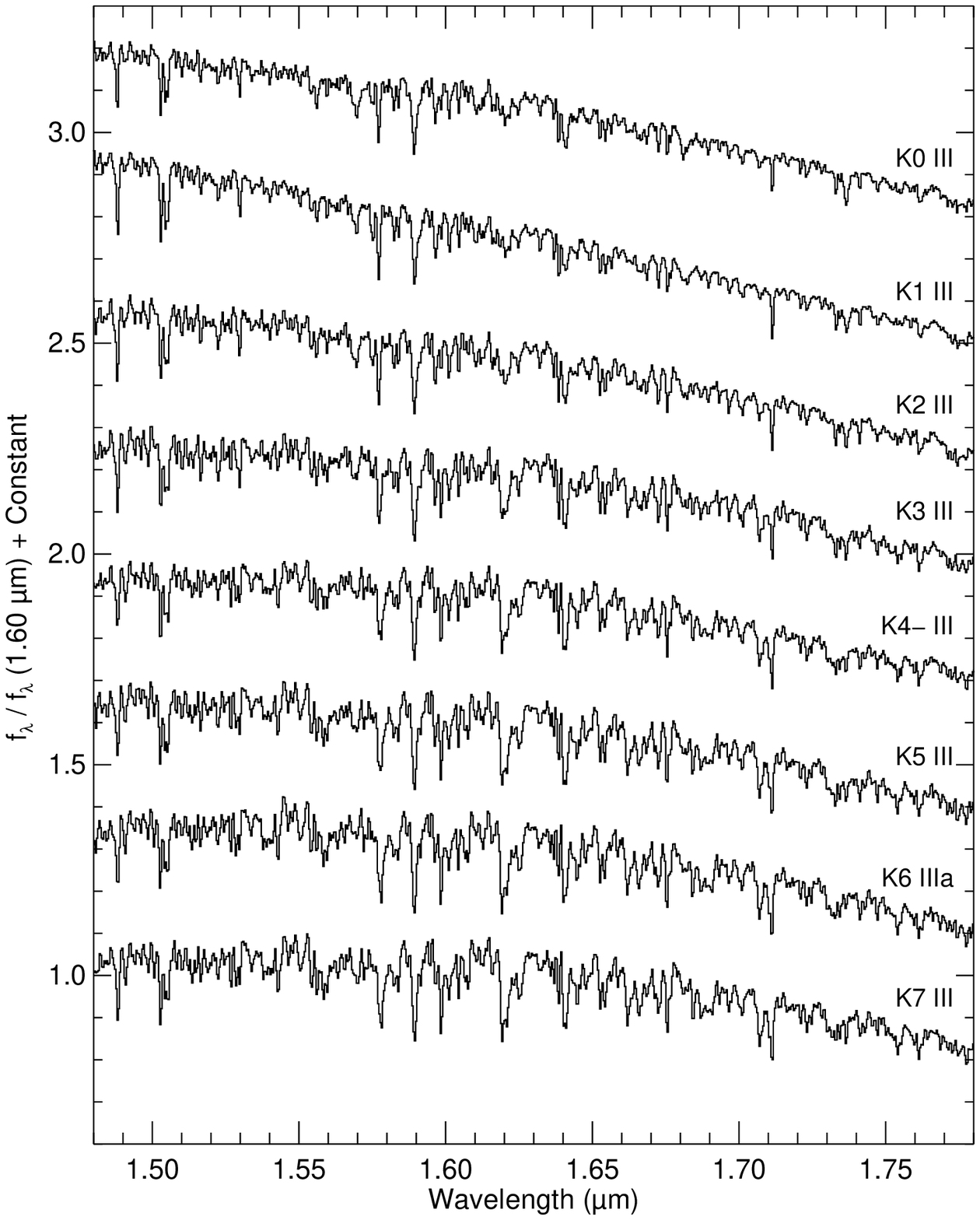}}
\caption{\label{fig:K_IIIH} A sequence of K giant stars plotted over the
  $H$ band (1.48$-$1.78~$\mu$m).  The spectra are of HD~100006 (K0~
  III), HD~25975 (K1~III), HD~137759 (K2~III), HD~221246 (K3~III),
  HD~207991 (K4-~III), HD~181596 (K5~III), HD~3346 (K6~IIIa), and
  HD~194193 (K7~III).  The spectra have been normalized to unity at
  1.60~$\mu$m and offset by constants.}
\end{figure}

\begin{figure}
\centerline{\includegraphics[width=6.0in,angle=0]{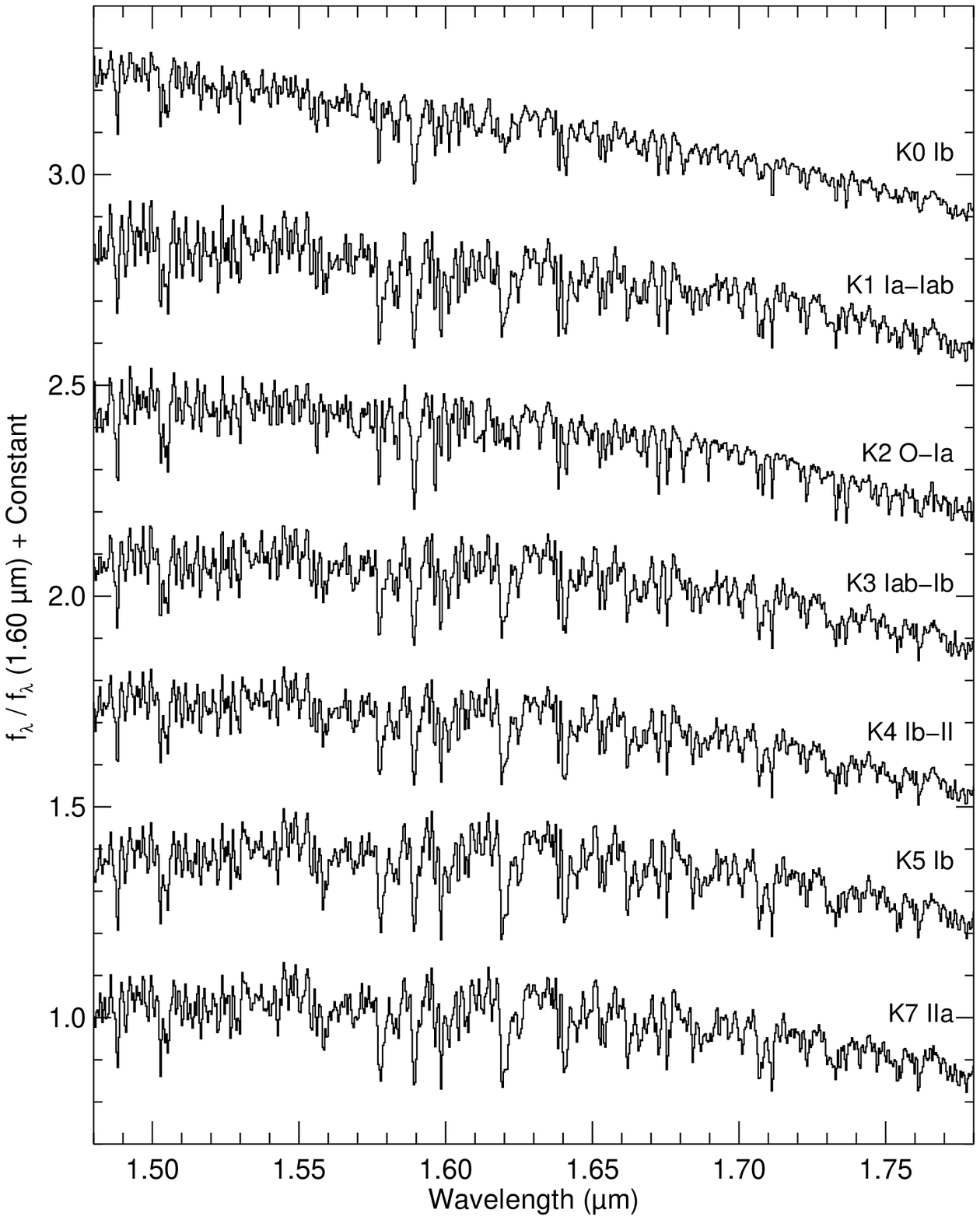}}
\caption{\label{fig:K_IH} A sequence of K supergiant stars plotted over
  the $H$ band (1.48$-$1.78~$\mu$m).  The spectra are of HD~44391
  (K0~Ib), HD~63302 (K1~Ia-Iab), HD~212466 (K2~O-Ia), HD~187238
  (K3~Iab-Ib), HD~201065 (K4~Ib-II), HD~216946 (K5~Ib), and HD~181475
  (K7~IIa).  The spectra have been normalized to unity at 1.60~$\mu$m
  and offset by constants.}
\end{figure}

\begin{figure}
\centerline{\includegraphics[width=6.0in,angle=0]{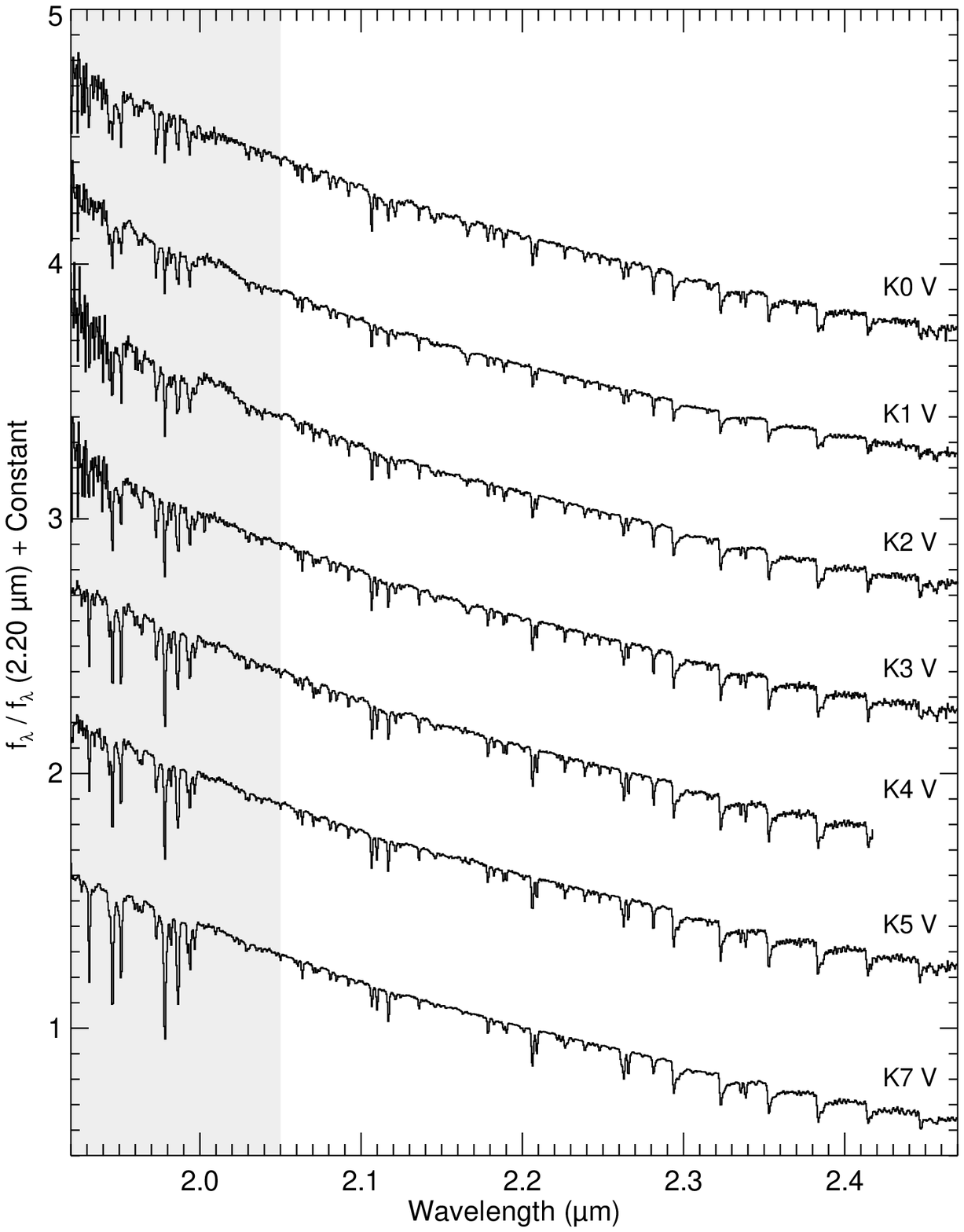}}
\caption{\label{fig:K_VK} A sequence of K dwarf stars plotted over the
  $K$ band (1.92$-$2.5~$\mu$m).  The spectra are of HD~145675 (K0~V),
  HD~10476 (K1~V), HD~3765 (K2~V), HD~219134 (K3~V), HD~45977 (K4~V),
  HD~36003 (K5~V), and HD~237903 (K7~V).  The spectra have been
  normalized to unity at 2.20~$\mu$m and offset by constants.}
\end{figure}

\begin{figure}
\centerline{\includegraphics[width=6.0in,angle=0]{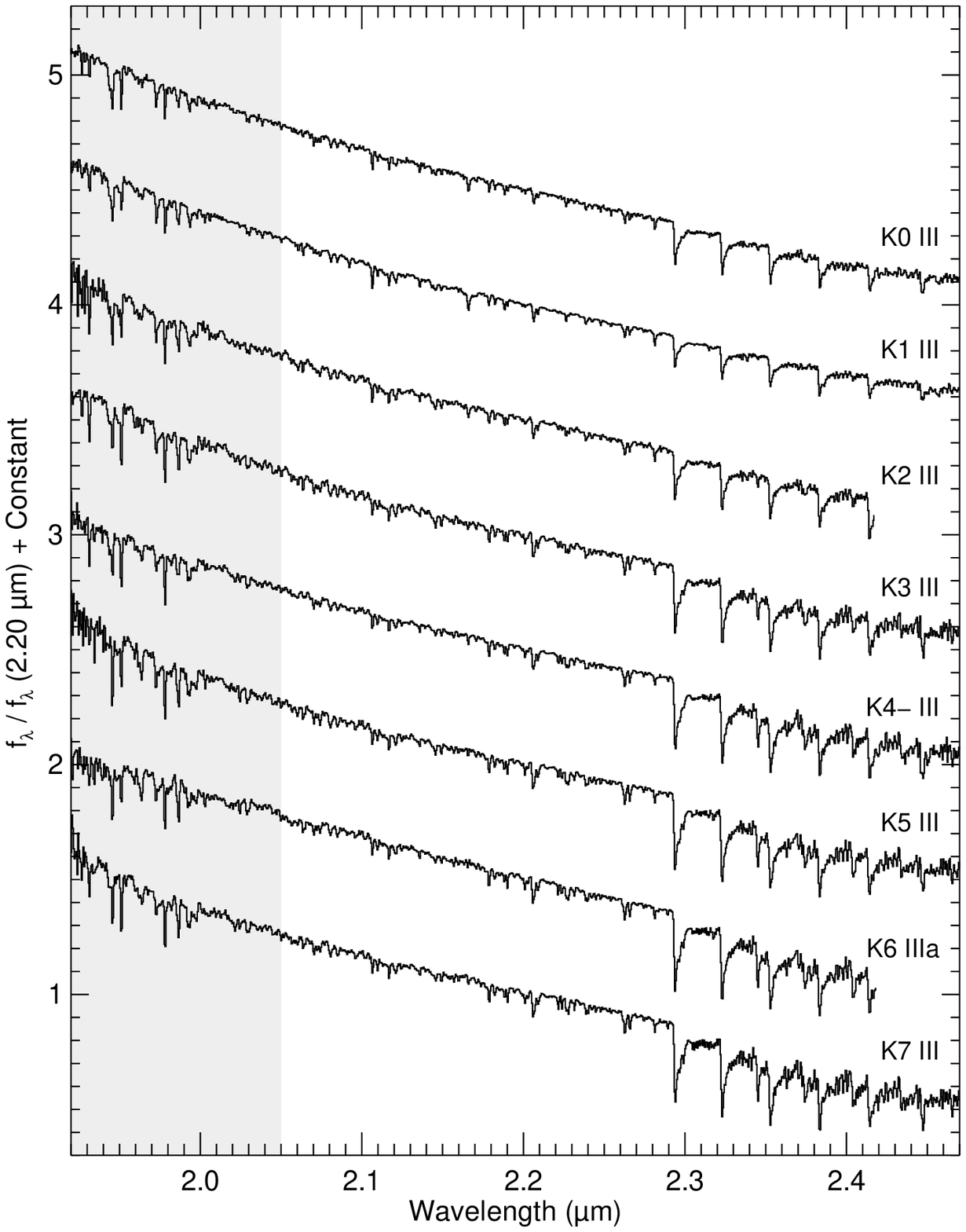}}
\caption{\label{fig:K_IIIK} A sequence of K giant stars plotted over the
  $K$ band (1.92$-$2.5~$\mu$m).  The spectra are of HD~100006 (K0~ III),
  HD~25975 (K1~III), HD~137759 (K2~III), HD~221246 (K3~III), HD~207991
  (K4-~III), HD~181596 (K5~III), HD~3346 (K6~IIIa), and HD~194193
  (K7~III).  The spectra have been normalized to unity at 2.20~$\mu$m
  and offset by constants.}
\end{figure}

\begin{figure}
\centerline{\includegraphics[width=6.0in,angle=0]{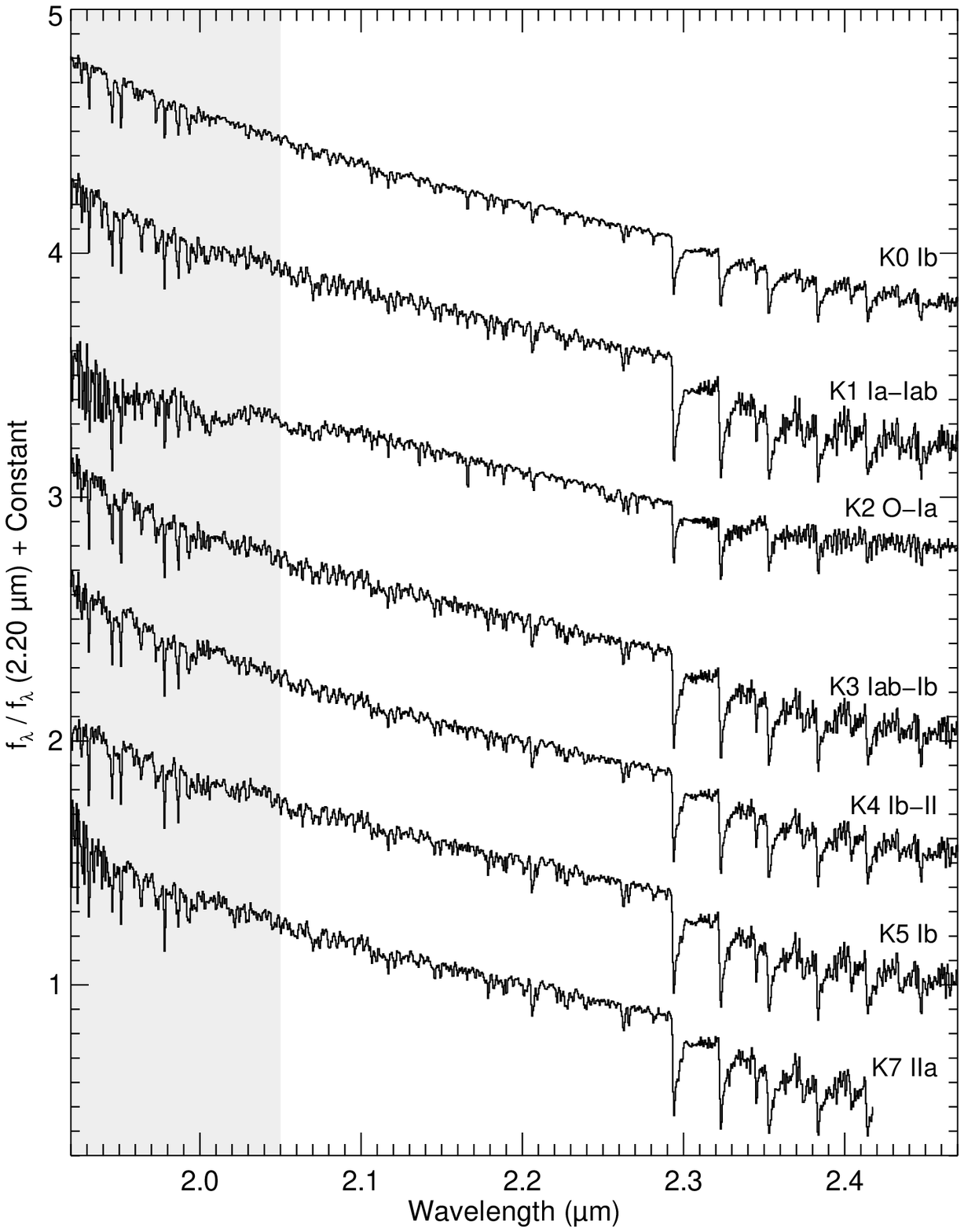}}
\caption{\label{fig:K_IK} A sequence of K supergiant stars plotted over
  the $K$ band (1.92$-$2.5~$\mu$m).  The spectra are of HD~44391
  (K0~Ib), HD~63302 (K1~Ia-Iab), HD~212466 (K2~O-Ia), HD~187238
  (K3~Iab-Ib), HD~201065 (K4~Ib-II), HD~216946 (K5~Ib), and HD~181475
  (K7~IIa).  The spectra have been normalized to unity at 2.20~$\mu$m
  and offset by constants.}
\end{figure}

\begin{figure}
\centerline{\includegraphics[width=6.0in,angle=0]{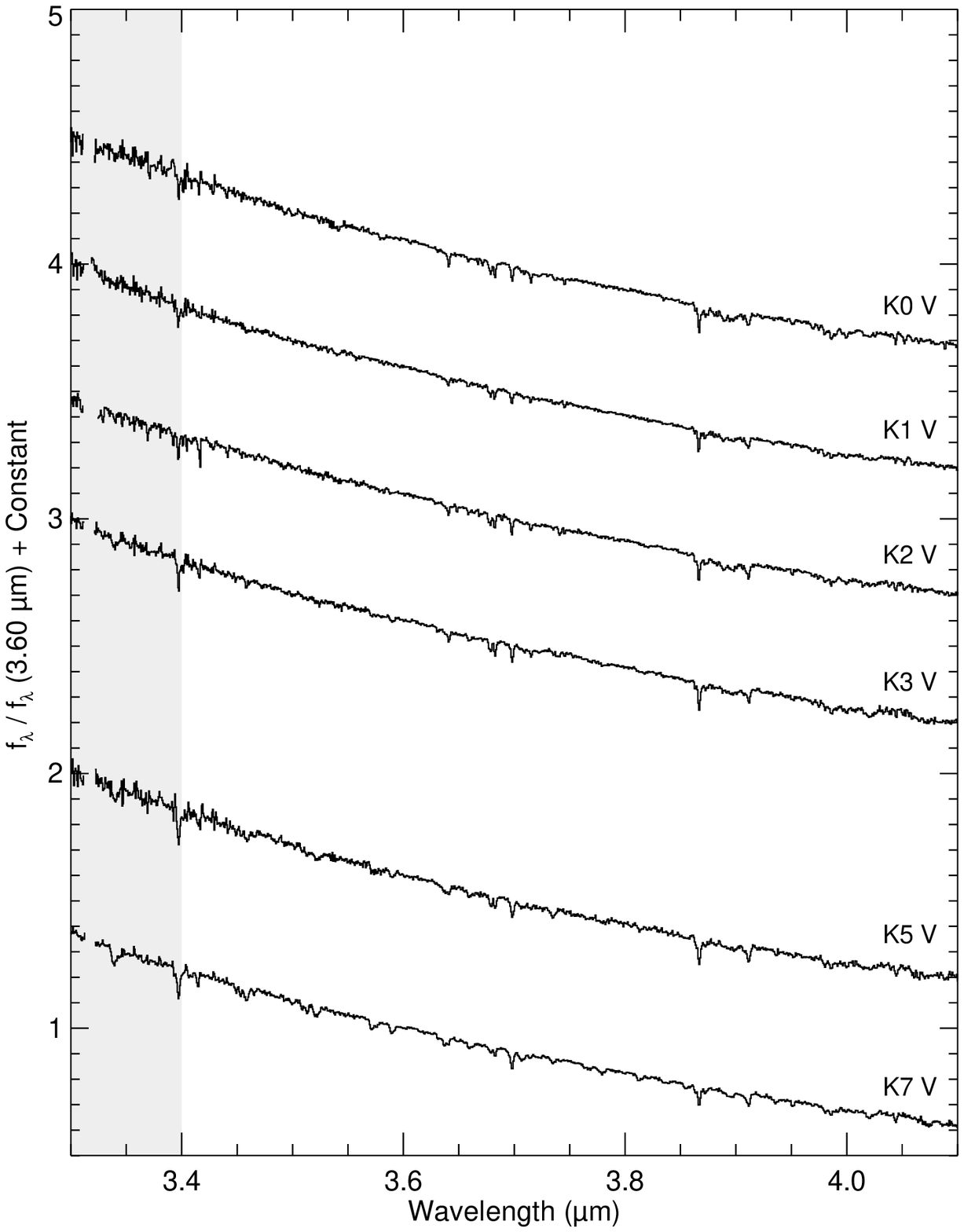}}
\caption{\label{fig:K_VL} A sequence of K dwarf stars plotted over the
  $L'$ band (3.6$-$4.1~$\mu$m).  The spectra are of HD~145675 (K0~V),
  HD~10476 (K1~V), HD~3765 (K2~V), HD~219134 (K3~V),
  HD~36003 (K5~V), and HD~237903 (K7~V).  The spectra have been
  normalized to unity at 2.20~$\mu$m and offset by constants.}
\end{figure}

\begin{figure}
\centerline{\includegraphics[width=6.0in,angle=0]{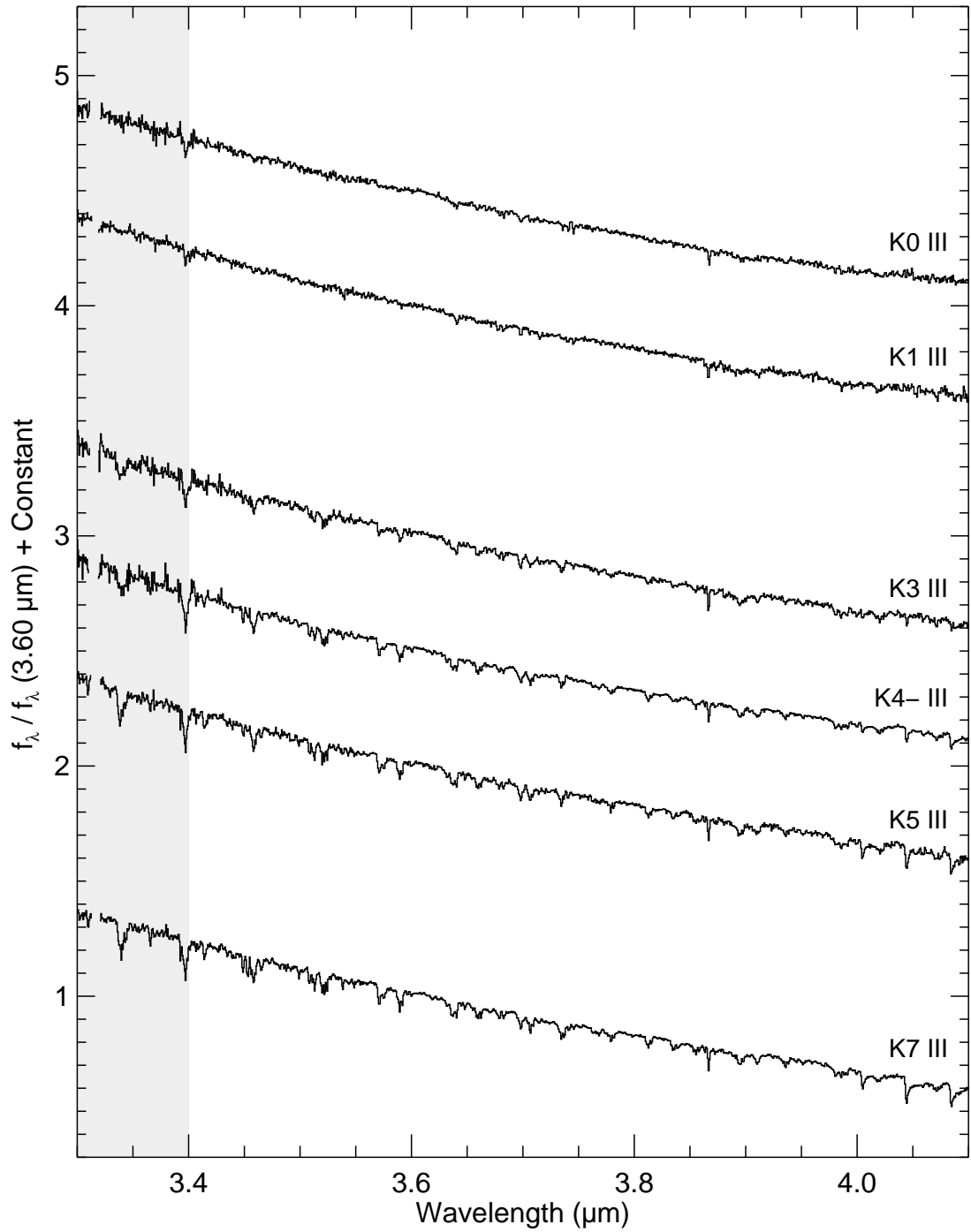}}
\caption{\label{fig:K_IIIL} A sequence of K giant stars plotted over the
  $L'$ band (3.6$-$4.1~$\mu$m).  The spectra are of HD~100006 (K0~ III),
  HD~25975 (K1~III), HD~137759 (K2~III), HD~221246 (K3~III), HD~207991
  (K4-~III), HD~181596 (K5~III), and HD~194193
  (K7~III).  The spectra have been normalized to unity at 3.6~$\mu$m
  and offset by constants.}
\end{figure}

\begin{figure}
\centerline{\includegraphics[width=6.0in,angle=0]{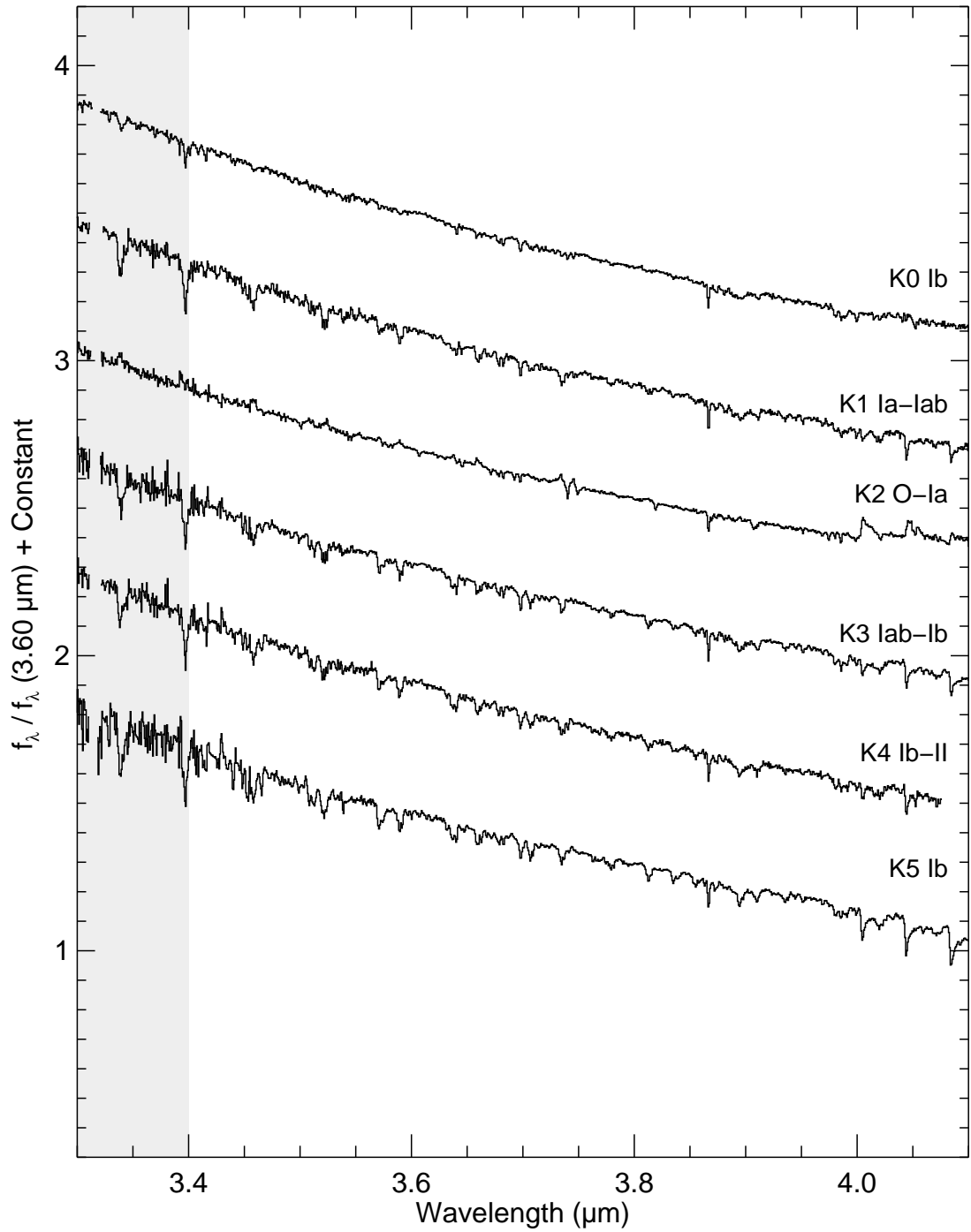}}
\caption{\label{fig:K_IL} A sequence of K supergiant stars plotted over
  the $L'$ band (3.6$-$4.1~$\mu$m).  The spectra are of HD~44391
  (K0~Ib), HD~63302 (K1~Ia-Iab), HD~212466 (K2~O-Ia), HD~187238
  (K3~Iab-Ib), HD~201065 (K4~Ib-II), and HD~216946 (K5~Ib).
  The spectra have been normalized to unity at 3.6~$\mu$m
  and offset by constants.}
\end{figure}

\begin{figure}
\centerline{\includegraphics[width=6.0in,angle=0]{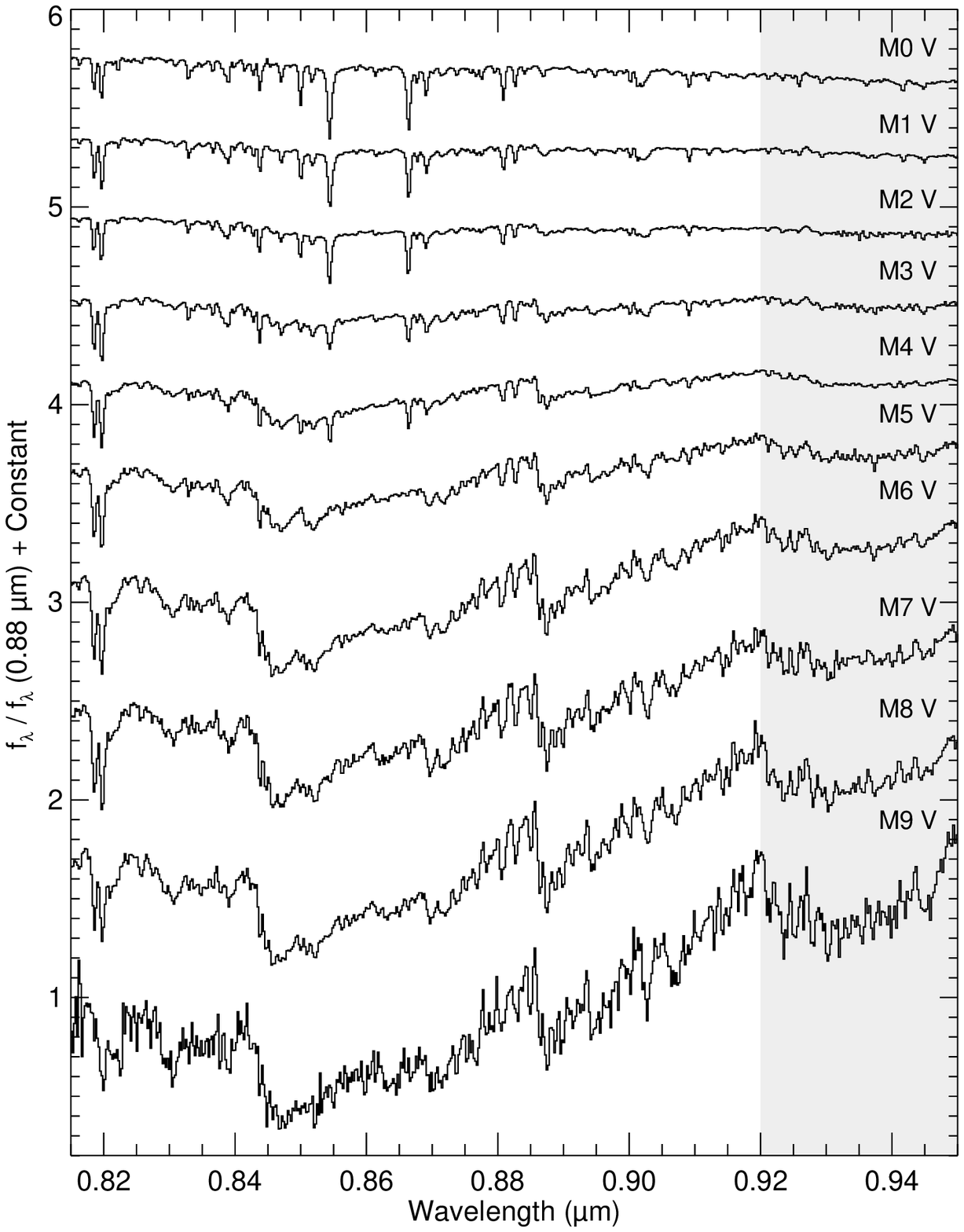}}
\caption{\label{fig:M_VI} A sequence of M dwarf stars plotted over the
  $I$ band (0.82$-$0.95~$\mu$m).  The spectra are of HD~19305 (M0~V),
  HD~42581 (M1~V), HD~95735 (M2~V), Gl~388 (M3~V), Gl~213 (M4~V), Gl~51
  (M5~V), Gl~406 (M6~V), Gl~644C (M7~V), Gl~752B (M8~V), and LP944-20
  (M9~V).  The spectra have been normalized to unity at 0.88~$\mu$m and
  offset by constants.}
\end{figure}

\begin{figure}
\centerline{\includegraphics[width=6.0in,angle=0]{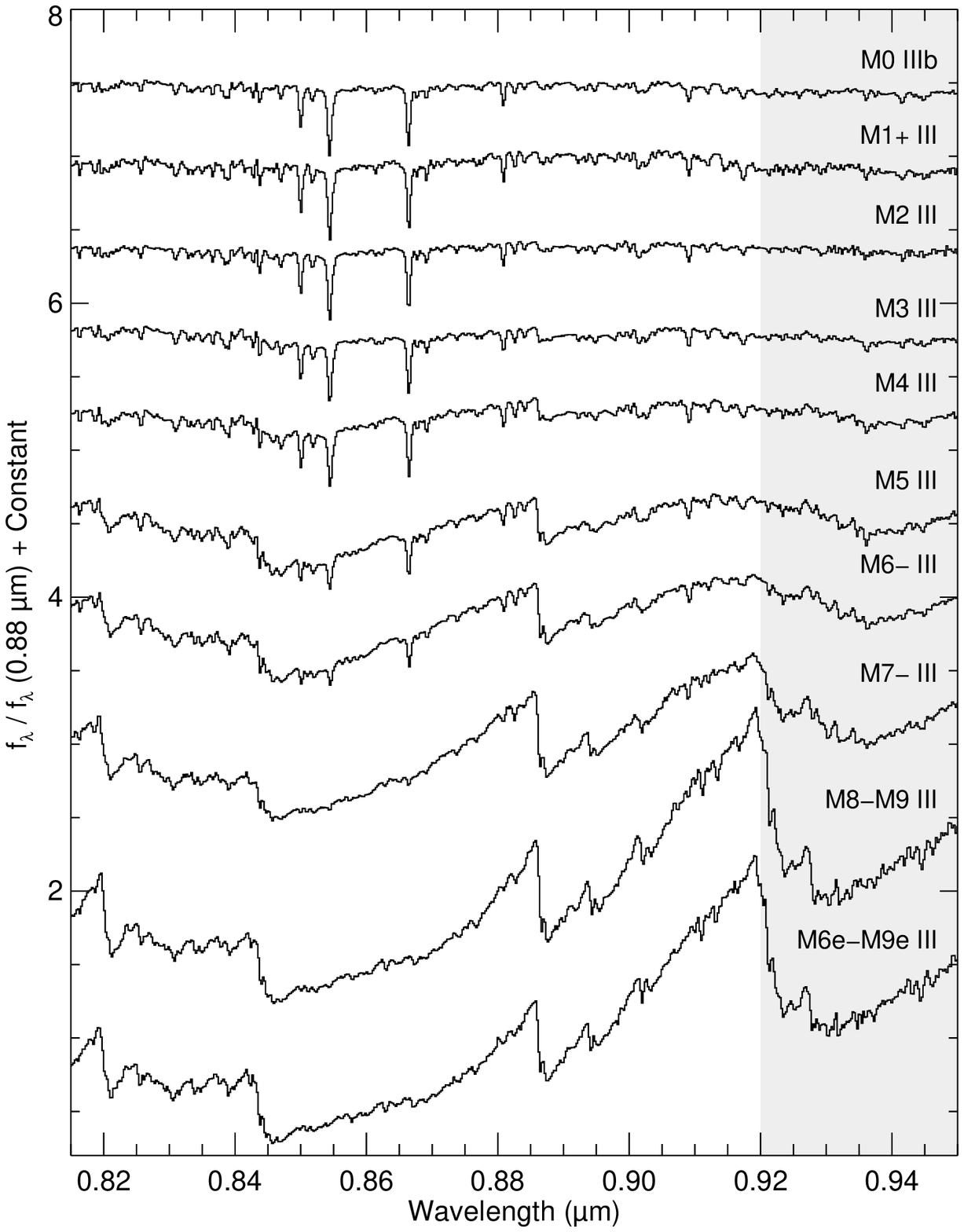}}
\caption{\label{fig:M_IIII} A sequence of M giant stars plotted over the
  $I$ band (0.82$-$0.95~$\mu$m).  The spectra are of HD~213893 (M0~IIIb),
  HD~204724 (M1+~III), HD~120052 (M2~III), HD~39045 (M3~III), HD~4408
  (M4~III), HD~175865 (M5~III), HD~18191 (M6-~III), HD~108849 (M7-~III),
  IRAS~21284-0747 (M8-M9~III), and HD~HD 69243 (M6e-M9e~III).  The
  spectra have been normalized to unity at 0.88~$\mu$m and offset by
  constants.}
\end{figure}

\begin{figure}
\centerline{\includegraphics[width=6.0in,angle=0]{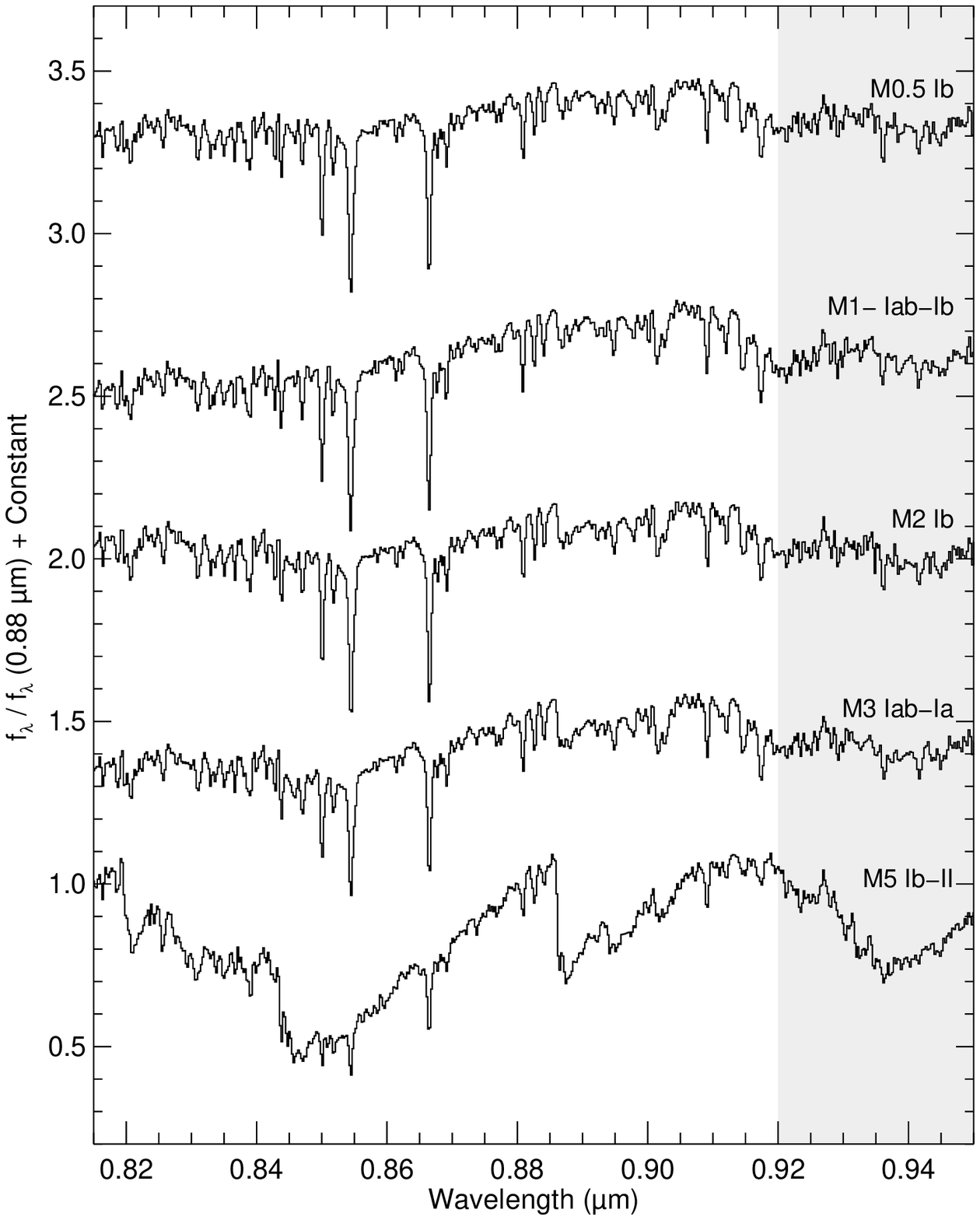}}
\caption{\label{fig:M_II} A sequence of M supergiant stars plotted over
  the $I$ band (0.82$-$0.95~$\mu$m).  The spectra are of HD~236697
  (M0.5~Ib), HD~14404 (M1-~Iab-Ib), HD~10465 (M2~Ib), CD~-31~4916
  (M3~Iab-Ia), and HD~156014 (M5~Ib-II).  The spectra have been
  normalized to unity at 0.88~$\mu$m and offset by constants.}
\end{figure}

\begin{figure}
\centerline{\includegraphics[width=6.0in,angle=0]{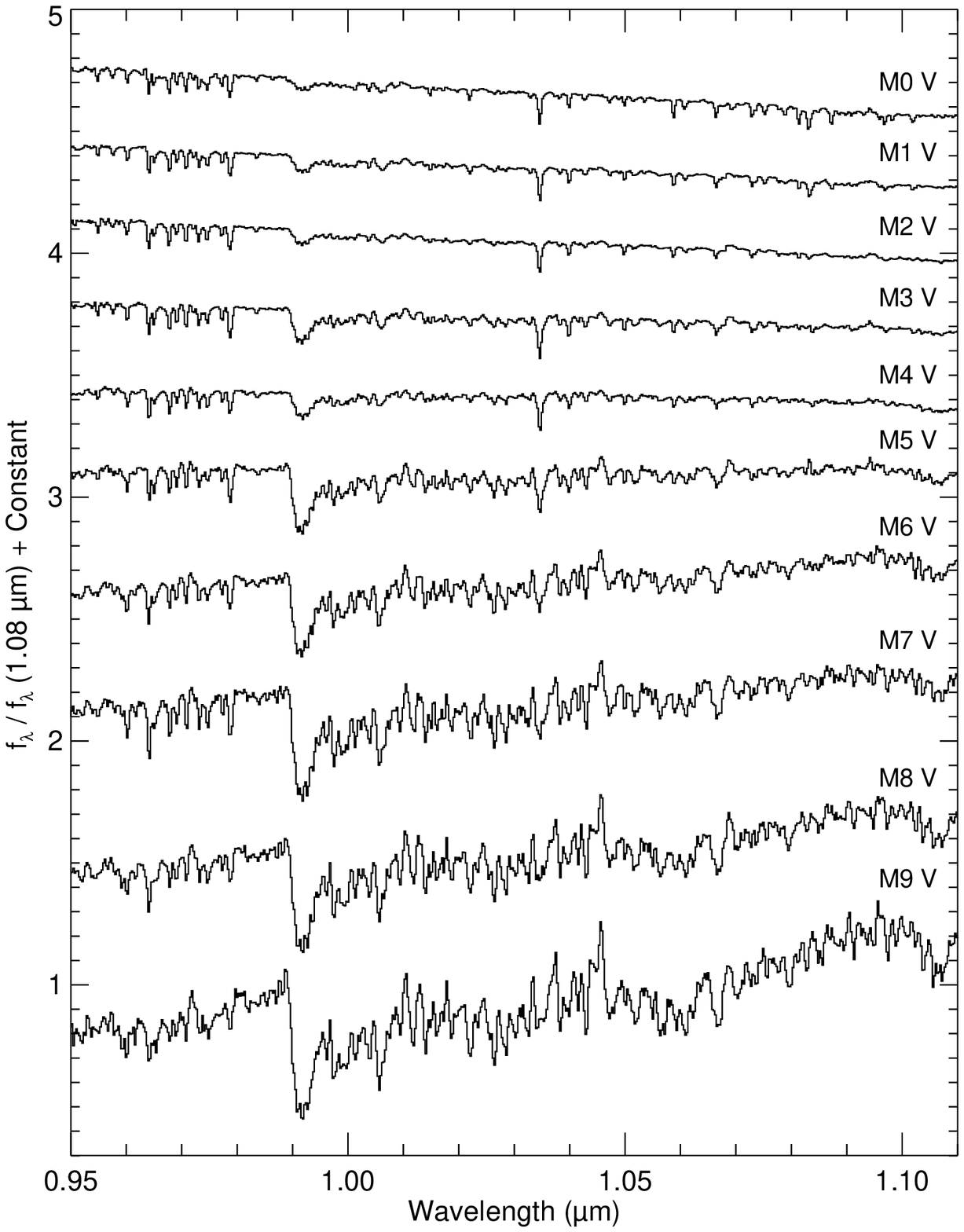}}
\caption{\label{fig:M_VY} A sequence of M dwarf stars plotted over the
  $Y$ band (0.95$-$1.10~$\mu$m).  The spectra are of HD~19305 (M0~V),
  HD~42581 (M1~V), HD~95735 (M2~V), Gl~388 (M3~V), Gl~213 (M4~V), Gl~51
  (M5~V), Gl~406 (M6~V), Gl~644C (M7~V), Gl~752B (M8~V), and LP944-20
  (M9~V).  The spectra have been normalized to unity at 1.08~$\mu$m and
  offset by constants.}
\end{figure}

\begin{figure}
\centerline{\includegraphics[width=6.0in,angle=0]{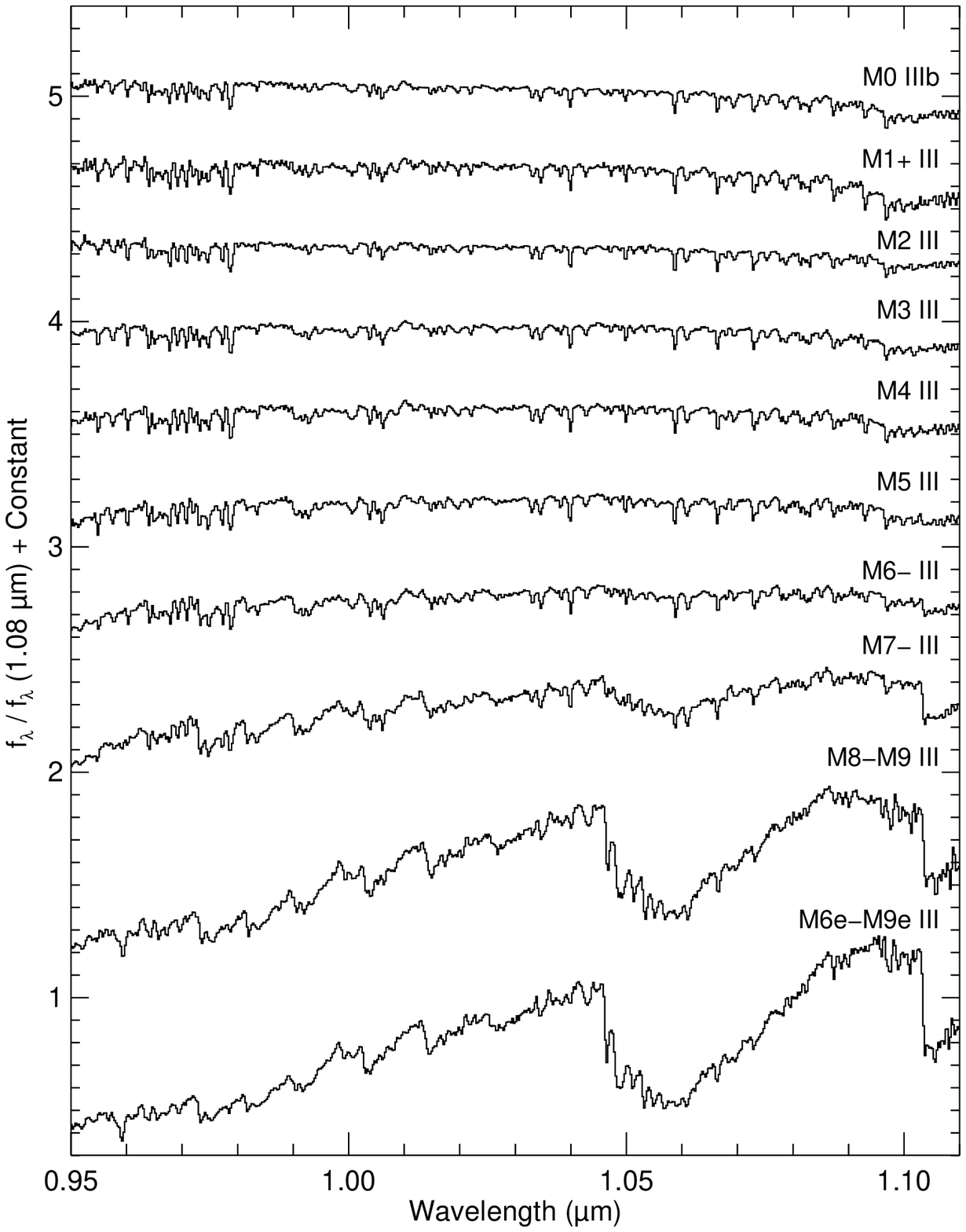}}
\caption{\label{fig:M_IIIY} A sequence of M giant stars plotted over the
  $Y$ band (0.95$-$1.10~$\mu$m).  The spectra are of HD~213893 (M0~IIIb),
  HD~204724 (M1+~III), HD~120052 (M2~III), HD~39045 (M3~III), HD~4408
  (M4~III), HD~175865 (M5~III), HD~18191 (M6-~III), HD~108849 (M7-~III),
  IRAS~21284-0747 (M8-M9~III), and HD~HD 69243 (M6e-M9e~III).  The
  spectra have been normalized to unity at 1.08~$\mu$m and offset by
  constants.}
\end{figure}

\begin{figure}
\centerline{\includegraphics[width=6.0in,angle=0]{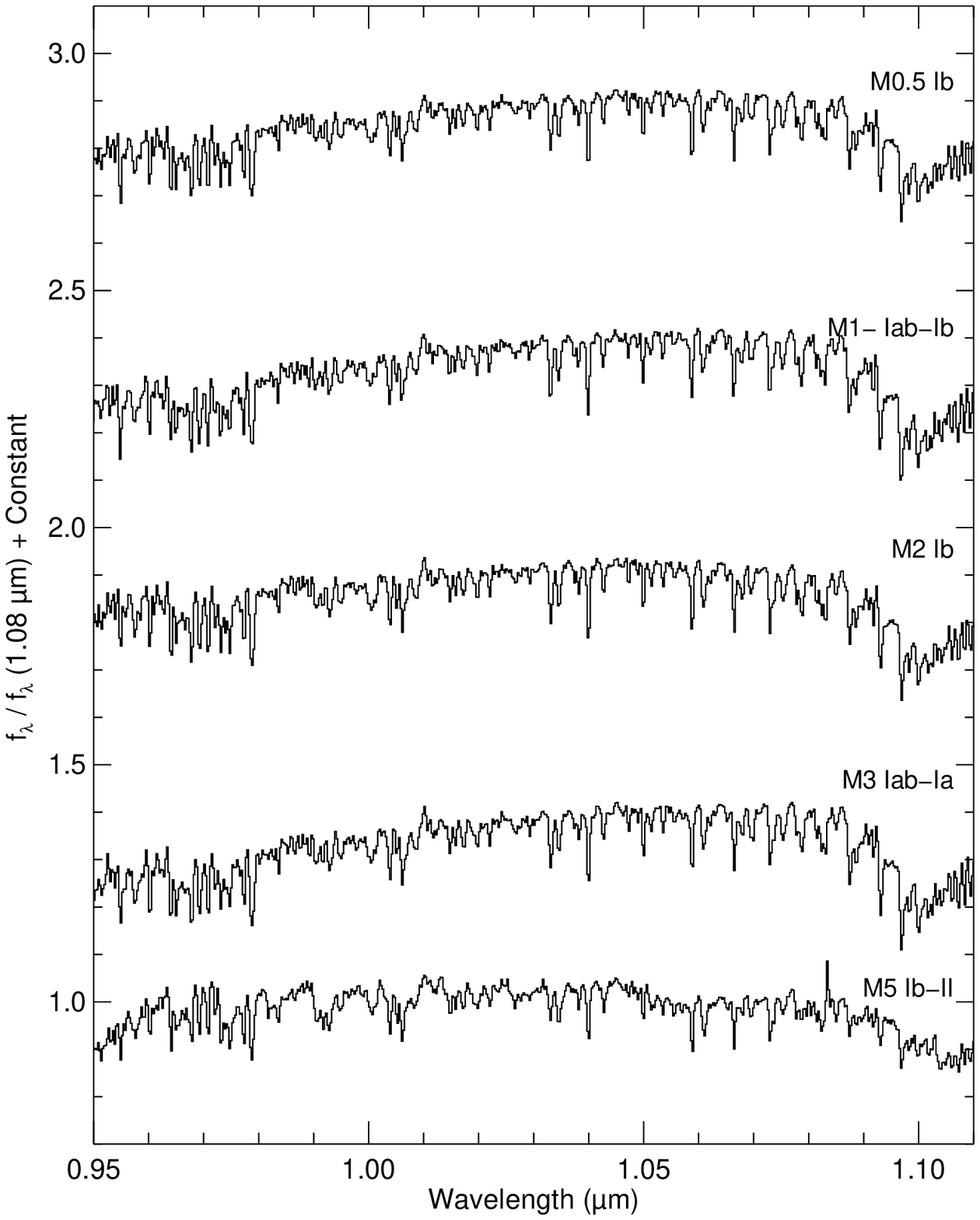}}
\caption{\label{fig:M_IY} A sequence of M supergiant stars plotted over
  the $Y$ band (0.95$-$1.10~$\mu$m).  The spectra are of HD~236697
  (M0.5~Ib), HD~14404 (M1-~Iab-Ib), HD~10465 (M2~Ib), CD~-31~4916
  (M3~Iab-Ia), and HD~156014 (M5~Ib-II).  The spectra have been
  normalized to unity at 1.08~$\mu$m and offset by constants.}
\end{figure}

\begin{figure}
\centerline{\includegraphics[width=6.0in,angle=0]{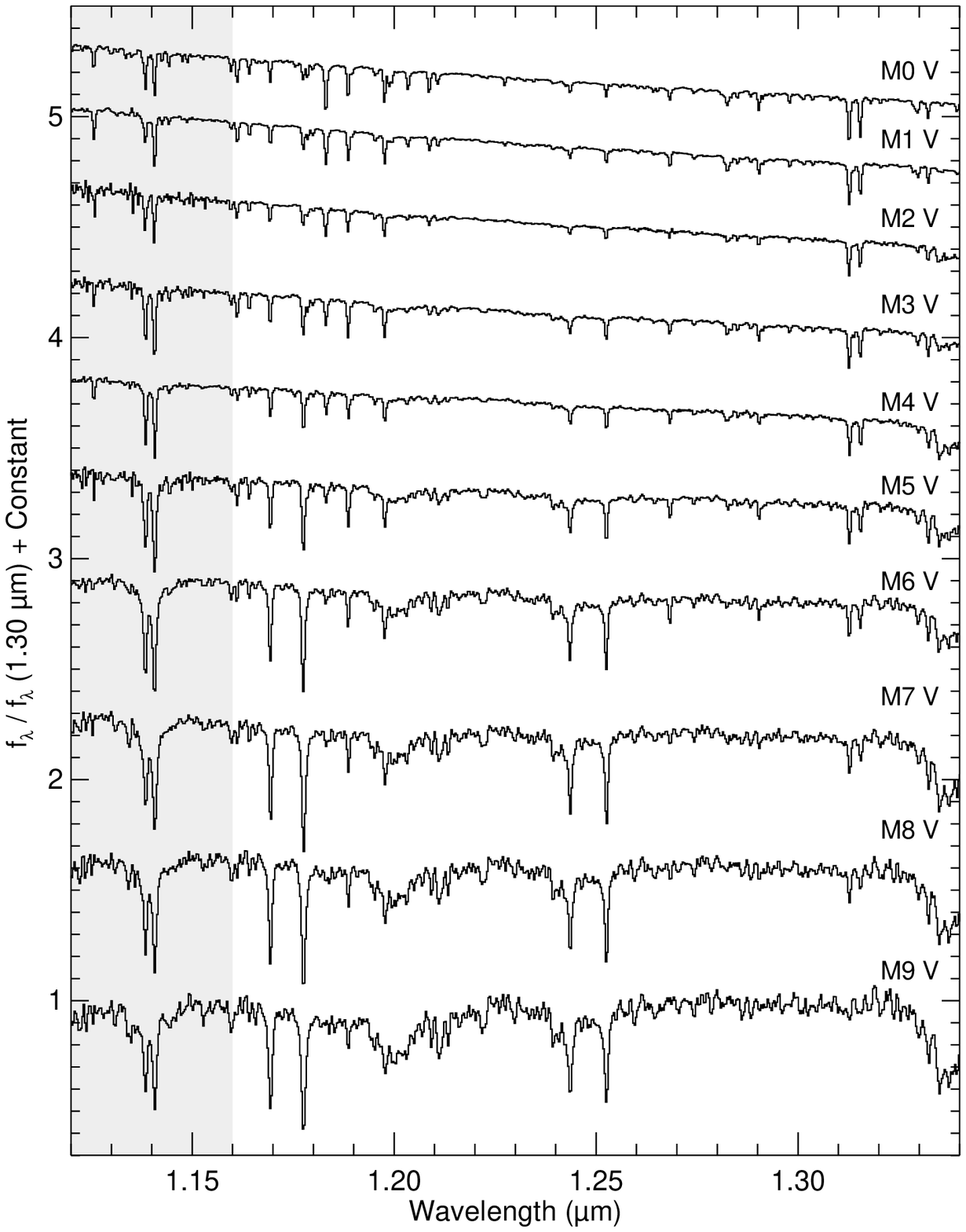}}
\caption{\label{fig:M_VJ} A sequence of M dwarf stars plotted over the
  $J$ band (1.12$-$1.34~$\mu$m).  The spectra are of HD~19305 (M0~V),
  HD~42581 (M1~V), HD~95735 (M2~V), Gl~388 (M3~V), Gl~213 (M4~V), Gl~51
  (M5~V), Gl~406 (M6~V), Gl~644C (M7~V), Gl~752B (M8~V), and LP944-20
  (M9~V).  The spectra have been normalized to unity at 1.30~$\mu$m and
  offset by constants.}
\end{figure}

\begin{figure}
\centerline{\includegraphics[width=6.0in,angle=0]{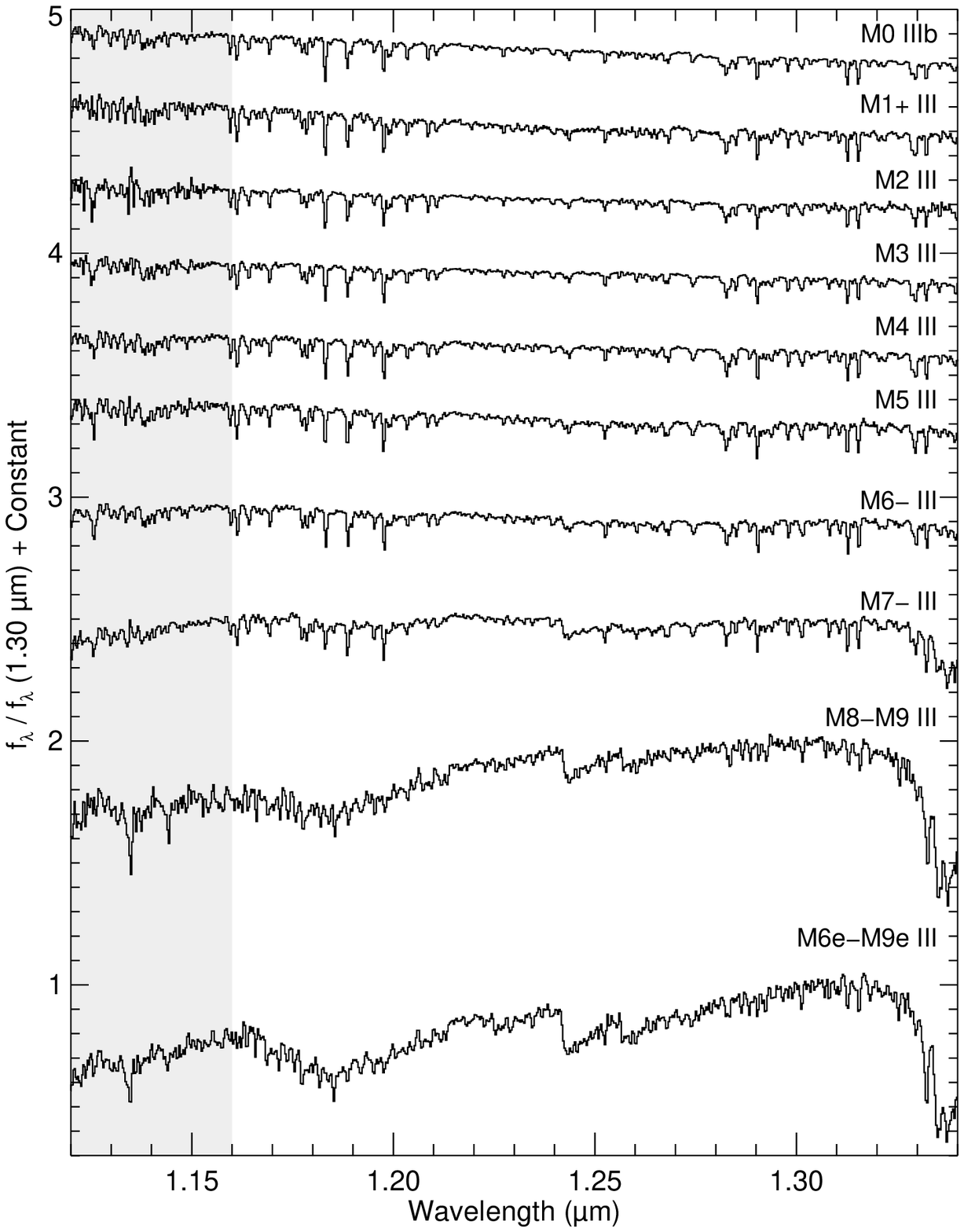}}
\caption{\label{fig:M_IIIJ} A sequence of M giant stars plotted over the
  $J$ band (1.12$-$1.34~$\mu$m).  The spectra are of HD~213893 (M0~IIIb),
  HD~204724 (M1+~III), HD~120052 (M2~III), HD~39045 (M3~III), HD~4408
  (M4~III), HD~175865 (M5~III), HD~18191 (M6-~III), HD~108849 (M7-~III),
  IRAS~21284-0747 (M8-M9~III), and HD~HD 69243 (M6e-M9e~III).  The
  spectra have been normalized to unity at 1.30~$\mu$m and offset by
  constants.}
\end{figure}

\begin{figure}
\centerline{\includegraphics[width=6.0in,angle=0]{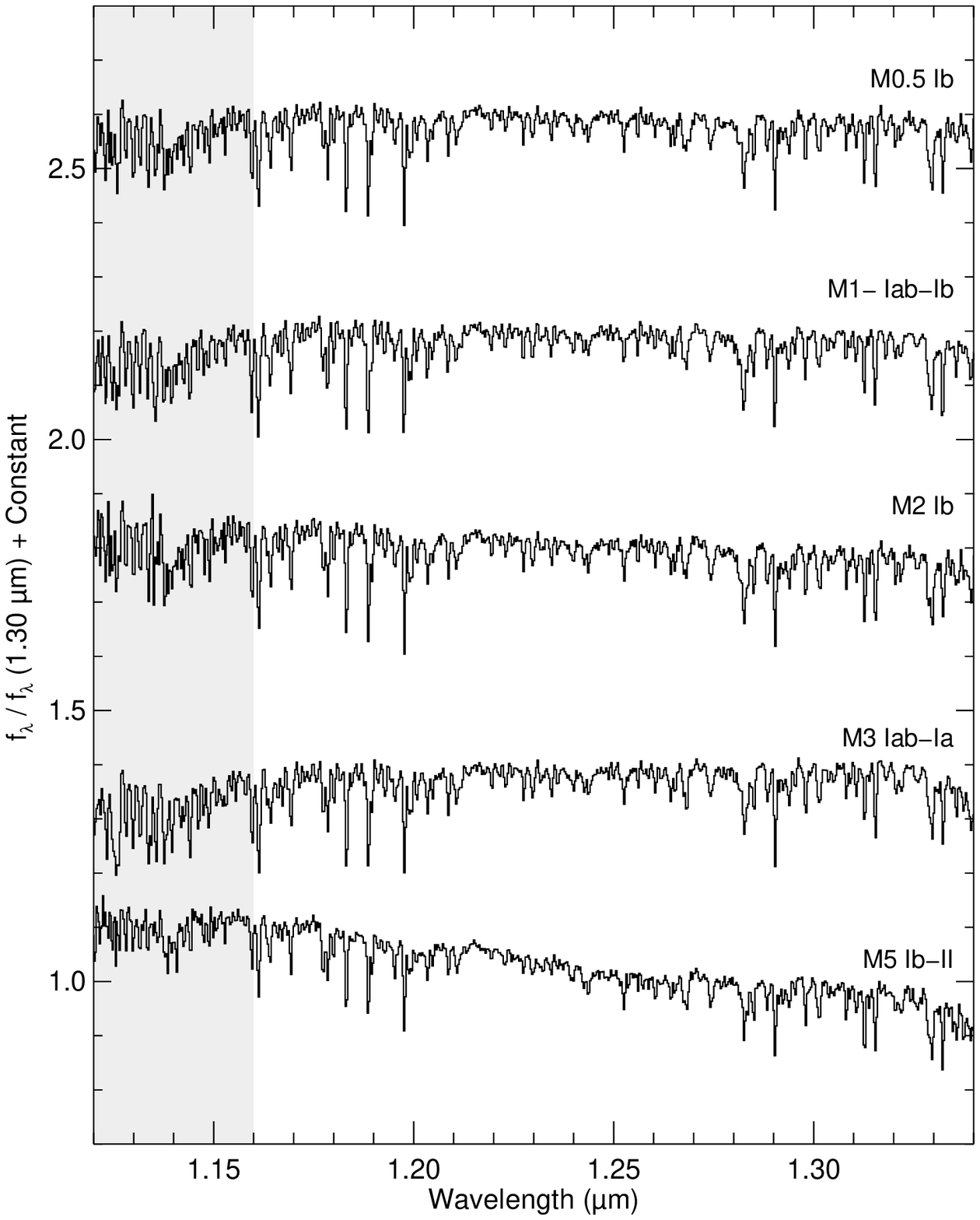}}
\caption{\label{fig:M_IJ} A sequence of M supergiant stars plotted over
  the $J$ band (1.12$-$1.34~$\mu$m).  The spectra are of HD~236697
  (M0.5~Ib), HD~14404 (M1-~Iab-Ib), HD~10465 (M2~Ib), CD~-31~4916
  (M3~Iab-Ia), and HD~156014 (M5~Ib-II).  The spectra have been
  normalized to unity at 1.30~$\mu$m and offset by constants.}
\end{figure}

\begin{figure}
\centerline{\includegraphics[width=6.0in,angle=0]{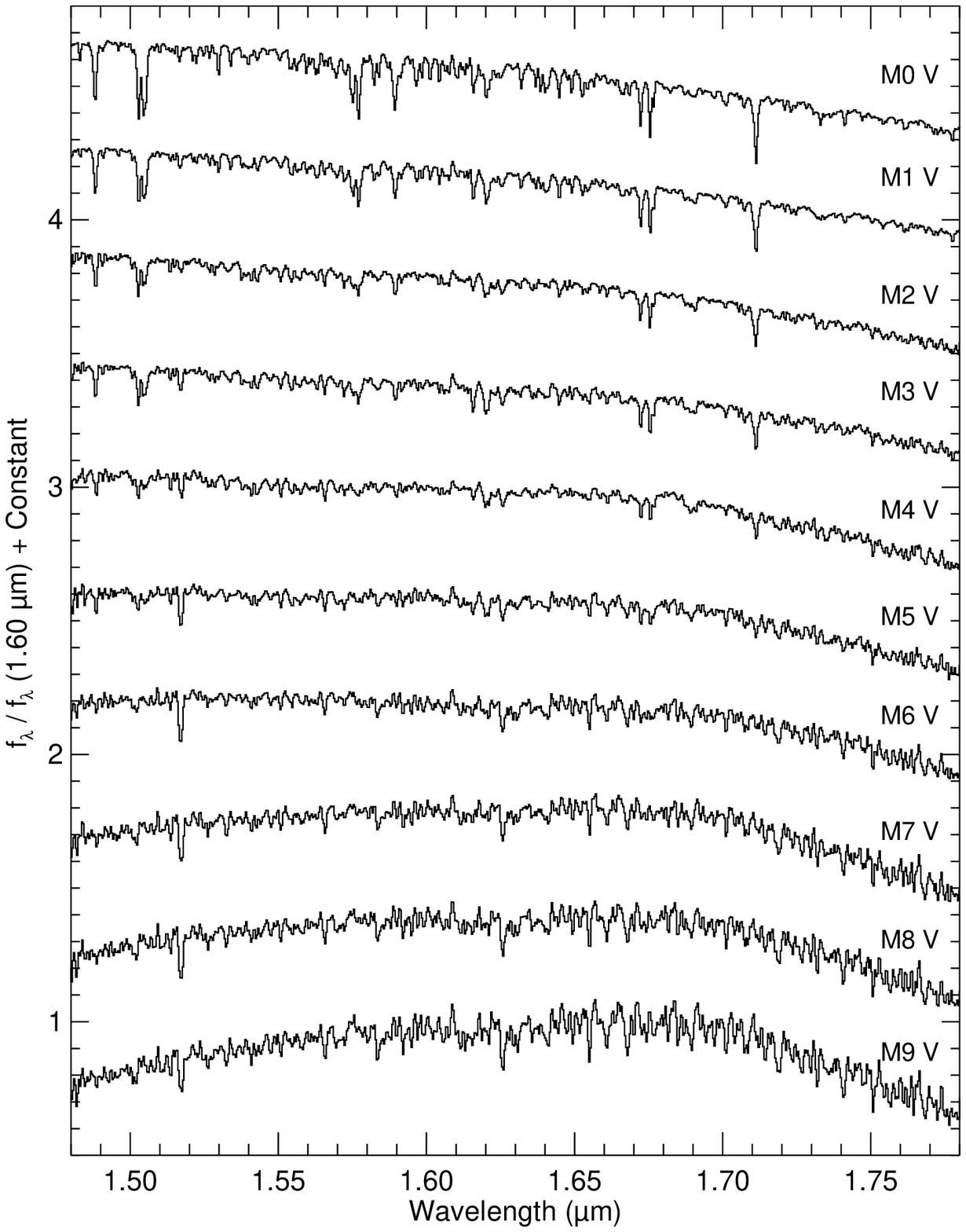}}
\caption{\label{fig:M_VH} A sequence of M dwarf stars plotted over the
  $H$ band (1.48$-$1.78~$\mu$m).  The spectra are of HD~19305 (M0~V),
  HD~42581 (M1~V), HD~95735 (M2~V), Gl~388 (M3~V), Gl~213 (M4~V), Gl~51
  (M5~V), Gl~406 (M6~V), Gl~644C (M7~V), Gl~752B (M8~V), and LP944-20
  (M9~V).  The spectra have been normalized to unity at 1.60~$\mu$m and
  offset by constants.}
\end{figure}

\begin{figure}
\centerline{\includegraphics[width=6.0in,angle=0]{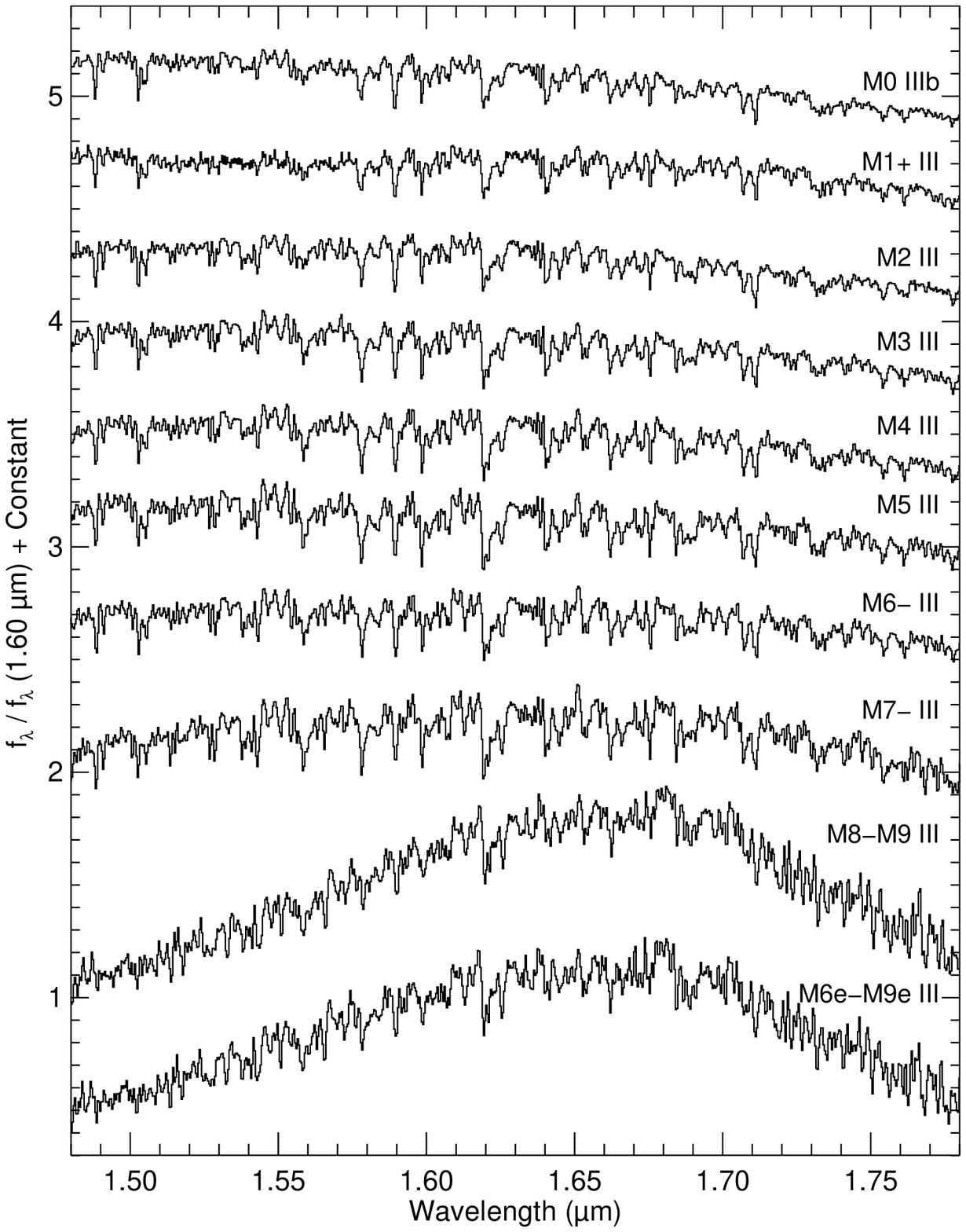}}
\caption{\label{fig:M_IIIH} A sequence of M giant stars plotted over the
  $H$ band (1.48$-$1.78~$\mu$m).  The spectra are of HD~213893 (M0~IIIb),
  HD~204724 (M1+~III), HD~120052 (M2~III), HD~39045 (M3~III), HD~4408
  (M4~III), HD~175865 (M5~III), HD~18191 (M6-~III), HD~108849 (M7-~III),
  IRAS~21284-0747 (M8-M9~III), and HD~HD 69243 (M6e-M9e~III).  The
  spectra have been normalized to unity at 1.60~$\mu$m and offset by
  constants.}
\end{figure}

\begin{figure}
\centerline{\includegraphics[width=6.0in,angle=0]{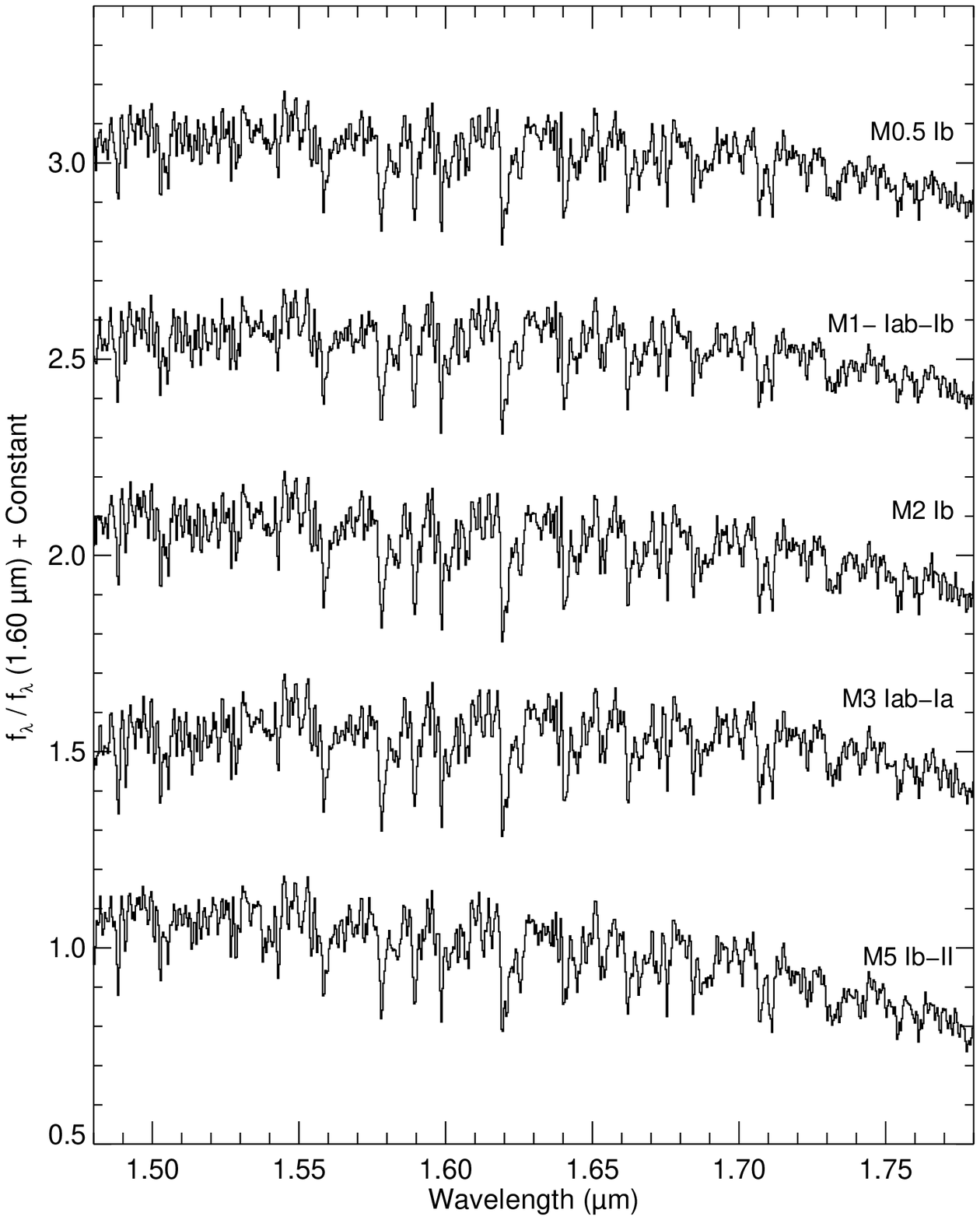}}
\caption{\label{fig:M_IH} A sequence of M supergiant stars plotted over
  the $H$ band (1.48$-$1.78~$\mu$m).  The spectra are of HD~236697
  (M0.5~Ib), HD~14404 (M1-~Iab-Ib), HD~10465 (M2~Ib), CD~-31~4916
  (M3~Iab-Ia), and HD~156014 (M5~Ib-II).  The spectra have been
  normalized to unity at 1.60~$\mu$m and offset by constants.}
\end{figure}

\begin{figure}
\centerline{\includegraphics[width=6.0in,angle=0]{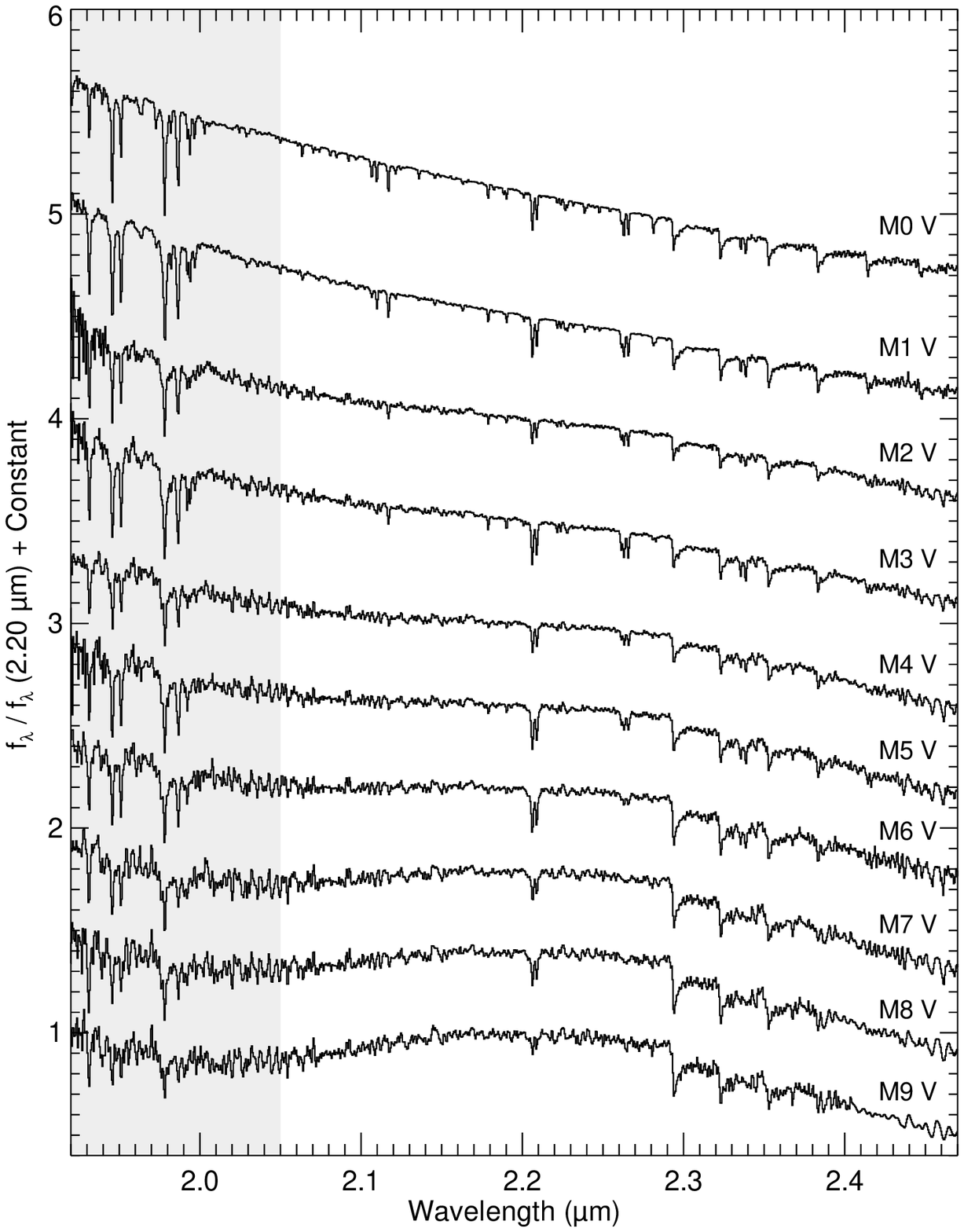}}
\caption{\label{fig:M_VK} A sequence of M dwarf stars plotted over the
  $K$ band (1.92$-$2.5~$\mu$m).  The spectra are of HD~19305 (M0~V),
  HD~42581 (M1~V), HD~95735 (M2~V), Gl~388 (M3~V), Gl~213 (M4~V), Gl~51
  (M5~V), Gl~406 (M6~V), Gl~644C (M7~V), Gl~752B (M8~V), and LP944-20
  (M9~V).  The spectra have been normalized to unity at 2.20~$\mu$m and
  offset by constants.}
\end{figure}

\begin{figure}
\centerline{\includegraphics[width=6.0in,angle=0]{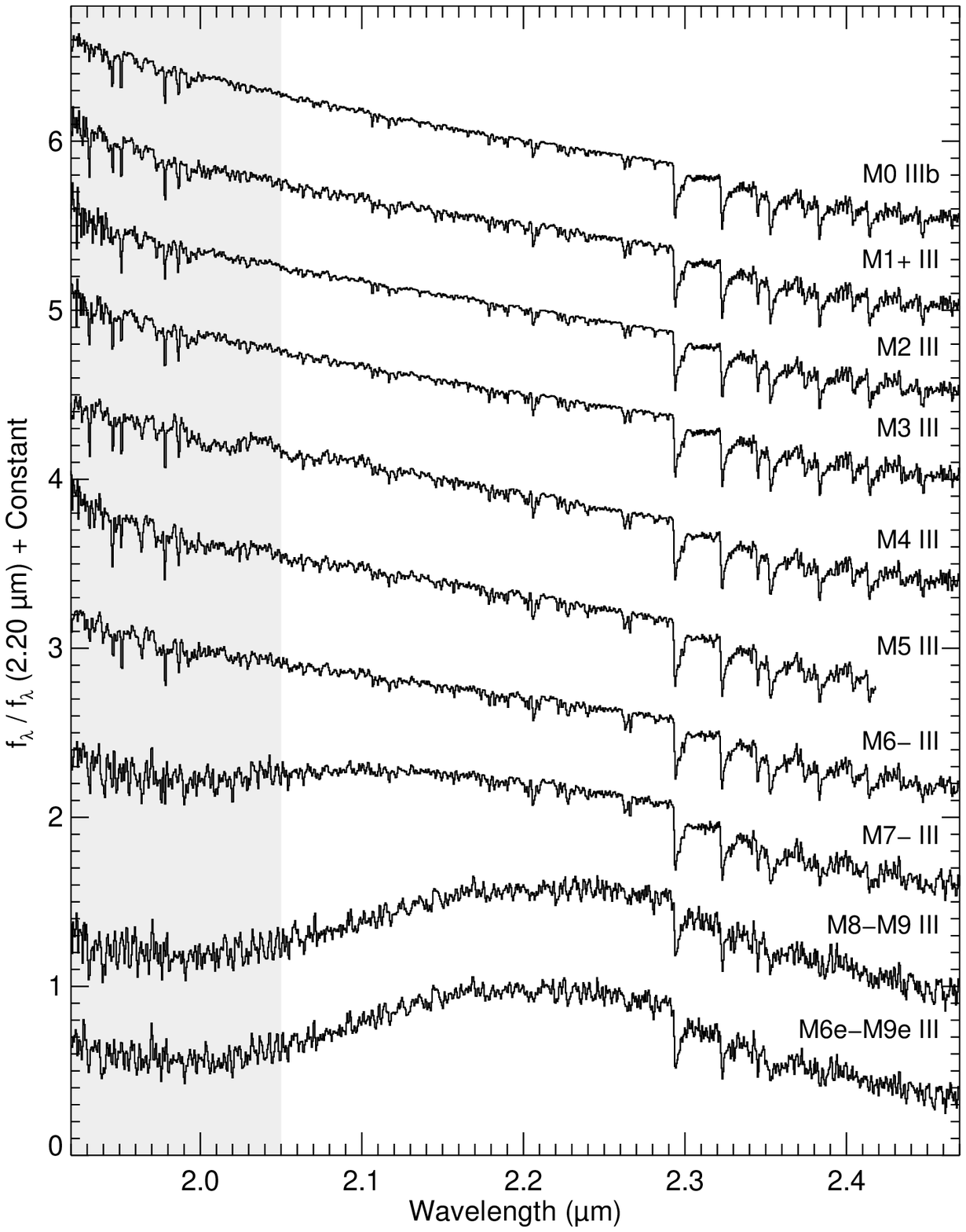}}
\caption{\label{fig:M_IIIK} A sequence of M giant stars plotted over the
  $K$ band (1.92$-$2.5~$\mu$m).  The spectra are of HD~213893 (M0~IIIb),
  HD~204724 (M1+~III), HD~120052 (M2~III), HD~39045 (M3~III), HD~4408
  (M4~III), HD~175865 (M5~III), HD~18191 (M6-~III), HD~108849 (M7-~III),
  IRAS~21284-0747 (M8-M9~III), and HD~HD 69243 (M6e-M9e~III).  The
  spectra have been normalized to unity at 2.20~$\mu$m and offset by
  constants.}
\end{figure}

\begin{figure}
\centerline{\includegraphics[width=6.0in,angle=0]{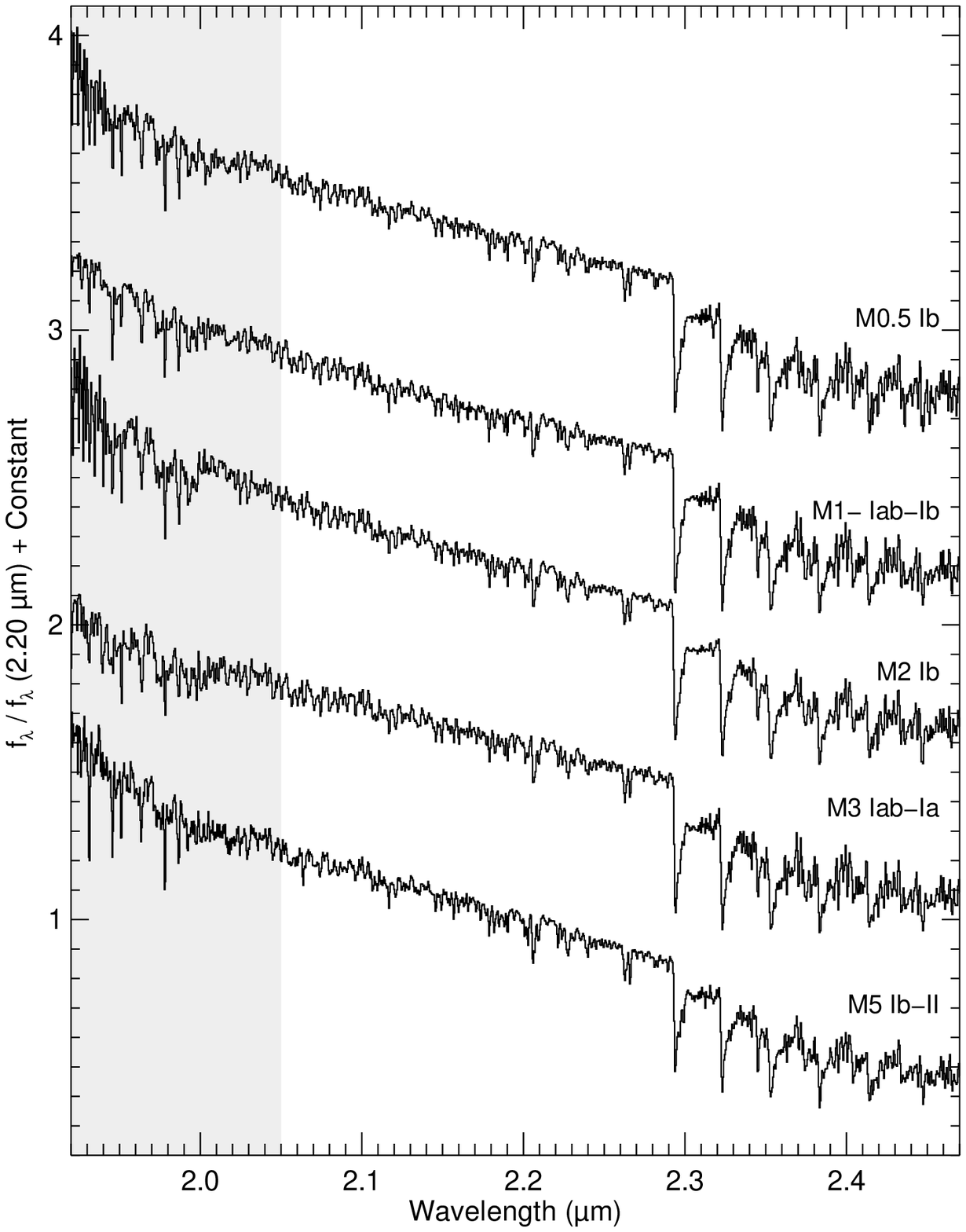}}
\caption{\label{fig:M_IK} A sequence of M supergiant stars plotted over
  the $K$ band (1.92$-$2.5~$\mu$m).  The spectra are of HD~236697
  (M0.5~Ib), HD~14404 (M1-~Iab-Ib), HD~10465 (M2~Ib), CD~-31~4916
  (M3~Iab-Ia), and HD~156014 (M5~Ib-II).  The spectra have been
  normalized to unity at 2.20~$\mu$m and offset by constants.}
\end{figure}

\begin{figure}
\centerline{\includegraphics[width=6.0in,angle=0]{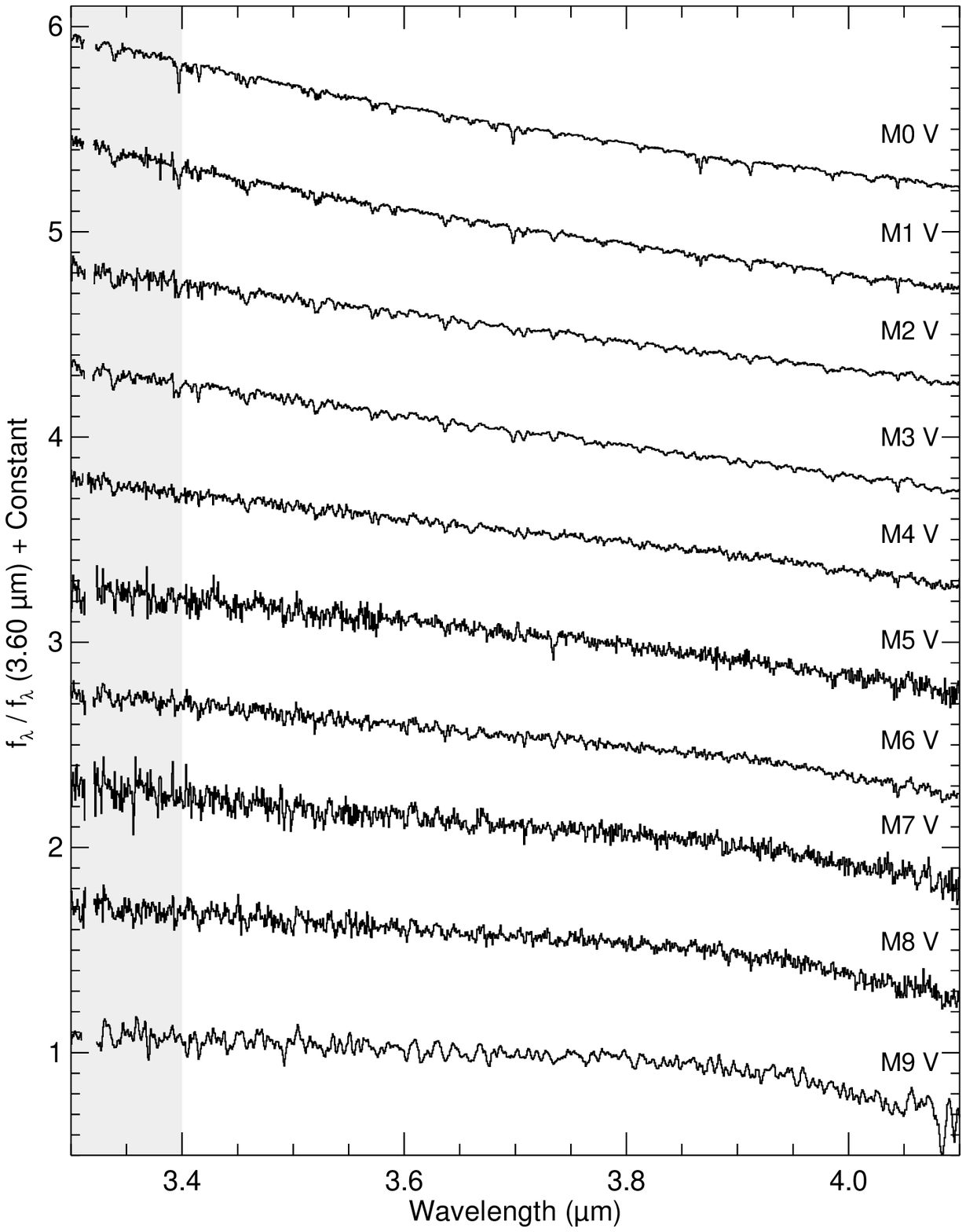}}
\caption{\label{fig:M_VL} A sequence of M dwarf stars plotted over the
  $L'$ band (3.6$-$4.1~$\mu$m).  The spectra are of HD~19305 (M0~V),
  HD~42581 (M1~V), HD~95735 (M2~V), Gl~388 (M3~V), Gl~213 (M4~V), Gl~51
  (M5~V), Gl~406 (M6~V), Gl~644C (M7~V), Gl~752B (M8~V), and LP944-20
  (M9~V).  The spectra have been normalized to unity at 3.6~$\mu$m and
  offset by constants.}
\end{figure}

\begin{figure}
\centerline{\includegraphics[width=6.0in,angle=0]{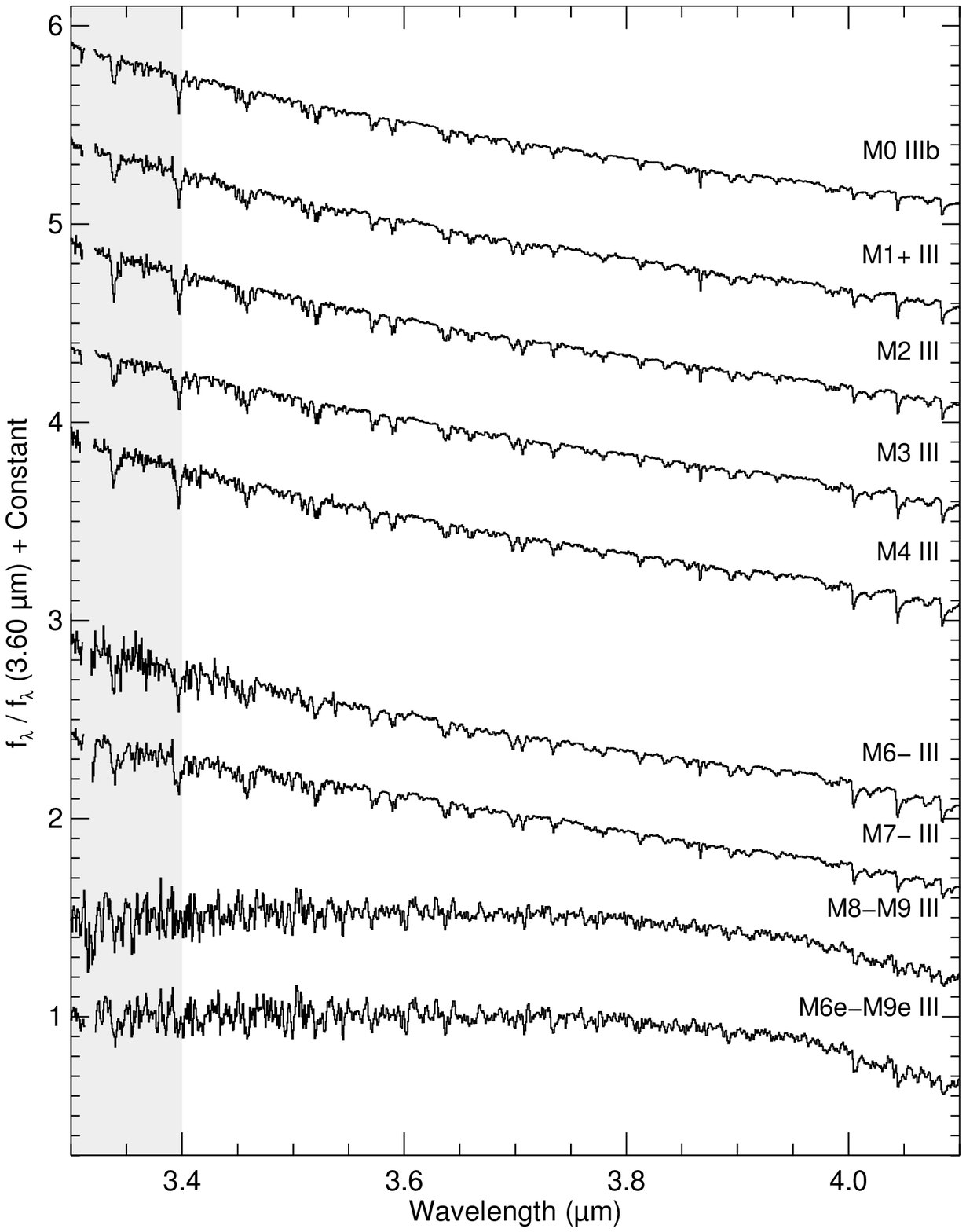}}
\caption{\label{fig:M_IIIL} A sequence of M giant stars plotted over the
  $L'$ band (3.6$-$4.1~$\mu$m).  The spectra are of HD~213893 (M0~IIIb),
  HD~204724 (M1+~III), HD~120052 (M2~III), HD~39045 (M3~III), HD~4408
  (M4~III), HD~18191 (M6-~III), HD~108849 (M7-~III),
  IRAS~21284-0747 (M8-M9~III), and HD~HD 69243 (M6e-M9e~III).  The
  spectra have been normalized to unity at 3.6~$\mu$m and offset by
  constants.}
\end{figure}

\begin{figure}
\centerline{\includegraphics[width=6.0in,angle=0]{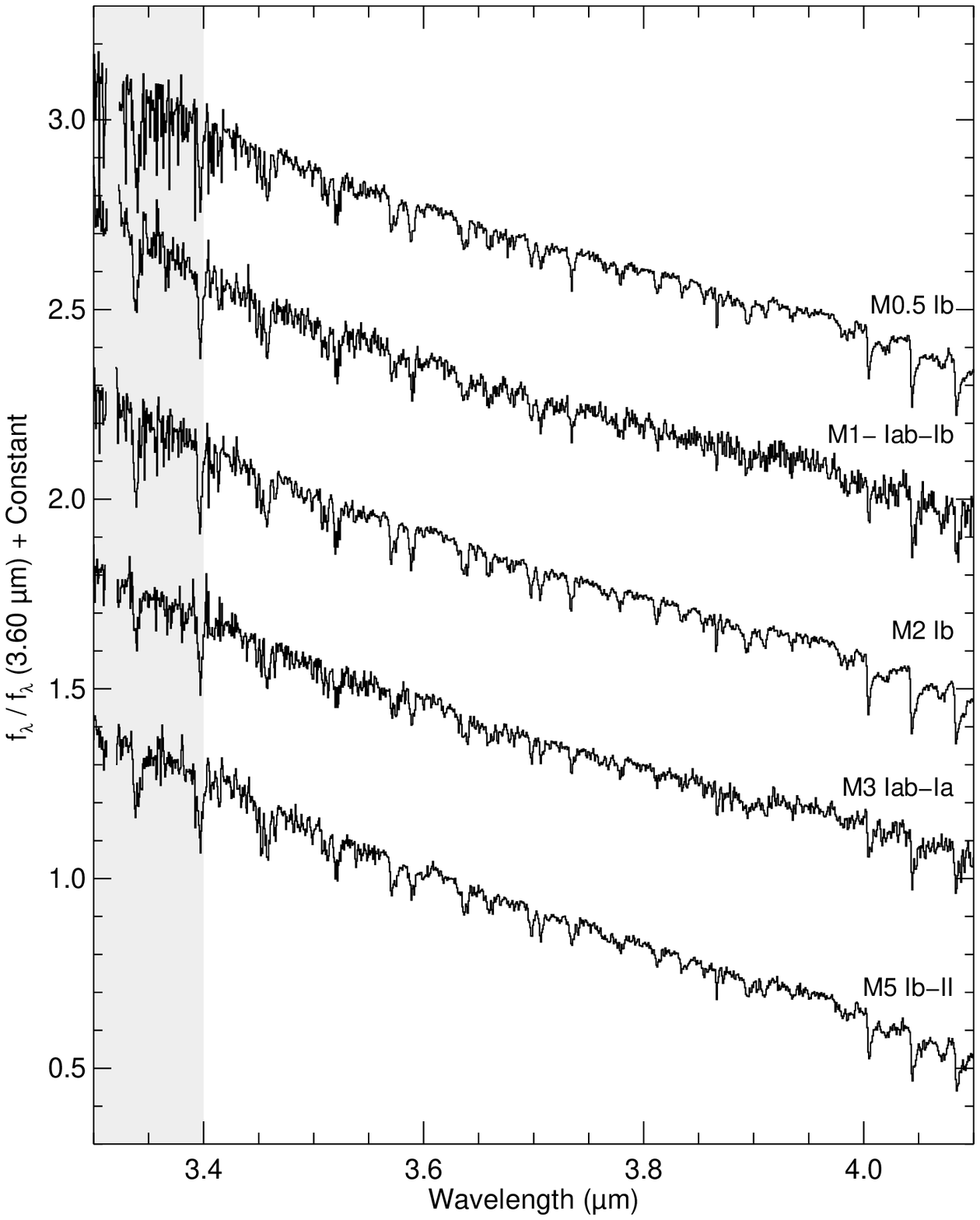}}
\caption{\label{fig:M_IL} A sequence of M supergiant stars plotted over
  the $L'$ band (3.6$-$4.1~$\mu$m).  The spectra are of HD~236697
  (M0.5~Ib), HD~14404 (M1-~Iab-Ib), HD~10465 (M2~Ib), CD~-31~4916
  (M3~Iab-Ia), and HD~156014 (M5~Ib-II).  The spectra have been
  normalized to unity at 3.6~$\mu$m and offset by constants.}
\end{figure}

More complete spectral sequences are plotted in this section: F stars
(bands $I$, $Y$, $J$, $H$, and $K$) in Figures \ref{fig:F_VI} to
\ref{fig:F_IL}, G stars (bands $I$, $Y$, $J$, $H$, and $K$) in Figures
\ref{fig:G_VI} to \ref{fig:G_IL}, K stars (bands $I$, $Y$, $J$, $H$, and
$K$) in Figures \ref{fig:K_VI} to \ref{fig:K_IL}, and M stars (bands
$I$, $Y$, $J$, $H$, and $K$) in Figures \ref{fig:M_VI} to
\ref{fig:M_IL}.  For feature identifications see the partial spectral
sequences plotted in Figures \ref{fig:FGKM_V} to \ref{fig:MSC}.

 
\clearpage

\clearpage

\clearpage

\clearpage

\clearpage

\clearpage

\clearpage

\clearpage

\clearpage

\clearpage

\clearpage

\clearpage

\clearpage

\clearpage

\clearpage

\clearpage

\clearpage

\clearpage

\clearpage
%
%

\clearpage

\clearpage

\clearpage

\clearpage

\clearpage

\clearpage

\clearpage

\clearpage

\clearpage

\clearpage

\clearpage

\clearpage

\clearpage

\clearpage

\clearpage

\clearpage

\clearpage

\clearpage
%
%

\clearpage

\clearpage

\clearpage

\clearpage

\clearpage

\clearpage

\clearpage

\clearpage

\clearpage

\clearpage

\clearpage

\clearpage

\clearpage

\clearpage

\clearpage

\clearpage

\clearpage

\clearpage
%

\clearpage

\clearpage

\clearpage

\clearpage

\clearpage

\clearpage

\clearpage

\clearpage

\clearpage

\clearpage

\clearpage

\clearpage

\clearpage

\clearpage

\clearpage

\clearpage

\clearpage

\end{document}